\documentclass[12pt,a4paper]{article}

\usepackage [unicode=true,bookmarksopen=true,colorlinks=true,urlcolor=blue, anchorcolor=blue,citecolor=blue,filecolor=blue,linkcolor=blue, menucolor=blue,pagecolor=blue,linktocpage=true]{hyperref}

\usepackage{amsmath,amsfonts,latexsym,amssymb,amscd}
\usepackage{pslatex}
\usepackage[latin1]{inputenc}
\usepackage[T1]{fontenc}
\usepackage{pspicture}
\usepackage{verbatim,amsthm,curves,graphics}
\usepackage{mathrsfs}

\def\ben{\begin{equation}}

\def\een{\end{equation}}

\def\bena{\begin{eqnarray}}

\def\eena{\end{eqnarray}}

\def\f(#1/#2){\frac{#1}{#2}}

\def\Frac(#1/#2){\left(\frac{#1}{#2}\right)}

\def\chris(#1-#2-#3){{\mit \Gamma}^{#1}{}_{{#2}{#3}} }

\def\tilchris(#1-#2-#3){\tilde{{\mit \Gamma}}^{#1}{}_{{#2}{#3}}}

\def\hatchris(#1-#2-#3){\hat{{\mit \Gamma}}^{#1}{}_{{#2}{#3}}}

\newcommand{\non}{\nonumber}

\theoremstyle{definition}

\newtheorem{thm}{Theorem}

\newtheorem{lemma}{Lemma}

\newtheorem{defn}{Definition}[section]

\newcommand{\Wn}{{\bf W}_{00}}

\newcommand{\myid}{{1 \!\!\! 1}}
\newcommand{\mr}{{\mathbb R}}
\newcommand{\mn}{{\mathbb N}}
\newcommand{\mc}{{\mathbb C}}

\newcommand{\e}{{\rm e}_\otimes}
\newcommand{\supp}{\operatorname{supp}}
\newcommand{\bv}{\operatorname*{B.V.}}
\newcommand{\btheta}{\theta}

\newcommand{\beps}{\epsilon}
\newcommand{\I}{{\mathcal J}_0}
\newcommand{\WF}{{\rm WF}}
\newcommand{\B}{{\mathcal B}}
\newcommand{\F}{{\mathcal F}}
\newcommand{\W}{{\bf W}_0}
\newcommand{\D}{{\mathcal D}}
\newcommand{\A}{{\mathcal A}}
\newcommand{\K}{{\bf K}}

\renewcommand{\H}{{\mathcal H}}
\newcommand{\N}{{\mathcal N}}
\newcommand{\st}{\star_\hbar}

\renewcommand{\O}{{\mathcal O}}
\renewcommand{\L}{{\bf L}}
\newcommand{\J}{{\bf J}}
\renewcommand{\P}{{\bf P}}

\newcommand{\eff}{{\rm eff}}

\renewcommand{\d}{{d}}

\newcommand{\h}{{\mbox{\tiny $\rm{H}$}}}
\newcommand{\subsc}{I}

\begin{document}

\title{Renormalized Quantum Yang-Mills Fields in Curved Spacetime}

\author{Stefan Hollands\thanks{\tt HollandsS@Cardiff.ac.uk
    }\\
    \\
{\it School of Mathematics} \\
{\it Cardiff University}\\
{\it UK}\\
and\\
{\it Institut f\" ur  Theoretische Physik}\\
{\it Universit\" at G\" ottingen}\\
{\it FRG}\\
\\ \\ \\
{ Dedicated to K. Fredenhagen on the occasion of his 60th birthday}
\\ \\ \\
}

\maketitle

\pagebreak

\begin{abstract}
We present a proof that quantum Yang-Mills theory can be consistently
defined as a renormalized, perturbative quantum field theory on an arbitrary
globally hyperbolic curved, Lorentzian spacetime. To this end, we construct
the non-commutative algebra of observables, in the sense of formal
power series, as well as a space of corresponding quantum states.
The algebra contains all gauge invariant, renormalized, interacting quantum
field operators (polynomials in the field strength and its derivatives),
and all their relations such as commutation
relations or operator product expansion.
It can be viewed as a deformation quantization of
the Poisson algebra of classical Yang-Mills theory equipped with the
Peierls bracket. The algebra is constructed as the cohomology of an auxiliary
algebra describing a gauge fixed theory with ghosts and anti-fields. A key technical
difficulty is to establish a suitable hierarchy of Ward identities at the
renormalized level that ensure
conservation of the interacting BRST-current, and that
the interacting BRST-charge is nilpotent. The algebra of physical
interacting field observables is obtained as the cohomology of this
charge. As a consequence of our constructions, we can prove that the
operator product expansion closes on
the space of gauge invariant operators. Similarly, the renormalization
group flow is proved not to leave the space of gauge invariant operators.
The key technical tool behind these arguments is a new universal
Ward identity that is formulated at the algebraic level, and that is proven to
be consistent with a local and covariant renormalization prescription. We also develop a new
technique to accomplish this renormalization process, and in particular give a
new expression for some of the renormalization constants in terms of cycles.
\end{abstract}

\pagebreak

\tableofcontents

\pagebreak

\section{Introduction}

The known interactions of elementary particles seem to be
well-described by quantized field theories with local gauge
invariance such as QCD. Such theories have been extensively investigated
in the context of flat Minkowski spacetime from a variety of different
angles. It has in particular been demonstrated that these quantum
field theories are internally consistent,
at least to all orders in the renormalized perturbation expansion. The early
Universe on the other hand is described by a strongly
curved spacetime, and important
new quantum field theory effects arise in this situation--- an
important example being
the generation of primordial fluctuations that have left an imprint
in the CMB as well as the large scale structure of the universe.
For this reason, it is obviously important to study quantum gauge
theories in curved Lorentzian spacetimes such as the expanding
Universe. The question how to consistently construct such theories in arbitrary curved,
globally hyperbolic spacetimes is an open problem.

As a first step in this direction, we will prove in this paper
that perturbative non-abelian pure
Yang-Mills theory can be consistently quantized
on any globally hyperbolic spacetime, to all orders in perturbation
theory, and any gauge group $G$ that is a direct product of $U(1)^l$
and a simple compact Lie group.
The essence of our proof is an inductive construction of an explicit
renormalization prescription for the perturbatively defined
interacting field quantities
that preserves gauge invariance, and that depends locally and
covariantly upon the spacetime metric.
The proof of this statement is rather complicated, and it relies
partly on auxiliary constructions that have been previously given in the
literature. Some of these constructions are not so widely known as the
renormalization techniques in flat spacetime, and there is at present
no comprehensive review. We therefore found it appropriate to present these
constructions in the form of a report.

\subsection{Generalities}
Quantum field theory in curved spacetime is a framework wherein one considers quantized fields
propagating on a rigidly fixed, non-dynamical, Lorentzian spacetime rather than
flat Minkowski spacetime. It is thus a generalization of the usual setting of quantum field theory. 
In order to have a well-defined propagation
of the fields in curved spacetime (even at the classical level), one usually assumes that
the spacetime does not have any gross causal pathologies such
as closed time-like curves,
(a typical assumption is that the spacetime is ``globally hyperbolic'') but otherwise no
restrictions on the metric are placed. In particular, one does not have
to (and does not want to) assume that the metric has any
isometries, or that it is a solution to a particular field equation.
As quantum field theory on flat spacetime, quantum field theory on curved
spacetime is in general only believed to be an effective theory with a limited
range of validity. It is expected to loose predictive power when the spacetime
curvatures become as large as the inverse Planck length, or in quantum states
where typical quantum field observables such as the quantum stress energy
operator have expectation values or variances (fluctuations) of the order of the Planck length.
On the other hand, the theory is expected to be a very good approximation
when the spacetime curvatures are of the order (or below) the scale of
elementary particle physics such as $\Lambda_{QCD}$, or even
the grand unification (GUT) scale, which is expected to be
the relevant scale during inflation. Naturally, it is also in this regime
(as well as in the case of black holes) that the most interesting physical effects
predicted by the theory occur.

Independent of those questions regarding the limits of physical applicability
of quantum field theory in curved spacetime, one may ask whether this theory,
in itself, has a consistent mathematical formulation or not---just as it is a relevant question
whether classical mechanics has a well-defined mathematical formulation even though
it clearly has a limited range of validity as a physical theory. Unfortunately, this question
is a very difficult one, which has not been answered in a satisfactory manner for
interacting quantum field theory models
even in flat spacetime (in 4 dimensions). Nevertheless, there exist perturbative approaches
to interacting quantum field theory in Minkowski spacetime, and it is by now well-understood
how to calculate, in principle, terms of arbitrary high order in the perturbation expansion. In particular,
one has a good understanding how to systematically deal with the problem of renormalization that
needs to be addressed at each order to get meaningful expressions, and it is known how to
calculate quantities of physical interest for, say, the purposes of collider physics. In fact, this
approach is at present by far the most powerful method to obtain theoretical predictions for particle
physics experiments, and to test quantum field theory.

In quantum field theories in curved spacetime, new conceptual problems arise because one no longer has a
preferred vacuum state in time-dependent spacetimes, as may be understood from the familiar fact
that time-dependent background fields tend to give rise to particle creation. Thus, a state that may be
thought of as a vacuum at one time may fail to be the vacuum at later time. This suggests to use an
S-matrix formulation of the theory, but such a formulation also does not make sense in general if the
spacetime does not have any asymptotically time-independent regions in the far past or future, or if the
metric approaches a time-independent metric too slowly. At the technical level, one no longer has a
clear cut relation between quantum field theory on Lorentzian spacetimes and Riemannian spacetimes, because
a general (even analytic) Lorentzian spacetime will not be a real section in a complexified manifold that also
has a real, Riemannian section. Furthermore, familiar flat space techniques such as momentum space,
dimensional regularization, or the Euclidean path integral, are not available on a generic curved Lorentzian manifold.

As had been realized for some time, these conceptual problems can in principle be overcome by
shifting the emphasis to the local quantum field operators, which can be unambiguously defined on
any (globally hyperbolic) Lorentzian spacetime. The key insight was that the algebraic relations between the
quantum fields (such as commutators, or the "operator product expansion") have an invariant meaning
for any such spacetime, even if there are no states with a definite particle interpretation.
Nevertheless, it remained an unsolved problem how to construct in practice interesting (non-free)
quantum field theories perturbatively on a general globally hyperbolic spacetime, mainly because of
the very complicated issues related to renormalization on a curved manifold. A fully satisfactory
construction of perturbative, renormalized quantum field theory on curved space was finally given in a series of papers~\cite{Brunetti2000, Brunetti1996, Hollands2000, Hollands2001,
Hollands2003} where it was shown that the algebras of local observables (interacting local fields) can always be constructed at the level of formal power series in the coupling, independent of the
asymptotic behavior of the metric at infinity. It was shown in detail
how to perform the renormalization process in a local and covariant way, and it was thereby seen
that the remaining finite renormalization ambiguities correspond
to the possibility of adding finite local terms (possibly with curvature couplings)
to the Lagrangian, and to the possibility of making finite field-redefinitions ("operator mixing with curvature").
These constructions also provided a completely new, geometrical understanding of the
nature of the singularities of multi-point operator products and their expectation values
in terms of "microlocal analysis"~\cite{Hormander,Bros,Sato}, and thereby provided a geometric
generalization of the usual spectrum condition in Minkowski spacetime quantum field theory to
curved manifolds.  By considering the behavior of the theory under a rescaling of the metric $g \to \mu^2 g$, a
definition of the renormalization group could be given~\cite{Hollands2003}, and detailed
results about the (poly-logarithmic) scaling behavior of products of interacting field
operators were thereby obtained.
It is also understood how to construct the operator product expansion from the
algebra of interacting fields in curved space, and this gives direct information
about the interplay between quantum field interactions and spacetime curvature
at small scales~\cite{Hollands2006}.

\subsection{Renormalization of theories without local gauge invariance}

The building blocks in the renormalized perturbation series for the interacting fields are the time-ordered products
$T_n(\O_1 \otimes \cdots \otimes \O_n)$ of composite fields in the underlying free field theory. In standard
approaches in flat spacetime, these objects are typically
viewed as operators on a Hilbert space (``Fock-space''), but in curved spacetime there is no preferred Hilbert-space
representation. In this context, it is more useful to view them instead as members of an abstract algebra, which may in the end be represented
on a Hilbert space (typically in infinitely many inequivalent ways). The first step in the renormalization program therefore is to define a
suitable abstract algebra, and this can indeed be done using the techniques of the ``wave front set.''  The next step is to actually
construct the time-ordered products as specific elements in this algebra. A naive definition leads to infinite meaningless
expressions, but one can show that it is possible to obtain meaningful objects by a process called ``renormalization''.
Conceptually, the best approach here is to first formulate a set of conditions (``renormalization conditions'') on the
time-ordered products to be constructed, and then show via an explicit construction that these properties can be satisfied.
It turns out that the conditions do not uniquely fix the time ordered products, but there remain certain
finite renormalization ambiguities. In curved spacetime, it is a major challenge to formulate sufficiently strong renormalization
conditions in order to guarantee that these ambiguities only consist in adding finite ``contact terms'' at each order
$n$, which are covariant expressions of the Riemann curvature and the fields of a suitable dimension. A key condition to
guarantees this is that the $T_n$ should themselves be local and covariant~\cite{Hollands2000},
and a precise formulation of that condition
naturally leads to a formulation of quantum field theory in the language of  category theory~\cite{Brunetti2003}.
The condition of locality and covariance is a rather strong one, and it is correspondingly non-trivial to
find a renormalization method that will ensure that this condition is indeed satisfied. Such a scheme was
found in~\cite{Hollands2000,Hollands2001} for interacting scalar field theory,
based on key earlier work of~\cite{Brunetti2000, Brunetti1996}, and also on the work~\cite{Duetsch2000,Duetsch2000b},
where an algebraic variant of perturbation theory in flat space was developed.  We will present these
constructions in section~3 of the paper. Here we follow the general steps proposed in these references, but
we develop a new technique to perform the actual renormalization (extension) step. Our new method (described
in the proof of Lemma~\ref{lemma4})
is more explicit than previous constructions, and also gives an interesting new formula
for some of the renormalization constants describing the departure from homogeneous scaling in terms of
an integral of a closed form of a cycle in $\mr^{4n}$, see Proposition~1.

In quantum field theory, one typically wants certain fields to have special properties. For example, an important observable in any
theory with a metric is the stress energy tensor, which is conserved at the classical level if the metric is the
only background field (as we assume). One would like the corresponding quantum field to be conserved as well. In perturbative quantum field theory, it is far from obvious that the corresponding
interacting quantum field quantity is also conserved, and indeed there exist theories where this fails to be the case~\cite{Witten}.
In general, one can formulate a set of renormalization conditions on the time-ordered products (the ``principle of perturbative
agreement''~\cite{Hollands2005}) that will guarantee conservation to all orders in the perturbation expansion. In~\cite{Hollands2005},
it was shown that the question whether or not these identities can be satisfied is equivalent to the question whether
a certain cohomological class on the space of all metric defined by the field theory is trivial or not. The obstruction
sometimes cannot be lifted, and then the renormalization condition is impossible to satisfy: There are anomalies.
Similarly, in gauge theories, one wants certain currents to be conserved at the quantum level and it is important
to ensure that there are no anomalies.

\subsection{The problem of local gauge invariance}

In fact, the perturbative construction of renormalized field theories on curved space
without local gauge invariance does not carry over
straightforwardly to theories with local gauge invariance, and the
construction of such models was therefore up to now an important open problem. The key
obstacle is that the field equations of local gauge theories, such as
e.g. the pure Yang-Mills theory studied in this paper, are not globally hyperbolic in nature even if
the underlying spacetime is globally hyperbolic. This, however, is a
basic assumption in the constructions~\cite{Brunetti1996,Brunetti2000,
Hollands2000,Hollands2001}. In theories with local gauge invariance,
the field equations fail to be hyperbolic in nature
precisely due to local gauge invariance, because it implies that solutions to
the field equations are not entirely determined by their initial data
on some Cauchy surface as required by hyperbolicity, but also on an
arbitrary choice of local gauge. At the classical level, this problem can be
dealt with by simply fixing a suitable gauge. However, at the quantum level, it is
problematical to base the theory on a gauge-fixed formulation, because
gauge fixing typically has non-local features. This causes severe
problems e.g. for the renormalization process. An elegant and very
successful approach circumventing these problems is the
BRST-method~\cite{Becci1975,Becci1976}.  This method consists in
replacing the original action by a new action containing additional
dynamical fields. That new action yields hyperbolic field equations,
and has an invariance under a nilpotent so-called ``BRST
transformation'', $s$, on field space.
Gauge invariant field observables are precisely those that are annihilated by $s$, or more
precisely, the cohomology classes of $s$. Furthermore, the
classical Poisson (or Peierls) brackets~\cite{Peierls1952,Marolf1994,
DeWitt2004,Duetsch2000}
of the gauge fixed theory are invariant under $s$. Thus,
as first suggested by~\cite{Duetsch1999} (based on~\cite{Kugo1980}),
one can try to proceed by first quantizing the brackets of the gauge
fixed action (in the sense of deformation quantization~\cite{Duetsch2000,Duetsch2000b,Bayen1977a,Bayen1977b}),
promote the differential $s$ to a graded derivation
at the quantum level leaving the quantized brackets invariant,
and then at the end define the algebra of physical observables
to be the kernel (or rather cohomolgy) of the quantum BRST-differential. As we will prove in
this paper, this program can be carried out successfully
for renormalized Yang-Mills theory in curved spacetime, at the level of formal
power series in the coupling constant.

Thus, the first step consists in finding an appropriate gauge fixed
and BRST invariant modified action, $S$, for pure Yang-Mills theory
in curved space involving the gauge field, and new auxiliary fields (``anti-fields'').
This step is completely analogous to Yang-Mills theory in flat space.
Next, one needs to ``quantize'' the brackets associated with the new
action $S$. It is not known presently how to do this
non-perturbatively even in flat space, but one can proceed in a
perturbative fashion as in theories without local gauge invariance.

The final step special to gauge theories is now to
define a quantum BRST derivation acting on the quantum interacting fields
This derivation should (a) leave the product invariant,
(b) square to $0$, and (c) go over
to the classical BRST transformation $s$ in the classical limit.
The natural strategy for constructing the quantum BRST transformation
is to consider the quantum Noether current corresponding to
the classical BRST-transformation. One then defines a
corresponding charge, and defines BRST-derivation via the graded commutator
in the star-product with this charge. While this definition automatically
satisfies (a), it is highly non-obvious that it would also satisfy
properties (b) and (c). In fact, it is even unclear whether
that the quantum Noether current operator associated with
the BRST-transformations  is conserved, as would be required in order to
yield a conserved charge.

The basic reason why it is a non-trivial challenge to establish
conservation of the quantum BRST current, as well as (b)
and (c), is that the construction of the
time ordered products $T_n$ used to define the interacting
quantum fields via the Bogoliubov formula involve renormalization.
It is far from obvious that a renormalization prescription exists such
that interacting BRST current will be conserved, and such that (b) and (c) will hold.
In fact, as we will show, these properties follow from
a new infinite hierarchy of Ward identities for the time-ordered products [see eq.~\eqref{WW} for
a generating functional of these identities], which are
violated for a generic renormalization prescription.
We will show that there nevertheless exists a renormalization prescription compatible
with locality and covariance such that
these Ward identities are satisfied in curved space, to all orders in
the renormalized perturbation expansion, when the gauge group is
a product of $U(1)^l$ and a semi-simple group. Thus, we can define an
algebra of interacting quantum fields as the cohomology of
the quantum BRST-differential, and this defines perturbative quantum
Yang-Mills theory. In a second step, we then define
quantum
states (i.e., representations) of this algebra by a deformation
argument. Here we rely on a construction invented in~\cite{Duetsch1999}.
As a by-product of our constructions, we can also show that the
operator product expansion in curved space~\cite{Hollands2006} closes among
gauge-invariant operators, and that the renormalization group flow likewise
closes among gauge-invariant operators.

Our approach has several virtues also in the context in flat spacetime. The key virtue
is that, since our constructions are entirely local, there is a clear separation between
issues related to the ultra-violet (UV) and infra-red (IR) behavior of the theory. In particular,
in our approach, the identities reflecting gauge invariance may be formulated and proved entirely
independently from the infrared behavior of the theory, while the infra-red cutoff is only
removed in the very end in an entirely well-defined manner at the algebraic level
(``algebraic adiabatic limit''~\cite{Brunetti2000}). In this way,
infra-red divergences are neither encountered at the level of the interacting
field algebras, nor in fact at the level of quantum states,
i.e., representations\footnote{However, we would encounter the
  familiar infra-red divergences if we were to try to construct scattering
  states. Actually, it is clear that those types of states cannot be defined in
  a generic curved spacetime anyway even for massive fields, so we do not see this as a
  problem.}. In this respect, our approach is
different from traditional treatments based on Feynman diagrams or effective actions, which
are only formal in as far as the treatment of the IR-problems are concerned.
We explain in some more detail the relation of our approach to those treatments in sec.~4.9.

A local approach that is similar to ours in spirit has previously been
taken in the context of QED on flat
spacetime in~\cite{Duetsch1999}, and in~\cite{Duetsch2001,Duetsch2002} for
non-abelian gauge theories on flat spacetime. Note, however,
that the ``Master Ward identity'' expressing the conditions for local gauge invariance in~\cite{Duetsch2002}
was taken as an axiom and has not been shown to be consistent yet\footnote{For recent
  progress in analyzing the validity of the Master Ward identity,
  see~\cite{MWD}.},
as opposed to the Ward identities of
our paper, which are shown to hold. Also, our Ward identities~\eqref{WW} appear
to be different from those expressed in the
Master Ward Identity of~\cite{Duetsch2002,Duetsch2001}.

\subsection{Summary of the report}

This report is organized as follows. In section~2, we first review basic notions from classical field theory,
including classical BRST-invariance and associated cohomological constructions. The material in this
secion is well-known and serves mainly to set up the notations and provide basic results that are needed
in later sections. In section~3, we review the perturbative construction of interacting quantum
field theory on curved spacetime. We focus on theories without local gauge invariance. We explicitly
describe scalar field theory, and we briefly mention the changes that have to be made for ghost and vector fields
(in the Lorentz gauge). We give a detailed renormalization prescription for the time-ordered products, their
renormalization ambiguities, and describe
how interacting fields may be constructed from them. We also show how the method works in some concrete examples.
The material presented in this section is to some extent taken from~\cite{Brunetti2000, Hollands2000, Hollands2001, Duetsch2000,
Duetsch2001, Duetsch2004}, but there are also some important
new developments. In section~4, we  construct perturbatively renormalized
quantum Yang-Mills theory. We first give an outline of the basic strategy, and then fill in the technical details in the
later sections. We present our new Ward-identities in subsection~4.3, and then prove them in section~4.4. We prove in
4.5 that our identities formally imply the BRST-invariance of the $S$-matrix, in 4.5 that they imply the conservation of the
interacting BRST-current, and in 4.6 that they imply the nilpotency of the interacting BRST-charge operator.
%In section 5
%we outline a proof that quantum Yang-Mills theory has a conserved quantum stress tensor, and we explain how the trace
%anomaly arises.
We conclude and name open problems in
section~6. Appendix A contains a treatment of free $U(1)$-theory
avoiding the introduction of the vector potential and an explanation of the
new superselection sectors arising in this context.
The appendices B--E contain definitions and various
constructions that are omitted from the main part of the paper.

\subsection{Guide to the literature}

A standard introduction to the theory of quantum fields on a curved space is~\cite{Wald},
which gives an in-depth discussion of the conceptual problems of the theory,
as well as the Hawking and Unruh-effect, at the level of free quantum
fields. The generalization of the latter effect to certain black-hole spacetimes---emphasizing
especially the role of the so-called ``Hadamard condition''---is
discussed in the review-style article~\cite{Kay}. Other
monographs are~\cite{Fulling,Birrell}.
The perturbative construction of interacting scalar quantum field theories on curved spaces was given in
the series of papers~\cite{Brunetti2000,Brunetti1996,Hollands2000,Hollands2001,Hollands2005}.
Important contributions to the understanding of Hadamard states in terms of microlocal analysis,
which were a key input in these papers, were made by
Radzikowski~\cite{Radzikowski1996a,Radzikowski1996b}. These results
are reviewed and extended in the very readable paper~\cite{Junker1}.
A complete characterization of the state space of perturbative
quantum field theory using microlocal analysis is given in~\cite{Ruan}.
A definition and analysis of the renormalization group in curved
space was given in~\cite{Hollands2003}. The generalization of the Wilson
operator product expansion in curved spacetime was constructed to all
orders in perturbation theory in~\cite{Hollands2006}.
Perturbative scalar quantum field theory on Riemannian spaces was treated in~\cite{Bunch1981} using the
BPHZ method, and by~\cite{Kopper} using the method of flow equations.
General theorems about quantum field theory in curved spacetime
within a model-independent setting were obtained in~\cite{HollandsPCT} (PCT-theorem),
and by~\cite{Verch2000} (spin and statistics theorem). The literature on the quantization of
gauge theory, and especially Yang-Mills theory in flat spacetime is huge. The use of ghost fields was
proposed first by~\cite{Fadeev}, and the early approaches to prove gauge invariance at the renormalized
level used the method of Feynman graphs, together with special regularization techniques~\cite{tHooft1,tHooft2,tHooft3}.
More recent discussions based on the Hopf-algebra structure behind renormalization~\cite{Kreimer1,Kreimer2,Kreimer3}
may be found in~\cite{Suijlekom1,Suijlekom2}.
With the discovery of the BRST-method~\cite{Becci1975,Becci1976}, cohomological methods were developed and
used to argue that gauge invariance can be maintained at the
perturbative level in flat spacetime.
Comprehensive reviews containing many
references are~\cite{deWitt1,Piguet1995,Henneaux1995,Barnich2000}, see
also
e.g.~\cite{Stora1, Stora2,Dubois-Violette1985,Dubois-Violette1990,Dubois-Violette1992}. There are also other approaches
to quantum gauge invariance in flat space,
based on the Epstein-Glaser method~\cite{Epstein1973} for renormalization. These are
described in the monographs~\cite{Scharf1,Scharf2} and
also in~\cite{Stora1990},
which also contain many references. For a related approach,
see~\cite{Steinmann1990}. The idea to formulate quantum gauge theory at the level of
observables, and to implement the gauge invariance in the operator
setting was developed in flat space
in~\cite{Duetsch1999,Duetsch2001,Duetsch2002},
building on earlier work of~\cite{Kugo1980}. A somewhat more detailed comparison
between the various approaches to the gauge invariance problem and our solution is given in Sec.~4.9, where
additional references are given.

\medskip
\noindent
{\bf Note added in proof:} It has been brought to our attention by M. D\" utsch 
that the original definition of the extension $u'$ in eq.~\eqref{shilovtrick} was incorrect. The present, 
corrected, version of formula~\eqref{shilovtrick} was suggested by him and has in fact already 
been considered in earlier  papers~\cite{garcia_bondia1,garcia_bondia2}. Correcting this formula has had 
no impact on the other conclusions or arguments of this paper. 

\section{Generalities concerning classical field theory}

\subsection{Lagrange formalism}

Most, though not all, known quantum field theories have a classical
counterpart that is described in terms of a classical Lagrangian field
theory. This is especially true for the gauge theories studied in this
paper, so we collect some basic notions and results from Lagrangian
field theory in this subsection that we will need later.
Not surprisingly, for perturbative quantum field
theories derived from a classical Lagrangian, many
formal aspects can be formulated using the language of classical
field theory, but we emphasize that, from the physical viewpoint,
quantum fields are really fundamentally different from classical
fields.

To specify a classical field theory on an $n$-dimensional manifold
$M$, we first need to specify its field content. We will generally divide the
fields into background fields, collectively denoted $\Psi$, and
dynamical fields, collectively denoted $\Phi$. Both background and dynamical fields
are viewed as sections in a certain fibre bundle, $B \to M$, over the
spacetime manifold. We will assume that the background fields always
comprise a Lorentzian metric $g = g_{\mu\nu} dx^\mu dx^\nu$ over
$M$ (which is a section in the bundle of non-degenerate
symmetric tensors in $T^*M \otimes T^*M$ of signature $(-++ \dots
+)$). More generally, the background fields may comprise a non-abelian
background gauge connection, or various external sources. We will also
admit Grassmann-valued fields, which are described in more detail
below. The dynamical fields will typically satisfy equations of motion, which
are derived from an action principle. By contrast, the background fields will
never be subject to any equations of motion.

To set up an action principle, we need to specify a Lagrangian. The
Lagrangians that we will consider have the property that they are
locally and covariantly constructed out of the dynamical fields
$\Phi$, and the background fields $\Psi$. In particular, they do not depend implicitly on
additional background structure such as the specification of a
coordinate system. Since such functionals will play an important role in
perturbation theory, it is worth defining
the notion that a quantity is locally and covariantly out of a set
of dynamical and non-dynamical fields $\Phi, \Psi$ with some care.
Let us denote by
$B \to M$ the ``total bundle'' in which the dynamical and
non-dynamical fields live. For example, in case all the fields
are tensor fields, the total bundle is simply the direct sum of
all the tensor bundles corresponding to the various types of
fields. If $x \in M$, we let $J^k_x(B)$ denote the space of
``$k$-jets'' over $M$. This is defined as the equivalence class of
all sections $\sigma=(\Phi, \Psi): M \to B$, with the equivalence relation
$\sigma_1 \sim \sigma_2$ if $\nabla^q \sigma_1|_x= \nabla^q \sigma_2 |_x$
for all $q \le k$,
where $\nabla$ is any affine connection in the bundle $B$, and
where we have put
\ben\label{ksym}
\nabla^k \sigma = dx^{\mu_1} \otimes \dots \otimes dx^{\mu_k}
\nabla_{(\mu_1} \cdots \nabla_{\mu_k)} \sigma \, .
\een
We say that
a $p$-form $\O = \O_{\mu_1 \dots \mu_p} dx^{\mu_1} \wedge \dots
\wedge dx^{\mu_k}$ is constructed out of $\sigma=(\Phi,\Psi)$ and its first
$k$ derivatives if $\O$ is a map
\ben\label{pform}
\O: J^k_x(B) \to \bigwedge^p T^*_x M
\een
for each $x \in M$, which we will also write as
$\O(x) = \O[\sigma(x), \nabla \sigma(x), \dots, \nabla^k
\sigma(x)]$.
Now let $\psi: M \to M'$ be an immersion that lifts to a
bundle map $B \to B'$ denoted by the same symbol,
and let $\sigma$ and $\sigma'$ be sections in $B \to M$ respectively
$B' \to M'$ such that $\sigma = \psi^* \sigma'$. We will say that
$\O$ is a $p$-form that is locally constructed out of the fields
$\sigma$ if we have
\ben
\O[\sigma(x), \nabla \sigma(x), \dots, \nabla^k \sigma(x)] =
\psi^* \O[\sigma'(x'), \nabla \sigma'(x'), \dots, \nabla^k \sigma'(x')]
\, , \quad  \psi(x) = x' \, ,
\een
for any $x$ and any such embedding $\psi$. This condition makes precise
the idea that $\O$ is only constructed out of $\sigma =(\Phi,\Psi)$ and finitely
many of its derivatives, but depends on ``nothing else''. For example,
if the fields are a background metric, $g$, and a set of dynamical
tensor or spinor fields $\Phi$, then one can show that
$\O$ can depend upon the metric only via the curvature, i.e., it
may be written in the form
\ben\label{thomas}
\O(x) =
\O[\Phi(x), \nabla \Phi(x), \dots,
\nabla^k \Phi(x), g(x),
R(x), \nabla R(x), \dots,
\nabla^{k-2} R(x)]
\een
where $\nabla$ is now the Levi-Civita (or spin-) connection associated with
$g$, and
$R = R_{\mu\nu\sigma\rho}(dx^\mu \wedge dx^\nu) \otimes (dx^\sigma
\wedge dx^\rho)$ is the curvature
tensor. This result is sometimes called the ``Thomas replacement
theorem,'' and a proof may be found in~\cite{Iyer1994} and in lemma \ref{trt} below. 

The second
example relevant to this work is when
the background fields contain in addition a background gauge connection $\bar \nabla$ in a
principal fibre bundle, such as $B = M \times G$. Then the lift of $\psi$ to a bundle map
$B \to B'$, with $B' = M' \times G$ incorporates the specification
of a map $\gamma: M \to G$ that provides the identification of the fibres, i.e.,
a local gauge transformation. The condition that $\nabla = \psi^* \nabla'$ then means that
$\bar \nabla' = \bar \nabla + \gamma^{-1} d\gamma$, and the condition of local covariance of
a functional $\O$ now implies that $\O$ can depend on the connection only via
its curvature $f$ and its covariant derivatives $\bar \nabla \bar f, \dots, \bar \nabla^{k-2} \bar f$
(Here $\bar \nabla$ acts as the Levi-Civita connection of $g$ on the tensorial structure).
More generally, if in addition there are dynamical fields $\Phi$ valued
in an associated bundle $B \times_G V$ (with $V$ a representation of $G$),
then $\O$ can only depend on gauge invariant combinations of $\Phi, \bar \nabla \Phi,
\dots, \bar \nabla^k \Phi$.

These statements can be proved by the same type arguments as in~\cite{Iyer1994}.
For completeness, we give a proof of this generalization of the
Thomas replacement theorem incorporating gauge fields in sec.~2.3 below.
In our later application to Yang-Mills theory,  the dynamical and background fields will be identified as follows, see below for more explanations:

\begin{enumerate}
\item ({\bf Dynamical fields} $\Phi$) Gauge connection $\mathcal D$ (usually decomposed as ${\mathcal D}=
\bar \nabla + i\lambda A$, where $\bar \nabla$ is a fixed background gauge connection $\lambda \in \mr$ is fixed and $A$ is the dynamical field in the adjoint representation), 
(anti-) ghost fields $C,\bar C$, auxiliary field $B$. $(A,B,C,\bar C)$ are collectively called $\Phi_i$ below. 
The gauge group $G$ acts on these fields via the adjoint 
representation.

\item ({\bf Background fields} $\Psi$) Metric $g$, background gauge connection $\bar \nabla$, ``anti-field sources'' of dynamical fields collectively called 
$\Phi^\ddagger_i$ below, see table \ref{table:anti-fields}. (The background connection is set to the trivial gauge connection in $B=M \times G$ in the body of the paper for simplicity). 
\end{enumerate}

We denote the space of all locally covariant $p$-form functionals~\eqref{pform}
by $\P^p(M)$, or simply by $\P^p$, and we define
\ben
\P(M) = \bigoplus_{p=0}^n \P^p(M) \, .
\een
We also assume for technical reasons that the expressions
in $\P$ have at most polynomial dependence upon the dynamical fields
$\Phi$, and an analytic dependence upon the background fields
$\Psi$.
These definitions can easily be generalized to the case
when $(\Phi,\Psi)$ are not ordinary fields valued in some bundle, but
instead Grassmann valued fields. A Grassmann valued field is by
definition simply a field that is valued in the infinite dimensional
exterior algebra $E$, which
is the graded vector space
\ben
E = {\rm Ext}(V) = \bigoplus_n E_n, \quad E_n = \bigwedge^n V
\een
with $V$ some infinite-dimensional complex vector space. The space $E$ is
equipped with the wedge product $\wedge:E_n \times E_m \to E_{m+n}$, which has the property
that $e_n e_m = (-1)^{nm} e_m e_n$ for $e_n \in E_n, e_m \in E_m$, and
$e_n e_m = 0$ for all $e_n$ if and only if $e_m = \lambda e_n$. The
elements $e_n$ in $E_n$ are assigned Grassmann parity $\epsilon(e_n) =
n$ modulo 2. Thus, when Grassmann valued field are present, expressions
$\O\in \P^p$ are no longer valued in the $p$-forms over $M$, but
instead in the set of $p$-forms over $M$, tensored with $E$. A Grassmann
valued field consequently has a formal expansion of the form
\ben
\Phi(x) = \sum_{n \ge 0} e_n \Phi_n(x), \quad e_n \in E_n,
\een
where each $\Phi_n$ is an ordinary $p$-form field.

A Lagrangian is a (possibly $E$-valued)
$n$-form $\L = \L[\Phi, \Psi]$ that is locally and
covariantly constructed out of the dynamical fields $\Phi$, the
background fields $\Psi$, and finitely many of its derivatives.
For manifolds $M$ carrying an orientation, which we shall assume to be
given from now on, one can define a canonical volume $n$-form $\beps =
\epsilon_{\mu_1 \dots \mu_n} dx^{\mu_1} \wedge \dots \wedge
dx^{\mu_n}$ by the standard formula
\ben
dx = \beps = \sqrt{-g} \, dx^0 \wedge \dots \wedge dx^{n-1}
\een
where $x^0, \dots, x^{n-1}$ is right handed, and where $\sqrt{-g}$ is
the square root of minus the determinant of $g_{\mu\nu}$. Using the
volume $n$-form, one defines the Hodge dual of a form by
\ben
*\alpha_{\mu_1 \dots \mu_{n-p}} = \frac{(-1)^p}{(n-p)!} \epsilon^{\nu_1
  \dots \nu_p}{}_{\mu_1 \dots \mu_{n-p}} \alpha_{\nu_1 \dots \nu_p}
\een
and it is thereby possible to convert the Lagrangian into a
scalar. This is more standard in the physics literature, but
for our purposes it will be slightly more advantageous to view $\L$
as an $n$-form. For
compactly supported field configurations, we may form an associated
action by integrating the Lagrangian $n$-form over $M$,
\ben
S = \int_M \L \, .
\een
We define the left and right variation, $\delta_L S/\delta \Phi(x)$ resp.
$\delta_R S/\delta \Phi(x)$ with respect to the
dynamical fields by the relation
\ben
\frac{d}{dt} S[\Phi_t; \Psi] \Bigg|_{t=0} = \int_M \frac{\delta_R
  S}{\delta \Phi(x)} \delta \Phi(x) = \int_M \delta\Phi(x) \frac{\delta_L S}{\delta \Phi(x)}, \quad
\delta \Phi(x) = \frac{d}{dt} \Phi_t(x) \Bigg|_{t=0} \, .
\een
The left and right derivatives may differ from each other only for
Grassmann-valued fields $\Phi$, and we adopt the convention
that the left derivative is meant by default if the subscript is suppressed.
In terms of the Lagrangian $n$-form, the variational derivative is
given by
\ben
\frac{\delta S}{\delta \Phi(x)} = \sum_{q=0}^k (-1)^q
\bar \nabla_{(\mu_1 \dots \mu_q)}
\left\{ \frac{\partial \L}{\partial( \bar \nabla_{(\mu_1 \dots \mu_q)} \Phi(x))}
\right\} \, ,
\een
where we use the abbreviation $\bar \nabla_{(\mu_1 \dots \mu_k)}$ for the
$k$-fold symmetrized derivative in eq.~\eqref{ksym}. The quantity
$\delta S/\delta \Phi(x)$ is an $n$-form that is locally and
covariantly constructed out of the dynamical fields and the background fields
and their derivatives, and may hence be viewed as
a differential operator acting on $\Phi$. Field configurations
$\Phi$ satisfying the differential equation
\ben\label{eom}
\frac{\delta S}{\delta \Phi(x)} = 0
\een
are said to satisfy the equations of motion associated with $S$, or
to be ``on shell.''

A symmetry is an infinitesimal field variation $s\Phi=\delta \Phi$ of
the dynamical fields such that $s\L = d{\bf B}$ for some locally
constructed $(n-1)$-form $\bf B$. The existence of symmetries implies
the existence of a conserved Noether current, $\J$, defined by
\ben
\J(\Phi) = \btheta(\Phi, s\Phi) - {\bf B}(\Phi) \, ,
\een
where $\btheta$ is the $(n-1)$ form defined by
\ben\label{thetadef}
\theta_{\nu_1 \dots \nu_{n-1}}(\Phi, \delta \Phi)
= \sum_{q=0}^{k-1}
\bar \nabla_{(\mu_1 \dots \mu_q)} \delta \Phi
\left\{ \frac{\partial \L_{\nu_1 \dots \nu_{n-1}\sigma} }{\partial(
    \bar \nabla_{(\mu_1 \dots \mu_q \sigma)} \Phi)}
\right\} \, ,
\een
where we are suppressing the dependence upon the background fields.
$\btheta$ is the boundary term that would arise if $\L$ is varied
under an integral sign. As a consequence of the definition,
we have
\ben
d\J = \sum s\Phi_i \frac{\delta S}{\delta \Phi_i} \, ,
\een
so $\J$ is indeed conserved on shell. In the context of perturbation
theory studied in this paper, the Lagrangian is a power series
\ben
\L = \L_0 + \lambda \L_1 + \lambda^2 \L_2 + \dots ,
\een
where $\L_0$ is called the ``free Lagrangian'' and contains only
terms at most quadratic in the dynamical fields $\Phi$, hence
giving rise to linear equations of motion. If the symmetry is also
a formal power series
\ben
s = s_0 + \lambda s_1 + \lambda^2 s_2 + \dots ,
\een
then there is obviously an expansion
\ben
\J = \J_0 + \lambda \J_1 + \lambda^2 \J_2 + \dots ,
\een
$s_0$ is a symmetry of the free Lagrangian $\L_0$
with corresponding conserved Noether current $\J_0$ when the equations
of motion hold for $\L_0$.

\medskip

The theories that we will deal
with in this paper all have the property that $\L_0$ contains
the highest derivative terms in the dynamical fields $\Phi$. In this case, it is natural to
assign a ``canonical dimension'' to each of the dynamical fields as
follows. Let us assume that the background fields consist of a
metric, $g$, and a covariant derivative operator, $\nabla$, which
acts like the Levi-Civita connection on tensors. Consider
a rescaling of the metric by a constant conformal factor,
$\mu^2 g$, where $\mu \in \mr$. Then there exists typically a unique rescaling
$\Phi_i \to \mu^{d(\Phi_i)} \Phi_i$, $\Psi_i \to \mu^{d(\Psi_i)} \Psi_i$ and $c_i \to \mu^{d(c_i)} c_i$
of the dynamical fields, the background fields, and the coupling constants in $\L_0$ such
that $\L_0 \to \L_0$. The numbers $d(\Phi_i), d(\Psi_i)$ and $d(c_i)$ are
called the ``engineering dimensions'' of the fields and the couplings,
respectively. The corresponding dimension of composite objects in
$\P$ is given by the counting operators $\N_f, \N_c, \N_r: \P(M) \to \P(M)$
\bena\label{Ndefs0}
\N_{f} &=& \sum (d(\Phi_i) + k) \, \partial^k \Phi_i
\frac{\partial}{\partial(\partial^k \Phi_i)} \, \\
\N_{c} &=& \sum d(c_i)  \, c_i
\frac{\partial}{\partial c_i} \, \\
\N_{r} &=& \sum (d(\Psi_i) + k) \, \partial^k \Psi_i
\frac{\partial}{\partial(\partial^k \Psi_i)} \, .
\eena
Not for all $S$, and not for all choices of the background fields
$\Psi$ do the equations of motion~\eqref{eom} possess a well posed initial value
formulation, which is a key requirement for a physically reasonable
theory. For first order differential equations one can formulate general
conditions under which the equations will posses a well-posed initial
value formulation. For example, for first order systems of so-called
``symmetric hyperbolic type,'' the initial value problem is well posed
in the sense that, given initial data for $\Phi$ on a suitably chosen
$n-1$-dimensional hypersurface, there exists a unique solution for
sufficiently short ``times,'' i.e., in some open neighborhood of
$\Sigma$. Furthermore, the propagation of disturbances is ``causal''
in a well-defined sense, see e.g.~\cite{Geroch2000}. Equations of
motion of higher differential order can always be reduced to ones of
first order by picking suitable auxiliary field variables, but it is
not obvious in a given example which choice will lead to a symmetric
hyperbolic system. Fortunately, the equations of motion that we will
study in this paper will all be of the form of a simple
wave-equation. Actually, since we only consider perturbation theory,
we will only be concerned with the existence of solutions for the
``free theory,'' defined by $S_0$. For the actions considered in this
paper, the corresponding equations are linear, and of the form
\ben\label{wave}
0 = \frac{\delta S_0}{\delta \Phi} =
\square \Phi + (\text{lower order terms})
\een
where $\square = g^{\mu\nu} \nabla_\mu \nabla_\nu$ is the wave
operator in curved space. Such equations do posses a well-posed
initial value formulation if the metric does not have any gross causal
pathologies, such as closed timelike curves.
A typical such equation
(for a real scalar field $\Phi = \phi$) is the Klein-Gordon equation
\ben\label{kgj}
(\square - m^2) \phi = j \, ,
\een
where $m^2$ is a constant. For that equation, the initial value problem is
well-posed globally for example
if the spacetime manifold $(M,g)$ is ``globally hyperbolic,'' meaning by
definition that
there exists a (necessarily spacelike)
``Cauchy-surface'', $\Sigma$, i.e., a surface which has
the property that any inextendible timelike curve hits $\Sigma$
precisely once. We will always assume in this work that
$(M,g)$ is globally hyperbolic. Then, given any
$f_0, f_1 \in C^\infty_0(\Sigma)$, there exists a unique
solution to eq.~\eqref{kgj} such that $\phi | \Sigma = f_0$, and
$n^\mu \nabla_\mu \phi | \Sigma = f_1$, where $n$ is the timelike
normal to $\Sigma$.

The well-posedness of the initial value problem for the Klein-Gordon
equation directly leads to the existence of advanced and retarded
propagators, which are the uniquely determined distributions
$\Delta_A, \Delta_R$ on $M \times M$ with the properties
\ben
(\square - m^2) \Delta_A(x,y) = \delta(x,y) =
(\square - m^2) \Delta_R(x,y)
\een
and the support properties
\ben
\supp \Delta_{A,R} \subset \{(x,y) \in M \times M \mid
\quad y \in J^\mp(x) \} \, ,
\een
where
$J^\pm(S)$ denotes the causal future/past of a set $S \subset M$ and
is defined as the set of points $x \in M$ with the property that there
is a future/past directed timelike or null curve $\gamma$ connecting
$x$ with a point in $S$.

\subsection{Yang-Mills  theories, consistency conditions, cohomology}

The theory that we are considering in this paper is
pure Yang-Mills theory, classically described by the action
\ben
S_{ym} =
-\frac{1}{2} \int_M F^I \wedge * F_I  \, .
\een
Here, $F_{\mu\nu} = (i/\lambda)[{\mathcal D}_\mu, {\mathcal D}_\nu]$
is the 2-form field strength tensor of a gauge
connection ${\mathcal D}$ in some principal
$G$-bundle over $M$, where $G$ is a direct product of $U(1)^l$ and
a semi-simple Lie group. $\lambda$ is a coupling constant 
that could be omitted at the classical level. For the sake of simplicity, we will assume
that the principal bundle is toplogically trivial,
i.e., of the form $M \times G$.
We denote the generators of the gauge Lie algebra
by $T_I, I = 1, \dots, dim(G)$, and we write $F = T_I
F^I_{\mu\nu}dx^\mu \wedge dx^\nu$ for the
components of the field strength and similarly for any other
Lie-algebra valued field. Lie algebra indices $I$ are raised an
lowered with the
Cartan-Killing metric $k_{IJ}$  defined by
${\rm Tr} \, ad(T_I) ad(T_J)$ for the generators of the
semi-simple part, and by 1 for the abelian factors.

The classical field equations for the connection ${\mathcal D}$ derived
from this action are
\ben
g^{\mu\nu}
[{\mathcal D}_\mu , [{\mathcal D}_\nu, {\mathcal D}_\sigma ]] = 0 \, ,
\een
or, written in more conventional form,
\ben
{\mathcal D}_{[\mu} *\! F_{\nu\sigma]} = 0 \, .
\een
As is particularly clear from the first formulation, the connection ${\mathcal D}$ is the
dynamical field variable in this equation. It is convenient to
decompose it into a fixed background connection $\bar \nabla$, plus $\lambda$ times
a Lie-algebra valued 1-form field $A = T_I A_\mu^I dx^\mu$,
\ben
\D = \bar \nabla + i\lambda A \, , \quad \quad \lambda \in \mr \, .
\een
The 1-form field $A$ is now the dynamical variable. The coupling constant
$\lambda$ is redundant at the classical level and may be absorbed in $A$, but 
it is useful as an explicit perturbation parameter when one wants to study the theory
perturbatively. The coupling constant $\lambda$ acquires a new role at the quantum level
due to renormalization effects as we will see below. 
It is convenient to define $\nabla$ on tensor fields
to be the standard Levi-Civita connection of the metric. The
background derivative operator then has the curvature tensor
\ben
[\bar \nabla_\mu, \bar \nabla_\nu] k_\sigma = R_{\mu\nu\sigma}{}^\rho k_\rho +
\bar f_{\mu\nu}^I {\rm R}(T_I) k_\sigma
\een
where $\rm R$ is the representation of the Lie-algebra associated with
$k_\mu$, and $\bar f=T_I \bar f_{\mu\nu}^I dx^\mu \wedge dx^\nu$ is the curvature of the background
gauge connection $\bar \nabla$. In Minkowski space, it is typically assumed
that $\bar \nabla = \partial$, implying that $\bar f=0$.

For simplicity, we will usually assume in the following that the background gauge connection $\bar \nabla$
has been chosen as the standard flat connection in the
bundle $M \times G$, so that $\bar f=0$ in our case. The advantage of
this choice is that we can effectively replace $\bar \nabla$ in all formulas involving 
Lie-algebra valued forms by the exterior differential $d$ without having to worry
about the background curvature $\bar f$. It is clear, however, that it would in principle be 
an advantage to have a formalism allowing an arbitrary background gauge connection. 
In such a formalism, one would be able to address the question to what extent the 
theory (after quantization) remains background independent with respect to the background gauge field, 
i.e. independent of the particular way of decomposing the full gauge connection into 
${\mathcal D} = \bar \nabla + i\lambda A$. 

With this choice of decomposition ${\mathcal D} = d + i\lambda A$, the curvature $F$ is given by
\ben
F^I=
dA^I
+ \frac{i\lambda}{2}  f^{I}{}_{JK} A^J\wedge A^K
\een
where $f^I{}_{JK}$ are the structure constants of the Lie-algebra defined
by $[T_I, T_J] = f_{IJ}{}^K T_K$.
The equations of motion, when written in terms of $A$, are not
hyperbolic, in the sense that the highest derivative term is not
of the form of a wave equation. Thus, the equations of motion for
Yang-Mills theory do not straightforwardly admit an initial value formulation.
This feature is a consequence of the fact that the Yang-Mills
Lagrangian and equations of motion is invariant under the group of local
gauge transformations acting on the dynamical fields by
${\mathcal D} \mapsto \gamma(x)^{-1} {\mathcal D}
\gamma(x)$, where $\gamma: M \to G$ is any smooth function valued in the group,
or equivalently by
\ben
\bar \nabla \mapsto \bar \nabla + \gamma^{-1} d \gamma \, , \quad A \mapsto \gamma^{-1} A \gamma \equiv Ad(\gamma) A \, 
\een
in case we have an arbitrary background connection $\bar \nabla$.
Since such local gauge transformations allow one to make
local changes to the dynamical field variables, it is clear that
those are not entirely specified by initial conditions.
However, the freedom of making local gauge transformation can be used to set
some components of $A$ to zero, so that the remaining components
satisfy a hyperbolic equation and consequently admit a well-posed initial
value formulation, as described
e.g. in~\cite{Choquet}. Later, we want to perturbatively
construct a quantum version of Yang-Mills theory, and for this
purpose, another approach seems to be much more convenient.
This approach consists in adding further fields to the theory
which render the equations of motion hyperbolic, and which can, at
a final stage, be removed by a symmetry called ``BRST-symmetry''.

In the BRST
approach, one introduces additional dynamical Grassmann
Lie-algebra valued fields $C=T_I C^I, \bar C= T_I \bar C^I$,
and a Lie-algebra valued
field $B=T_I B^I$, and one defines a new theory with action $S_{tot}$ by
\ben\label{Sdef}
S_{tot} = S_{ym} + S_{gf} + S_{gh} \, ,
\een
where $S_{gf}$ is a ``gauge fixing'' term defined by
\ben
S_{gf} = \int_M B^I (i{\mathcal G}_I + \frac{1}{2} B_I)
\een
with a local covariant
``gauge fixing'' functional ${\mathcal G}$ of the field
$A$, and where $S_{gh}$ is the ``ghost'' term, defined by
\ben
S_{gh} = i \int_M {\mathcal D}_\mu C^J \frac{\delta ({\mathcal G^I}
    \bar C_I)}{\delta A_\mu^J} \, \beps \, .
\een
The total set of dynamical fields is denoted
$\Phi = (A^I, C^I, \bar C^I, B^I)$, and their assignment of ghost number,
Grassmann parity,  dimension, and form degree is summarized in the
following table
\begin{table}
\label{table:fields}
\begin{center}
\begin{tabular}{|c||cccc|}
\hline
$\Phi$& $A^I$ & $C^I$ & $\bar C^I$ & $B^I$\\
\hline
Dimension & 1 & 0 & 2 & 2\\
Ghost Number & 0 & 1 & $-1$ & 0\\
Form Degree & 1 & 0 & 0 & 0\\
Grassman Parity & 0 & 1 & 1 & 0\\
Star Parity & + & + & + & $-$\\
\hline
\end{tabular}
\end{center}
\end{table}
The assignments of the dimensions are given for the case
when the spacetime $M$ is 4-dimensional, to which we now
restrict attention for definiteness. The ``star parity'' property concerns 
how the given field behaves under the *-operation. It will be 
referred to later when the algebras of fields is equipped with a $*$-operation.

To state the relation between the auxiliary theory
and the original Yang-Mills theory,
one first observes that the action $S_{tot}$ of the auxiliary theory
is invariant under the following so-called BRST-transformations~\cite{Becci1975,Becci1976}:
\bena
s A^I &=& d C^I + i\lambda f^I{}_{JK}A^J C^K , \\
s C^I &=& -\frac{i\lambda}{2} f^I{}_{JK} C^J C^K , \\
s \bar C^I &=& B^I \, ,  \\
s B^I &=& 0\, .
\eena
The assignment of the various gradings to the fields are done in such
a way that $s$ has dimension 0, ghost number +1, Grassmann parity +1,
and form degree 0.
It is declared on arbitrary local covariant functionals $\O \in \P(M)$
of the dynamical fields $A, C, \bar C, B$ and the background fields
by the rules $\nabla \circ s - s \circ \nabla = 0 = dx^\mu \circ s + s \circ dx^\mu$, and on (wedge) products
via the graded Leibniz rule, $s(\O_p \wedge \O_q) =
s\O_p \wedge \O_q + (-1)^{p+\epsilon(\O_p)} \O_p \wedge s\O_q$ (here $\epsilon$ gives the Grassmann parity of a field).
With these definitions, it follows that
\ben
s^2 = 0 \, , \quad sd + ds = 0 \, .
\een
The key equation is
\ben
sS_{tot} = 0 \, ,
\een
which one may verify by writing $S_{tot}$ in the
form
\ben
S_{tot} = S_{ym} + s\varPsi
\een
where
\ben
\varPsi = \int_M \bar C^I (\frac{1}{2} B_I + i{\mathcal G}_I) \beps \, .
\een
Indeed, $S_{ym}$ is invariant because $s$ just acts like an ordinary
infinitesimal gauge transformation on $A$, while $s$ annihilates
the second term because $s^2=0$. In this paper, we choose the
gauge fixing functional as
\ben\label{lorgauge}
{\mathcal G}^I = \nabla^\mu A_\mu^I \, .
\een
Then the equation of motion for $B^I$ is algebraic, $B^I = -i\nabla^\mu
A_\mu^I$. Inserting this into the equation of motion for $A^I_\mu$, one
sees that this equation is of the form~\eqref{wave}. Indeed, this special
choice of the gauge fixing function effectively eliminates a term
of the form $\nabla^\mu \nabla_\nu A^I_\mu$ (which would spoil
hyperbolicity) from the equations of
motion for the gauge field, thus leaving only the wave operator.
The remaining equations for $C^I, \bar C^I$ are also of the
form~\eqref{wave}. Thus, the equations of motion for the total action
$S_{tot}$ are of wave equation type. They consequently possess a well-posed
initial value formulation at the linear level, which is
sufficient for perturbation theory, and in fact also at the non-linear
level~\cite{Choquet}.

Given that $S_{tot}$ defines a classical theory with a well-posed initial
value formulation, we may define an associated graded
Peierls
bracket~\cite{Duetsch2002,Duetsch2004,Peierls1952,DeWitt2004,Marolf1994}
$\{\O_1, \O_2\}_{\rm P.B.}$,
for any pair of local\footnote{The Peierls bracket may also be
defined for certain non-local functionals. The consideration of
such functionals is necessary in
order to contain a set of functionals that is stable under
the bracket.}
functionals $\O_1, \O_2 \in \P$.
Since the action $S_{tot}$ is invariant under $s$,
it follows that the (graded) Peierls bracket is also invariant
under $s$, in the sense that
\ben\label{derivation}
s \{\O_1, \O_2\}_{\rm P.B.} = \{s\O_1, \O_2\}_{\rm P.B.} +
(-1)^{\epsilon(\O_1)+deg(\O_1)} \{\O_1, s\O_2\}_{\rm P.B.} \, ,
\een
$(-1)^{\epsilon(\O_1)}$ denoting the Grassmann parity of a functional
of the fields, and $deg(\O_1)$ the form degree. The connection between
the classical auxiliary
theory associated with $S_{tot}$, and Yang-Mills theory with
action $S_{ym}$ is based on the following key Lemma:
\begin{lemma}\label{thm1}
Let $\O \in \P$ be a local covariant functional of the background
connection, the background metric, and the fields
$\Phi=(A, C, \bar C, B)$. Let $s \O = 0$. Then, up to
a term of the form $s\O'$, $\O$ is a linear combination of
elements of the form
\ben\label{pqd}
\O = \prod_k r_{t_k}(g, R, \nabla R, \dots, \nabla^k R) \prod_i p_{r_i}(C) \,
\prod_j \Theta_{r_j}(F, {\mathcal D} F, \dots, {\mathcal D}^l F) \, ,
\een
where $p_r,\Theta_s$ are invariant polynomials of the Lie-algebra of $G$,
where $F = F_{\mu\nu} dx^\mu \wedge dx^\nu$, and where $r_t$ is
a local functional of the metric $g$, and the Riemann tensor $R$ and its derivatives.
\end{lemma}
The lemma is essentially a standard result in BRST-cohomolgy,
see e.g.~\cite{Barnich2000} and
the references cited there.
The only difference to the formulation given in~\cite{Barnich2000} is that,
in the present setting, the coefficients $r_t$
can only depend locally and covariantly
upon the metric (as opposed to being an arbitrary form on spacetime).
The fact that
$r_t$ then has to be a functional of the Riemann tensor and its derivatives follows again
from the ``Thomas replacement argument'', see e.g.~\cite{Iyer1994} and the next subsection.
Thus, at zero ghost number, the local and covariant functionals
in the kernel of $s$ are precisely the local gauge invariant
observables of Yang-Mills theory modulo an element in the image of
$s$, so the equivalence classes of the kernel of $s$ modulo the
image of $s$ at zero ghost number,
\ben
\{\text{class. gauge. inv. fields} \}
= \frac{{\rm Kernel} \, s}{{\rm Image} \, s}
\quad (\text{at zero ghost number}),
\een
are in one-to-one correspondence with the gauge invariant
observables. Furthermore, by~\eqref{derivation}, the
brackets are well-defined on the cohomology classes, and the
Yang-Mills equations of motion hold modulo $s$. Thus, the theory
whose observables are defined by the equivalence classes of
$s$ (at zero ghost number), and whose bracket is defined by the
Peierls bracket may be viewed as a definition of
classical Yang-Mills theory.

\medskip

The BRST-transformation $s$ plays a crucial role also in the
perturbative quantum field theory associated with Yang-Mills theory,
where its role is among other things to derive certain consistency conditions on
the terms in the renormalized perturbation
series. We therefore discuss some of the relevant facts about the
BRST-transformation in some more detail. Since $s^2 = 0$, the BRST
transformation defines a ``differential'', or, more precisely, a differential
complex
\ben
s: \P_0 \to \P_1 \to \dots \to \P_N \to \dots
\een
where a subscript denotes the grading of the functionals
in $\P$ by the ghost number, defined by the ghost number
operator $\N_g$ counting the ghost number of an element in $\P$
by the formula
\ben\label{Ngdef}
\N_{g} = \sum \nabla^k C^I \frac{\partial}{\partial (\nabla^k C^I)}
               - \nabla^k \bar C^I \frac{\partial}{\partial (\nabla^k
                 \bar C^I)} \, .
\een
Thus, $\P$ is doubly graded space, by the form degree and ghost number, and we
write $\P^q_p$ for the subspace of elements with form degree $q$ and ghost number $p$.
We define the cohomology ring $H^p(s, \P^q)$ to be the set of
all local covariant $q$-form functionals $\O$ of ghost number $p$,
and $s \O = 0$, modulo the set of a $q$-form
functionals $\O = s\O'$ with ghost number $p$, i.e.,
\ben
H^p(s, \P^q(M)) =
\frac{\{ {\rm Kernel} \, s: \P^q_p \to \P^q_{p+1} \}}{\{ {\rm Image} \, s : \P^q_{p-1} \to \P^q_p \} } \, .
\een
The above lemma may be
viewed as the determination of the space $H^q(s, \P^p)$ for all
$q,p$. We will also encounter another cohomology ring, consisting of all
$s$-closed local covariant functionals modulo exact local covariant
functionals. To describe this ring more precisely, it is useful to
know the following result, sometimes called ``algebraic Poincare
Lemma'', or ``fundamental Lemma of the calculus of variations'':

\begin{lemma} (Algebraic Poincare lemma)\label{lemma2}
Let $\alpha=\alpha[\Phi, \Psi]$
be a $p$-form on an $n$-dimensional manifold $M$, which
is locally and covariantly constructed out of a number of
dynamical fields $\Phi$, and background fields $\Psi$.
Assume that $d\alpha[\Phi, \Psi] = 0$ for all $\Psi$, and that
each $\Psi$ is pathwise connected to a reference $\Psi_0$ for
which $\alpha[\Phi, \Psi_0] = 0$. Then $\alpha = d\beta$ for
some $\beta = \beta[\Phi, \Psi]$ which is locally constructed out
of the fields.
\end{lemma}

The proof is given for convenience in the next subsection. Consider
now a $\O_q \in \P^q$ such that $s \O_q = d \O_{q-1}$, i.e.,
$\O_q$ is $s$-closed modulo $d$. Then, by $s^2=0$ and $d s + s d = 0$,
the form $s \O_{q-1}$ is $d$-closed, and hence $d$-exact by the
fundamental lemma, so $s\O_{q-1} = d\O_{q-2}$. We can now repeat this
procedure until we have reached the forms of degree $0$, thereby
arriving at what is called a ``decent-equation'', or a ``ladder'':
\bena
s \O_q &=& d \O_{q-1} \\
s \O_{q-1} &=& d \O_{q-2} \\
&&\dots\\
s \O_1 &=& d \O_0\\
s \O_0 &=& 0 \, .
\eena
Note that, within each ladder, the form degree plus the ghost number
is constant. We denote the space of $\O_q$ that are $s$-closed modulo
$d$ at ghost number $p$, factored by elements that are $s$-exact modulo $d$ by
$H^p(s|d, \P^q) \equiv H^p(s, H^q(d, \P))$. In practice, ladders can
be used to determine the cohomology of $s$ modulo $d$.

For the purpose of perturbative quantum field theory, it will be convenient to
consider yet another cohomology ring related to $s$ that incorporates also
the equations of motion. Let us add to the theory a further set of
background fields (``BRST sources,'' or ``anti-fields''~\cite{Batalin1981})
$\Phi^\ddagger = (A_I^\ddagger,
C_I^\ddagger, \bar C_I^\ddagger, B_I^\ddagger)$
corresponding to the dynamical fields $\Phi=(A^I,C^I,\bar C^I,B^I)$:
\begin{table}
\label{table:anti-fields}
\begin{center}
\begin{tabular}{|c||cccc|}
\hline
$\Phi^\ddagger$& $A_I^\ddagger$ & $C_I^\ddagger$ & $\bar C_I^\ddagger$
& $B_I^\ddagger$\\
\hline
Dimension & 3 & 4 & 2 & 2\\
Ghost Number & $-1$ & $-2$ & 0 & $-1$\\
Form Degree & 3 & 4 & 4 & 4\\
Grassmann Parity & 1 & 0 & 0 & 1\\
Star Parity & $+$ & $-$ & $-$ & $+$ \\
\hline
\end{tabular}
\end{center}
\end{table}
Consider now the action
\ben\label{fullaction}
S[\Phi, \Phi^\ddagger]
= S_{ym} + S_{gf} + S_{gh} + S_{sc}, \quad S_{sc} = -\int_M s\Phi_i \wedge
\Phi^{\ddagger i}
\een
The new action is still BRST-closed, $sS=0$, because it
is given by the sum of $S_{tot}$ and a BRST-exact term, and it satisfies in
addition $(S,S)=0$, where the ``anti-bracket'' $( \, . \, , \, . \,)$
is defined by the equation~\cite{Batalin1981}
\ben\label{bvh}
(F_1, F_2) = \int_M \left[\frac{\delta_R F_1}{\delta \Phi_i(x)}
\wedge \frac{\delta_L F_2}{\delta \Phi^{\ddagger i}(x)} - 
%(-1)^{deg(\Phi_i^\ddagger)}
\frac{\delta_R F_1}{\delta \Phi^{\ddagger i}(x)}
\wedge \frac{\delta_L F_2}{\delta \Phi_i(x)} \right] \, .
\een
The local anti-bracket satisfies the graded Jacobi-identity
\ben\label{Jacobiid}
(-1)^{\epsilon_3 \epsilon_1}
((F_1, F_2), F_3)+
(-1)^{\epsilon_2 \epsilon_1}
((F_2, F_3), F_1)+
(-1)^{\epsilon_3 \epsilon_2}
((F_3, F_1), F_2)=0
\een
and as a consequence $(F,(F,F))=0$ for any $F$. The differential incorporating the
equations of motion is defined by
\ben
\hat s F = (S, F) \, .
\een
It satisfies $\hat s^2 = 0$ as a consequence of $(S,S)=0$ and the
Jacobi identity, as well as, $\hat s d + d \hat
s = 0$. It differs from the BRST-differential $s$ by the ``Koszul-Tate-differential'' $\sigma$
\ben
\hat s = s + \sigma \, ,
\een
where it can be checked explicitly that $-\sigma^2 = \sigma s + s \sigma$. It acts on the fields by
\ben
\sigma \Phi_i = 0, \quad \sigma \Phi^{\ddagger i} = \frac{\delta_R
  S}{\delta \Phi_i} \, .
\een
Thus, acting with $\sigma$ on a monomial in $\P$
containing an anti-field automatically gives an
expression containing a factor of the equations of motion, i.e.,
an on-shell quantity. This will be useful in the context of perturbative
quantum field theory in order to keep track of such terms.
Starting from the differential $\hat s$, one can again define
cohomology rings $H^p(\hat s, \P^q)$ and $H^p(\hat s|d, \P^q)$. The
ring $H^0(\hat s, \P^q)$ is still described by Lemma 2, because
one can prove in general that $H^p(s,\P^q)$ and $H^p(\hat s, \P^q)$
are isomorphic, see e.g.~\cite{Barnich2000}. The relative cohomology rings
$H^p(\hat s|d, \P^q)$ appear in the analysis of gauge invariance
in quantum Yang-Mills theory. They
are also known, but they
depend somewhat upon the choice of the gauge group $G$.
They are described by the following theorem, see e.g.~\cite{Barnich2000}:

\begin{thm}\label{thm2}
Let the Lie-group $G$ be a simple or semi-simple compact Lie group with no abelian factors,
and let $n=dim(M)$. Then each class in $H(\hat s|d, \P^n)$ is a linear
combination of expressions $\O$ of the form~\eqref{pqd}, and
representatives $\O'$ of the form
\ben\label{chsim}
\O' = \text{$n$-form part of} \quad
\prod_k r_{t_k}(R, \nabla R, \dots, \nabla^{n_k} R)
\prod_i q_{r_i}(C+A, F) \prod_j f_{s_j}(F) \, ,
\een
where $q_{r_i}(A+C,F)$ are the Chern-Simons forms,
\ben
q_r(A+C,F) = \int_0^1 {\rm Tr} \bigg(
(C+A)[tF + \lambda t(t-1) (C+A)^2]^{m(r)-1}
\bigg) \, dt \, .
\een
where $f_s$ are strictly gauge-invariant monomials of $F$ containing
only the curvature, $F$, but not its derivatives. The numbers $m(r)$
are the degrees of the independent Casimir elements of $G$, and the
trace is in some representation. The $r_t$
are taken to be a basis of closed forms $dr_t = 0$
that are analytic functions of the metric and
the covariant derivatives of the Riemann tensor.
For $p<n$, a basis of $H(\hat s|d, \P^p)$ is given by the
$\O'$ at form degree $p$,
together with all elements $\O$ of the form \eqref{pqd},
for any Lie-group $H=U(1)^l \times G$, with $G$ a semi-simple compact Lie group.
\end{thm}
{\bf Remarks}:
1) The statement of the theorem given in~\cite{Barnich2000} only
asserts that the $r_t$ are closed forms on $M$. To obtain that
the $r_t$ in fact have to be analytic functions
of $R, \nabla R, \nabla^2 R, \dots$, one has to use that, as we are assuming,
the elements in $\P^q$ are locally and covariantly constructed out of
the metric in the sense described above, with an analytic dependence
upon the spacetime metric. It then follows from the ``Thomas
replacement argument'' (see~\cite{Iyer1994} and lemma \ref{trt} below) that the $r_t$ have to be analytic functions
of the curvature tensor and its derivatives. It
furthermore follows that the $r_t$ may be chosen
to be characteristic classes
\ben
r_t = {\rm Tr}\Big(  R \wedge \cdots \wedge R \Big),
\een
where Tr is the trace in a representation of the Lie-algebra of $SO(n-1,1)$, and where
$R = T_{ab} R^{ab}_{\mu\nu} dx^\mu \wedge dx^\nu$ is the curvature 2-form
of the metric, identified with a 2-form valued in the Lie-algebra of $SO(n-1,1)$
via a tetrad field $e^a_\mu dx^\mu$.

2) There are more elements
in $H(\hat s|d, \P^n)$ when the group $G$ has abelian
factors, see e.g.~\cite{Henneaux1995} for a discussion.
In pure Yang-Mills theory, abelian factors decouple and hence
can be treated separately.

\medskip

In perturbation theory, we expand $S$ as
\ben
S = S_0 + \lambda S_1 + \lambda^2 S_2 \, ,
\een
and we correspondingly expand the Lagrangian as
\bena
\L_0 &=& \frac{1}{2} dA^I\wedge *\! dA_I - i d\bar C^I \wedge *\! dC_I
+ B^I(id *\! A_I + \frac{1}{2} *\! B_I)
+ s_0 A_{I} \wedge A^{\ddagger I} \non\\
&& + B_I \wedge \bar C^{\ddagger I}\, , \\
\L_1 &=&  \frac{1}{2} f_{IJK} *\! dA^I \wedge A^J \wedge A^K + f_{IJK}
\bar C^I \wedge A^J \wedge *\! dC^K \non\\
&& + s_1 A_I \wedge A^{\ddagger I}
   + s_1 C_I \wedge C^{\ddagger I}
   + s_1 \bar C_I \wedge \bar C^{\ddagger I}\\
\L_2 &=& \frac{1}{4} f^I{}_{JK} f_{ILM} A^J \wedge A^K *\! (A^L \wedge A^M)
\eena
in our choice of gauge~\eqref{lorgauge}.
We correspondingly have an expansion of the Slavnov Taylor differential as
$\hat s = \hat s_0 + \lambda \hat s_1 + \lambda^2 \hat s_2$,
and similarly of the Koszul Tate differential as
$\sigma = \sigma_0 + \lambda \sigma_1 + \lambda^2
\sigma_2$. The zeroth order parts of these expansions still define
differentials. The free Slavnov Taylor differential
$\hat s_0 \O = (S_0, \O)$, decomposed as
\ben\label{freeKT}
\hat s_0 = s_0 + \sigma_0
\een
will play an important role in
perturbative quantum field theory. Its action is given explicitly by
\ben\label{hs01}
\hat s_0 A^I = dC^I, \quad
\hat s_0 C^I = 0, \quad
\hat s_0 \bar C^I = B^I, \quad
\hat s_0 B^I = 0 \,
\een
on the fields, where it coincides with that of $s_0$.
Its action on the anti-fields is given by
\ben\label{hs02}
\hat s_0 A^\ddagger_I = \frac{\delta S_0}{\delta A^I}, \quad
\hat s_0 C^\ddagger_I = \frac{\delta S_0}{\delta C^I}, \quad
\hat s_0 \bar C^\ddagger_I = \frac{\delta S_0}{\delta \bar C^I}, \quad
\hat s_0 B^\ddagger_I = \frac{\delta S_0}{\delta B^I} \,
\een
where it coincides with that of $\sigma_0$. The actions of
$s_0$ and $\sigma_0$ are summarized in the following table:
\begin{center}
\begin{tabular}{|c||c|c|}
\hline
Field & $s_0$ & $\sigma_0$ \\
\hline
$A^I$ & $dC^I$ & 0\\
$B^I$ & 0 & 0\\
$C^I$ & 0 & 0\\
$\bar C^I$ & $B^I$ & 0\\
\hline
$A_I^\ddagger$ & 0 &$- d*dA_I-i*dB_I$\\
$B_I^\ddagger$ & 0 &$B_I -id*A_I+\bar C^\ddagger_I$\\
$C_I^\ddagger$ & 0 &$id*d\bar C_I - dA_I^\ddagger$\\
$\bar C_I^\ddagger$ & 0 & $i d*dC_I$\\
\hline
\end{tabular}
\end{center}
%%%%%%%%%%%%%%%%%%%
In perturbation theory, if $F = F_0 + \lambda F_1 +
\lambda^2 F_2 + \dots$, equations like $sF=0$ are understood in the
perturbative sense, as the hierarchy of identities obtained by
expanding the terms out in $\lambda$. This makes no difference with
regard to the above two cohomological lemmas, which now also have to be
interpreted in
the sense of formal power series (in fact, the proof of those lemmas
is in some sense perturbative). We finally mention a few identities
satisfied by the BRST-current $\J$ defined above that we will need
later. First, from the expression for the differential of the BRST
current, we have
\ben\label{sjofx}
d\J(x) = \sum_i (S, \Phi_i(x))(\Phi^{\ddagger i}(x), S) \, .
\een
Applying the differential $\widehat s=(S, \, . \,)$ and using the Jacobi identity
for the anti-bracket as well as $(S,S)=0$, we get
\ben
d \hat s \J = 0 \, ,
\een
so by lemma~\ref{lemma2}, we have the identity
\ben
\hat s \J = d\K \,  \ , 
\een
for the $(n-2)$-form $\K$.
The free BRST-current and its non-zero perturbations $\J_0, \J_1, \J_2, ...$ are given by
\bena\label{j0def}
\J_0 &=& * dA^I \wedge dC_I -iB^I * dC_I = \hat s_0(dA^I \wedge A_I
-i\bar C^I \wedge * dC_I)\, ,\\
\J_1 &=& if_{IJK}[
C^J A^K \wedge *  dA^I + \frac{i}{2} *d\bar C^I  C^J C^K + iB^I C^J * A^K -
\frac{1}{2} C^J C^K A^{\ddagger I} + \frac{1}{2} dC^I \wedge *(A^J \wedge A^K)
] \, , \non \\
\J_2 &=&  f^I_{JK}f_{ILM} *(A^J \wedge A^K) \wedge A^L C^M \, 
\non
\eena
and the non-zero $\K_0, \K_1, \K_2, ...$ are given by
\bena
\K_1&=& \frac{-i}{2} f_{IJK} dA^I C^J C^K \, \\
\K_2 &=&  f_{IJK} f^I{}_{MN} A^J \wedge A^K C^M C^N \, .
\eena
The forms $\L, \J, \K$ are in fact the elements of a ladder, 
\ben
\hat s \L = d \J \, , \quad 
\hat s \J = d \K \, ,  \quad 
\hat s \K = ... \ , 
\een
in the space $H^0(\hat s|d, {\bf P}^4)$. To show the first relation 
$\hat s \L = \d \J$, we collectively denote by $\psi = (\Phi, \Phi^\ddagger)$  the fields and
anti-fields, and  we recall that $\J[\psi] = \btheta[\psi, s\Phi]$, where
$\btheta$ was defined in eq.~\eqref{thetadef}. Also, $\hat s = s + \sigma$, so
\ben
\hat s \L[\psi] = \sigma \L[\psi] + \frac{\delta S[\psi]}{\delta \Phi} \wedge s\Phi + d\btheta[\psi, s\Phi]
= \sigma \L[\psi] + \sigma \Phi^\ddagger \wedge s\Phi + d\btheta[\psi, s\Phi]
= d\btheta[\psi, s\Phi] \, ,
\een
using in the last line that $\sigma \L = -\delta S/\delta \Phi \wedge s\Phi$. The
equation we have just derived may be expanded in powers of $\lambda$, leading for instance to the relations
\bena
\hat s_0 \L_0 &=& d\J_0  \, ,\\
\hat s_0 \L_1 + \hat s_1 \L_0 &=& d\J_1  \, , \\
\hat s_0 \L_2 + \hat s_1 \L_1 + \hat s_2 \L_0 &=& d\J_2 \ , 
\eena
and
\bena\label{jladder}
\hat s_0 \J_0 &=& d\K_0 \, , \\
\hat s_1 \J_0 + \hat s_0 \J_1 &=& d\K_1 \, , \\
\hat s_2 \J_0 + \hat s_1 \J_1 + \hat s_0 \J_2 &=& d\K_2 \ .
\eena

\subsection{Proof of the Algebraic Poincare Lemma, and the Thomas Replacement Theorem}

\begin{lemma} (Algebraic Poincare Lemma)\label{lemma2}
Let $\alpha=\alpha[\Phi, \Psi]$
be a $p$-form on an $n$-dimensional manifold $M$, which
is locally and covariantly constructed out of a number of
dynamical fields $\Phi$, and background fields $\Psi$.
Assume that $d\alpha[\Phi, \Psi] = 0$ for all $\Psi$, and that
each $\Psi$ is pathwise connected to a reference $\Psi_0$ for
which $\alpha[\Phi, \Psi_0] = 0$. Then $\alpha = d\beta$ for
some $\beta = \beta[\Phi, \Psi]$ which is locally constructed out
of the fields.
\end{lemma}

The algebraic Poincare lemma has been rediscovered many times,
and different proofs exist in the literature. Here we follow the proof
given in~\cite{Wald1990}, for other accounts see e.g.~\cite{Barnich2000}.

\noindent
{\em Proof:}
One first considers the case when
$\alpha[\Phi,\Psi]$ is linear in $\Psi$, i.e., of the form
\ben
\alpha_{\mu_1 \dots \mu_p} = \sum_{i=0}^k A^i{}_{\mu_1\dots\mu_p}{}^{\nu_1
  \dots \nu_i}(\Phi) \nabla_{(\nu_1} \cdots \nabla_{\nu_i)}\Psi \,,
\een
where we may assume that $A^i$ is totally symmetric in the upper
indices, and totally anti-symmetric in the lower indices.
The condition that $d\alpha = 0$ implies the condition
\ben
A^k{}_{[\mu_1 \dots \mu_p}{}^{\nu_1 \dots \nu_k}
\delta_{\gamma]}{}^\delta \nabla_{(\delta} \nabla_{\nu_1} \dots
\nabla_{\nu_k)} \Psi = 0 \, .
\een
At each $x \in M$, $\nabla_{(\nu_1} \dots
\nabla_{\nu_k)} \Psi |_x$ can be chosen to be an arbitrary
totally symmetric tensor, so we must have
\ben
A^k{}_{[\mu_1 \dots \mu_p}{}^{(\nu_1 \dots \nu_k}
\delta_{\gamma]}{}^{\delta)} = 0 \, .
\een
Contracting over $\delta, \gamma$ and using the symmetries of $A^k$,
one finds
\bena
&&\bigg[
\frac{n}{(k+1)(p+1)} + \frac{k}{(k+1)(p+1)} - \frac{p}{(k+1)(p+1)}
\bigg] A^k{}_{\mu_1 \dots \mu_p}{}^{\nu_1 \dots \nu_k} -\\
&&-\frac{kp}{(k+1)(p+1)} A^k{}_{\gamma[\mu_2 \dots
  \mu_p}{}^{\gamma(\nu_2 \dots \nu_k} \delta_{\mu_1]}{}^{\nu_1)} = 0 \non
\eena
and therefore that
\ben\label{71}
A^k{}_{\mu_1 \dots \mu_p}{}^{\nu_1 \dots \nu_k} =
\frac{kp}{(k+1)(p+1)} A^k{}_{\gamma[\mu_2 \dots
  \mu_p}{}^{\gamma(\nu_2 \dots \nu_k} \delta_{\mu_1]}{}^{\nu_1)} \, .
\een
For $k=0$, this condition simply reduces to $A^0=0$ and hence
$\alpha=0$, thus proving that the lemma is trivially fulfilled when $k=0$ and
when $\alpha$ depends linearly on $\Psi$.
For $k>0$, one may proceed inductively. Thus, assume that the
statement has been shown for all $k\le m-1$. Define
\ben
\tau_{\mu_2 \dots \mu_m}=
 \frac{mp}{(m+1)(p+1)} A^m{}_{\gamma[\mu_2 \dots
  \mu_p]}{}^{\gamma\nu_2 \dots \nu_m} \nabla_{(\nu_2} \cdots
\nabla_{\nu_m)} \Psi \, ,
\een
and let
\ben
\alpha' = \alpha - d\tau \, .
\een
Then $\alpha'$ is still closed and locally constructed from
$\Phi,\Psi$, linear in $\Psi$, but by~\eqref{71},
it only contains terms with
a maximum number $m-1$ of derivatives on $\Phi$. For such
$\alpha'$, we inductively know that $\alpha'=d\gamma$ for a
locally constructed $\gamma$. Thus, $\alpha = d(\gamma+\tau)$,
thereby closing the induction loop. Thus, we have proved the lemma
when $\alpha$ depends linearly upon $\Psi$.

Consider now the case when $\alpha[\Phi,\Psi]$ is non-linear in
$\Psi$. Let $\tau \mapsto \Psi_\tau$ be a smooth path in field space
with $\Psi_0=\Psi$.
Putting $\frac{d}{d\tau} \Psi |_{\tau=0} = \delta
\Psi$, we have
\ben
d \bigg\{ \frac{d}{d\tau} \alpha[\Phi, \Psi_\tau] \Big|_{\tau=0} \bigg\}=
d \bigg\{ \sum_{i=1}^k
\frac{\partial \alpha[\Phi, \Psi]}{\partial
    (\nabla_{(\mu_1} \dots  \nabla_{\mu_i)} \Psi)} \,
\, \nabla_{(\mu_1} \dots  \nabla_{\mu_i)}
\delta \Psi \bigg\}= 0 \, .
\een
Since this must hold for all paths, the identity holds for all
$\Phi, \Psi, \delta \Psi$. Thus, since this expression is linear
in $\delta \Psi$ and must hold for all
$\delta \Psi$, we can find a $\gamma$ such that
\ben
\frac{d}{d\tau} \alpha[\Phi, \Psi_\tau] \Bigg|_{\tau=0}
= d\gamma[\Phi, \Psi, \delta \Psi] \, .
\een
where $\gamma$ is constructed locally out of the fields. Thus, for any
path in field space, we have
\ben
\alpha[\Phi, \Psi_\tau] = \alpha[\Phi, \Psi_0]+ d  \bigg\{ \int_0^\tau
\gamma\bigg[\Phi, \Psi_t, \frac{d}{dt} \Psi_t \bigg] \, dt \bigg\}
\, .
\een
Consequently, for any field configuration $\Psi$ that can be reached by
a differentiable path from a reference configuration $\Psi_0$
for which $\alpha(\Phi, \Psi_0)$, we can write $\alpha(\Phi, \Psi) =
d\beta(\Phi, \Psi)$. \qed

\medskip
\noindent

We next give the precise statement and proof of the Thomas replacement theorem in the
case that we have a background gauge field, a metric background field, other non-specified background fields and some dynamical fields.
Thus, we consider spacetime manifolds $(M,g)$, and $G$ principal fibre bundles $B \to M$
over $M$ with an arbitrary but fixed structure group $G$. On $B$, we consider
gauge connections $\bar \nabla$ which in applications would be the ``background gauge connection''. As above, if we have any section $k$ in
$(T^*M)^{\otimes m} \otimes (TM)^{\otimes n}$ times $B \times_G V$, where
$V$ is a vector space with an action of $G$, then we let $\nabla$ act on
the "tensor part" of $k$ by the Levi-Civita connection $\nabla$ of $g$, and on the
"fibre bundle part" by $\bar \nabla$. We denote by $j_x^p(g, \bar \nabla, \Phi)$ the $p$-jet
of the metric, the gauge connection some other fields $\Phi$ which are sections in suitable 
associated bundles $B \times_G V$. 

We would like to investigate the possible dependence of functionals of these fields on some other, perhaps, implicit 
background structure. In this section, we mean by this e.g. an implicitly chosen, fixed choice of coordinates 
(i.e. a set of suitably independent functions $x^\mu$ on $M$) or an implicitly chosen gauge (i.e. a 
global section $s$ of $B$). We denote by $\Psi = (x^\mu, s)$ this perhaps implicitly used information. Unlike in the 
rest of this paper $\Psi$ does not stand for anti-fields etc. 

The functionals we consider are thus of the type
\ben
\O(x) = \O[j^p_x(g, \bar \nabla, \Psi, \Phi)] \, .
\een
Let $\psi: B \to B$ be a bundle
morphism, i.e., a diffeomorphism of $B$ which is compatible with the
$G$-action on $B$ in the sense that $\gamma\psi(y) = \psi(\gamma y)$, where $\gamma \in G$.
Let $\psi^* g, \psi^* \bar \nabla, \psi^* \Phi$ be the pulled-back/gauge transformed metric/connection/fields.
We say that $\O$ depends locally and covariantly upon the metric, connection, and fields if
we have
\ben\label{Thomascond}
\psi^* \O[j^p(g, \bar \nabla, \Psi, \Phi)] = \O[j^p(\psi^* g, \psi^* \bar \nabla, \psi^* \Phi, \Psi)] \, ,
\een
where we note that $\psi^*$ does {\em not} act on the background structure $\Psi$ on the right side.
This equation is to hold for {\em all} $g, \nabla, \Phi$, and {\em some} choice of the background structure $\Psi=(x^\mu, s)$.

If the bundle is trivial, $B = M \times G$, then
the above condition can be stated somewhat more explicitly as follows. We may identify
the $p$-jet of the background structure $\Psi$ with a collection of
tensor fields on $M$, which
we again denote by $\Psi$ for simplicity. Let us introduce an
arbitrary {\em background} derivative operator $\partial$ (e.g. a coordinate derivative operator if a global coordinate system exists, as the notation suggests), 
and consider first the case when $\psi$ is a "pure diffeomorphism", i.e., $\psi = f \times id_G$, with $f$ a diffeomorphism
of $M$.
Let us decompose $\bar \nabla$ as $\bar \nabla = d + \bar A$, with $\bar A$ a Lie-algebra valued 1-form on $M$.
Then the above condition can be written as
\bena\label{thomasc1}
%\begin{split}
&f^* \O[g, \dots, \partial^p g, \bar A, \dots, \partial^p \bar A, \Phi, \dots, \partial^p \Phi, \Psi] \non \\
&=
\O[ f^* g, \dots, \partial^p f^* g, f^* \bar A, \dots, \partial^p f^* \bar A, f^* \Phi, \dots, \partial^p f^*\Phi, \Psi] \, ,
%\end{split}
\eena
where as usual we denote by $\partial^k = dx^{\mu_1} \otimes \dots \otimes dx^{\mu_k} \partial_{(\mu_1} \dots \partial_{\mu_k)}$
the symmetrized k-fold derivative. Note that, in the above expression, $f^*$ does not act on
any of the background fields $\Psi$, nor on $\partial$. Secondly, let $\psi$ be a "pure gauge transformation",
i.e., a transformation of the form $\psi = id_M \times \gamma$, where $\gamma: M \to G$ is a local gauge
transformation. Let $\bar A^\gamma = \gamma^{-1} \bar A \gamma + \, \gamma^{-1} d\gamma$, and let 
$\Phi^\gamma = {\rm R}(\gamma) \Phi$, where $\rm R$ is some finite dimensional representation of $G$ (so that 
$\Phi$ is a section of $B \times_G V$, with $V$ the representation space of $\rm R$). Then
the above condition~\eqref{Thomascond} becomes
\ben\label{thomasc2}
\O[g, \dots, \partial^p g, \bar A, \dots, \partial^p \bar A, \Phi, \dots, \partial^p \Phi, \Psi] =
\O[g, \dots, \partial^p g, \bar A^\gamma, \dots, \partial^p \bar A^\gamma, \Phi^\gamma, \dots, \partial^p \Phi^\gamma, \Psi] \, .
\een

\begin{lemma}\label{trt}
(Thomas Replacement Theorem) If $\O$ is a functional satisfying eq.~\eqref{Thomascond}
[or equivalently eqs.~\eqref{thomasc1} and~\eqref{thomasc2} when $B=M \times G$],
then it can be written as
\ben
\O(x) = \O[g(x), R(x), \dots, \nabla^{p-2} R(x), \bar f(x), \dots, \bar \nabla^{p-2} \bar f(x), \Phi, \dots, \bar \nabla^p \Phi] \, ,
\een
where $\bar f$ is the curvature of $\bar \nabla = \nabla + \bar A$, $\nabla$ is the Levi-Civita connection of $g$ and $R$ is the Riemann tensor of $g$. 
In particular, there cannot be any dependence upon the background fields $\Psi$.
\end{lemma}

{\bf Remark:} In the following sections, we will use this result with $\bar \nabla$ equal to the standard flat connection on the trivial bundle $M \times G$, 
so in this case $\bar f=0$. The general case is relevant if we want to generalize our constructions to arbitrary background gauge fields, as recently done by 
\cite{zahn}. 

\medskip
\noindent
{\em Proof:} The proof follows \cite{Iyer1994}, with a slight generalization due to
the presence of gauge fields ($\bar A$) that were not considered in that reference. For simplicity, 
we assume that the fields $\Phi$ are absent; the general case is dealt with in a very similar manner.
We first consider the case $B = M \times G$. Then our covariance condition
implies the conditions
\ben\label{poundsO}
\pounds_\xi \O = \sum_{k=0}^p \frac{\partial \O}{\partial(\partial^k g)} \partial^k \pounds_\xi g
+ \sum_{k=0}^p \frac{\partial \O}{\partial(\partial^k \bar A)} \partial^k \pounds_\xi \bar A
\een
for any vector field $\xi$ on $M$, and
\ben
0 = \sum_{k=0}^p \frac{\partial \O}{\partial(\partial^k \bar A)} \partial^k \bar \nabla h
\een
for any Lie-algebra valued function $h$ on $M$. We first analyze the first of
these conditions, following~\cite{Iyer1994}. First, we rewrite all
$\partial$-derivatives of $A$ in terms of $\nabla$-derivatives (where $\nabla$ is the
Levi-Civita connection of $g$ and not $\bar \nabla = \nabla + \bar A$), plus additional terms involving
$\partial$-derivatives of $g$. Thus, we write
\ben
\O(x) = \O[g(x), \dots, \partial^p g(x), \bar A(x), \dots, \nabla^p \bar A(x), \Psi(x)] \, .
\een
Next, we eliminate $\partial^k g$ in favor of $C$ and its $\partial$-derivatives, where
$C$ is the tensor field defined by
\ben
C^\mu{}_{\nu \sigma} = -\frac{1}{2} g^{\mu\alpha}(\partial_\alpha g_{\nu\sigma} - 2\partial_{(\nu} g_{\sigma)\alpha}) \, .
\een
We thereby obtain
\ben
\O(x) = \O[g(x), C(x), \dots, \partial^{p-1} C(x), \bar A(x), \dots, \nabla^p \bar A(x), \Psi(x)] \, .
\een
Next, we observe that the symmetrized derivatives of $C$ can be rewritten as\footnote{If $\partial$ has no curvature, as the notation suggests, 
then the symmetrization is superfluous.}
\bena
\partial_{(\alpha_1} \cdots \partial_{\alpha_l)} C^{\mu}{}_{\gamma\delta} &=& \partial_{(\alpha_1} \cdots \partial_{\alpha_l} C^{\mu}{}_{\gamma\delta)} \non\\
&+& \frac{l+3}{4(l+1)(l+2)} \sum_{i} \nabla_{(\alpha_1} \cdots \widehat \nabla_{\alpha_i} \cdots \nabla_{\alpha_l} (R^\mu{}_{\gamma\alpha_i \delta} +
R^\mu{}_{\delta \alpha_i \gamma}) \non\\
&+&\frac{3l+4}{8(l+1)(l+2)} \sum_{i \neq j} \bigg( \nabla_{(\gamma} \nabla_{\alpha_1} \cdots \widehat \nabla_{\alpha_i}
\widehat \nabla_{\alpha_j} \cdots \nabla_{\alpha_l)} R^\mu{}_{\alpha_i \delta \alpha_j} \non\\
&+& \nabla_{(\delta} \nabla_{\alpha_1} \cdots \widehat \nabla_{\alpha_i}
\widehat \nabla_{\alpha_j} \cdots \nabla_{\alpha_l)} R^\mu{}_{\alpha_i \gamma \alpha_j}
\bigg) \non\\
&+& \text{terms with less than $l$ derivatives on $C$.}
\eena
By iterating this substitutions, we can achieve that all derivatives of $C^\mu{}_{\nu\sigma}$
in $\O$ only appear in totally symmetrized form $\partial_{(\alpha_1} \dots \partial_{\alpha_l} C^\mu{}_{\nu\sigma)}$,
at the expense of possibly having an additional dependence upon the curvature tensor $R^\mu{}_{\alpha\beta\gamma}$
of the metric and its covariant derivatives. In other words, we may assume that $\O$ is given as
\bena
\O &=& \O\bigg[
g_{\mu\nu}, C^\mu{}_{\nu\sigma}, \dots, \partial_{(\alpha_1} \dots \partial_{\alpha_{p-1}} C^\mu{}_{\nu\sigma)}, \non\\
&& R^\mu{}_{\nu\sigma\rho}, \dots, \nabla_{(\alpha_1} \dots \nabla_{\alpha_{p-2})}
R^\mu{}_{\nu\sigma\rho}, \bar A_\mu, \dots,  \nabla_{(\alpha_1} \dots \nabla_{\alpha_{p})} \bar A_\mu; \Phi \bigg] \, .
\eena
We now apply the condition~\eqref{poundsO} to this expression. We find
\bena\label{derivativered}
&&\sum_{k=0}^{p-1} \frac{\partial \O}{\partial(\partial_{(\alpha_1} \dots \partial_{\alpha_{k}} C^\mu{}_{\nu\sigma)})}
\pounds_\xi \partial_{(\alpha_1} \dots \partial_{\alpha_{k}} C^\mu{}_{\nu\sigma)} +
\frac{\partial \O}{\partial \Psi} \pounds_\xi \Psi \non\\
&=&
\sum_{k=0}^{p-1} \frac{\partial \O}{\partial(\partial_{(\alpha_1} \dots \partial_{\alpha_{k}} C^\mu{}_{\nu\sigma)})}
\partial_{(\alpha_1} \dots \partial_{\alpha_{k}} \delta C^\mu{}_{\nu\sigma)}
\eena
where $\delta C^\mu{}_{\nu\sigma}$ is the variation arising from the variation $\delta g_\mu = \pounds_\xi g_{\mu\nu}
= 2 \nabla_{(\mu} \xi_{\nu)}$ under an infinitesimal diffeomorphism,
\ben
\delta C^\alpha{}_{\beta\gamma} = g^{\alpha\delta} (\partial_{(\beta} \partial_{\gamma)} \xi_{\delta} - {\mathcal R}_{\delta(\beta\gamma)\rho} \xi^\rho) - 2\partial^{(\alpha} \xi^{\delta)} g_{\delta \rho} C^\rho{}_{\beta\gamma} \, ,
\een
with ${\mathcal R}_{\mu\nu\sigma\rho}$ the curvature of $\partial$ (if any).
The terms in the above equation arising from an infinitesimal variation of $g_{\mu\nu},A_\mu, R_{\mu\nu\sigma\rho}$
and their $\nabla$-derivatives cancel out. The key point about the above equation is now that, on the left side, there
appears no more than one derivative of $\xi^\mu$, while on the right side there can appear as many as
$p+1$ symmetrized derivatives of $\xi^\mu$. Since the symmetrized derivatives of $\xi^\mu$ can be chosen
independently at each given point $x$ in $M$, it follows that a necessary condition for eq.~\eqref{derivativered}
to hold is that
\ben
\frac{\partial \O}{\partial(\partial_{(\alpha_1} \dots \partial_{\alpha_{k}} C^\mu{}_{\nu\sigma)})} = 0
\een
for $k=0, \dots, p-1$. Thus, our expression for $\O$ must have the form
\ben
\O = \O\bigg[
g_{\mu\nu}, R^\mu{}_{\nu\sigma\rho}, \dots, \nabla_{(\alpha_1} \dots \nabla_{\alpha_{p-2})}
R^\mu{}_{\nu\sigma\rho}, \bar A_\mu, \dots,  \nabla_{(\alpha_1} \dots \nabla_{\alpha_{p})} \bar A_\mu; \Psi \bigg] \, .
\een
We also get the condition that $\partial \O/\partial \Psi \cdot \pounds_\xi \Psi = 0$. If $\Psi$ only
consists of scalar fields, then it follows immediately that $\O$ cannot have any dependence on
$\Psi$. If $\Psi$ contains tensor fields, then we may reduce this to the situation of only scalar fields
by picking a coordinate system, and by treating the coordinate components of $\Psi$ as scalars.

We finally use the condition~\label{gaugecondition} to show that the $\bar A$-dependence of
$\O$ can only be through the field strength tensor $\bar f$ and its covariant derivatives $\bar \nabla^p \bar f$. To show this, we
rewrite
\bena
\nabla_{(\alpha_1} \dots \nabla_{\alpha_{l})} \bar A_\mu &=&
\nabla_{(\alpha_1} \dots \nabla_{\alpha_{l}} \bar A_{\mu)} + \frac{l}{l+1}
\bar \nabla_{(\alpha_1} \dots \bar \nabla_{\alpha_{l-1}} \bar f_{\alpha_l)\mu}\\
&+& \text{terms with no more than $l-1$ derivatives of $\bar A$.} \non
\eena
By repeatedly substituting this relation into $\O$, we can rewrite it as
\bena
\O &=& \O\bigg[
g_{\mu\nu}, \bar f_{\mu\nu}, \dots, \bar \nabla_{(\alpha_1} \dots \bar \nabla_{\alpha_{p-1})} \bar f_{\mu\nu}, \non\\
&& R^\mu{}_{\nu\sigma\rho}, \dots, \nabla_{(\alpha_1} \dots \nabla_{\alpha_{p-2})}
R^\mu{}_{\nu\sigma\rho}, \bar A_\mu, \dots,  \nabla_{(\alpha_1} \dots \nabla_{\alpha_{p}} \bar A_{\mu)} \bigg] \, .
\eena
We now substitute this into the (infinitesimal version) of our condition~\eqref{gaugecondition}, to get
\bena\label{derivativered}
0&=&\sum_{k=0}^{p} \frac{\partial \O}{\partial(\nabla_{(\alpha_1} \dots \nabla_{\alpha_{k}} \bar A_{\mu)})}
\nabla_{(\alpha_1} \dots \nabla_{\alpha_{k}} \bar \nabla_{\mu)} h \non\\
&+&
\sum_{k=0}^{p-2} \frac{\partial \O}{\partial(\bar \nabla_{(\alpha_1} \dots \bar \nabla_{\alpha_{k})} \bar f_{\mu\nu})}
[h,\bar \nabla_{(\alpha_1} \dots \bar \nabla_{\alpha_{k})} \bar f_{\mu\nu}] \, ,
\eena
for all Lie-algebra valued functions $h$. Note that, in the second sum, we have no derivatives of $h$,
while in the first sum we have at least one symmetrized derivative of $h$. Since the symmetrized
derivatives of $h$ are independent at each point, the above equation can only hold if
\ben
\frac{\partial \O}{\partial(\nabla_{(\alpha_1} \dots \nabla_{\alpha_{k}} \bar A_{\mu)})} = 0
\een
for all $k$. This proves the Thomas replacement theorem in the case when $B=M \times G$. But, since
it is a local statement and any principal fibre bundle is locally trivial, it must in fact hold
for any principal fibre bundle. \qed

\section{Quantized field theories on curved spacetime:
         Renormalization}

\subsection{Definition of the free field algebra $\W$ for scalar field theory}

Consider a classical scalar field $\phi$ described by the
quadratic Lagrangian
\ben\label{L0}
\L_0 = \frac{1}{2}(d\phi \wedge * d\phi - m^2 * \!\phi^2) \, \, .
\een
The quantity $m^2$ is a real parameter (we do not assume $m^2 \ge 0$).
In this section, we explain how to quantize such a theory in curved
spacetime, and how to define Wick powers and time-ordered products of
$\phi$ at the quantum level. We assume only that $(M,g)$ is globally
hyperbolic and we assume for the rest of the paper that the
spacetime dimension is 4. We do not
assume that $(M,g)$ has any symmetries. As discussed above, if
$(M,g)$ is globally hyperbolic, then the Klein-Gordon equation has
a well-posed initial value formulation and unique retarded and advanced
propagators $\Delta_R$ and $\Delta_A$. A fundamental object in the
quantization of $\phi$ is the commutator function,
\ben
\Delta = \Delta_A-\Delta_R
\een
which is antisymmetric, $\Delta(x,y)=-\Delta(y,x)$. We want to define
a non-commutative product $\st$ between classical field observables such that
\ben\label{com}
\phi(x) \st \phi(y) - \phi(y) \st \phi(x) = i\hbar\Delta(x,y) \myid \, .
\een
This formula is motivated by the fact that, as $\hbar \to 0$, we would
like the above commutator divided by $i\hbar$ to go to the classical
Peierls bracket. The classical Peierls bracket for a linear scalar
field with Lagrangian $\L_0$, however, is given by $\{\phi(x), \phi(y)\}_{\rm P.B.} =
\Delta(x,y)$, see e.g.~\cite{Duetsch1999}.

To define the desired ``deformation quantization'', we proceed as follows.
We first consider the free *-algebra generated by the expressions
$\phi(f)$, where $f$ is any smooth compactly supported testfunction,
to be thought of informally as the integral expressions
$\int \phi(x) f(x) \, dx$. We now simply factor this free algebra by the
relation~\eqref{com}. This defines the desired deformation
quantization algebra $\Wn$. Evidently, the construction of $\Wn$
only depends upon the spacetime $(M,g)$ and its orientations, because
these data uniquely determine the retarded and advanced propagators.

The algebra $\Wn$ by itself is too small to serve as an arena for
renormalized perturbation theory. It does not, for example, even
contain the Wick-powers of the free field, or other quantized
composite fields, which are a
minimal input to even define interactions at the quantum level. More generally, to do
perturbation theory we need an algebra that also contains the time-ordered products
of composite fields, and these are, of course, not
contained in $\Wn$ either. Thus, our first task is to define an
algebra that is sufficiently big to contain such quantities.
The key input in the construction of such an algebra is
an arbitrary, but fixed 2-point function $\omega(x,y)$ on $M \times M$
of ``Hadamard type'' which serves to define a
suitable completion of $\Wn$. This is by definition
a distribution on $M \times M$ which is
(a) a bisolution to the equations of motion, that is,
\ben\label{bisol}
(\square -m^2)_x \omega(x,y) = (\square-m^2)_y \omega(x,y) = 0\, ,
\een
which (b) satisfies
\ben
\omega(x,y) - \omega(y,x) = i\Delta(x,y)
\een
and which (c) has a wave front set~\cite{Hormander}
of ``Hadamard type''~\cite{Radzikowski1996a}
\bena\label{wfs2}
\WF(\omega) &=& \{(x_1, k_1, x_2, k_2) \in T^* M \times T^*M ; \non \\
                   &&\text{$x_1$ and $x_2$ can be joined by
                     null-geodesic $\gamma$} \non\\
                   &&\text{$k_1 = \dot \gamma(0)$ and $k_2 = -\dot
                     \gamma(1)$, and $k_1 \in \bar V^+$} \} \, .
\eena
The wave front set completely characterizes the singularity structure of
$\omega$, and its definition and properties are recalled in appendix~C.
It can be shown that, on any globally hyperbolic spacetime $(M,g)$,
there exist infinitely many distributions $\omega$ of Hadamard
type~\cite{Junker2003,FNW,Koehler}.
Using $\omega$, we now define the following set of generators of $\Wn$, where
$u=f_1\otimes\dots\otimes f_n$:
\bena\label{generators}
F(u)&=&\int_M \dots \int_M f_1(x_1) \cdots f_n(x_n) :\phi(x_1) \cdots
\phi(x_n):_\omega \, dx_1 \dots dx_n \non\\
&=& \frac{d^n}{i^n d\tau_1 \dots d\tau_n}
{\rm exp}_{\st} \bigg(i \sum_j \tau_j \phi(f_j) + \frac{\hbar}{2}
\sum_{i,j} \tau_i \tau_j \omega(f_i, f_j) \bigg) \Bigg|_{\tau_i=0} \, .
\eena
The commutator property of $\omega$ implies that the
quantities $:\phi(x_1) \dots \phi(x_n):_\omega$ are symmetric
in its arguments. In fact, these quantities are nothing but the
``normal ordered field products'' (with respect to $\omega$), but we
note that we do not think of these objects as operators defined on a
Hilbert space as is usually done when introducing normal ordered expressions.

So far, we have done nothing but to introduce a new set of expressions
in $\Wn$ that generate this algebra. We can express the product
between to elements $F(u),F(v)$ of the form~\eqref{generators}
as
\ben\label{star}
F(u) \st F(v)
= \sum_k \hbar^k F(u \otimes_k v)
\een
where $u \otimes_k v$ is the $k$-times contracted tensor product of
distributions $u,v$ in $n$ resp. $m$ spacetime variables. It
is defined by
\begin{multline}\label{contr}
(u \otimes_k v)(x_1, \dots, x_{n+m-2k}) = \\
\frac{n!m!}{k!} \sum_\pi \int u(x_{\pi(1)}, \dots, y_1, \dots)
v(x_{\pi(n-k+1)}, \dots, y_{k+1}, \dots) \prod_{i=1}^k
\omega(y_i,y_{k+i}) dy_1 \dots dy_{2k} \, ,
\end{multline}
where the sum is over all permutations of $n+m-2k$ elements.
A somewhat more symbolic, but more compact and suggestive
way to write the product is
\ben\label{starproduct}
F(u) \st F(v) =
: F(u) \, {\rm exp} \bigg(  \hbar \, {}_< {\mathcal D}_>
\bigg) F(v) :_\omega
\een
where ${}_< {\mathcal D}_>$
is the bi-differential operator defined by
\ben\label{derivative}
{}_< {\mathcal D}_> =
\int
\frac{\delta_L}{\delta \phi(x)}
\omega(x, y)
\frac{\delta_R}{\delta \phi(y)}  \, dxdy \, .
\een
The superscripts on the functional derivatives indicate that the
first derivative acts to the left (as a left derivative in case we have a theory with anti-commuting fields), and the second one to the
right factor in a tensor product (as a right derivative). These functional derivatives
are to be understood to act on an expression like
$:\phi(x_1) \dots \phi(x_n):_\omega$ a classical
product of classical fields in $\P(M)$. The point is now that the
product can still be defined on a much larger class of expressions.
These expressions are of the form
\ben\label{generators1}
F(u) = \int u(x_1, \dots, x_n) :\phi(x_1) \cdots
\phi(x_n):_\omega \, dx_1 \dots dx_n
\quad (n \ge 1) \, ,
\een
where $u$ is now a {\em distribution} on $M^n$, rather than
the product of $n$ smooth functions on $M$ as above in
eq.~\eqref{generators}. To make the product well defined,
we only need to impose a mild wave-front set condition on the $u$~\cite{Duetsch1999}:
\ben\label{wfs1}
\WF(u) \cap \bigcup_{x\in M}
[(\bar V^+_x)^{\times n} \cup
(\bar V^-_x)^{\times n}]
 = \emptyset
\, ,
\een
with $\bar V^\pm_x$ denoting the closure of the future/past lightcone at $x$.
The reason for imposing this condition is that it ensures, together
with~\eqref{wfs2}, that the distributional products in the
contracted tensor products that arise when
carrying out the product $F \st G$ of two expressions of the type
\eqref{generators1} make sense. The point is that in such a product, there
appear distributional products of $u,v,\omega$ in the contracted tensor product
of $u,v$, see eq.~\eqref{contr}. Normally, the product of distributions does not
make sense, but due to our wave front set conditions on $u,v,\omega$, the relevant
products exist due to the fact that vectors in the wave front set
of $\omega,u,v$ can never add up to 0, see appendix C for details.
We define the desired enlarged algebra, $\W$,
to be the algebra generated by~\eqref{generators1}, with the product $\st$.
It can be viewed in a certain sense as the closure of $\Wn$, because
the distributions $u$ in eq.~\eqref{generators1} can be approximated, to
arbitrarily good precision by sums of smooth functions of the form
$f_1 \otimes \dots \otimes f_n$ as in~\eqref{generators} (in the H\"ormander
topology~\cite{Hormander}). The algebra $\W$ will turn out to be big
enough to serve as an arena for perturbation theory.
For example, it can be seen immediately that $\W$ contains normal
ordered Wick-powers of $\phi(x)$: Namely, since the wave-front set of
the delta-distribution on $M^n$ is
\ben
\WF(\delta) = \{(x,k_1, \dots, x, k_n); \quad x \in M, k_i \in T^*_x
M, \sum k_i = 0\}
\een
it follows that $u(y,x_1, \dots, x_n) = f(y) \delta(y,x_1, \dots, x_n)$
satisfies the wave front condition~\eqref{wfs1}. The corresponding
generator $F$ as in~\eqref{generators1} may be viewed as the normal ordered
Wick power \\
$:\phi^n(x):_\omega$, smeared with $f(x)$.

As it stands, the Klein-Gordon equation
is not implemented in the algebra $(\W, \st)$. This
could easily be incorporated by factoring $\W$ by an appropriate
ideal (i.e., a linear subspace that is stable under $\st$-multiplication
by any $F \in \W$). The ideal for the field equation is simply the
linear space
\bena\label{ideal}
\I &=& \Bigg\{ F=\int u(x_1, \dots, x_n) :\phi(x_1) \cdots \frac{\delta
S_0}{\delta \phi(x_i)} \cdots \phi(x_n):_\omega  \, dx_1 \dots dx_n ,\non\\
  &&\text{for some $u$ of compact support, $\WF(u) \cap \bigcup_{x\in M}
[(\bar V^+_x)^{\times n} \cup
(\bar V^-_x)^{\times n}]
 = \emptyset$ } \Bigg\}
\eena
of generators containing a factor of the wave equation. This space is
stable under the adjoint operation and $\st$-products with any $F \in \W$ by eq.~\eqref{bisol} and so
indeed an ideal. If we consider the factor algebra
\ben
pr: \W \to \F_0 = \W/\I \, ,
\een
then within $\F_0$,  the field equation $(\square -m^2) \phi(x)=0$ holds.
The factor algebra $\F_0$ is the algebra of physical interest for
free field theory. For physical applications, one is interested in
representations of $\F_0$ as operators on a Hilbert space, ${\mathcal H}_0$, and
in $n$-point functions of observables in $\F_0$ in physical states.
However, in the context of perturbation theory, it will be much
more useful to work with the algebra $\W$ at intermediate stages.

To make physical predictions, one finally needs to represent
the algebra of observables $\F_0$ as linear operators with a dense,
invariant domain on a Hilbert space ${\mathcal H}_0$. A vector state
$|\Psi\rangle$ in ${\mathcal H}_0$ is said to be of Hadamard form
if its $n$-point functions
\ben
G_n^\Psi(x_1, \dots, x_n) = \langle \Psi | \pi_0(\phi(x_1)) \dots
\pi_0(\phi(x_n)) |\Psi \rangle
\een
are of "Hadamard form". By this
one means that the 2-point function has a wave front set of
Hadamard form~\eqref{wfs2}, and that its truncated $n$-point
functions\footnote{The truncated $n$-point functions of
a hierarchy of $n$-point distributions $\{h_n\}$ are defined by
the generating functional $h^c(\e^f) = \log h(\e^f)$, where
$h(\e^f) = \sum_n h_n(f,f, \dots, f)/n!$.}
are smooth for
$n \neq 2$. A Hadamard representation is
a representation containing a dense, invariant domain of Hadamard states.
Hadamard representations may be constructed on any
globally hyperbolic spacetime as one may show using the
deformation argument of~\cite{Fulling,Koehler} (or the
construction of~\cite{Junker2003}, and combining these with those of~\cite{Ruan}).
We describe the deformation construction below in sec.~4.2 in the context of
gauge theories.

It is clear that, since $\W(M,g)$ was obtained
as the completion of the algebra $\Wn(M,g)$, also
$\W(M,g)$ depends locally and covariantly
upon the metric. Because this fact will be of key importance when we
formulate the local and covariance condition of renormalized
time-ordered products, we now explain more formally what exactly we
mean by this statement. Consider two oriented and time-oriented
spacetimes $(M,g)$ and $(M',g')$ and a map $\psi: M \to M'$ which is an
orientation and causality preserving\footnote{An isometric embedding may
be such that the intrinsic notion of causality is not the same as the
notion of causality inherited from the ambient space. Examples of this
sort may be constructed by embedding suitable regions of Minkowski spacetime
into Minkowski space with periodic identifications in one or more spatial directions.}
 isometric embedding. Then
there is a corresponding isomorphism
\ben\label{alphapsi}
\alpha_\psi: \W(M,g) \to \W(M',g')\, ,
\een
which behaves naturally under composition of embeddings. This map is
simply defined on $\Wn(M,g)$ by setting $\alpha_\psi(\phi_{M,g}(f)) =
\phi_{M',g'}(\psi_* f)$, where $\psi_* f(x') = f(x)$ for
$x=\psi(x')$. Since, as explained above, $\W(M,g)$ is essentially the
closure of $\Wn(M,g)$, we can define $\alpha_\psi$ on $\W(M,g)$ by continuity.
The action of $\alpha_\psi$ on $F$ of the form~\eqref{generators} may be
calculated straightforwardly from the
definition. However, we note that its form will depend on
the choices $\omega$ and $\omega'$ for the Hadamard bidistributions on
$M$ respectively $M'$, and will look somewhat involved if $\omega$ and
$\omega'$ are such that $\psi^* \omega' \neq \omega$. These
expressions are given in~\cite{Hollands2000}, but will not be needed here.

\subsection{Renormalized Wick products and their time-ordered products}

In the previous section we have laid the groundwork for the
construction of linear quantum field theory in curved spacetime by
giving the definition of an algebra $\W(M,g)$ associated with a
free Lagrangian $\L_0$ that can be viewed as a deformation
quantization of the algebra of classical observables with the Peierls
bracket. In this section we shall identify, within $\W(M,g)$, the
various objects that have the interpretation of the various Wick
powers in the theory, and their time-ordered products. Those objects
will be the quantities of prime interest in the perturbative constructions
in the subsequent sections. For simplicity,
we first address the case when $\L_0$ describes a linear, hermitian
scalar field $\phi$, see eq.~\eqref{L0}.

Actually, for reasons that we will explain below, it is
convenient to adopt a unified viewpoint on the Wick products and their
time-ordered products. We define a
time-ordered  product with $n$ factors (where $n\ge 1$) to be a linear map
\ben
T_n: \P^{k_1}(M) \otimes \dots \P^{k_n}(M) \to \D'\bigg(M^n; \wedge^{k_1} T^*M \times \dots \times \wedge^{k_n} T^* M \bigg) \otimes \W \, ,
\een
taking values in the distributions over $M^n$ with target space $\W$.
Thus, the linear map $T_n$ takes as arguments the tensor product of
$n$ local covariant classical forms $\O_1, \dots, \O_n$, and
it gives an expression $T_n(\O_1(x_1) \otimes \dots \otimes
\O_n(x_n))$,
which is itself a distribution in $n$ spacetime variables
$x_1, \dots, x_n$, with values in $\W$, i.e.,
$T_n(\O_1(x_1) \otimes \dots \otimes \O_n(x_n))$ is itself a map
that needs to be smeared with $n$-test forms $f_1(x_1), \dots,
f_n(x_n)$, where the $i$-th test form is an element in the set of compactly
supported smooth forms $f_i \in \Omega_0^{4-k_i}(M)$ over $M$. The set
$\D'(M^n; \wedge^{k_1} T^*M \times \dots \times \wedge^{k_n} T^* M)$ denotes the dual space (in the
standard distribution topology~\cite{Hormander}) of the
space of forms $\Omega^{4-k_1}_0(M) \times \dots \times \Omega^{4-k_n}_0(M)$.

The time-ordered products $T_n$ are
characterized abstractly by certain properties which we
will list. We define the Wick powers of a field to be the time-ordered
products with 1 factor, i.e., $n=1$. We will formulate the properties
of the time-ordered products in the form of axioms in this section,
but we will see in the following section that one can turn these
properties into a concrete constructive algorithm for these
quantities. In fact, as we will see, the properties
that we wish the time-ordered products to have do not {\em uniquely}
characterize them, but leave a certain ambiguity. This ambiguity corresponds
precisely to the renormalization ambiguity in other approaches in flat
spacetime, with the addition of couplings to curvature. However, we
note that our time-ordered products are rigorously defined, by
contrast to the corresponding quantities in other approaches to
renormalization in flat spacetime, where they are a priori only formal (i.e., infinite) objects.

\medskip
\noindent
\paragraph{T1 Locality and covariance}
The time ordered products are locally and covariantly
constructed in terms of the metric. This means that, if
$\psi:M \to M'$ is a causality preserving isometric embedding
between two spacetimes preserving the causal structure,
and $\alpha_\psi$ denotes the
corresponding homomorphism $\W(M,g) \to \W(M',g')$, see eq.~\eqref{alphapsi}, then we have
\ben
\alpha_\psi \circ T_n = T_n' \circ \bigotimes^n \psi_*
\een
where $T_n$ denotes the time-ordered product on $(M,g)$,
while $T_n'$ denotes the time-ordered product on $(M',g')$.
The mapping $\psi_*: \P(M) \to \P(M')$ is the natural push-forward
map. Thus, the local and covariance condition imposes a relation
between the construction of time-ordered products on locally isometric
spacetimes.Written more explicitly (in the case of scalar operators),
the local covariance condition is
\ben\label{lcc}
\alpha_\psi \bigg[ T_n(\phi^{k_1}(x_1) \otimes \dots \phi^{k_n}(x_n)) \bigg]=
T_n'(\phi^{k_1}(x_1') \otimes \dots \phi^{k_n}(x_n'))
\quad \psi(x_i) = x_i'\, .
\een
In particular, if $n=1$,
then the Wick products $T_1(\O(x))$ are local covariant fields in one
variable. As we will see
more clearly in the next subsection, the requirement of locality and
covariance is a non-trivial renormalization
condition already in the case of 1 factor.

It is instructive to consider the local covariance requirement for the
special case where $M = M'$ is Minkowksi spacetime, with $g = g'$
the Minkowski metric $-dt^2 +dx^2 + dy^2 + dz^2$. In that case, the
causality and orientation preserving isometric embeddings are
just the proper, orthochronous Poincare transformations $\psi = (\Lambda, a)
\in P^\uparrow_+$, while the map $\alpha_\psi$ may be implemented
by $Ad(U_0(\Lambda,a))$ in the vacuum Hilbert space representation $\pi_0$ of
the algebra $\W$ (we need to assume $m^2 \ge 0$ to have that representation),
with $U_0(\Lambda,a)$ the unitary representative of the proper orthochronous
Poincare transformation $(\Lambda, a)$ on the Hilbert space of the representation
$\pi_0$. The local covariance condition~\eqref{lcc} reduces
in that case to
 \ben\label{lccp}
Ad[U_0(\Lambda,a)]
\pi_0 \bigg( T_n\Big(\phi^{k_1}(x_1) \otimes \dots \phi^{k_n}(x_n)\Big) \bigg)
=
\pi_0 \bigg( T_n \Big(
\phi^{k_1}(\Lambda x_1-a) \otimes \dots \phi^{k_n}(\Lambda x_n- a) \Big) \bigg)
\een
which is the standard transformation law for the time ordered product (and in fact any
relativistic field) in Minkowski spacetime.

\paragraph{T2 Scaling.}
We would like the time-ordered products to satisfy a certain
scaling relation. For distributions $u(x), x \in \mr^n$
on flat space, it is natural to
consider the scaled distribution $u(\mu x), \mu \in \mr_+$. Such a
distribution is then said to scale homogeneously with
degree $D$ if $u(\mu x) = \mu^D u(x)$, in the sense of distributions, which
is equivalent to the differential relation
\ben
\bigg(\mu \frac{\partial}{\partial \mu} -D \bigg) u(\mu x) = 0 \, .
\een
More generally, it is said to scale ``polyhomogeneously'' or
``homogeneously up to logarithms'' if instead only
\ben
\bigg(\mu \frac{\partial}{\partial \mu} -D \bigg)^N u(\mu
x) = \frac{\partial^N}{\partial (\log \mu)^N} \left[ \mu^D
  u(\mu x) \right] =
0 \, .
\een
holds for some $N \ge 2$, which gives the highest power $+1$ of the
logarithmic corrections.

For the quantities in the quantum field
theory associated with the Lagrangian $\L_0$ on a generic curved
spacetime without dilation symmetry, we do not expect
a simple scaling behavior under rescalings in an arbitrarily chosen
coordinate system. However, we know that the Lagrangian $\L_0$ has an
invariance under a rescaling
\ben
g \mapsto \mu^2 g, \quad m^2 \mapsto \mu^{-2} m^2,
\quad \phi \mapsto \mu^{-1} \phi \, .
\een
It is therefore natural to expect that the time-ordered products can be
constructed so as to have a simple scaling behavior under such a
rescaling. However, due to quantum effects, one cannot expect
an exactly homogeneous scaling, but only a homogeneous scaling
behavior that is modified by logarithms. To describe this behavior, we
must first take into account that the time-ordered products associated
with the spacetime metric $g$ live in a different algebra than the
time-ordered products associated with $\mu^2 g$, so we must first
identify these algebras.  This is achieved by the
linear map $\sigma_\mu: \phi \mapsto \mu \phi$, which
may be checked to define an isomorphism between
$\W(M, g, m^2)$ and $\W(M, \mu^2 g, \mu^{-2} m^2)$. The desired polyhomogeneous scaling behavior is then
formulated as follows. Let
\ben
T_n[\mu] = \sigma^{-1}_\mu \circ T_n \circ
\bigotimes^n {\rm exp}(
\ln \mu \cdot \N_d)
\een
where $\N_d$ is the dimension counter, defined as $\N_d:=\N_c+\N_f+\N_r$, where
$\N_c, \N_f, \N_r: \P(M) \to \P(M)$ are the number counting operators for
the coupling constants, fields, and curvature terms, defined for Klein-Gordon theory
in 4 spacetime dimensions by
\bena\label{Nedefs}
\N_f &:=& \sum_k (1+k) (\nabla^k \phi) \frac{\partial}{\partial(\nabla^k \phi)} \, ,\\
\N_c &:=&  2 m^2 \frac{\partial}{\partial m^2} \, ,\\
\N_r &:=& \sum_k (k+2) (\nabla^k R) \frac{\partial}{\partial(\nabla^k R)} \, .
\eena
For example
\ben
T_n[\mu](\phi^{k_1}(x_1) \otimes \dots \otimes \phi^{k_n} (x_n)) =
\mu^{k_1+\dots+k_n} \sigma^{-1}_\mu T_n
(\phi^{k_1}(x_1) \otimes \dots \otimes \phi^{k_n} (x_n)) \, .
\een
Because we have put the identification map $r_\mu$ on the right
side, $T_n[\mu]$ defines a new time ordered product in the algebra
associated with the unscaled metric, $g$, and coupling constants. In the
absence of scaling anomalies, this would be equal to the original
$T_n$ for all $\mu \in \mr_+$. As we have said, it is not
possible to achieve this exactly homogeneous scaling behavior, so we
only postulate the polyhomogeneous scaling behavior
\begin{equation}
\frac{\partial^N}{\partial (\log \mu)^N} \, T_n[\mu] = 0.
\end{equation}

\paragraph{T3 Microlocal Spectrum condition.}
Consider a time ordered product $T_n(\O_1(x_1) \otimes \cdots
\otimes \O_n(x_n))$ as an $\W$ valued distribution on $M^n$.
Then we require that
\begin{equation}
\label{microcond}
\WF(T_n) \subset C_T(M, g),
\end{equation}
where the set $C_T(M, g) \subset T^*M^n \setminus 0$
is described as follows (we
use the graph theoretical notation introduced in
\cite{Brunetti1996,Brunetti2000}):
Let $G(p)$ be a ``decorated embedded graph''
in $(M, g)$. By this we mean an embedded graph $\subset M$ whose
vertices are points $x_1, \dots, x_n \in M$
and whose edges, $e$, are oriented null-geodesic curves. Each such null
geodesic is equipped with a coparallel, cotangent covectorfield $p_e$.
If $e$ is an edge in $G(p)$ connecting the points $x_i$ and $x_j$
with $i < j$, then $s(e) = i$ is its source
and $t(e) = j$ its target. It is required that
$p_e$ is future/past directed if $x_{s(e)} \notin J^\pm(x_{t(e)})$.
With this notation, we define
\begin{eqnarray}
\label{gamtdef}
C_T(M, g) &=&
\bigg\{(x_1, k_1; \dots; x_n, k_n) \in T^*M^n \setminus 0 \mid
\exists \,\, \text{decorated graph $G(p)$ with vertices} \nonumber\\
&& \text{$x_1, \dots, x_n$ such that
$k_i = \sum_{e: s(e) = i} p_e - \sum_{e: t(e) = i} p_e
\quad \forall i$} \bigg\}.
\end{eqnarray}

\paragraph{T4 Smoothness.}
The functional dependence of the time ordered products
on the spacetime metric, $g$, is such that
if the metric is varied smoothly, then the time ordered
products vary smoothly, in the sense described in~\cite{Hollands2000}.

\paragraph{T5 Analyticity.}
Similarly, we require that, for an analytic family of analytic
metrics (depending analytically upon a set of parameters),
the expectation value of the time-ordered products in an
analytic family of states\footnote{As explained in remark (2) on P. 311
of \cite{Hollands2000}, it suffices to consider a suitable analytic family of
linear functionals on $\W$ that do not necessarily satisfy the positivity
condition required for states.} varies analytically in the same sense as in
T4.

\paragraph{T6 Symmetry.}
The time ordered products are symmetric under a permutation of
the factors,
\ben
T_n(\O_1(x_1) \otimes \cdots \otimes \O_n(x_n))
= T_n(\O_{\pi 1}(x_{\pi 1}) \otimes \cdots \otimes \O_{\pi n}(x_{\pi n}))
\een
for any permutation $\pi$.

\paragraph{T7 Unitarity.}
Let $\bar T_n(\otimes_i \O_i(x_i)) =
[T_n(\otimes_i \O_i(x_i)^*)]^*$ be
the ``anti-time-ordered'' product. Then we require
\begin{equation}
\bar T_n \bigg( \bigotimes_{i=1}^n \O_i(x_i) \bigg) =
\sum_{I_1 \sqcup \dots \sqcup I_j = \underline{n}}
(-1)^{n + j} T_{|I_1|}\bigg(
\bigotimes_{i \in I_1} \O_i(x_i) \bigg) \st \dots \st
T_{|I_j|}\bigg(\bigotimes_{j \in I_j} \O_j(x_j) \bigg),
\label{atoprod}
\end{equation}
where the sum runs over all partitions of the set $\{1, \dots, n\}$ into
pairwise disjoint subsets $I_1, \dots, I_j$.

\paragraph{T8 Causal Factorization.}
The ``product'' $T_n$ is time ordered in the sense that the following causal
factorization property is to be satisfied. Let $\{x_1, \dots, x_i\}
\cap J^-(\{x_{i+1},\dots,x_n\}) = \emptyset$. Then we have
\begin{multline}
T_n(\O_1(x_1) \otimes \dots \otimes \O_n(x_n)) \\=
T_i(\O_1(x_1) \otimes \dots \otimes \O_i(x_i)) \st
T_{n-i}(\O_{i+1}(x_{i+1}) \otimes \dots \otimes \O_n(x_n)) \, .
\end{multline}
For the case of 2 factors, this means
\ben
T_2(\O_1(x_1) \otimes \O_2(x_2)) =
\begin{cases}
T_1(\O_1(x_1)) \st T_1(\O_2(x_2)) & \text{when $x_1 \notin J^-(x_2)$;}\\
T_1(\O_2(x_2)) \st T_1(\O_1(x_1)) & \text{when $x_2 \notin J^-(x_1)$.}
\end{cases}
\een

\paragraph{T9 Commutator.}
The commutator of a time-ordered product with a free field is given by
lower order time-ordered products times suitable commutator functions,
namely
\ben
\left[T_n \bigg( \bigotimes_i^n \O_i(x_i) \bigg), \phi(x) \right]_{\st} =
i\hbar\sum_{k=1}^n T_n\bigg(\O_1(x_1) \otimes \dots \int \Delta(x,y)
\frac{\delta \O_k(x_k)}{\delta \phi(y)} \otimes \dots \O_n(x_n) \bigg),
%\quad \text{mod $\I$}
\label{T9}
\een
where $\Delta$ is the
causal propagator.

\paragraph{T10 Field equation.}
The free field equation $\delta S_0/\delta \phi$ holds in the sense
that
\ben
T_{n+1}\bigg(
\frac{\delta S_0}{\delta \phi(x)} \otimes \bigotimes_i^n \O_i(x_i)
\bigg)
= \sum_i
T_n
\bigg(
\O_1(x_1) \otimes \cdots \frac{\delta \O_i(x_i)}{\delta
    \phi(x)} \otimes \cdots \O_n(x_n)
\bigg) \quad \text{mod $\I$.}
\een

\paragraph{T11 Action Ward identity}
If $d_k = dx_k^\mu \wedge \frac{\partial}{\partial x_k^\mu}$ is the
exterior differential acting on the $k$-th spacetime variable, then we
have
\ben
T_n(\O_1(x_1)\dots \otimes d_k\O(x_k) \dots \otimes \O_n(x_n)) =
d_k \, T_n  (\O_1(x_1)\otimes  \dots \otimes \O(x_n)) \, .
\een
Thus, derivatives can be freely pulled inside the time-ordered
products.

Condition T11 can be stated as saying that
$T_n$ may alternatively be viewed as a linear map
$T_n:  A^{\otimes n} \to \W$ for each $n$, where
$A$ is the space of all local action functionals, i.e., all expressions
of the form $F=\int \O \wedge f$, where $f \in \Omega_0^p(M)$ is any $p$-form of
compact support, and where
$\O \in \P^{4-p}$. To explain how this comes about, consider the integrated field
polynomial $F= \int f \wedge d\O$. It may equivalently be written
as $-\int (df) \wedge \O$, so the time ordered product should
give the same result for either choice. T11 means that
the time ordered products
$\int f(x_i) \, T_n(\dots \otimes d_i\O(x_i) \otimes \dots)$ and
$-\int d_i f(x_i) \, T_n  (\dots \otimes \O(x_i) \otimes \dots)$ are equal,
where the exterior derivative
$d_i=dx_i^\mu \wedge \partial/\partial x_i^\mu$
acts on the $i$-th spacetime argument. This means that $T_n$
may be viewed as a functional taking as arguments the
integrated functionals (or "actions") in $A$, because
it does not matter how $F$ is represented. This is
the origin of the name ``action Ward identity'' for
T11. The action Ward identity also means that
we may apply the Leibniz rule for derivative of
quantum Wick powers, i.e., time ordered products
with one factor, which is why the same condition was
called ``Leibniz rule'' in~\cite{Hollands2005}.

\subsection{Inductive construction of time-ordered products}

In the previous subsection, we have given a list of
properties of the local Wick powers and their time-ordered
products. We now present an algorithm showing how these can be
constructed, and thus in particular demonstrating that axioms T1 through
T11 are not empty. We shall reduce the problem to successively simpler
problems by a series of reduction steps. These steps are as follows:
\begin{enumerate}
\item First, construct the time-ordered products with one factor.
\item Assuming inductively that time-ordered products with $n$ factors have been
  constructed, we show, following the ideas of ``causal perturbation
  theory''~\cite{Epstein1973,Bogoliubov1952,Stora1990,Steinmann1990}
  that the time-ordered products with $n+1$ factors are
  already uniquely fixed, apart from points on the total diagonal, by
  the lower order time-ordered products.
\item The problem of extending the time-ordered products at order
  $n+1$ to the total diagonal
  is reduced to that of extending certain scalar distributions to the
  total diagonal.
\item The problem of reducing the scalar functions on $M^{n+1}$ to the
  diagonal is reduced to that of extending a set of distributions on
  the $(n+1)$-fold Cartesian power of Minkowski space via a curvature
expansion.
\item The extension of the Minkowski distributions is performed. This
step corresponds to renormalization.
\end{enumerate}
Thus, we shall proceed inductively in the number
of factors, $n$, appearing in the time ordered product
$T_n(\O_1(x_1) \otimes \cdots \otimes \O_n(x_n))$.
To keep our discussion as simple as possible,
we now restrict attention to the case when the fields $\O_i \in \P$ in the
time ordered product contain no spacetime
derivatives, i.e., $\O_i = \phi^{k_i}$ for some natural numbers
$k_i$. We will also assume for simplicity that external potential
$v$ in the Klein-Gordon equation vanishes, so that there are no
coupling parameters to consider. We briefly explain how to deal with the
general case in the end.

{\bf Time-ordered products with 1 factor:}
For $n=1$ the time ordered products are just the
local covariant Wick powers, i.e., $T_1(\phi^k(x))$ is a local covariant
field in one spacetime variable, interpreted as the $k$-th local
covariant Wick power of $\phi$. These Wick powers may be constructed
as follows. Let $H(x,y)$ be the ``local Hadamard parametrix,'' for
the Klein-Gordon operator, given by
\ben\label{Hdef}
H(x,y) = \frac{1}{2\pi^2} \bigg(\frac{u(x,y)}{\sigma + it 0} + v(x,y) \, \log
(\sigma + it0) \bigg) \, .
\een
Here, $\sigma(x,y)$ is the signed squared geodesic distance between two
points $x,y$ in a convex normal neighborhood of $M$, and
$u, v$ are smooth kernels
that are locally constructed in terms of the metric, which are
determined by the Hadamard recursion relations~\cite{DeWitt1962},
which are obtained by demanding that $H$ be a bi-solution
(modulo a smooth remainder) of the Klein-Gordon equation.
Their construction is recalled in Appendix~D.
The quantity $t(x,y)=T(x)-T(y)$ is defined in terms of an arbitrary global time
coordinate $T$.

Consider now, for any $k \ge
1$, the ``locally normal ordered expressions''
\begin{multline}\label{locnorm}
: \phi(x_1) \cdots \phi(x_k) :_\h \, \\
=
\frac{\delta^k}{i^k \delta f(x_1) \dots \delta f(x_k)}
\exp_{\st} \bigg(
i\int_M f(x) \phi(x) + \frac{\hbar}{2} \int_{M \times M} H(x,y) f(x) f(y)
\bigg) \Bigg|_{f=0} \, .
\end{multline}
Because $H$ is defined locally and covariantly in terms of the metric,
it follows that $:\phi(x_1) \dots \phi(x_k):_\h$
are local and covariant
fields that are defined in a convex normal neighborhood of the
diagonal $\Delta_k$, where
\ben
\Delta_k = \{(x,x,\dots,x) \mid \quad x \in M\} \subset M^k \, .
\een
The following lemma shows that the normal ordered
quantities~\eqref{locnorm} differ from the quantities
$:\phi(x_1) \dots \phi(x_n):_\omega$ only by
a smooth function (valued in $\W$).

\begin{lemma} Let $\omega(x,y)$ be a 2-point function of Hadamard form,
  i.e., the wave front set $\WF(\omega)$ is given by~\eqref{wfs2}.
Then locally (i.e., where $H$ is defined), $\omega-H$ is smooth, i.e.,
\ben
\omega(x,y) = \frac{1}{2\pi^2}
\bigg( \frac{u(x,y)}{\sigma + it 0} + v(x,y) \, \log
(\sigma + it0) \bigg) + \quad (\text{smooth function in $x,y$}).
\een
Furthermore, any two Hadamard states can at most differ by a globally
smooth function in $x,y$.
\end{lemma}
The proof is given in Appendix~E.

Because the normal ordered products may be smeared with a $\delta$-function
(or derivatives thereof), we may define
\ben\label{wickprod}
T_1\Big( \phi^k(x) \Big)
= \, : \phi^k(x) :_\h \,
%= \int_{M^n} :\phi(y_1)\cdots\phi(y_n):_\h \,
%\nabla^{k_1}_{y_1} \cdots \nabla_{y_n}^{k_n} \delta(x,y_1,\dots,y_n)
\een
which is a well defined element in $\W$ after smearing with
any testfunction $f \in C^\infty_0(M)$.
This defines our time-ordered products with one factor.
It follows from the definition of $H$ that $T_1(\phi^k(x))$ is a local
covariant field, i.e., it satisfies T1
for $n=1$. The other properties T2---T11 are also seen to be
satisfied using the properties of $H$ described in Appendix~D.

\medskip

{\bf Time-ordered products with $n>1$ factors:}
We have defined the time-ordered products with $n=1$ factor,
and we may inductively assume that time ordered products with
properties T1--T11
have been defined for any number of factors $\le n$. The key idea of
causal perturbation
theory~\cite{Epstein1973,Bogoliubov1952,Steinmann1990,Stora1990}
is that the time ordered
products with $n+1$ factors are already uniquely determined as
algebra-valued distributions on the manifold $M^{n+1}$ minus its total
diagonal $\Delta_{n+1}=\{(x,x,\dots,x) \in M^{n+1}\}$
by the causal factorization requirement T8, once
the time ordered products with less than or
equal to $n$ factors are given. The construction of the time ordered
products at order $n+1$ is then equivalent to the task of extending
this distribution in a suitable way compatible with the other
requirements T1--T10. In order to perform this task in an efficient
way, it is useful to derive a number of properties that hold at all
orders $m \le n$ as a consequence of T1--T10.

The first property is a
local Wick expansion for time ordered products~\cite{Hollands2001}. This is a key
simplification, because it will enable one to reduce the problem of
extending algebra valued quantities to one of finding an extension of
c-number distributions. In the simplest case, when none of the $\O_i$
contain derivatives of $\phi$, we have  in an open neighborhood of $\Delta_m$
\begin{multline}\label{wickexp00}
T_m\Big( \phi^{k_1}(x_1) \otimes \cdots \otimes \phi^{k_m}(x_m) \Big)
\\=
\sum_{0 \le j_i \le k_i} \prod_i
\left(
\begin{matrix}
k_i \\
j_i
\end{matrix}
\right)
t_{j_1, \dots, j_m}(x_1, \dots, x_m) \,
: \phi^{k_1-j_1}(x_1) \cdots \phi^{k_m-j_m}(x_m) :_\h
%\sum_{l_{ij}}
%t\Bigg(
%$\bigotimes_{i=1}^m
%\Bigg\{
%\prod_j
%\frac{\partial^{l_{ij}}}{\partial^{l_{ij}}[\nabla^j \phi(x_i)]}
%\Bigg\}
%\O_i(x_i)
%\Bigg) \,
%:\prod_{i=1}^m \nabla^{l_{ij}} \phi(x_i) :_\h,
%\quad \forall m \le n
\end{multline}
for all $1< m \le n$,
where $t_{j_1, \dots, j_m}$ are c-number distributions. The Wick expansion when derivatives are present
is analogous. The Wick expansion formula can be proved
from axiom~T9. Because the time-ordered products are
local and covariant, the
c-number distributions in the Wick expansion have the same property,
in the sense that if $\psi: (M',g') \to (M,g)$ is an isometric, causality and
orientation preserving embedding, so that if $\psi^* g = g'$, then
\ben\label{tloc}
t_{j_1, \dots, j_m}\Big[ \psi^* g; x_1, \dots, x_m \Big] =
t_{j_1, \dots, j_m}\Big[ g; \psi(x_1) , \dots, \psi(x_m) \Big] \, .
\een
Because
$H$ and the local normal ordered products are
in general only defined in a neighborhood of the diagonal, 
it follows that also the c-number distributions are only defined on a
neighborhood of the diagonal, but this will turn out to be
sufficient for our purposes.

It follows from the scaling property T2 and the corresponding
scaling properties of $H$ that
\ben\label{littletscale}
\frac{\partial^N}{\partial(\log \mu)^N} \left\{ \mu^{j_1+\dots+j_m}
t_{j_1, \dots, j_m}\Big[\mu^{-2} m^2, \mu^2 g; x_1, \dots, x_m \Big]
\right\} = 0
\een
for some $N$. This relation, together with the condition of
locality and covariance and the analytic dependence of
the time ordered products on the metric, can be used to derive a
subsequent "scaling-" or "curvature expansion"~\cite{Hollands2001}
of each of the
distributions $t_{j_1, \dots, j_m}$ in powers of
the Riemann tensor and the coupling constants (in our case only $m^2$) at
a reference point:

\paragraph{\bf Proposition 0:}\label{scalingexpansion}
The distributions $t:=t_{j_1, \dots, j_m}$ have the asymptotic expansion
\ben\label{scalexp}
t(\exp_y \xi_1, \dots, \exp_y \xi_{m-1}, y)
=
\sum_{k=0}^S C^k_{\mu_1 \dots \mu_t}(y)
\, u_k^{\mu_1 \dots \mu_t}(\xi_1,
\dots, \xi_m) + r^S(y, \xi_1, \dots, \xi_{m-1}) \, .
\een
in an open neighborhood of the diagonal $\Delta_m$. The terms
have the following properties:
\begin{enumerate}
\item[(i)] The remainder $r^S$ is a
distribution of scaling degree (see Appendix C for the
mathematical definition of this concept)
strictly lower than the scaling degree of any term in the sum.
\item[(ii)]
Each $u_k$ is a Lorentz invariant distribution on
$(\mr^4)^{m-1}$, i.e.,
\ben
u_k^{\mu_1\dots\mu_t}(\Lambda \xi_1, \dots, \Lambda\xi_m)
= \Lambda^{\mu_1}_{\nu_1} \cdots \Lambda^{\mu_t}_{\nu_t}
u_k^{\nu_1\dots\nu_t}(\xi_1, \dots, \xi_m)
\quad \forall \Lambda \in {\rm SO}_0(3,1) \, .
\een
\item[(iii)]
Each distribution $u_k$ scales almost
homogeneously under a coordinate rescaling, i.e.,
\ben\label{scalingcon}
\frac{\partial^N}{\partial(\log \mu)^N} \left[ \mu^{\rho}
u_k^{\mu_1 \dots \mu_t}(\mu\xi_1, \dots, \mu\xi_{m-1})
\right] = 0
\een
with $\rho \in \mn$. The scaling condition can be rewritten equivalently as
\ben\label{scale}
\bigg(\sum_{i=1}^{m-1} \xi_i^\nu \frac{\partial}{\partial \xi_i^\nu}
-\rho
\bigg)^N u_k^{\mu_1 \dots \mu_t}(\xi_1, \dots, \xi_{m-1}) = 0 \, .
\een
\item[(iv)] Each
term $C^k$ is a polynomial in $m^2$ and the covariant
derivatives of the Riemann tensor ,
\ben
C^k_{\mu_1 \dots \mu_t}(y) =
C^k_{\mu_1 \dots \mu_t}[m^2, R(y), \nabla R(y), \dots, \nabla^l R(y)] \, .
\een
\item[(v)] The scaling degree $\rho = sd(u_k)$ is
given by
\ben
sd(u_k) = \sum_i j_i - {\mathcal N}_{r}(C^k) \, ,
\een
where $\N_r$
is the dimension counting operator for curvature terms and dimensionful
coupling constants (in our case only $m^2$), see eq.~\eqref{Nedefs}.
\end{enumerate}

\medskip

By the above proposition, we see that, by
including sufficiently (but finitely many) terms
in the scaling expansion~\eqref{scalexp} (i.e., choosing $S$ sufficiently large), one can achieve that the
remainder $r^S$
has arbitrarily low scaling degree. It does {\em not} mean that the
sum is convergent in any sense (it is not).

Having stated the detailed properties of the time ordered products with
$\le n$ factors, we are now resume the main line of
the argument and perform the construction of the
time-ordered products with $n+1$ factors.
Let $I$ be a proper subset of $\{1, 2, \dots, n+1\}$, and let
$U_I$ be the subset of $M^{n+1}$ defined by
\begin{equation}\label{ci}
U_I = \{(x_1, x_2, \dots, x_{n+1}) \mid x_i \notin J^-(x_j)
\quad \text{for all $i \in I, j \notin I$} \} \, .
\end{equation}
It can be seen~\cite{Brunetti2000} that the sets $U_I$ are open and that the collection
$\{U_I\}$ of these sets covers the manifold $M^{n+1} \setminus \Delta_{n+1}$.
We can therefore define an algebra valued distribution $T_{n+1}$ on
this manifold by declaring it for
each $(x_1, \dots, x_{n+1}) \in U_I$ by
\begin{multline}\label{Tdef}
T_{n+1}\Big( \phi^{k_1}(x_1) \otimes \cdots \otimes
             \phi^{k_{n+1}}(x_{n+1}) \Big) = \\
T_{|I|}\Big( \otimes_{i \in I} \phi^{k_i} (x_i) \Big)
\st T_{n+1-|I|}\Big(\otimes_{j \in \underline{n+1} \setminus I} \phi^{k_j}
  (x_j) \Big) \quad
\forall (x_1, \dots, x_{n+1}) \in U_I \, .
\end{multline}
To avoid a potential inconsistency in this definition for
points in $U_I \cap U_J \neq \emptyset$ for different $I,J$, we must
show that the definition agrees for different $I,J$. This can be
achieved using the causal factorization property T8 of the time ordered
products with less or equal than $n$ factors~\cite{Epstein1973,Brunetti2000}. Property T8 applied to the
time ordered products with $n+1$ factors also implies that the
restriction of $T_{n+1}$ to $M^{n+1} \setminus \Delta_{n+1}$ must agree with
\eqref{Tdef}. Thus, property T8 alone determines the time ordered
products up to the total diagonal,
as we desired to show, see~\cite{Brunetti2000} for details.

In fact---assuming that time ordered products with less or
equal than $n$ factors have been defined so as to satisfy properties
T1--T11 on $M^{n}$---one can argue in a relatively straightforwardly way
that the fields defined by eq.~\eqref{Tdef} with $n+1$ factors
automatically satisfy\footnote{Of course, if any $T_{n+1}$ failed to
satisfy any of these properties on $M^{n+1} \setminus \Delta_{n+1}$,
we would have a proof that no definition of time ordered products
could exist that satisfies T1--T9.} the restrictions of properties
T1--T9 to $M^{n+1} \setminus \Delta_{n+1}$, while T10 and T11 are empty in the
present case for time ordered products without derivatives.

Our remaining task is to find an extension of each of the
algebra-valued distributions $T_{n+1}$ in $n+1$ factors from $M^{n+1}
\setminus \Delta_{n+1}$ to all of $M^{n+1}$ in such a way that properties
T1--T9 continue to hold for the extension. This step, of course,
corresponds to renormalization. Condition T8 does not impose any
additional conditions on the extension, so we need only satisfy
T1--T7 and T9. However, it is not difficult to see that if an
extension $T_{n+1}$ is defined that satisfies T1--T5 and T9, then that
extension can be modified, if necessary, so as to also satisfy the
symmetry and unitarity conditions, T6 and T7, see~\cite{Hollands2000}.

Thus, we have reduced the problem of defining time ordered products to
the problem of extending the distributions $T_{n+1}$ defined by
\eqref{Tdef} from $M^{n+1} \setminus \Delta_{n+1}$ to all of
$M^{n+1}$ so that properties T1--T5 and T9 continue to hold for the
extension. To find that extension, we now make a Wick expansion
of $T_{n+1}$, which follows from the Wick expansion at lower orders.
That Wick expansion will contain c-number distribution coefficients,
$t$, that are defined as distributions on a neighborhood of $\Delta_{n+1}$ in
$M^{n+1} \setminus \Delta_{n+1}$. They possess a scaling expansion analogous
to~\eqref{scalexp}, with distributions $u_k$ that are defined on
$(\mr^4)^n \setminus 0$.
As we have just argued, time ordered products satisfying all of our
conditions will exist if and only if the c-number distributions $t$
defined away from $\Delta_{n+1}$ appearing in the Wick expansion for
$T_{n+1}$ analogous \eqref{wickexp00} can be extended to distributions
defined on an open neighborhood of $\Delta_{n+1}$
in such a way that the distribution $T_{n+1}$ defined by
\eqref{wickexp00} continues to satisfy properties T1--T5. It is
straightforward to check that this will be the case if and only if the
extensions $t$ satisfy the following five corresponding conditions:

\paragraph{t1 Locality/Covariance.}
The distributions $t=t_{j_1, \dots, j_{n+1}}$
are locally constructed from the metric in a covariant manner in the
following sense. Let $\psi: M \to M'$ be a causality-preserving
isometric embedding, so that $\psi^* g'=g$.
Then eq.~\eqref{tloc} holds for $m=n+1$.

\paragraph{t2 Scaling.}
The extended distributions $t$
scale homogeneously
up to logarithmic terms,  in the sense that
there is an $N \in \mn$ such that~\eqref{littletscale} holds for
$m=n+1$.

\paragraph{t3 Microlocal Spectrum Condition.}
The extension satisfies the wave front set condition
that the restriction of $\WF(t)$  to the
diagonal $\Delta_{n+1}$ is contained in $\{(x, k_1, \dots,
x,k_{n+1}) \mid \sum k_i =  0\}$.

\paragraph{t4 Smoothness.}
$t$ depends smoothly on the metric.

\paragraph{t5 Analyticity.}
For analytic spacetimes $t$ depends analytically on the metric.

\medskip
\noindent
In summary, we have reduced the problem of defining
time ordered products to the following question: Assume that time
ordered products involving $\leq n$ factors have been constructed so
as to satisfy our requirements T1--T9. Define $T_{n+1}$ by \eqref{Tdef}
and define the distributions $t$ on $M^{n+1} \setminus
\Delta_{n+1}$ by the analogy of \eqref{wickexp00} for $T_{n+1}$,
in a neighborhood of the
diagonal. Can each $t$ be extended to
a distribution defined on a neighborhood of $\Delta_{n+1}$
so as to satisfy requirements
t1--t5?

The answer to this question is ``yes,'' and we shall now show how the
desired extension of $t(x_1, \dots, x_{n+1})$ may be found.
The idea is that, since the remainder in the scaling
expansion~\eqref{scalexp} for $t$ has an arbitrary low scaling
degree for sufficiently large
$m$ by item (v), it can be extended to the diagonal $\Delta_{n+1}$ by
continuity~\cite{Brunetti2000}, i.e., there is no need to
``renormalize'' the remainder for sufficiently large but finite $S$.
In fact, by Thm.~5.3 of~\cite{Brunetti2000}, it is sufficient to choose
any $S \ge d-4n$ for this purpose. Furthermore, each term in the sum
in the scaling expansion~\eqref{scalexp} can be written as
$C^k(y) \cdot u_k(\xi_1, \dots, \xi_n)$ by (i). Each $u_k$ is an almost
homogeneous, Lorentz invariant $n$-point distribution on $(\mr^4)^n
\setminus 0$. As we will see presently in lemma~\ref{lemma4}~\cite{Hollands2001}, this Minkowski distribution can be
extended to a distribution on $(\mr^4)^n$ with the same properties
[possibly with a higher $N$ than that appearing~\eqref{scale}], by
techniques in Minkowski space.
It is this step that corresponds to the renormalization. As a consequence of
the properties satisfied by the extension $u$, the corresponding
extension $t$ can be seen to satisfy t1)---t5), thus solving the
renormalization problem for the time ordered products
$T_{n+1}(\otimes_{i=1}^{n+1} \phi^{k_i}(x_i))$ with $n+1$ factors.

\begin{lemma}\label{lemma4}
Let $u \equiv u_{\mu_1 \dots \mu_l}(\xi_1, \dots,
\xi_n)$ be a Lorentz invariant tensor-valued distribution on $\mr^{4n}
\setminus 0$ which scales almost homogeneously with degree $\rho \in \mc$
under coordinate rescalings, i.e.,
\begin{equation}\label{sscale}
S_{\rho}^N u = 0 \quad \text{
for some natural number $N$.}
\end{equation}
where
\ben
S_\rho = \sum_{i=1}^n \xi_i^\mu \partial/\partial \xi_i^\mu +
\rho \, .
\een
Then $u$ has a Lorentz invariant extension, also denoted $u$, to a distribution on
$\mr^{4n}$ which also scales almost homogeneously with degree $\rho$
under rescalings of the coordinates. Moreover:
\begin{enumerate}
\item If $\rho \in {\mathbb Z}$, $\rho < 4n$, then $u$ can be extended
  by continuity, the extension is
  unique, and $S_\rho^N u = 0$.
\item If $\rho \in \mc \setminus {\mathbb Z}$ then
  the extension is unique, and $S_\rho^N u = 0$.
\item If $\rho \in {\mathbb Z}$, $\rho \ge 4n$, then the extension is
  not unique, and $S^{N+1}_\rho u = 0$. Two different extensions can
  differ at most by a distribution of the form $L \delta$, where $L$
  is a Lorentz-invariant partial differential operator in $\xi_1,
  \dots, \xi_n$ containing derivatives of degree $\rho-4n$.
\end{enumerate}
\end{lemma}

\noindent
{\it Proof:}
A proof of this important lemma was given first in~\cite{Hollands2001}. In this paper, we 
choose to give a somewhat different, alternative, proof, parts of which are closely related 
also to the `improved Epstein-Glaser renormalization' of~\cite{garcia_bondia1, garcia_bondia2}.
The proof given here has the advantage that it is somewhat more constructive and explicit.
We will first construct an extension that satisfies the almost
homogeneous scaling property.
This extension need not satisfy the Lorentz
invariance properties. However, we will show that the extension can be
modified, if necessary, so that the desired Lorentz-invariance
property is satisfied, while retaining the desired almost homogeneous
scaling behavior. The proof of the theorem given here differs
from that given in~\cite{Hollands2001}, and thereby provides an
alternative construction of the extension. A less general result of
a similar nature for distributions with an {\it exactly}
homogeneous scaling has previously been obtained in
\cite[Thms. 3.2.3 and 3.2.4]{Hormander}. Thus, our theorem generalizes this
result to the case of almost homogeneous scaling.
To simplify the notation, we set $x=(\xi_1, \dots, \xi_n) \in \mr^{4n}$
throughout this proof.

The almost homogeneous scaling property of $u$, eq.~\eqref{sscale}, or
the equivalent form of this condition~\eqref{scalingcon} implies that
$u(r x)$ can be written in the form
\ben\label{umux}
u(r x) = r^{-\rho} \sum_{k=0}^{N-1} \frac{(\log r)^k}{k!} v_k(x)
\quad\quad r>0 \, ,
\een
where $v_k$ are the distributions defined on $\mr^{4n} \setminus 0$ by
\ben\label{vkdef}
v_k = S_\rho^k u \, .
\een
Choose an arbitrary compact $4n-1$-dimensional surface
$\Sigma \subset \mr^{4n}$ homeomorphic
to the sphere $S^{4n-1}$ around the origin of $\mr^{4n}$ that intersects
each orbit of the scaling map $x \mapsto \mu x$ transversally and
precisely once\footnote{For example, we may choose $\Sigma$ to be the
sphere $S^{4n-1}$ defined relative to some auxiliary Euclidean metric
on $\mr^{4n}$.}. The
first aim is to show that the distributions $v_k$ can be restricted
to $\Sigma$. To prove this, it is convenient to use
the methods of microlocal analysis, in particular the following
result~\cite{Hormander}:
If $\varphi$ is a distribution on a manifold $X$ with a
submanifold $Y$, then $\varphi$ can be restricted to $Y$ if its
wave front set~(see Appendix~C) satisfies
$\WF(\varphi) |_Y \cap N^* Y = \emptyset$, where $N^* Y$ is the
``conormal bundle,'' defined as
\ben
N^* Y = \{(y,k) \in T^*_y X ; \,\, y \in Y, k_i w^i =0
\,\, \forall w \in T_y Y \} \, .
\een
We would like to apply this result to the situation $\Sigma=Y, \mr^{4n} \setminus
0 = X$, and $v_k = \varphi$.
To estimate the wave front set of the distributions $v_k$, we use
another result from microlocal analysis~\cite{Hormander}. Suppose $A$ is a
differential operator on $X$ such that $A\varphi$ is smooth.
Then $\WF(\varphi) \subset {\rm char}(A) \setminus 0$, where the
characteristic set of $A$ is defined by ${\rm char}(A) = \{(x,k) \in
T^*_x X ; \,\, a(x,k) = 0\}$, where $a$ is the principal symbol of
$A$. In our case, we have $S_\rho^{N-k} v_k = 0$, so
\ben
\WF(v_k) \subset {\rm char}(S_\rho^{N-k}) \setminus 0 =
\bigg\{(x,k) \in T^* \mr^{4n}; \,\, \sum_i \xi_i \cdot k_i = 0, k \neq
0 \bigg\}
\een
because the principal symbol of $S_\rho$ is given by
$s(x,k) = \sum \xi_i \cdot k_i$, where we recall the notation
$x=(\xi_1, \dots, \xi_n)$, and where we have set $k=(k_1, \dots, k_n)
\in (\mr^{4n})^*$. Assume now that $(x,k) \in N^* \Sigma$, and at the same
time $(x,k) \in \WF(v_k) |_\Sigma$. Then, from the first condition, we
have $w \cdot k=0$ for all $w \in T_x \mr^{4n}$ that are tangent to
$S$, while from the second condition, we have $x \cdot k=0$ and $k\neq
0$. Since $\Sigma$ is transverse to the scaling orbits, it follows that
$k=0$, a contradiction. Hence $\WF(v_k)|_\Sigma \cap N^* \Sigma = \emptyset$,
and $v_k$ can be restricted to $\Sigma$. We denote points in $\Sigma$ by
$\hat x$, and we denote the restriction simply by $v_k(\hat x)$,
by the usual abuse of notation.

Let $\Sigma \subset \mr^{4n}$ a
submanifold of dimension $4n-1$ as above, and define, for $r > 0$
\ben
\Sigma_r = \{ r\hat x \in \mr^{4n}; \,\, \hat x \in \Sigma \} \, .
\een
We let $d^{4n} x$ be the usual $4n$-form on $\mr^{4n}$ with the orientations
induced from $\mr^4$, i.e.,
\ben
d^{4n} x = d^4 \xi_1 \wedge \dots \wedge d^4 \xi_n, \quad d^4 \xi = d\xi^0 \wedge \dots \wedge d\xi^3 \, ,
\een
where we have put again $x = (\xi_1, \dots, \xi_n)$ to lighten the notation.
We also define the $3$-form $w$ on $\mr^4$ and the $4n-1$ form $\Omega$ on
$\mr^{4n}$ by
\bena\label{Odef}
&&w(\xi) = \sum_{\mu=0}^3 \xi^\mu
d\xi_1 \wedge \dots \widehat{d \xi^{\mu}} \wedge \dots d\xi^3, \\
&& \Omega(x) = \sum_{i=1}^n d^4 \xi_1 \wedge \dots w(\xi_i) \wedge \dots d^4 \xi_n
\eena
where a caret denotes omission. Because we are assuming that the surface $\Sigma$ is
transverse to the orbits of dilations in $\mr^{4n}$, the
map $(r, \hat x) \in \mr_+ \times \Sigma \mapsto r\hat x \in \mr^{4n} \setminus 0$
is an diffeomorphism. If $i_r : \Sigma_r \to \mr^{4n}$ is the natural inclusion,
then we may write
\ben
d^{4n} x = \frac{dr}{r} \wedge i_r^* \Omega \, .
\een

Now let $f$ be a test function of compact support on $\mr^{4n}
\setminus 0$, i.e., $f$ is smooth, vanishes outside a compact set, and
vanishes in an open neighborhood of $0$.
From the equation for $d^{4n}x$, and from eq.~\eqref{umux}, we
then get the following representation for $u(f)$:
\bena
u(f) &=& \int_{\mr^{4n}} u(x) f(x) \, d^{4n} x \non\\
     &=& \int_0^\infty \left( \int_{\Sigma_r} u(x) f(x) \Omega(x) \right) \frac{dr}{r} \non\\
     &=& \int_0^\infty r^{4n-1} \left( \int_{\Sigma_1} u(rx) f(rx) \Omega(x) \right) dr \non\\
     &=& \int_0^\infty \sum_{k=0}^{N-1} r^{4n-1-\rho}
\frac{(\log r)^k}{k!} \left( \int_\Sigma  v_k(x) f(r x) \, \Omega(x) \right) dr \, .
\eena
The terms in the sum
may be written as residue using the equality
\ben
r^a = \sum_k \frac{a^k (\log r)^k }{k!} \, , 
\een
where we have introduced a complex number $a \in \mc$ close to $0$. 
To get the desired residue formula, let $f_r(\hat x)$ be the
function on $\Sigma$ defined by $f(r\hat x)$. Then we may write
\ben
v_k(f_r) = \int_\Sigma v_k(x) f_r(x) \, \Omega(x) \, ,
\een
and we have
\ben\label{extdef}
u(f) = {\rm Res}_{a=0} \sum_{k=0}^{N-1} \frac{1}{a^{k+1}}
\int_0^\infty r^{a+4n-1-\rho}
v_k(f_r) \, dr \, ,
\een
This formula is well defined because, since the support of $f$ is bounded away from
the origin in $\mr^{4n}$, the distribution $r \mapsto v_k(f_r)$ is in fact a smooth
test function on $\mr_+$ whose
support is compact and bounded away from $r=0$, showing that the 
integral is an analytic function of $a \in \mc$. We would like to
define an extension $u'$ of $u$ by generalizing
formula~\eqref{extdef} to arbitrary test functions $f$ on $\mr^{4n}$ whose support
is not necessarily bounded away from the origin.
If $f$ is an arbitrary test function then $r \mapsto v_k(f_r)$ vanishes
for sufficiently large $r>r_0$, but it no longer vanishes near $r=0$.
In that case, it is not obvious that the right side
of~\eqref{extdef} is still well-defined. Finding a well-defined replacement amounts to 
finding the desired extension $u'$ of $u$. For this, we let
\ben\label{hrdef}
h_k(r):=v_k(f_r)=\int_\Sigma v_k(x) f(rx) \, \Omega(x)\, ,
\een
and we define $u'(f)$ as
\bena\label{shilovtrick}
u'(f) &:=& {\rm Res}_{a=0} \sum_{k=0}^{N-1} \frac{1}{a^{k+1}} \int_0^\infty
r^{a-\rho+4n-1} \bigg( h_k(r) \, - \,
\sum_{j=0}^{m-1} \frac{r^j}{j!}
\frac{d^j h_k(0)}{dr^{j}}
\non\\
&& \hspace{2cm} - \Theta(1-r) \frac{r^m}{m!}
\frac{d^m h_k(0)}{dr^{m}} \bigg) \, dr \ , 
\eena
where $\Theta$ is the step function, and where $m = \lfloor {\rm Re} \, \rho - 4n \rfloor$. 
We claim that $u'$ is an extension of $u$. We split the integral into a contribution from $r >1$ and one from $r \le 1$. 
Firstly, for large $r >1$, the $r$-integral is 
absolutely convergent. This is clear because $h_k(r)$ and $\Theta(1-r)$ are of compact support, and because we may assume in order to 
take the residue that $|a| < \delta \ll 1$, so that the power of $r$ of the terms under the sum over $j$ is 
at most $r^{-2+\delta}$, making the $r$-integral therefore absolutely convergent for large $r$. 
Secondly, the $r$-integral is also well defined in the range $r \le 1$. To see this, note that 
$h_k(r) - \sum_{j \le m} r^j (d^j h_k(0)/dr^j)/j!$ is formally the Taylor remainder at order $m$. 
Looking at eq.~\eqref{hrdef}, one sees that this Taylor remainder corresponds to replacing $f(rx)$
by its $m$-th order Taylor remainder, which is of order $O(r^{m+1})$. Thus, the integrand
in~\eqref{shilovtrick} is of order $r^{{\rm Re} a}$ for small $r$ and hence the integral is convergent for 
$r \le 1$ and in fact defines an analytic function of $a$ for $|a| < \delta$. Thus, the integral on the 
right side is convergent for all $r$ and defines an analytic function of $a$ near $a=0$, so that the 
expression under the residue is meromorphic in $a$ there. It can be shown
using the methods described in chapter~I, paragraph~3 of~\cite{C}
that $u'(f)$ is not just a linear functional on the space of test-functions, but defines in fact a
distribution on $\mr^{4n}$. 

Furthermore, if $f$ has its support away from
$0$, then $h_k(r) = 0$ in an open neighborhood of $r=0$, and
we have $u'(f)=u(f)$. Consequently,~\eqref{shilovtrick}
defines an extension $u'$ of the distribution $u$ in all 
cases 1), 2) and 3) of the lemma.

We next need to analyze the scaling behavior of this extension $u'$.  
A straightforward calculation using eq.~\eqref{shilovtrick}
shows that
\bena
&&(S^N_\rho u')(f) = \\
&&-{\rm Res}_{a=0}\Bigg\{
\frac{\partial^N}{\partial (\log \mu)^N} \sum_{k=0}^{N-1}
\frac{\mu^a}{a^{k+1}} \Bigg[ \frac{r^{4n-\rho+a} h_k(0)}{4n-\rho+a}+
\dots + \frac{r^{4n-\rho+a+m} \tfrac{d^m}{dr^m} h_k(0)}{m!(4n-\rho+a+m)}
\Bigg]_{r=1}^{r=1/\mu}
\Bigg\}_{\mu=1} \, .\non
\eena
If we now assume that we are in case 3), i.e.,
$\rho \in {\mathbb N}_0+4n$, then $m=\rho-4n$, and the expression
is shown to be equal to
\ben
(S^N_\rho u')(f) = \frac{\tfrac{d^{\rho - 4n}}{dr^{\rho-4n}}
h_{N-1}(0)
}{(\rho-4n)!} \, .
\een
The terms on the right side can be evaluated as follows using the definition
of $h_{N-1}(r)$ and $v_{N-1}(x)$, see eqs.~\eqref{hrdef} and~\eqref{vkdef}:
\ben
\frac{d^{\rho - 4n}}{dr^{\rho-4n}} h_{N-1}(0) = \sum_{|\alpha|=\rho-4n}
\left( \int_{\Sigma} x^\alpha S^{N-1}_\rho u(x) \, \Omega(x) \right) (\partial_\alpha f)(0) \, ,
\een
where $\alpha = (\alpha_1, \dots, \alpha_{4n}) \in \mn_0^{4n}$ is a multi-index, and we are
using the usual multi-index notation
\ben
\partial_\alpha = \frac{\partial^{|\alpha|}}{\partial x_1^{\alpha_1} \dots \partial x_{4n}^{\alpha_{4n}}},
\quad |\alpha| = \sum_i \alpha_i, \quad x^\alpha = x_1^{\alpha_1} \cdots x_{4n}^{\alpha_{4n}} \, .
\een
Alternatively, we may write
\ben\label{uprdef}
S^N_\rho u'(x) = \sum_{|\alpha|=\rho-4n} c^\alpha \partial_\alpha \delta(x)
\een
in terms of the usual $\delta$-function on $\mr^{4n}$ concentrated at the
origin. The numerical constants $c^\alpha \in \mc$ are given, in fact, by the
formula
\ben
c^\alpha = \int_\Sigma F^\alpha(x) \, ,
\een
with $F^\alpha$ the (distributional) $(4n-1)-forms$ on $\Sigma$ defined by
\ben\label{Falphadef}
F^\alpha(x) := \frac{(-1)^{\rho-4n}}{(\rho-4n)!} \, x^\alpha S^{N-1}_\rho u(x) \cdot \Omega(x)
\quad \in
{\mathcal D}'\Big( \Sigma; \wedge^{4n-1} T^* \Sigma \Big) \, .
\een
Since the
delta-function is a homogeneous distribution of degree $-4n$, we have
$S_\rho \partial_\alpha \delta = \partial_\alpha S_{4n} \delta = 0$,
and therefore $S^{N+1}_\rho u' = 0$ by eq.~\eqref{uprdef}. Thus our extension
$u'$ is again an almost homogeneous distribution.

One may repeat this argument also for case 1) and 2) of the lemma.
In those cases, one finds $S_\rho^N  u'=0$. Thus, summarizing,
eq.~\eqref{shilovtrick} defines a distributional extension $u'$ of $u$ that
is almost homogeneous.
To simplify the notation, we will from now on denote this extension
again by $u$. 

\medskip

We now investigate the Lorentz transformation properties of
$u$. Our construction of the extension $u$ given above involved a
choice of a suitable $\Sigma$ transverse to the orbits of the
dilations. Since no $\Sigma$ with the
above properties exists that is at the same time
invariant under the Lorentz group,
the extension $u$ just constructed
will in general fail to be Lorentz invariant.
Restoring the tensor indices on $u$, we find by a calculation
using eq.~(\ref{shilovtrick}) that for any test function $f \in
C^\infty_0(\mr^{4n})$ and any Lorentz transformation, $\Lambda$, we have
\begin{equation}
\label{71}
{u}_{\mu_1 \dots \mu_l}(f) -
\Lambda^{\nu_1}_{\mu_1} \dots \Lambda^{\nu_l}_{\mu_l}
{u}_{\nu_1 \dots \nu_l}(R(\Lambda)f)
= \sum_{|\alpha| \le \rho- 4n} b^\alpha_{\mu_1 \dots \mu_l}(\Lambda)
\partial_\alpha
\delta(f),
\end{equation}
where $(R(\Lambda)f)(x) = f(\Lambda^{-1} x)$ and the $b^\alpha_{\mu_1 \dots
\mu_l}(\Lambda)$ are complex constants, which would vanish if and only
if the distribution $u$ were Lorentz invariant. We now apply the
differential operator $S_{\rho}^{N+1}$ to both sides of the above
equation. Since $S_{\rho}$ is itself a Lorentz invariant operator, we
have $R(\Lambda) S_{\rho} = S_{\rho} R(\Lambda)$. Therefore,
since $S_{\rho}^{N+1}
{u} = 0$, the
operator $S_\rho^{N+1}$
annihilates the left side of eq.~\eqref{71}, so we obtain
\begin{equation}
0 = S_{\rho}^{N+1}
\sum_{|\alpha| \le {\rm Re}(\rho)- 4n} b^\alpha_{\mu_1 \dots \mu_l}(\Lambda)
\partial_\alpha \delta =
\sum_{|\alpha| \le {\rm Re}(\rho)- 4n} (\rho- 4n - |\alpha|)^{N+1}
b^\alpha_{\mu_1 \dots \mu_l}(\Lambda)
\partial_\alpha \delta.
\end{equation}
It follows immediately that $b^\alpha_{\mu_1 \dots \mu_l}(\Lambda)
= 0$, except possibly when $|\alpha| = \rho- 4n$, which evidently can only
happen when $\rho$ is an integer. Thus, focussing on that case, we have
\begin{equation}
{u}_{\mu_1 \dots \mu_l}(f) -
\Lambda^{\nu_1}_{\mu_1} \dots \Lambda^{\nu_l}_{\mu_l}
{u}_{\nu_1 \dots \nu_l}(R(\Lambda)f)
= b_{\mu_1 \dots \mu_l}^{\nu_1 \dots \nu_{\rho-4n}}(\Lambda)
\partial_{\nu_1} \dots \partial_{\nu_{\rho-4n}}
\delta(f)
\end{equation}
for all $f$ and all Lorentz-transformations $\Lambda$. Using this equation,
one finds the following transformation property for $b(\Lambda)$,
\begin{equation}\label{btrafo}
0=b(\Lambda_1 \Lambda_2) - b(\Lambda_1) - {\rm D}(\Lambda_1)b(\Lambda_2)
\equiv (\delta b)(\Lambda_1,\Lambda_2),
\end{equation}
where we have now dropped the tensor-indices and where ${\rm D}$ denotes the
tensor representation of the Lorentz-group on the
space ${\rm D}=(\otimes^l \mr^4)^*
\otimes (\otimes^{\rho-4n} \mr^4)$. This relation is of cohomological nature.
To see its relation to cohomology, one defines the following group-cohomology
rings, see e.g.~\cite{Varadarajan}:

\begin{defn}
Let $G$ be a group, $\rm D$ a representation of $G$ on a vector space $V$,
and let $c^n$ be the space of functionals $\xi_n: G^{\times n} \to V$. Let
$\delta: c^n \to c^{n+1}$ be defined by
\bena
(\delta \xi_n)(g_1, \dots, g_{n+1}) &=& {\rm D}(g_1) \xi_n(g_2, \dots, g_{n+1})
+ \sum_{i=1}^n (-1)^i \xi_n(g_1, \dots, g_i g_{i+1}, \dots, g_{n+1})\non\\
&& +(-1)^{n+1} \xi_n(g_1, \dots, g_n) \, .
\eena
Then $\delta^2 = 0$. The corresponding cohomology rings are defined as
\ben
H^n(G; {\rm D}) = \frac{ \{ {\rm Kernel} \, \delta: c^n \to c^{n+1} \} }{\{ {\rm Image} \, \delta: c^{n-1} \to c^{n} \}} \, .
\een
\end{defn}

According to this definition, eq.~\eqref{btrafo} may be viewed~\cite{pr} as
saying that $b \in H^1(SO_0(3,1); {\rm D})$.
It is a classical result of
Wigner~\cite{Wigner} that this ring is trivial for
the Lorentz group and any finite-dimensional $\rm D$. It follows
that there is an $a$ such that $b = \delta a$, or
\begin{equation}
b(\Lambda) = (\delta a)(\Lambda) \equiv a - {\rm D}(\Lambda)a \quad \forall \Lambda,
\end{equation}
where $a$ is an element in $H^0(SO_0(3,1); {\rm D}) =
{\rm D}=(\otimes^l \mr^4)^* \otimes
(\otimes^{\rho-4n} \mr^4)$.
This enables us to define a modified extension $\hat u$ by
\begin{equation}\label{uprime}
u'{}_{\mu_1 \dots \mu_l} :=
{u}_{\mu_1 \dots \mu_l} -
a_{\mu_1 \dots \mu_l}^{\nu_1 \dots \nu_{\rho-4n}}
\partial_{\nu_1} \dots \partial_{\nu_{\rho-4n}}
\delta,
\end{equation}
where we have now restored the tensor indices. It is easily checked
that $u'$ is Lorentz invariant and satisfies $S_{\rho}^{N+1} u' = 0$.
In cases 1) and 2), $u$ actually even satisfies $S_\rho^N u=0$, so the
modified extension~\eqref{uprime} even satisfies $S_\rho^N u' = 0$.
We have therefore accomplished the goal of constructing the desired
extension of $u$ in cases 1), 2) and 3).

The uniqueness statement immediately follows from the fact that the
difference between any two extensions has to be a Lorentz-invariant
derivative of the delta-function, $L\delta$, such that $S_\rho^{N+1}
L\delta = 0$. Thus, $L$ can be non-zero only when $\rho$ is an
integer, and $L$ must have degree of precisely
$\rho-4n$.
\qed

\medskip

From the proof of the lemma, we get the following interesting proposition:

\paragraph{Proposition 1:}\label{prop1}
Let $u(x)$ be a Lorentz invariant (possibly tensor-valued) distribution on $\mr^{4n}
\setminus 0$ which scales almost homogeneously with degree $\rho \in 4n+{\mathbb N}_0$
under coordinate rescalings, i.e.,
\begin{equation}\label{sscale}
S_{\rho}^N u(x) = 0 \quad \text{
for some natural number $N$, $x \neq 0$,}
\end{equation}
Then $u$ has a Lorentz invariant extension, also denoted $u$, to a distribution on
$\mr^{4n}$ which also scales almost homogeneously with degree $\rho$
under rescalings of the coordinates. We have
$S^{N+1}_\rho u = 0$, and
\ben\label{uprdef}
S^N_\rho u(x) = \sum_{|\alpha|=\rho-4n} c^\alpha \partial_\alpha \delta(x)
\een
in terms of the usual $\delta$-function on $\mr^{4n}$ concentrated at the
origin. The numerical constants $c^\alpha \in \mc$ are Lorentz-invariants,
and are given by the
formula
\ben\label{cal}
c^\alpha = \int_\Sigma F^\alpha(x) \, ,
\een
where $\Sigma \subset \mr^{4n}$ is {\em any} closed $(4n-1)$ submanifold
enclosing the origin $0 \in \mr^{4n}$ which is transverse to the orbits of
to the dilations of $\mr^{4n}$. Here,
the distributional $(4n-1)$-forms $F^\alpha \in \D'(\Sigma; \wedge^{4n-1} T^* \Sigma)$
on $\Sigma$ are defined in eq.~\eqref{Falphadef}, and are closed,
\ben
dF^\alpha = 0 \, .
\een

\medskip
\noindent
{\em Proof:} That the $c^\alpha$ are Lorentz invariants is obvious because the extension $u$
is Lorentz-invariant and $S_\rho$ commutes with Lorentz transformations. So we only need to show that the $(4n-1)$-forms $F^\alpha$ are closed. We first compute
\ben
d\Omega(x) = 4n \, d^{4n} x \,
\een
using the definition of the $(4n-1)$-form $\Omega$, see eq.~\eqref{Odef}.
By a straightforward computation using the definition of $\Omega$, we
also have
\ben
d[x^\alpha S^{N-1}_\rho u(x)] \wedge \Omega(x) =
x^\alpha (S_0+|\alpha|) [(S_0-\rho)^{N-1} u(x)] \, d^{4n} x \, \, .
\een
Using next the fact that $|\alpha| = \rho-4n$, and that $S_\rho^N u = (S_0-\rho)^N u = 0$,
we find
\ben
d[x^\alpha S^{N-1}_\rho u(x)] \wedge \Omega(x) = -4n \, x^\alpha S^{N-1}_\rho u(x) \,
d^{4n} x \,
\een
so
\bena
dF^\alpha(x) &=& \frac{(-1)^{\rho-4n}}{(\rho-4n)!} \,
d[x^\alpha S_\rho^{N-1} u(x) \, \Omega(x)] \\
&=& \frac{(-1)^{\rho-4n}}{(\rho-4n)!} \left\{  d[x^\alpha S_\rho^{N-1} u(x)] \wedge \Omega(x) +
x^\alpha S^{N-1}_\rho u(x) \, d\Omega(x) \right\} = 0 \, .
\eena
\qed

\medskip
\noindent
{\bf Remark:}
If we did not already know that $u$ was Lorentz invariant, it would at first sight appear from eq.~\eqref{cal}
somewhat surprising that the $c^\alpha$ have these properties, given that the surface $\Sigma$ appearing on
the right side must be compact, and thus cannot possibly be Lorentz invariant. To see explicitly that this is 
nevertheless the case, one can proceed as follows. We would like to see explicitly that
$\Lambda^\alpha_\beta c^\beta = c^\alpha$ for any Lorentz transformation. Indeed, 
\bena
\Lambda^\alpha_\beta c^\beta &=& \int_\Sigma \Lambda^\alpha_\beta F^\beta(x)\non\\
                             &=& \int_{\Lambda^* \Sigma} \Lambda^\alpha_\beta F^\beta(\Lambda^{-1} x)\non\\
                             &=& \int_{\Lambda^* \Sigma} F^\alpha(x) \non\\
                             &=& \int_\Sigma F^\alpha(x) + \int_U dF^\alpha(x) \non\\
                             &=& c^\alpha \, .
\eena
Here we have used in the first step the definition of $c^\alpha$, in the second step we have
used the standard transformation formula of an integral under a diffeomorphism, denoting by
$\Lambda^* \Sigma$ image of $\Sigma$ under the natural action of $\Lambda$ on $\mr^{4n}$.
In the third step we have used that $F^\alpha$ itself is Lorentz invariant, and in the
fourth step we have used Stoke's theorem for the open set $U \subset \mr^{4n}$
such that $\partial U = -\Sigma \cup \Lambda^* \Sigma$, and in the
fifth step we used $dF^\alpha = 0$.

\medskip

In summary, we have now described how to construct the time ordered products
$T_n(\otimes_{i=1}^n \phi^{k_i})$ of Wick
monomials without derivatives. These construction can in principle be
generalized to time ordered products of Wick monomials $\O_i$
containing derivatives by generalizing the Wick expansion to fields
with derivatives. A non-trivial new renormalization condition
now arises from T10, because $S_0$ contains derivatives. This
condition is not automatically satisfied, but it is
not difficult to see that we can change, if necessary, our construction of the time
ordered products, so as to also satisfy T10~\cite{Hollands2005}.

We finally have to consider condition T11. This condition is
satisfied by our construction for $T_1$, but not in general for $T_n$
when $n>1$. The operational meaning
of this requirement is that ``derivatives can be freely pulled through
the time-ordering symbol''. This identity is a non-trivial requirement
because both sides of the equation mean quite different things a
priori: The first expression means the time ordered product of fields,
one of which contains a total derivative, the second expression denotes
the derivative, in the sense of distributions, of the algebra valued
distribution given by the time ordered product of the fields without
the total derivative. That these two quantities are actually the same
is not obvious from the above construction, and is therefore an
additional renormalization condition, called the ``action Ward
identity'' in~\cite{Duetsch2004}, and the ``Leibniz rule''
in~\cite{Hollands2005}. It is shown in these two references how,
starting from a prescription that satisfies T1---T10 but
possibly does not satisfy this
renormalization condition, one can go to a prescription which does.

The action Ward identity is at odds
with conventions often found in standard textbooks on field theory
in Minkowski spacetime~\cite{Weinberg1996}, where
the derivative is not taken to commute with $T_n$. To illustrate this
difference in point of view, consider the time ordered product
$T_2(\phi(x) \otimes \phi(y))$. According to condition T11,
we have $(\square_x-m^2) T_2(\phi(x) \otimes \phi(y))
= T_2((\square_x - m^2)\phi(x) \otimes \phi(y))$. In our approach, the time
ordered products need not vanish when acting on a factor of the wave
equation, so this quantity does not need to vanish. In fact, one can
see that the time-ordered product under consideration
is uniquely determined by the properties T1---T10, and
we have $T_2((\square_x - m^2)\phi(x) \otimes \phi(y)) = i\hbar\delta(x,y)
\myid$. In standard approaches, on the other hand, it is assumed that
the time ordered product vanishes when acting on $(\square_x -m^2) \phi(x)$,
because the time-ordering symbol is viewed as on operation
acting on on-shell quantized fields, rather than just
classical polynomial expressions in $\P$. On the other hand,
in most standard approaches, it is not assumed that derivatives commute with $T_2$. In this way,
one reaches the same conclusion for the example just considered, and
both viewpoints are consistent for that example.
However, the standard viewpoint gets very awkward in general when
considering more complicated time ordered products of fields with
derivatives, for a discussion see e.g.~\cite{Duetsch2002}. This is because it is in
general inconsistent to assume that
a time ordered product containing a factor $\O \square \phi$ vanishes,
because of possible anomalies. On the other
hand, the Leibniz rule can always be satisfied, and possible anomalies
can thereby be analyzed consistently.

\subsection{Examples}

Here we illustrate the above general construction of the time-ordered product
by some simple examples. The simplest non-trivial example of a time
ordered product with one factor is
$T_1(\phi^2(x))=:\phi^2(x):_\h$. Using the definition of the locally
normal ordered product, this may be viewed as a ``point-splitting''
definition, see e.g.~\cite{DeWitt1962}.
%\ben
%T_1(\phi^2(x)) = \lim_{y \to x} :\phi(x)\phi(y):_\h =
%\lim_{y \to x} [ \phi(x)\phi(y) - H(x,y) \myid ] \, .
%\een
Consider next the time ordered product $T_2(\phi^2(x_1) \otimes
\phi^2(x_2))$. By T8, it is defined
for non-coincident points $x_1 \neq x_2$ by the prescription
\ben
T_2(\phi^2(x_1) \otimes \phi^2(x_2)) =
\begin{cases}
:\phi^2(x_2):_\h \st :\phi^2(x_1):_\h
& \text{when $x_1 \notin J^+(x_2)$;}\\
:\phi^2(x_1):_\h \st :\phi^2(x_2):_\h
& \text{when $x_1 \notin J^-(x_2)$.}
\end{cases}
\een
In order to extend the definition to coincident points
$x_1=x_2$, i.e., to make the time-ordered product a well
defined distribution on the entire product manifold $M^2$, we now
use the expansion procedures described in general in the previous
section. Using the definition of the product $\st$, and of the locally normal
ordered products, we have
\begin{multline}
:\phi^2(x_1):_\h \st :\phi^2(x_2):_\h = \\
: \phi^2(x_1) \phi^2(x_2) :_\h
- 2 \hbar H(x_1,x_2) : \phi(x_1) \phi(x_2) :_\h +
\hbar^2 H(x_1,x_2)^2 \, \myid \, ,
\end{multline}
for points $x_1,x_2$ that are sufficiently close to each other so that
the local Hadamard parametrix $H(x_1,x_2)$ is well-defined.
Using furthermore the definition of the local Feynman parametrix
$H_F$ (see eq.~\eqref{Hfdef}) and
\ben
\Theta(T(x)-T(y)) \, H(x,y) + \Theta(T(y)-T(x)) \, H(y,x) = iH_F(x,y)
\een
with $\Theta$ the step function, 
we can write the time ordered product under consideration as
\begin{multline}
T_2(\phi^2(x_1) \otimes \phi^2(x_2)) = \\
: \phi^2(x_1) \phi^2(x_2) :_\h
+2(\hbar/i) H_F(x_1,x_2) : \phi(x_1) \phi(x_2) :_\h +
(\hbar/i)^2 H_F(x_1,x_2)^2 \, \myid \, ,
\end{multline}
for non-coinciding points $x,y$. This is the desired local
Wick-expansion. Comparing with eq.~\eqref{wickexp00}, we read off
\ben
t_{0,0}(x_1,x_2) = 1, \quad t_{1,1}(x_1,x_2) = (\hbar/i) H_F(x_1,x_2),
\quad t_{2,2}(x_1,x_2) = (\hbar/i)^2 H_F(x_1,x_2)^2
\een
for the coefficients in the Wick expansion. The coefficients
$t_{0,0}, t_{1,1}$ may be extended to coincident points $x=y$
by continuity, because their scaling degree is 0 resp. 2, which is
less than 4, but the distribution $t_{2,2}$ has scaling degree 4
and therefore cannot be extended to the diagonal by continuity, but
must instead be extended non-trivially. Actually, since $t_{2,2}$
is the square of the distribution $H_F$ with singularities on the
lightcone, it is instructive to check explicitly
that it is even defined for non-coincident points that are on the lightcone. This can be done
using the wave front set: For $x_1 \notin J^+(x_2)$, the pair
$(x_1, k_1; x_2, k_2) \in T^*(M^2)$ is in the wave front set of $H_F$ (see appendix C) if and
only if $x_1$ and $x_2$ can be joined by a null-geodesic $\gamma:
(0,1) \to M$, with $\dot \gamma(0) = k_1$ and $\dot \gamma(1) = -k_2$,
with $k_1 \in V^*_+$. Similarly, for $x_1 \notin J^-(x_2)$, the pair
$(x_1, k_1; x_2, k_2) \in T^*(M^2)$ is in the wave front set if and
only if $x_1$ and $x_2$ can be joined by a null-geodesic $\gamma:
(0,1) \to M$, with $\dot \gamma(0) = k_1$ and $\dot \gamma(1) = -k_2$,
with $k_1 \in V^*_-$. It follows that, when $x_1 \neq x_2$, elements
$(x_1, k_1, x_2, k_2) \in \WF(H_F)$ can never add up to the zero
element. Thus, by the general theorems about the wave front set
summarized in appendix~C, arbitrary powers $H_F(x_1,x_2)^n$
exist in the distributional sense, i.e., as distributions on $M^2
\setminus \Delta_2$. On the other hand, when $x_1=x_2$, arbitrary
elements of the form $(x_1, k, x_2, -k)$ are in $\WF(H_F)$. Thus,
for coincident points, the elements in the wave front set can add up
to zero, and the product $H_F(x_1,x_2)^n$ is therefore not defined as
a distribution on all of $M^2$, i.e., including coincident points.

In order to extend $t_{2,2}$ to a well-defined distribution to
all of $M^2$, we now need to perform the scaling expansion of $t_{2,2}$,
which in turn can be obtained from the scaling expansion of $H_F$. The
latter can be found using expansions for the recursively defined
coefficients in the
local Hadamard parametrix, see e.g.~\cite{DeWitt1962}. Up to numerical prefactors,
it is given by (we assume for simplicity that $m^2=0$)
\ben
H_F(\exp_y \xi,y) \sim \frac{1}{\xi^2 + i0} + R_{\mu\nu}(y) \bigg(
-\frac{1}{6} \frac{\xi^\mu \xi^\nu}{\xi^2 + i0} + \frac{1}{12}
\eta^{\mu\nu} \log(\xi^2 + i0)
\bigg) + \dots ,
\een
where the dots stand for a remainder with scaling degree $<2$, where $\xi \in T_y M$
has been identified with a vector in $\mr^4$ via a tetrad, and where
$\xi^2 = \eta_{\mu\nu} \xi^\mu \xi^\nu$. From this we obtain
the first terms in the scaling expansion of $t_{2,2}$ up to numerical
prefactors as
\ben
t_{2,2}(\exp_y \xi,y) \sim u(\xi) + R_{\mu\nu}(y) \, u^{\mu\nu}(\xi) + \dots
\een
where the dots stand for terms of scaling degree less than 2. The
distributions $u$ and $u^{\mu\nu}$ are
defined on $\mr^4 \setminus 0$ and is given
there by
\ben\label{uu}
u(\xi) =
\frac{1}{(\xi^2 + i0)^2} \, , \quad u^{\mu\nu}(\xi)=
-\frac{1}{3} \frac{\xi^\mu \xi^\nu}{(\xi^2 + i0)^2} + \frac{1}{6}
\frac{\eta^{\mu\nu} \log(\xi^2 + i0)}{\xi^2 + i0} \, .
\een
$u$ has
scaling degree 4, while $u^{\mu\nu}$ has scaling degree 2. Thus, by
lemma~\ref{lemma4}, we
need to extend non-trivially only $u$, while $u^{\mu\nu}$ and
the remainder (i.e., the dots in the scaling expansion of
$t_{2,2}$) can be extended by continuity. An extension to all of $\mr^4$
(i.e., including $\xi=0$) of $u$ can easily be guessed, but we here
prefer to give a systematic method, which is needed anyway in
more complicated examples. A constructive method to obtain an
extension of $u$ is provided by lemma~\ref{lemma4}. However, that
has the disadvantage of being somewhat complicated because it
involves a non-Lorentz invariant surface $S$ at intermediate
steps, which is awkward in concrete calculations\footnote{Note,
  however, that this is not an obstacle in the corresponding
``Euclidean situation'', where one may take $S$ simply to be a
Euclidean sphere.}. Instead we here
present a different method, that is more practical and works in a
wide class of examples. That method is based upon the fact that,
for complex scaling degree, there is a unique extension of
a homogeneous distribution by lemma~\ref{lemma4}.
The method has also appeared in the
context of BPHZ-renormalization in momentum space under the name
``analytic renormalization''~\cite{Speer0,Speer1,Speer2}.

Consider instead of $u$ the distribution
given by
\ben
u_a(\xi) = \frac{1}{(\xi^2 + i0)^{2-a}}, \quad a \in \mc \setminus
{\mathbb Z} \, .
\een
By contrast to $u$, this is well defined on all of $\mr^4$, see e.g.~\cite{C}, and
also~\cite{GB} for a treatment of such so-called ``Riesz-distributions''.
An extension $u'$ of $u$ can now be obtained by taking the residue
of the meromorphic function $a \mapsto u_a(f)/a$,
\ben\label{EX}
u'(f) = {\rm Res}_{a=0} \frac{u_a(f)}{a} \, .
\een
Indeed, if the support of $f$ excludes $0$, then $u'(f)$ obviously
must coincide with $u(f)$, because we may then use formula~\eqref{uu}
to get $u(f)$. The almost homogeneous scaling property of $u'(f)$
under rescalings of $f(\xi) \to f(\mu \xi)$ also immediately follows from the
definition. To get a more explicit formula for the extension, we
compute the fourier transform of $u_a$, given up to numerical factors by~\cite{GB}
\ben\label{ua}
\hat u_a(p) = 4^a \frac{\Gamma(a)}{\Gamma(2-a)}(p^2-i0)^{-a} \, .
\een
We expand this expression around $a=0$ using the well-known residue of the
$\Gamma$-function at 0 and substitute the resulting expression
into eq.~\eqref{EX}. We obtain, up to numerical prefactors
\ben
\hat u'(p) = \ln [l^2 (p^2 -i0)]
\een
where $l$ is some constant. Taking an inverse
fourier transform then gives the desired extension
\ben
u'(\xi) = -\frac{1}{2}
\partial^2 \bigg( \frac{\log [l^{-2} (\xi^2 +i0)]}{\xi^2 + i0}\bigg)
\, .
\een
where $\partial^2 = \eta^{\mu\nu} \partial^2/\partial \xi^\mu \partial
\xi^\nu$. Note that the extension has acquired a
logarithm, which is a general phenomenon according to
lemma~\ref{lemma4}. Different choices of $l$ change the extension
by a term proportional to $\delta^4(\xi)$, and thus correspond to
the different extensions of $u(\xi)$.
Thus, inserting this extension into the scaling
expansion of $t_{2,2}$, we obtain
the desired extension of $T_2(\phi^2(x_1) \otimes
\phi^2(x_2))$.

Our last example is the time ordered product $T_3(\phi^3(x_1) \otimes
\phi^3(x_2) \otimes \phi^4(x_3))$ with 3 factors. The terms in the
Wick expansion of this quantity that need to be extended
non-trivially from $M^3 \setminus \Delta_3$ to $M^3$ are
\bena
t_{3,3,2}(x_1,x_2,x_3) &=& t_{1,1}(x_1,x_2) t_{1,1}(x_2,x_3)
t_{2,2}(x_1,x_3), \\
t_{3,3,4}(x_1,x_2,x_3) &=& t_{1,1}(x_1,x_2) t_{2,2}(x_2,x_3)
t_{2,2}(x_1,x_3) \, .
\eena
All other terms are either already
well-defined as distributions on all of $M^3$ (assuming the
corresponding time ordered products with 2 factors have been defined),
or can be extended by continuity. We focus on the last term
$t_{3,3,4}$. Again, for the sake of illustration of the
general construction, we first verify explicitly that this
distribution is indeed well-defined on $M^3 \setminus \Delta_3$.
Consider a point $(x_1, x_2, x_3) \notin \Delta_3$. Then it must be
possible to separate one point, from the remaining two points by a Cauchy
surface. For definiteness, let us assume that this point is $x_3$, and
that $x_1, x_2 \notin J^+(x_3)$. Then $(x_1, k_1; x_3, k_3)$ is in the
wave front set of $t_{2,2}(x_1,x_3)$ if and only if $k_1 \sim -k_3$,
and if $k_1 \in V^*_+$. Likewise, $(x_2, p_2; x_3, p_3)$ is in the
wave front set of $t_{2,2}(x_2,x_3)$ if and only if $p_2 \sim -p_3$,
and if $p_2 \in V^*_+$. Finally $(x_1,q_1; x_2, q_2)$ is in the wave
front set of $t_{1,1}(x_1,x_2)$ iff $q_1 \sim -q_2$ and
$q_1 \in V_{\pm}^*$ when $x_1 \notin J^\pm(x_2)$, or
iff $q_1=-q_2$ when $x_1=x_2$. We now add up these wave front set
elements, viewed in the obvious way as elements in $T^*_{x_1} M \times
T^*_{x_2} M \times T^*_{x_3} M$. We obtain the set
\ben
S = \{(x_1, k_1+q_1; x_2, p_2+q_2; x_3, k_3+p_3)\} \, .
\een
Assume first that $x_1=x_2$.
Clearly, if e.g. $k_1+q_1=0$, then $q_1 \in V^*_-$, so $p_2+q_2=p_2-q_1 \neq
0$, because $p_2 \in V^*_+$. Thus, $S$ cannot contain the zero
element, and the product defining $t_{3,3,4}$ is well-defined near
$(x_1,x_2,x_3)$ by thm.~\ref{thm8}. Similarly, if $x_1 \notin J^-(x_2)$, then $q_2 \in
V^*_+$, and again $p_2+q_2 \neq 0$, and again, $S$ cannot contain the
zero element.
The same type of argument can be made for all other
configurations of the points, except the configuration $x_1=x_2=x_3$.
Thus, by the general existence theorem~\ref{thm8} for products of
distributions, $t_{3,3,4}$ is indeed well-defined as a distribution on $M^3
\setminus \Delta_3$.

We next would like to construct an extension of $t_{3,3,4}$ along the
lines of our general construction. Thus, we must determine the scaling
expansion of $t_{3,3,4}$. It can be obtained
from the expansions
of the (extended) distributions $t_{2,2}$ and of $t_{1,1}$ that were
constructed above. We
focus on the terms that require a non-trivial extension
(up to numerical prefactors):
\ben\label{162}
t_{3,3,4}(\exp_y \xi_1, \exp_y \xi_2, y) \sim
u(\xi_1, \xi_2)
+ R_{\mu\nu}(y) u^{\mu\nu}(\xi_1, \xi_2)
+R_{\mu\nu\sigma\rho}(y) u^{\mu\nu\sigma\rho}(\xi_1,\xi_2) +
\dots \, ,
\een
where $u$ is the distribution defined on $(\mr^4)^2 \setminus
0$ given by
\ben
u(\xi_1, \xi_2) =
\frac{1}{4} \partial^2_1
\bigg( \frac{\log [l^{-2}(\xi^2_1 + i0)]}{\xi^2_1 + i0} \bigg)
\partial^2_2
\bigg( \frac{\log [l^{-2}(\xi^2_2 + i0)]}{\xi^2_2 + i0} \bigg)
  \frac{1}{(\xi_1-\xi_2)^2+i0}\non\\
\een
where $u^{\mu\nu}$ is the distribution defined on $(\mr^4)^2 \setminus
0$ given by
\begin{multline}
u^{\mu\nu}(\xi_1, \xi_2) = \\
-\frac{1}{2}
\partial^2_1 \bigg( \frac{\log [l^{-2} (\xi^2_1 + i0)]}{\xi^2_1 + i0}
\bigg)\bigg(
-\frac{1}{3} \frac{\xi^\mu_2 \xi^\nu_2}{(\xi^2_2 + i0)^2} + \frac{1}{6}
\frac{\eta^{\mu\nu} \log [l^{-2} (\xi^2_2 + i0)]}{\xi^2_2 + i0}
\bigg) \frac{1}{(\xi_1-\xi_2)^2+i0}
+ (\xi_1 \leftrightarrow \xi_2)
\end{multline}
and where $u^{\mu\nu\sigma\rho}$ is the distribution on $(\mr^4)^2
\setminus 0$ defined by
\begin{multline}
u^{\mu\nu\sigma\rho}(\xi_1, \xi_2) =
\frac{1}{4}
\partial^2_1 \bigg( \frac{\log [l^{-2} (\xi^2_1 + i0)]}{\xi^2_1 + i0}
\bigg)
\partial^2_2 \bigg( \frac{\log [l^{-2} (\xi^2_2 + i0)]}{\xi^2_2 + i0}
\bigg) \\
\cdot \bigg(
-\frac{1}{6} \frac{\xi^\mu_1 \xi^\sigma_1 \xi^\nu_2 \xi^\rho_2}{[
(\xi_1-\xi_2)^2+i0]^2}
-\frac{1}{12} \frac{\eta^{\mu\sigma}(
\xi^\nu_1 \xi^\rho_2+2\xi^\nu_1 \xi_1^\rho)}{(\xi_1-\xi_2)^2+i0}
+ \frac{1}{24}
\eta^{\mu\sigma} \eta^{\nu\rho} \log \{l^{-2} [(\xi_1-\xi_2)^2+i0] \}
\bigg) \\
+ (\xi_1 \leftrightarrow \xi_2)
\end{multline}
The dots in eq.~\eqref{162} again represent a remainder. This now has
scaling degree 6 and can thus be extended by continuity, while the
3 terms in the scaling expansion that are explicitly given have
scaling degree 10 for the first term respectively 8 for the second and
third term. They must thus be extended non-trivially. The extension of the
corresponding distributions $u,u^{\mu\nu},u^{\mu\nu\sigma\rho}$ now
can no longer be found by trial and error, but one must use a
constructive method, such as that given in the proof of lemma~\ref{lemma4}.
We will again not use this method here, but instead use a variant
of the method given above. For this, we consider the
distribution
\ben
u_{a,b,c}(\xi_1, \xi_2) = \frac{1}{(\xi_1^2+i0)^{2-a}(\xi_2^2+i0)^{2-b}
[(\xi_1-\xi_2)^2+i0]^{2-c}} \, .
\een
It can be checked using wave-front arguments similar to that given
above that this distributional product is well-defined
on $(\mr^4)^2 \setminus 0$ for $a,b,c \in \mc \setminus {\mathbb Z}$.
Furthermore, by Lemma~\ref{lemma4}, if $a+b+c \notin {\mathbb Z}$ this
distribution has a unique extension to all of $(\mr^4)^2$.
We define the desired extension of $u$ by the expression
\ben\label{uext}
u'(f) = {\rm Res}_{c=1} {\rm Res}_{b=0} {\rm Res}_{a=0} \,
\frac{u_{a,b,c}(f)}{a b (c-1)} \, .
\een
This is an extension, because one can check that $u'(f)$ conicides
with $u(f)$ for any $f$ whose support excludes $\xi_1=\xi_2=0$, and
it is also clearly Lorentz invariant and has the
desired almost homogeneous scaling behavior. To
get a more explicit expression for $u'$, we perform a fourier
transformation of $u_{a,b,c}$
using eq.~\eqref{ua} and eq.~(23) of~\cite{Davidychev}. This gives,
up to numerical factors
\begin{eqnarray}
\hat u_{a,b,c}(p_1,p_2) =
\frac{4^{a+b+c}}{\Gamma(4-a-b-c)\Gamma(2-a)\Gamma(2-b)\Gamma(2-c)}
I_{a,b,c}(p_1, p_2)
\end{eqnarray}
where
\begin{eqnarray*}
&&I_{a,b,c}(p_1,p_2) = \\
&&
[(p_1+p_2)^2-i0]^{2-a-b-c}
\Gamma(c)\Gamma(a+b+c-2)\Gamma(2-a-c)\Gamma(2-c-b)
\times\\
&&F_4 \Bigg(c,a+b+c-2,a+c-1,b+c-1 \Bigg|
\frac{p_1^2}{(p_1+p_2)^2-i0},
\frac{p_2^2}{(p_1+p_2)^2-i0} \Bigg)+\\
&&
[(p_1+p_2)^2-i0]^{-a} (p_2^2-i0)^{2-b-c}
\Gamma(a)\Gamma(2-b)\Gamma(2-a-c)\Gamma(b+c-2)
\times\\
&&F_4 \Bigg( a,2-b,a+c-1,3-b-c \Bigg|
\frac{p_1^2}{(p_1+p_2)^2-i0},
\frac{p_2^2}{(p_1+p_2)^2-i0} \Bigg)+\\
&&
[(p_1+p_2)^2-i0]^{-b} (p_1^2-i0)^{2-a-c}
\Gamma(b)\Gamma(2-a)\Gamma(a+c-2)\Gamma(2-c-b)
\times\\
&&F_4 \Bigg(b,2-a,3-a-c,b+c-1 \Bigg|
\frac{p_1^2}{(p_1+p_2)^2-i0},
\frac{p_2^2}{(p_1+p_2)^2-i0} \Bigg)+\\
&&
[(p_1+p_2)^2-i0]^{c-2} (p_1^2-i0)^{2-a-c} (p_2^2-i0)^{2-b-c}\times\\
&&\Gamma(4-a-b-c)\Gamma(2-c)\Gamma(a+b-2)\Gamma(b+c-2)
\times\\
&&F_4 \Bigg( 4-a-b-c,2-c,3-a-c,3-b-c \Bigg|
\frac{p_1^2}{(p_1+p_2)^2-i0},
\frac{p_2^2}{(p_1+p_2)^2-i0} \Bigg) \, .
\end{eqnarray*}
Here, $F_4$ is the Appell function, defined by
\ben
F_4(\alpha,\beta,\gamma,\delta|z_1,z_2) = \sum_{j_1,j_2=0}^\infty
\frac{(\alpha)_{j_1+j_2}(\beta)_{j_1+j_2}}{(\gamma)_{j_1}(\delta)_{j_2}}
z_1^{j_1} z_2^{j_2} \, ,
\een
with $(\alpha)_j$ the Pochhammer symbol.
The fourier transform of the extension is then given by
\ben
\hat u'(p_1,p_2) = {\rm Res}_{c=1} {\rm Res}_{b=0} {\rm Res}_{a=0}
\frac{\hat u_{a,b,c}(p_1,p_2)}{ab(c-1)} \, ,
\een
which may be evaluated readily using the Laurent expansion of the
Gamma-function.
It is worth noting that the
extension $u'$ given by expression~\eqref{uext}
now implicitly contains third powers of the logarithm, thus again
confirming the general theorem that there are logarithmic corrections
to the naively expected homogeneous scaling behavior.

\subsection{Ghost fields and vector fields}

The above algebraic construction of Wick-powers and their time-ordered
products may be generalized to a multiplet
of scalar or tensor fields satisfying a system of wave equations on $M$ with local covariant
coefficients or to Grassmann valued fields. In the BRST approach to gauge theory, the relevant fields are
(gauge fixed) vector fields, and ghost fields.

Classical ghost fields are valued in the
Grassmann algebra $E$. For gauge theory, the relevant ghost fields
are described, at the free level, by the Lagrangian
\ben
\L_0 = -i d\bar C \wedge * dC \, .
\een
The fields $C,\bar C$ are independent and
take values in the Grassmann algebra $E$.
In particular, the ``bar'' over $\bar C$ is a purely conventional notation and
is {\em not} intended to mean any kind of conjugation. The
non-commutative *-algebra $\W$ corresponding to this classical
Lagrangian is described as follows. As above, we consider
a bi-distribution $\omega^{\rm s}(x,y)$ on $M \times M$ of Hadamard form
(we put a superscript ``s'' for ``scalar''), and
we consider distributions $u$ on $M^{n}$ which are
{\em anti}-symmetric in the variables, and which satisfy the
wave-front condition~\eqref{wfs1}. With each such distribution,
we associate a generator $F(u)$, which we (purely formally) write as
\begin{multline}
F(u) = \\
\int u(x_1, \dots, x_n; y_1, \dots, y_m) :C(x_1) \cdots C(x_n)
\bar C(y_1) \cdots \bar C(y_m):_\omega
\, dx_1 \dots dx_n \, dy_1 \dots dy_n \, .
\end{multline}
We now define a $\st$-product between such generators.
This is again defined by
eq.~\eqref{star}, where the derivative operator~\eqref{derivative}
is now given by
\ben
{}_< {\mathcal D}_> =
-i \int
\frac{\delta_L}{\delta C(x)}
\omega^{\rm s}(x,y)
\frac{\delta_R}{\delta \bar C(y)}  -
\frac{\delta_L}{\delta \bar C(x)}
\omega^{\rm s}(x,y)
\frac{\delta_R}{\delta C(y)}
\, dx dy \, .
\een
Here, as above, it is understood that a functional derivative acting
on $F(u)$ is executed by formally treating the fields in the normal
ordered expression as classical fields, i.e., by formally identifying
$:C(x_1) \cdots C(x_n)\bar C(y_1) \cdots \bar C(y_m):_\omega$ with
the classical field expression. The
operation * of conjugation is defined as $C(x)^* = C(x)$ and
$\bar C(x)^* = \bar C(x)$. This is consistent with the product.
It leads to the anti-commutation relations for the ghost fields,
\bena\label{ccomm}
\bar C(x) \st C(y) + C(y) \st \bar C(x) &=& \hbar
\Delta^{\rm s}(x,y) \myid \,
, \\
C(x) \st C(y) + C(y) \st C(x) &=&
\bar C(x) \st \bar C(y) + \bar C(y) \st \bar C(x) = 0 \, ,
\eena
where we have put a superscript on ``s'' the scalar causal propagator
$\Delta^{\rm s}$ to distinguish it from the vector propagator
below. The field equations may be implemented, as in the scalar case, by dividing $\W$ by the ideal $\I$
generated by $\square C(x)$ and $\square \bar C(x)$.
Time-ordered products of Grassmann fields are also defined in the same
way as above, the only minor difference being that they are not
symmetric in the tensor factors, but have graded symmetry
according to the Grassmann parity of the arguments. For example,
T6 reads instead
\ben\label{Antisym}
T_n( \dots \otimes \O_1(x_j) \otimes \O_2(x_{j+1}) \otimes \dots )
=
(-1)^{\epsilon_j \epsilon_{j+1}}
T_n( \dots \otimes \O_2(x_{j+1}) \otimes \O_1(x_{j}) \otimes \dots ) \, .
\een
There are similar signs also in T9.

\medskip

We next consider 1-form (or vector) fields, $A$. In the Lorentz gauge,
their classical dynamics is described by the Lagrangian
\ben
\L_0 = \frac{1}{2}(dA \wedge *dA +  \delta A \wedge *\delta A) \, .
\een
where $\delta = * d *$ is the co-differential (divergence).
Their equation of motion is the canonical wave equation for vectors,
$(d\delta + \delta d)A=0$, or
\ben\label{vec}
(g_{\mu\nu} \square + R_{\mu\nu}) A^\nu = 0 \,
\een
in component notation.
It is seen from the component form of the equation that it is
hyperbolic in nature, and hence has unique
fundamental retarded and advanced solutions,
$\Delta^{\rm v}_A$ and $\Delta^{\rm v}_R$, where we have
put a superscript ``v'' in order to distinguish them from
their scalar counterparts.

To define the corresponding quantum algebra of observables,
we proceed by analogy with the scalar case. For this, we pick an arbitrary
distribution $\omega^{\rm v}$ taking values in $T^*M \times T^*M$ of
Hadamard form. Thus, $\omega^{\rm v}(x,y)$ satisfies the
vector equations of motion~\eqref{vec} in $x$ and $y$, its
anti-symmetric part is given by $i\Delta^{\rm v}(x,y)$, where
$\Delta^{\rm v}$ is the difference between the fundamental advanced
and retarded vector causal propagators, and its wave-front set
is given by eq.~\eqref{wfs2}. The algebra $\W$ is generated by
expressions of the form
\ben
F(u) = \int u(x_1, \dots, x_n) \, :A(x_1) \dots A(x_n):_\omega
\, dx_1 \dots dx_n \, ,
\een
where $u(x_1, \dots, x_n)$ is a distribution with wave front
set~\eqref{wfs1}, now taking values in the bundle
$TM \times \dots \times TM$, and the *-operation is declared
by $A(x)^* = A(x)$. The $\st$-product is again defined by
eq.~\eqref{star}, where the derivative operator~\eqref{derivative}
is now given by
\ben
{}_< {\mathcal D}_> =
%\frac{1}{2}
\int
\frac{\delta_L}{\delta A(x)}
\omega^{\rm v}(x,y)
\frac{\delta_R}{\delta A(y)}
\, dx dy \, .
\een
From this, we can calculate the commutation relations for the field
$A(x) = :A(x):_\omega$,
\ben\label{acomm}
A(x) \st A(y) - A(y) \st A(x) = i\hbar \,
\Delta^{\rm v}(x,y) \, \myid \, .
\een
The construction of Wick powers and their time-ordered products is
completely analogous to the scalar case, the only difference is that
the Hadamard scalar parametrix $H$ must be replaced by a vector
Hadamard parametrix, whose construction is described in
Appendix~D.2.

\subsection{Renormalization ambiguities of the time-ordered products}

In the previous section, we have described the construction of local
and covariant renormalized time ordered products in globally
hyperbolic Lorentzian curved spacetimes. We now address the issue
to what extend the time ordered products are unique. Thus, suppose
we are given two
prescriptions, called $T =\{T_n\}$ and $\hat T = \{\hat T_n\}$, satisfying the conditions T1---T11. We would like to know how they can differ.
To characterize the difference, we introduce a hierarchy $D=\{D_n\}$ of linear functionals with the following properties. Each $D_n$ is a linear map
\ben
D_n: \P^{k_1}(M) \otimes \dots \otimes \P^{k_n}(M) \to
\P^{k_1/\dots/k_n}(M^n)[[\hbar]] \, ,
\een
where we denote by $\P^{k_1/\dots/k_n}(M^n)$ the space of all
distributional local, covariant functionals of $\phi$ and its covariant
derivatives $\nabla^k \phi$, of $m^2$, of the metric, and of
the Riemann tensor and its
covariant derivatives $\nabla^k R$,
which are supported on the total diagonal, and which take values in
the bundle
\ben\label{bundle}
\bigwedge^{k_1} T^* M \times \dots \times \bigwedge^{k_n} T^*M
\subset
\bigwedge^{k_1+\dots+k_n} T^*M^n
\een
of antisymmetric tensors over $M^n$. Thus, if $\O_i \in \P^{k_i}(M)$, then
$D_n(\otimes_i \O_i) \in \P^{k_1/\dots/k_n}(M^n)$,
and $D_n$ is a (distributional) polynomial,
local, covariant functional of $\phi$, the mass, $m^2$, and
the Riemann tensor and its derivatives
taking values in the $k_1+\dots+k_n$ forms over $M^n$,
which is supported on the total diagonal, i.e.,
\ben
\supp \, D_n(\O_1(x_1) \otimes \cdots \otimes \O_n(x_n))
= \{x_1=x_2=\dots=x_n\} = \Delta_n \, .
\een
 It is a $k_1$-form
in the first variable $x_1$, a $k_2$-form in the second variable
$x_2$, etc.

The difference between two prescriptions $T$ and $\hat T$ for time
ordered products satisfying T1---T11 may now be expressed in
terms of a hierarchy $D=\{D_n\}$ as follows.
Let $F= \int f \wedge \O$ be an integrated local
functional $\O \in \P(M)$, and formally combine the
time-ordered functionals into a generating functional
written
\ben
T(\e^F) := \sum_{n=0}^\infty \frac{1}{n!} T_n(F^{\otimes n}),
\een
where $\exp_\otimes$ is the
standard map from the vector space  of local actions to the tensor algebra
(i.e., the symmetric Fock space) over the space of local action functionals.
We similarly write $D(\e^F)$ for the corresponding generating functional
obtained from $D$. The difference between the time-ordered
products $T$ and $\hat T$ may now be expressed in the following way~\cite{Hollands2000}:
\ben
\label{unique1}
\hat T\bigg(\e^{iF/\hbar}  \bigg) = T \bigg(
\e^{i[F + D({\rm exp}_\otimes F)]/\hbar}\bigg)  \, .
\een
where $D = \{D_n\}$ is a hierarchy of functionals of the type
just described. Each $D_n$ is a formal power series
in $\hbar$, and if each $\O_i = O(\hbar^0)$, then it can be shown
that $D_n(\otimes \O_i) = O(\hbar)$, essentially because there are no
ambiguities of any kind in the underlying classical theory.
The expression $D(\e^F)$ may be viewed as being equal to
the finite counterterms that characterize the difference between the
two prescriptions for the time ordered products. Note that in curved
space, there is even an ambiguity in defining time-ordered products
with one factor (the Wick powers), so even $D_1$ might be
non-trivial.

The counterterms, i.e., the maps $D_n$, satisfy a number of
properties corresponding to the properties T1---T11 of
the time ordered products~\cite{Hollands2000}. As we have already said, the $D_n$
are supported on the total diagonal, and this corresponds to the
causal factorization property T8. The $D_n$ are local and covariant
functionals of the field $\phi$,
the metric, and $m^2$, in the following sense: Let $\psi:M \to M'$ be
any causality and orientation preserving isometric embedding, i.e.,
$\psi^* g' = g$. If $D_n$ and $D_n'$ denote the
functionals on $M$ respectively $M'$, then we have
that $\psi^* \circ D_n' =
D_n \circ (\psi^* \otimes \dots \otimes \psi^*)$. This follows from T1. It follows from the
smoothness and analyticity properties T4, T5 and the scaling property T2
that the $D_n$ depend only polynomially on the Riemann curvature tensor,
the mass parameter $m^2$, and the field $\phi$.
Since there is no ambiguity in defining the identity
operator, $\myid$, or the basic field, $\phi$, we must have
\ben\label{norm}
D_1(\myid) = D_1(\phi) = 0 \, .
\een
As a consequence of the symmetry of the time-ordered products T6, the maps
$D_n$ are symmetric (respectively graded symmetric when
Grassmann valued fields would be present), and as a
consequence of the field independence property T9, they must satisfy
\ben\label{fieldindep}
\frac{\delta}{\delta \phi(y)}
D_n \bigg( \O_1(x_1) \otimes \cdots \otimes \O_n(x_n) \bigg) =
\sum_k D_n \bigg( \O_1(x_1) \otimes \cdots \frac{\delta
\O_k(x_k)}{\delta \phi(y)}
  \otimes \cdots \O_n(x_n) \bigg) \, .
\een
In particular, the $D_n$
depend polynomially upon the field
$\phi$.
As a consequence of the scaling
property T2 of  time-ordered products, the
engineering dimension of each term appearing in $D_n$
must satisfy the following constraint. As above,
let $\N_{r}$ be the counter of Riemann curvature tensors,
let $\N_{f}$ be the dimension
counter for the fields, and let $\N_c$ be the counter for
the coupling constant (in this case $m^2$), see eq.~\eqref{Nedefs}. Let the dimension
counter $\N_d: \P \to \P$ be defined as above by
$\N_{d} = \N_c + \N_{r} + \N_{f}$
Then we must have
\ben\label{dimension}
(\N_{d} + sd)  D_n \bigg( \O_1(x_1) \otimes \cdots \otimes \O_n(x_n) \bigg)
= \sum_{i=1}^n
D_n \bigg( \O_1(x_1) \otimes \cdots \N_{d} \O_i(x_i) \otimes \dots
\O_n(x_n) \bigg) \, .
\een
where $sd$ is the scaling degree, see appendix C.
The unitarity requirement T7 on the time-ordered products
yields the constraint
\ben\label{unitary}
D_n \bigg( \O_1(x_1) \otimes \cdots \otimes \O_n(x_n) \bigg)^*
=
-D_n \bigg( \O_1(x_1)^* \otimes \cdots \otimes \O_n(x_n)^* \bigg) \, .
\een
and the action Ward identity T11 implies that one can freely
pull an exterior derivative $d_i = dx_i^\mu \wedge \frac{\partial}{\partial x_i^\mu}$
into $D_n$,
\ben\label{Dactionward}
d_i \, D_n \bigg( \O_1(x_1) \otimes \cdots \O_i(x_i) \otimes \dots O_n(x_n) \bigg)=
D_n \bigg( \O_1(x_1) \otimes \cdots d_i \, \O_i(x_i) \otimes \dots O_n(x_n) \bigg) \, .
\een

The meaning of the above restrictions on $D_n$ is maybe best illustrated in
some examples. The dimension of the
coupling is $d(m^2)=+2$, and the dimension of the
field is $d(\phi)=+1$. Consider the composite field $\phi^2 \in
\P$. In curved spacetime, there is an ambiguity $D_1(\phi^2)$ in defining
$T_1(\phi^2)$, given by
\ben
\hat T_1(\phi^2) = T_1(\phi^2) + (\hbar/i) \, T_1(D_1(\phi^2)) \, .
\een
By properties~\eqref{fieldindep} and \eqref{norm}, we must have
$\frac{\delta}{\delta \phi} D_1(\phi^2)=0$, so
$D_1(\phi^2)$ must be a multiple of the identity operator, so
$D_1(\phi^2) = ic \myid$. By the local and covariance property and
the dimensional constraint~\eqref{dimension}, $c=aR+bm^2$, where $a,b$ are
constants that must be real in view of~\eqref{unitary}. Thus, we have
the familiar result that the Wick power $T_1(\phi^2)$ is unique only
up to curvature/mass terms. Consider next the ambiguity in defining
the time ordered product of two factors of $\phi^2$, given by
\ben
\hat T_2(\phi^2 \otimes \phi^2) = T_2(\phi^2 \otimes \phi^2) +
(\hbar/i)^2 \,
T_1(D_2(\phi^2 \otimes \phi^2))
\een
(here we are assuming that $D_1(\phi^2)=0$ for simplicity).  By the same reasoning as above,
this must now be given by
\ben
D_2(\phi^2(x) \otimes \phi^2(y)) = c\delta(x,y)
\een
for some real constant $c$, because the
scaling degree of the delta function in 4 dimensions is $+4$.
If $\phi^2$ in this formula would be replaced by
$\phi^3$, then the right side could be a constant times the wave operator $\square$ of the
delta function, or by a real linear combination of $m^2, R$ and $\phi^2$, times the delta-function.

We summarize the renormalization ambiguities again in the
``main-theorem of renormalization theory:''

\begin{thm}\label{Tuniqueness}\cite{Hollands2000,Hollands2001}
Time ordered products $T$ with the above properties T1-T11 exist. If
$T=\{T_n\}$ and $\hat T = \{\hat T_n\}$ are two different time ordered products
satisfying conditions T1--T11,
then their difference is given by
\bena\label{tren}
&&\hat T_n \bigg( \O_1(x_1) \otimes \dots \otimes \O_n(x_n) \bigg) = \\
&&\sum_{I_0 \cup I_1 \cup \dots I_r \subset \underline{n}}
T_{r+1} \bigg( \bigotimes_{j \in I_0} \O_j(x_j) \otimes \bigotimes_k (\hbar/i)^{|I_k|} D_{|I_k|}\bigg[ \bigotimes_{i \in I_k} \O_i (x_i)\bigg]\bigg) \, .\non
\eena
Here, the sum runs over all partitions $I_0 \cup \dots \cup I_r = \underline{n}$
of $\underline{n}=\{1,\dots,n\}$,
and $D =\{D_n\}$ is a hierarchy of counterterms described above. Conversely, if $D$ is as above, then
$\hat T$ defines a new hierarchy of time-ordered products with the properties T1---T11.
\end{thm}

\subsection{Perturbative construction of interacting quantum fields}

In the previous sections we have given the construction of Wick powers and their
time-ordered products in a theory that is classically described by a
Lagrangian $\L_0$ at most quadratic in the field, with associated classical field
equations of wave-equation type. Those quantities may be used to give a
definition of an interacting quantum field theory via a perturbation expansion. For
definiteness, consider a scalar field described by the classical Lagrangian
$\L = \L_0 + \lambda \L_1$,
\ben
\L = \frac{1}{2}(d\phi \wedge * d\phi + m^2 * \! \phi^2) + \lambda \, *\! \phi^N = \L_0 + \lambda \L_1 \, .
\een
We would like to construct quantities in the interacting quantum field theory as
formal power series in $\lambda$. Even in flat spacetime, one may encouter infra-red divergences
if one tries to define the terms in such expansions, but such infra-red divergences are
absent if one considers, instead of the interaction $I= \int \lambda \L_1$, a cutoff interaction,
$F=\int \lambda f \L_1$, where $f$ is a smooth cutoff function of compact support that is one
in a globally hyperbolic subregion of the original spacetime $(M,g)$. The perturbative
formula for the interacting fields associated with this interaction  is
then
\ben
\O(x)_F =  T\bigg(\e^{iF/\hbar} \bigg)^{-1} \st
\frac{\delta}{\delta j(x)} T \bigg( \e^{iF/\hbar + \int j \wedge \O} \bigg) \bigg|_{j=0} \, .
\een
This formula is called ``Bogoliubov's
formula,''~\cite{Bogoliubov1952}. Each term in the formal power series for
$\O(x)_F$ is a well-defined element in $\W$, due to the infra-red
cutoff in the interaction $F$. The subscript ``$F$'' indicates throughout this
paper an ``interacting field'' defined by $F$, which is an
element in the ring\footnote{
The fact that, implicit in the notation ``$\mc[[\hbar]]$'', the
interacting field only contains non-negative powers of $\hbar$,
is not so obvious and follows from the fact that $R_n$ itself is
of order $\hbar^n$, see~\cite{Duetsch2000b}.}
$\W \otimes \mc[[\lambda, \hbar]]$, as opposed to the classical field
expression $\O \in \P$.
The expansion coefficients in $\lambda$ of the interacting fields define the
so-called ``retarded products,''~\cite{Kallen1950}
\ben\label{bogoliubov}
\O(x)_F = \sum_{n=0}^\infty \frac{i^n}{\hbar^n n!} R_n(\O(x); F^{\otimes n}) =: R\Big( \O(x); \e^{iF/\hbar} \Big) \,.
\een
The retarded products are maps $R_n: \P^{\otimes(n+1)} \to \D'(M^{n+1}) \otimes \W$
with properties similar to the properties T1---T11 of the time-ordered products. The
symmetry property only holds with respect to the $n$-arguments separated by the semicolon.
Their definition in terms of time-ordered products is
\bena\label{retardeddef}
&& R_n\bigg(\Psi(y); \O_1(x_1) \otimes \dots \otimes \O_n(x_n) \bigg) \\
&=&
\sum_{I_1 \cup \dots \cup I_j = \underline{n}} (-1)^{n+j+1}
T_{|I_1|}\bigg( \bigotimes_{k \in I_1} \O_k(x_k) \bigg)
\st \dots \st
%T_{|I_{j-1}|}\bigg( \bigotimes_{k \in I_{j-1}} \O_k(x_k) \bigg)
%\st
T_{|I_j|}\bigg(\Psi(y) \otimes \bigotimes_{k \in I_j} \O_k(x_k) \bigg) \, , \non
\eena
where the sum runs over all partitions $I_1 \cup \dots \cup I_j$
of $\underline{n} = \{1, \dots, n\}$.
An important property of the retarded product is that their support is restricted to the
set
\ben
\supp \, R_n(\Psi(y); \O_1(x_1) \otimes \dots \otimes \O_n(x_n)) \subset
\{ (y, x_1, \dots, x_n) \in M^{n+1} \mid x_i \in J^-(y) \quad \forall i \} \, .
\een
The support property follows from the causal factorization property of the time-ordered products.
A useful combinatorial identity for the retarded products is the
Glaser-Lehmann-Zimmermann (GLZ) relation, which states that~\cite{Duetsch2002}
\bena\label{GLZ}
&&
R_n\bigg( \Psi_1(y_1); \Psi_2(y_2) \otimes
\bigotimes_{i=1}^{n-1} \O_i(x_i)
\bigg) -
R_n\bigg( \Psi_2(y_2); \Psi_1(y_1) \otimes
\bigotimes_{i=1}^{n-1} \O_i(x_i) \bigg)=\non\\
&&
\sum_{I \cup J = \underline{n}}
\bigg[
R_{|I|}\bigg(\Psi_1(y_1);
\bigotimes_{i \in I}
\O_i(x_i)
\bigg),
R_{|J|}\bigg(\Psi_2(y_2);
\bigotimes_{j \in J}
\O_j(x_j)
\bigg)
\bigg]
\eena
The GLZ-relation may be used to express the commutator of two interacting fields in terms of
retarded products as follows:
\ben
[\Psi_1(x_1)_F, \Psi_2(x_2)_F] = \sum_{n=0}^\infty \frac{i^n}{\hbar^n n!} \bigg[
R_{n+1}(\Psi_1(x_1);
\Psi_{2}(x_2) \otimes F^{\otimes n} ) - (1 \leftrightarrow 2)
\bigg] \, .
\een
As a consequence of the GLZ-relation and
the support properties of the retarded products, any two interacting fields
located at spacelike separated points commute\footnote{In case when
  Grassmann valued fields are present, the commutator is replaced by
the graded commutator, and the minus sign on the right side is
replaced by $-(-1)^{\epsilon_1\epsilon_2}$, where $\epsilon_i$ are
the Grassmann parities of $\Psi_i$.}.
Thus, we have constructed interacting fields as formal
power series in the coupling constant via the
time-ordered products in the underlying free field theory. If one changes the definition of the
time-ordered products along the lines described in the previous subsection, then
there is a corresponding change in the interacting theory, affecting both the interaction Lagrangian,
as well as resulting in general in a multiplicative redefinition of the interacting fields.
To describe this in more detail, we introduce the linear map ${\bf Z}_F: \P(M) \to \P(M)[[\lambda,\hbar]]$
by
\ben\label{zfdef}
{\bf Z}_F(\O(x)) := \O(x) + D(\O(x) \otimes \e^F) \, ,
%= \O(x) + \sum_{n \ge 0} \frac{1}{n!} D_{n+1}(\O(x) \otimes F^{\otimes n}) \, ,
\een
where $D = \{D_n\}$ is the hierarchy of distributions encoding the
difference between two prescriptions $T$ and $\hat T$ for time ordered products. We may introduce
a basis in $\P(M)$, and represent this map by its matrix
\ben
{\bf Z}_F(\O_i(x)) = \sum_j Z^j_i \, \O_j(x) \, .
\een
For renormalizable interactions $(\N_f F \le 4)$, ${\bf Z}_F$ leaves each finite dimensional
subspace of $\P$
invariant, but this is no longer the case
for non-renormalizable interactions. Now, if $\hat O(x)_F$ is the definition of the
interacting field using the time ordered products $\hat T$, and $\O(x)_F$ that
using $T$, then the two are related by
\ben\label{Och}
\hat \O(x)_F = {\bf Z}_F[\O(x)]_{F+D({\rm exp}_\otimes \, F)} \, .
\een

We now explain how one can remove the cutoff
implemented by the cutoff function $f$ in the interaction $F=\int \lambda f \O$ at the algebraic level.
The key identity~\cite{Brunetti2000} in this construction is
\ben
V_{F_1,F_2} \st \O(x)_{F_2} \st V_{F_1,F_2}{}^{-1} = \O(x)_{F_1}
\een
where $F_1, F_2$ are any two local interactions as above that are equal in an open
neighborhood of $x$, and where $V_{F_1,F_2} \in \W \otimes \mc[[\hbar, \lambda]]$
are unitaries that can be written in terms of retarded products. They satisfy the
cocycle condition
\ben
V_{F_1,F_2} \st V_{F_2,F_3} \st V_{F_3,F_1} = \myid \, .
\een
To construct
the limit of the interacting fields as $f\to 1$, one can now proceed as follows.
For simplicity, let us assume that
$M=\mr \times \Sigma$, with $\Sigma$ compact. The cutoff function may
then be chosen to be of compact support in a ''time-slice''
$M_{2\tau} = \Sigma \times (-2\tau,2\tau)$, and to be equal to one in a somewhat smaller time-slice,
say $M_{\tau}$. To indicate the dependence upon the cutoff $\tau$, let us write the cutoff function as
$f_\tau$, and let us correspondingly write.
Let $F_\tau = \int \lambda f_\tau\L_1$ and $\O_{F_\tau}$ for the
corresponding interacting field defined using $F_\tau$ as the
interaction. Finally, let $U_\tau = V_{F_\delta,F_\tau}$, for some fixed $\delta$.
The interacting fields defined with respect to the true interaction
$I=\int \lambda \L_1$ may now defined as the limit
\ben\label{OIdef}
\O(x)_I = \lim_{T \to \infty} U_{\tau} \st \O(x)_{F_\tau} \st U_{\tau}{}^{-1} \, .
\een
The sequence on the right side is trivially convergent, because it only contains a finite number of
terms for each fixed $x$, by the cocycle condition. More precisely, the terms in the sequence
will remain constants once $\tau$ has become so large that $x \in M_{\tau}$. It is important to note that this
would not be the case if we had not inserted the unitary operators under the limit sign. In that case,
our notion of interacting field would have coincided with the naive ``adiabatic limit'' which intuitively
corresponds to the situation where the interacting field is fixed at $\tau=-\infty$. By contrast, our limit
corresponds intuitively to fixing the field during``finite time interval'' corresponding to the
neighborhood $\Sigma \times (-\delta, \delta)$. Actually,
one can see that the defining formula for $U_\tau$ and the interacting
field will still make sense also for spacetimes with non-compact
Cauchy surface. We can now define the algebras of interacting field
observables as
\ben
\F_I(M,g) = {\rm Alg}\bigg\{ G_I \,\,\, \bigg| \,\,\, G = \int g \wedge \O \bigg\} \bigg/ \I \, .
\een
We note that these are subalgebras of $\F_0[[\lambda, \hbar]]$. While the embedding
of this algebra as a subalgebra of $\F_0[[\lambda, \hbar]]$ depends upon the choice of
the cutoff function $f$, it can be proved~\cite{Brunetti2000,Hollands2003} that the definition
of $\F_I$ as an abstract algebra is independent of our choice of
the sequence of cutoff functions $\{f_\tau\}$.
Another important consequence of our definition of the interacting
fields is that, if we want to investigate properties of the
interacting field near a point $x$, we only have to work in practice with the cutoff interaction $F$ where $f$ is equal
to 1 on a sufficiently large neighborhood containing $x$. For example, if we want to check whether an interacting
current $\J(x)_I$ is conserved, we only need to check whether $d\J(x)_F = 0$ for any cutoff function $f$
which is equal to 1 in an open neighborhood of $x$.

The effect of changing the renormalization conditions may also be discussed at the level
of the interacting fields $\O_I$ and the associated interacting field algebra $\F_I$. For this,
consider again two prescriptions $T$ and $\hat T$ for defining the time-ordered products,
and let us denote by $\O_I$ and $\hat O_I$ the respective interacting fields, and by
$\F_I$ and $\hat \F_I$ the interacting field algebras.
Let us denote
by ${\bf Z}_I: \P \to \P[[\lambda,\hbar]]$ the limit of the map ${\bf Z}_F$ as the cutoff implicit in $F$
is removed. This limit exists, because all the functionals $D = \{D_n\}$ in the defining
relation~\eqref{zfdef} for ${\bf Z}_F$ are supported only on the total diagonal. Then one can derive from
eq.~\eqref{Och} that there exists an algebra isomorphism
\ben
\rho: \hat \F_{I} \to \F_{\hat I}, \quad \rho(\hat \O_{I}) = {\bf Z}_I(\O)_{\hat I} \, ,
\een
with $\hat I = I + D(\e^I)$. The algebra isomorphism map $\rho$ is needed in order to compensate
for the difference between the unitaries $U_\tau$ and $\hat U_\tau$ in the two prescriptions,
see eq.~\eqref{OIdef}, and see~\cite{Hollands2003} for details.
A particular case of this map again arises when the prescription $\hat T$
is defined in terms of a change of scale (see T2) from the time ordered product $T$.
Then we obtain, for each scale $\mu \in \mr^+$, a map $\rho_\mu$, which
depends polynomially on $\mu$ and $\ln \mu$. This map
defines the renormalization group flow in curved spacetime~\cite{Hollands2003} together
with the corresponding ``mixing matrices,'' i.e., the matrix components $Z^i_j(\mu)$ of the maps
$Z_I(\mu)$.

\section{Quantum Yang-Mills theory}

\subsection{General outline of construction}\label{genoutline}

\subsubsection{Free fields}

We now construct quantum Yang-Mills theory along the lines
outlined in the introduction. As our starting point, we take the
auxiliary theory described classically by the auxiliary action $S$ with ghosts
and anti-fields, see eq.~\eqref{Sdef}. Thus, the set of dynamical and
background fields is
\begin{center}
\begin{tabular}{|c|c|}
\hline
background fields & dynamical fields\\
\hline
spacetime metric $g$ & \\
anti-ghost $C^\ddagger, \bar C^\ddagger$ & ghost $C, \bar C$\\
anti-vector $A^\ddagger$ & vector $A$ \\
anti-auxiliary $B^\ddagger$ & auxiliary $B$ \\
\hline
\end{tabular}
\end{center}
We assume that the group $G$ is a direct product of a
semi-simple group and $U(1)^l$, and that the dimension of
spacetime is 4. We split the action $S$ into a
free part $S_0$ containing only expressions at most quadratic
in the dynamical fields, and an interaction part, $\lambda S_1 + \lambda^2
S_2$. The action $S_0$ describes the free classical auxiliary theory. Its
field equations are hyperbolic. As we shall describe
in more detail below, we can thus define an algebra $\W$ that
represents a deformation quantization of the free field theory
associated with the free auxiliary action $S_0$, and this algebra
contains all local covariant Wick-powers, and their time-ordered
products.

As in the classical case, the so-obtained auxiliary theory is by
itself not equivalent to (free) Yang-Mills theory, because it contains
gauge-variant observables and observables with non-zero ghost number.
To obtain a quantum theory of (free) Yang-Mills theory, we pass from
the algebra of observables, $\W$, to the cohomology algebra
constructed from the (free) quantum BRST-charge $Q_0$. For this,
we consider first the (free) classical BRST-current $\J_0$, which defines
a quantum Wick power $T_1(\J_0)$, which we denote again by $J_0$ by
abuse of notation. Let us assume for simplicity that
the spacetime $(M,g)$ has a compact Cauchy surface $\Sigma$. Then there is
a closed compactly supported 1-form $\gamma$ on $M$ such
that $\int_M \gamma \wedge \alpha =
\int_\Sigma \alpha$ for any closed 3-form $\alpha$, i.e., $[\gamma]
\in H^1_0(M,d)$ is dual to the cycle $[\Sigma] \in H_3(M,\partial)$. We
can then define the {\em free} BRST-charge by
\ben\label{Qdef}
Q_0 = \int_M \gamma \wedge \J_0
\een
As we will show below, the local covariant quantum BRST current $\J_0
:= T_1(\J_0)$ can be defined so that it is closed $d\J_0 =0$
modulo $\I$, so evidently $Q_0$ is independent, modulo $\I$, of
the choice of the representer $\gamma$ in $H^1(M,d)$. We will also
show that $Q_0$ is nilpotent, $Q_0^2 = 0$ modulo $\I$.
It follows from this fact that the linear quotient space
\ben\label{hatF}
\hat \F_0 = \frac{{\rm Kernel} \, [Q_0, \, . \,] \cap \F_0 \cap {\rm
    Kernel} \, \N_g}{{\rm Image}
  \, [Q_0, \, . \,] \cap \F_0 \cap {\rm Kernel} \, \N_g} \, , \quad
\F_0 = \W/\I
\een
is well defined, and that it
is again an algebra. Above, we have explained that
$\F_0$ is a deformation
quantization of the classical theory associated with
$S_0$ in the sense that, when $\hbar \to 0$,
the commutator divided by $\hbar$ goes over to the
Peierls bracket of the classical observables. In particular,
the commutator divided by $\hbar$ with $Q_0$
goes to the classical BRST-variation, $\hat s_0$. Furthermore,
as we explained above, the cohomology of $\hat s_0$ is in
1-1 correspondence with classical gauge-invariant observables, so
that, in the classical limit, the algebra $\hat \F_0$ is the Poisson algebra
of physical, gauge-invariant observables.
Thus, it is natural to define $\hat \F_0$ to be the algebra of physical
observables also in the quantum case.

Consider now a representation $\pi_0$ of the free algebra
$\F_0$ on an inner product space ${\mathcal H}_0$. For simplicity,
let us denote representer $\pi_0(Q_0)$ of the BRST-charge in this representation
again by $Q_0$. We require $Q_0$ to be hermitian with respect to
the (necessarily indefinite) inner product. We would like to know under which condition this
representation induces a Hilbert-space representation $\hat \pi_0$ on the factor
algebra $\hat {\mathcal F}_0$. Following~\cite{Duetsch1999}, let us suppose that the representation
fulfills the following additional

\medskip
\noindent
\paragraph{Positivity requirement:}
A representation is called positive if the following hold: (a)
if $|\psi\rangle \in {\rm Kernel} \, Q_0$, then $\langle \psi |
\psi \rangle \ge 0$, and (b) if
$|\psi\rangle \in {\rm Kernel} \, Q_0$, then $\langle \psi |
\psi \rangle = 0$ if and only if
$|\psi\rangle \in {\rm Image} \, Q_0$.

\medskip
\noindent
It is elementary to see that if the positivity requirement is
fulfilled, then the representation $\pi_0$
induces a representation $\hat \pi_0$ of the physical observables
$\hat \F_0$ on the inner product space
\ben\label{Hphy0}
\hat {\mathcal H}_0
= \frac{{\rm Kernel}\, Q_0}{{\rm Image} \, Q_0} \, ,
\een
which is in fact seen to be a pre-Hilbert space, i.e., carries a
positive definite inner product. As we will see below, when $G$ is
compact, there do indeed exist representations satisfying the
above positivity requirement if we restrict ourselves to the
ghost number 0 subalgebra of $\F_0$. As we will also see,
in static spacetimes $(M,g)$ or in spacetimes with
static regions, the states in
$\hat {\mathcal H}_0$ (in the ground state representation)
can be put into one-to-one correspondence with $\pm$-helicity
particle states of the
electromagnetic field, and $\hat {\mathcal H}_0$ contains
a dense set of Hadamard states. However, in generic time-dependent spacetimes,
such an interpretation in terms of particles states is not possible.

When the Cauchy surfaces of $M$ are not compact, the charge $Q_0$ is
in general not defined as stated. The reason is that the 1-form field
$\gamma$ is no longer of compact support, but has non-compact support in
spatial directions. Nevertheless, we can see that if we formally consider the
graded commutator $[Q_0, \O(x)]$ with a local quantum
Wick-power, denoted $\O(x) := T_1(\O(x))$, then
there will be only contributions in the formal integral defining $Q_0$
(see~\eqref{Qdef}) from the portion of
the support of $\gamma$ that is contained in $J^+(x) \cup J^-(x)$.
All other contributions vanish due to the (graded) commutativity property, T9.
Since the intersection of the support of $\gamma$ and $J^+(x) \cup J^-(x)$
is compact for a suitable choice of $\gamma$, it follows that the commutator
of any local observable in $\F_0$ with $Q_0$ is always defined. Thus, while
$Q_0$ itself is undefined, the graded commutator still defines a graded
derivation. The definition of the algebra of gauge invariant
observables can then be given in terms of this graded
derivation. However, the construction of representations explicitly
used (the representer of) $Q_0$ itself, and not just the graded
commutator. Thus, it is not straightforward to obtain Hilbert space
representations on manifolds with non-compact Cauchy surfaces.

\subsubsection{Interacting fields}

A similar kind of construction as for free Yang-Mills theory can also
be given in order to perturbatively construct quantized interacting
Yang-Mills theory. The starting point is now the classical auxiliary
interacting field theory described by the auxiliary action
$S=S_0+\lambda S_1 + \lambda^2 S_2$. Thus,
the interaction is
\ben
I = \int (\lambda \L_1 + \lambda^2 \L_2) = \lambda S_1 + \lambda^2 S_2 \, .
\een
The first step is to
construct a quantum theory associated with this auxiliary action.
For simplicity, we again assume that $M$ has compact Cauchy-surfaces---the
general situation can again be treated by complete analogy with the free
field case as just described.
Following the general procedure described in Sec.~3.7, we first
introduce an infra-red cutoff for the interaction supported in
a compact region of spacetime, and construct the interacting
theory in that region.
To define the desired infra-red cutoff, we consider a compactly supported
cutoff function, $f$, which is equal to $1$ on the submanifold
$M_\tau = (-\tau,\tau) \times \Sigma$. We define a cutoff interaction, $F$, by
$F=\int\{f\lambda\L_1+f^2\lambda^2\L_2\}$, and we define corresponding
interacting fields $\O_F$ by Bogoliubov's formula. We then send the
cutoff $\tau$ to infinity at the algebraic level as described in sec.~3.7, and
get a corresponding algebra $\F_I$ of interacting fields $\O_I$.
This algebra of interacting fields is not
equivalent to quantum Yang-Mills theory, as it contains gauge variant
fields and fields of non-zero ghost number. As in the free case,
we obtain the algebra of physical field observables by considering the
cohomology of the (now interacting) BRST-charge operator,
$Q_I$.

To define this object, consider the interacting BRST-current with cutoff interaction, defined
by the Bogoliubov formula [see eq.~\eqref{bogoliubov}]
\bena
\J(x)_F &=& \frac{\delta}{\delta \gamma(x)} T(\e^{iF/\hbar})^{-1} \st
T(\e^{iF/\hbar+\int \gamma \wedge \J}) \Bigg|_{\gamma=0}\non\\
&=& \sum_{n\ge 0} \frac{1}{n!} \bigg( \frac{i}{\hbar} \bigg)^n
    R_n(\J(x); F^{\otimes n}) \, .
\eena
As in our general definition of interacting fields, we can then remove the
cutoff at the algebraic level by defining an interacting current
$\J(x)_I$.
We will show below that the interacting BRST-current $\J_{\subsc}(x)$ is conserved in
$M$, so we can define a corresponding interacting BRST-charge
by $Q_{\subsc} = \int \gamma \wedge \J_{\subsc}$, [compare eq.~\eqref{Qdef}].

We will furthermore show that the
so-defined charge is nil-potent, $Q_{\subsc}^2 = 0$. Thus, we can define
the physical observables as in the free field theory by the cohomology
of the interacting BRST-charge, i.e., the algebras of interacting
fields are defined by
\ben
\hat \F_I = \frac{{\rm Kernel} \, [Q_{\subsc}, \, . \,] \cap \F_{I} \cap {\rm Kernel} \, \N_g}{{\rm Image}
  \, [Q_{\subsc}, \, . \,] \cap \F_I \cap {\rm Kernel} \, \N_g} \, .
\een
Next, one would like to define representations of the algebra of
observables on a Hilbert space. Such representations can be obtained
from those of the free theory by a deformation
process~\cite{Duetsch1999}. For this, consider a state $|\psi_0 \rangle
\in {\mathcal H}_0$ in a representation $\pi_0$ of the underlying
free theory satisfying the above positivity requirement. Let also
$|\psi_0\rangle \in {\rm Kernel} \, Q_0$. Then, using $Q^2_I = 0$, and
$Q_I = Q_0 + \lambda Q_1 + \lambda^2 Q_2 + \dots$ one first shows that there
exists a formal power series
\ben\label{deformationp}
|\psi_I\rangle = |\psi_0\rangle
+ \lambda |\psi_1\rangle + \lambda^2 |\psi_2\rangle + \dots \in \H_I =
\H_0[[\lambda]]
\een
such that $Q_I |\psi_I\rangle = 0$, where $Q_I$ has been
identified with its representer in the representation $\pi_I$ that is
induced from the representation of the underlying free theory. In
order to construct the vectors $|\psi_i \rangle$, we proceed
inductively. We write the condition that $|\psi_I \rangle$ is in the kernel
of $Q_I$ and that $Q_I^2 = 0$ as
\ben\label{qqcon}
0 = \sum_{k=0}^m Q_k |\psi_{m-k} \rangle, \quad 0 = \sum_{k=0}^m Q_k
Q_{m-k} \, ,
\een
for all $m$. For $m=0$, the first equation is certainly satisfied, as
we are assuming $Q_0 |\psi_0 \rangle = 0$. Assume now that
$|\psi_0 \rangle, |\psi_1 \rangle, \dots, |\psi_{n-1} \rangle$ have
been constructed in such a way that the first
equation is satisfied up to $m=n-1$, and put
\ben
|\chi_m \rangle = \sum^{n-1}_{k=0} Q_{m-k} |\psi_k \rangle \, .
\een
Then, using the second equation in~\eqref{qqcon}, we see that
\ben
0=\sum_{k=0}^m Q_{m-k} | \chi_k \rangle \, , \quad
0=\sum_{k=0}^m \langle \chi_m | \chi_{m-k} \rangle \, ,
\een
for all $m$. We now use the inductive assumption that $|\chi_m\rangle =
0$ for $m \le n-1$, from which we get that $Q_0 | \chi_{n} \rangle =
0$, putting $m=n$ in the first equation. Putting $m=2n$ in the second
equation, we get $\langle \chi_n | \chi_n \rangle = 0$. In view of the
positivity requirement, we must thus have $|\chi_n\rangle =
-Q_0|\psi_n \rangle$ for some $|\psi_n \rangle$. We take this as the
definition of the $n$-th term for the deformed
state~\eqref{deformationp}. This then satisfies the induction
assumption at order $n$, thus closing the induction loop.

Thus, by the above deformation argument,
one sees that ${\rm Kernel} \, Q_I \subset \H_I$ is a non-empty
subspace. One furthermore shows that the representation $\pi_I$
satisfies an analog of the positivity requirement\footnote{
Since we are working over the ring
$\mc[[\lambda]]$ of formal power series in $\lambda$ in the
case of interacting Yang-Mills theory, the positivity requirement
needs to be formulated appropriately by specifying what it means
for a formal power series to be positive. For details, see~\cite{Duetsch1999}.
} for the interacting
theory. Thus, we obtain, as in the free case, a representation $\hat
\pi_I$ on the inner product space
\ben\label{Hphy}
\hat {\mathcal H}_I
= \frac{{\rm Kernel}\, Q_I}{{\rm Image} \, Q_I} \, ,
\een
and this space is again shown to be a pre-Hilbert space. For details
of these constructions, see sec.~4.3 of~\cite{Duetsch1999}.

\subsubsection{Operator product expansions and RG-flow}

As we have just described, a physical gauge invariant, interacting
field is an element in the algebra $\hat \F_0$, i.e.,
an equivalence class of an interacting field operator
$\O_{\subsc}(x)$ satisfying
\ben
[Q_{\subsc}, \O_{\subsc}(x)] = 0 \quad \forall x \in M \, ,
\een
modulo the interacting fields that can be written as
\ben\label{trivial}
\O_{\subsc}^{}(x) = [Q_{\subsc}^{}, \O'_{\subsc}(x)] \quad \forall x \in M \, ,
\een
for some local field $\O'$ (as usual, $[ \, , \, ]$ means the graded
commutator). Our constructions of the interacting BRST-charge do not
imply that the action of
$Q_{\subsc}$ on a local covariant interacting field is not
equivalent to $\hat s$. But it follows from general arguments that
\ben
[Q_{\subsc}, \O_{\subsc}(x)] = (\hat q \O)_{\subsc}(x) \quad \forall x \in M
\een
where $\hat q$ is a map
\ben
\hat q: \P^p(M) \to \P^p(M)[[\hbar]], \quad \hat q=\hat s + \hbar \hat q_1 +
\hbar^2 \hat q_2 + \dots \, .
\een
Because $Q_\subsc^2=0$, the map $\hat q$
is again a differential (the ``quantum BRST-differential''), $\hat q^2 = 0$, whose action on general
elements in $\P$ is different from that of $\hat s$. An exception of
this rule are the exactly gauge invariant elements $\O=\Psi$ at zero
ghost number, which by lemma~\ref{thm1} are of the form $\Psi = \prod
\Theta_{s_i}(F,{\mathcal D} F, {\mathcal D}^2 F, \dots)$, with
$\Theta_s$ invariant polynomials of the Lie-algebra. For such
elements, we shall show that
we have $\hat q \Psi = \hat s \Psi = 0$.
Thus,
\ben\label{qpsi}
[Q_{\subsc}, \Psi_{\subsc}(x)] = 0 \quad \forall x \in M
\een
and the corresponding interacting fields $\Psi_{\subsc}(x)$ are always observable.

Given $n$ local fields $\O_{j_1}, \dots, \O_{j_n} \in \P$, we can construct
the operator product expansion of the corresponding interacting quantum
fields,
\ben\label{ope}
\O_{j_1}(x_1)_{\subsc} \st \dots \st \O_{j_n}(x_n)_{\subsc}  \sim
\sum_k C_{j_1 \dots j_n}^k (x_1, \dots, x_n, y) \, \O_k(y)_{\subsc} \, .
\een
The operator product expansion is an asymptotic expansion
for $x_1, \dots, x_n \to y$, see \cite{Hollands2006}, where the
construction and properties of the expansion are described. Because
the action $S$ of the auxiliary theory has zero ghost number, the OPE
coefficients are non-vanishing only when
\ben
\sum_r \N_{g}(\O_{j_r}) = \N_{g}(\O_k) \, .
\een
Now assume that all operators $\O_{j_1}, \dots, \O_{j_n}$
are physically observable fields.
Then, since the graded commutator with $Q_{\subsc}$ respects the
$\st$-product, also all local operators $\O_k$ appearing on the
right side must be in the kernel of $Q_{\subsc}$. By the
same argument, if one of the operators on the left side
is of the trivial from~\eqref{trivial}, then it follows that each operator
in the expansion on the right side is of that form, too.
Thus, we conclude that the OPE closes on gauge invariant operators in the following sense:
Let $\O_{i_1}, \dots, \O_{i_n} \in \P$ be in the kernel of $\hat s$,
with vanishing ghost number, as characterized by thm.~\ref{thm1}.
Then $C_{i_1 \dots i_n}^k$ is non-vanishing only for $\O_k \in \P$
of vanishing ghost number that are in the kernel of $\hat s$. If one
$\O_{i_r}$ is in the image of $\hat s$, then $C_{i_1 \dots i_n}^k$
is non-vanishing only for $\O_k \in \P$
of vanishing ghost number that are in the image of $\hat s$. If one
drops the restriction to the 0-ghost number sector, then the same
statement is true with $\hat s$ replaced by $\hat q$.

\medskip
\noindent
By the same kind of
argument, one can also show that the renormalization group flow
closes on physical operators. The renormalization flow in curved
spacetime was defined in subsec.~3.7 as the behavior of the interacting
fields under a conformal change of the metric, $g \to \mu^2 g$. In
general we have $\rho_\mu (\O_i(x)_\subsc) = Z^j_i(\mu) \cdot
\O_j(x)_{I_\mu}$ for all $x \in M$, where $I_\mu$ is the renormalized
interaction, and
where $\rho_\mu: \F_\subsc(g) \to \F_{I_\mu}(\mu^2 g)$ is an algebraic
isomorphism implementing the conformal change of
the metric. Now, in the perturbative quantum field theory associated
with the auxiliary action $S$, we have
\ben
\rho_\mu (\J(x)_I) = Z(\mu) \cdot \J(x)_{I_\mu} + \sum_i \zeta_i(\mu) \cdot \O_i(x)_{I_\mu}
\quad \forall x \in M \, ,
\een
for some $Z(\mu), \zeta_i(\mu) \in \mc[[\lambda,\hbar]]$, and operators $\O_i \in \P^3(M)$
of dimension three not equal to the BRST-current and not equal to 0.
If we take the exterior derivative $d$ of this equation
and use that the interacting BRST-currents themselves are conserved, we obtain
$\sum \zeta_i(\mu) \cdot d\O_i(x)_{I_\mu} = 0$. Let $k$ be the largest
natural number such that $\zeta_i(\mu)$ is of order $\hbar^k$ for all $i$,
and let $z_i(\mu)$ be the $\hbar^k$-contribution to $\zeta_i(\mu)$. We can then
divide this relation by $\hbar^k$, and take the classical limit $\hbar \to 0$.
Because the classical limit of the interacting fields gives the corresponding
perturbatively defined classical interacting fields and because $I_\mu \to I$ as $\hbar \to 0$,
it follows that $\sum z_i(\mu) \cdot d\O_i(x)_I = 0$ for the corresponding on-shell
classical interacting fields. This means that $d \O_i(x)_I = 0$ for those $i$ such that $z_i(\mu) \neq 0$.
But there are no such 3-form fields of dimension three at the classical
level by the results of~\cite{Barnich2000} except for the zero field and the BRST-current. Thus,
we have found that $z_i(\mu) = 0$ for all $i$. By repeating this type of
argument for the higher orders in $\hbar$ in $\zeta_i(\mu)$, we can
conclude that $\zeta_i(\mu) = 0$ to all orders in $\hbar$.

Thus, we have found that BRST-current does not mix with other operators
under the renormalization group flow, from which it follows that
\ben
\rho_\mu (Q_I) = Z(\mu) \cdot Q_{I_\mu}\, .
\een
Hence, if $[Q_I, \O_i(x)_I]=0$ for all $x\in M$, then, by
applying $\rho_\mu$ to this relation, it also follows that
\ben
Z^i_j(\mu) [Q_{I_\mu}^{}, \O_i(x)_{I_\mu}] = 0 \, .
\een
Because $Z^i_j(\mu)$ is invertible (it is a formal power
series in $\lambda$ starting with $\delta^i_j$), we thus obtain the following result,
which states that the
RG-flow does not leave the sector of physical observables:
\begin{thm}\label{rgflow}
Let $\O_{i} \in \P$ be in the kernel of $\hat s$,
with vanishing ghost number, as characterized by thm.~\ref{thm1}.
Then $Z^j_i(\mu)$ is non-vanishing only for $\O_j \in \P$
of vanishing ghost number that are in the kernel of $\hat s$. If
$\O_{i}$ is in the image of $\hat s$, then $Z_i^j(\mu)$
is non-vanishing only for $\O_j \in \P$
of vanishing ghost number that are in the image of $\hat s$. If one
drops the restriction to the 0-ghost number sector, then the same
statement is true with $\hat s$ replaced by $\hat q$.
\end{thm}

\paragraph{Remark}: An interesting corollary to this theorem arises when
one considers the particular case when $\O$ is the Yang-Mills Lagrangian.
Since it is the only gauge invariant field at ghost number 0 of
this dimension, it does not mix with other field up to $Q_I$-exact terms
under the renormalization group flow. The corresponding {\em constant}
$Z_I(\mu)$ describing the field renormalization for the interacting field
corresponding to the Yang-Mills Lagrangian then defines the flow of the coupling
constant $\lambda$. Since our flow is local and covariant, it follows that this
flow automatically must be exactly the same as in Minkowski spacetime!

A similar remark would apply to more complicated gauge theories with
additional matter fields, as long as there cannot arise any additional couplings
to curvature of engineering dimension 4 (such as e.g. $R \, {\rm Tr} \, \Phi^2$ if the
gauge field is coupled to a scalar field $\Phi$ in some representation of the gauge group).
Even if there can arise such couplings, the above argument can still be used to directly infer the vanishing of all
$\beta$-functions in curved spacetimes with $R=0$ if the corresponding $\beta$-functions
vanish in flat spacetime.

\subsection{Free gauge theory}

We now describe in more detail the construction of free gauge theory
outlined in the previous section~4.1. As explained, our starting point is
the auxiliary theory that is classically described by the free action
$S_0$. 

\medskip
\noindent
{\bf Deformation quantization
algebra $\W$}: The theory contains the dynamical fields
$\Phi=(A^I, B^I, C^I, \bar C^I)$, as well as the background fields
$\Phi^\ddagger = (A_I^\ddagger, B_I^\ddagger, C_I^\ddagger, \bar
C_I^\ddagger)$. Of the dynamical fields, $B^I$ is only an auxiliary
field with no kinetic term in $S=0$, while for the vector field $A^I$ and
the ghost fields $C^I, \bar C^I$ the corresponding deformation quantization algebra 
was defined already above in sect.~3.5 by analogy with the model case of a scalar field 
described in sect.~3.1. Thus, the desired $\W$ will essentially be a tensor product
of the algebras for the vector and ghost fields, with additional ``commuting'' generators for the background 
fields. We now describe the construction in detail.

We first consider a vector Hadamard 2-point function $\omega^{\rm
  v}(x,y)$, and a scalar Hadamard 2-point function $\omega^{\rm
  s}(x,y)$. These quantities by definition satisfy the hyperbolic equations
\bena\label{wveq}
(d\delta + \delta d)_x \omega^{\rm v}(x,y) = 0
=(d\delta + \delta d)_y \omega^{\rm v}(x,y) \quad
(d\delta)_x \omega^{\rm s}(x,y) = 0
= (d\delta)_y \omega^{\rm s}(x,y) ,
\eena
the commutator property~\eqref{com}, and the wave front condition~\eqref{wfs2}.
Below, we will show that we can at least locally always choose them so that they
additionally satisfy the consistency relation
\ben\label{sv}
d_x \omega^{\rm s}(x,y) = -\delta_y \omega^{\rm v}(x,y), \quad
d_y \omega^{\rm s}(x,y) = -\delta_x \omega^{\rm v}(x,y),
\een
where $d_x = dx^\mu \wedge \frac{\partial}{\partial x^\mu}$, and
where $\delta_x = * d_x *$ is the co-differential, etc. So we assume 
\eqref{sv} in addition to \eqref{wveq}.

As a linear space, we define the
desired deformation quantization algebra $\W$ to be the vector space
generated by formal expression of the form
\ben\label{Fu}
F(u) = \int u_{i_1 \dots i_m}^{k_1 \dots k_n}
(x_1, \dots, x_n; y_1, \dots, y_m)
:\Phi^{i_1}(y_1) \dots \Phi^{i_m}(y_m)
 \Phi^\ddagger_{k_1}(x_1)
\dots  \Phi^\ddagger_{k_n}(x_n):_\omega \, ,
\een
where $u$ is a distribution subject to the wave front set
condition~\eqref{wfs1} in the variables $y_1, \dots, y_m$, but
not subject to any wave front set condition in the variables
$x_1, \dots, x_n$. Furthermore, $u$ is required to have symmetry properties in its 
argument compatible with the Grassmann parities summarized in tables~\ref{table:fields},~\ref{table:anti-fields}.

We define the $\st$-product as in eq.~\eqref{starproduct}, where the 
the differential operator is now
\ben\label{freegst}
{}_< {\mathcal D}_> = \int \frac{\delta_L}{\delta \Phi_k(x)}
\omega_{jk}(x,y)
\frac{\delta_R}{\delta \Phi_j(y)} \, dxdy
\een
where $j,k=(A^I,B^I,C^I,\bar C^I)$, and where
\ben
\Big( \omega_{jk}(x,y) \Big) = (k_{IJ}) \otimes
\left(
\begin{matrix}
\omega^{\rm v}(x,y) & -i\delta_y \omega^{\rm v}(x,y) & 0 & 0\\
-i\delta_x \omega^{\rm v}(x,y) & 0 & 0 & 0\\
0 & 0 & 0 & i\omega^{\rm s}(x,y)\\
0 & 0 & -i\omega^{\rm s}(x,y) & 0
\end{matrix}
\right) \, .
\een
Our definitions imply the commutation relations~\eqref{ccomm}, \eqref{acomm}
(with obvious modification to accommodate the Lie-algebra indices on
the fields $A^I, C^I, \bar C^I$), as well as
\ben
A^I(x) \st B^J(y) - B^J(y) \st A^I(x) = \hbar k^{IJ} \,
\delta_y \Delta^{\rm v}(x,y)
\, \myid
\, .
\een
The (graded) commutators of all other fields, in particular those involving
any of the background fields $A_I^\ddagger, B^\ddagger_I,
C^\ddagger_I, \bar C^\ddagger_I$, vanish. In this sense the background
fields are $\mc$-numbers, and their product is not deformed.

Finally, we define the *-operation in $\W$ by declaring declaring the *-operation 
to act on the fields as in the last row in tables~\ref{table:fields},~\ref{table:anti-fields}. 
This operation is compatible with $\st$-product and gives $\W$ the structure of a *-algebra.
This completes our construction of the quantization algebra $\W$ of free gauge theory.

We complete the discussion with a result on the existence of compatible scalar and 
vector Hadamard 2-point functions. 

\begin{thm}
Let $U \subset \Sigma$ be an open domain in a Cauchy surface $\Sigma$, with smooth boundary $\partial U$ and compact closure and 
vanishing first deRahm cohomology $H^1(U, d)$. Let $D(U) \subset M$ its domain of dependence. Then there exist within $D(U)$ 
Hadamard 2-point functions $\omega^{\rm s}, \omega^{\rm v}$ satisfying the compatibility 
condition~\eqref{sv}. 
\end{thm}
\noindent
{\bf Remark:} The fact that our theorem only guarantees the existence locally is not a problem for our later constructions, 
which are also local. The conditions on $U$ in the theorem arise because we need to exclude the existence of zero-modes.

\medskip
\noindent
{\it Proof:} 
It is relatively easy to prove the existence of a pair $(\hat \omega^{\rm s},
\hat \omega^{\rm v})$ satisfying~\eqref{sv}, the Hadamard
condition~\eqref{wfs2}, the commutator property,
and field equations~\eqref{wveq} on a spacetime $(\hat M,
\hat g)$, when $\hat M = \hat D(U)$ is the domain of dependence 
of $U$ inside the non-globally hyperbolic spacetime ${\mathbb R} \times U$
with metric 
\ben
\hat g = -dt^2 + h
\een
there, where $h = h_{ij} dx^i dx^j$
is a Riemannian metric on $U$ that does not depend upon $t$.

This can be shown as follows by improving a construction
by~\cite{Fewster2003}, which in turn builds on results
of~\cite{Junker2003}:  On the 3-dimensional compact
Riemannian spacetime $(U , h)$, we consider the Laplace-deRahm
operator $\Delta_h = d_U \delta_U + \delta_U d_U$ acting on 
$p$-forms with domain 
\ben\label{domain}
{\mathcal D}(\Delta_h) = \{ \xi \in H^2(U, \wedge^p T^* U) \mid 
\xi_{\rm tan} = 0 = \pounds_n \xi_{\rm nor} \ \ \text{on $\partial U$} \} \ . 
\een
Here $n$ is the normal to $\partial U$, ``tan'' resp. ``nor'' refer to the 
normal and tangential components of the form $\xi$ on $\partial U$ and 
$H^2$ indicates a Sobolov space of order $2$. Thus, we have ``Neumann'' conditions 
for the normal conditions and ``Dirichlet'' conditions for the tangential components. $\Delta_h$ is
self-adjoint on this domain by standard theory of elliptic partial differential operators on bounded domains,
see e.g.~\cite{Schechter}. Its spectrum is discrete, the eigenvalues are non-positive, 
and the eigenfunctions are smooth. Furthermore, one can see that the vanishing of the first deRahm cohomology 
implies the absence of zero eigenvalues for 1-forms ($p=1$), whereas the boundary conditions trivially 
imply the absence of zero eigenvalue for 0-forms ($p=0$). 

We denote a complete set of normalized
eigenfunctions of the scalar Laplace-operator $\Delta_h
= d_U \delta_U$, by $\varphi_{\bf k}$
with negative eigenvalues $-\nu(S, {\bf k})^2$,
labelled by an index ${\bf k} \in J(S)$ in a corresponding index
set. One defines $x=(t,{\bf x}) \in {\mathbb R} \times U$ and
\ben
u_{\bf k}^{\rm s}(t,{\bf x}) = e^{-i\nu(S, {\bf k})t} \varphi_{\bf
  k}({\bf x}), 
\een  
 as well as the ``scalar'' and ``longitudinal''
mode 1-forms on $M$ by
\bena
u_{S,{\bf k}}^{\rm v}(t,{\bf x}) &=& e^{-i\nu(S, {\bf k})t} \varphi_{\bf
  k}({\bf x}) \, dt \\
u_{L,{\bf k}}^{\rm v} (t,{\bf x}) &=&
\frac{1}{\nu(L, {\bf k})} e^{-i\nu(S, {\bf k})t} \, d \varphi_{\bf
  k}({\bf x})\,
\eena
with $\nu(L,{\bf k}) = \nu(S, {\bf k})$.
One next chooses an orthonormal set of eigenmodes for the
Laplacian $\Delta_h$ on $(U, h)$ acting on 1-forms with domain~\eqref{domain}. Using $H^1(U, d_U)=0$,
these can be uniquely decomposed into
ones in the kernel of $\delta_U$ and those in the image of the domain~\eqref{domain} under
$d_U$. We denote those in the image of $\delta_U$ by
$\xi_{{\bf k}}$ and their eigenvalues\footnote{Note that the
scalar and transversal eigenvalues need not coincide.}
by $-\nu(T, {\bf k})^2<0$,
where $\bf k$ is now an index from a set $J(T)$. We define the
corresponding ``transversal'' mode 1-forms on ${\mathbb R} \times U$ by
\ben\label{long}
u^{\rm v}_{T,{\bf k}}(t,{\bf x}) = {\rm e}^{-i\nu(T, {\bf k})t}
\, \xi_{\bf k}({\bf x}) \,\, .
\een
At this point, 
we define the vector Hadamard 2-point distribution by
\ben
\hat
\omega^{\rm v}(x,y) = - \sum_{\lambda} \sum_{{\bf k} \in
J(\lambda)} \frac{s(\lambda)}{2\nu(\lambda, {\bf k})}
\overline{ u^{\rm v}_{\lambda, {\bf k}}(x) }
u^{\rm v}_{\lambda, {\bf k}}(y) 
\een
where $s(S)=1, s(L)=-1=s(T)$, and $\lambda \in \{S,L,T\}$. It was proved
in~\cite{Fewster2003} that this is of Hadamard form within $\hat D(U)$ and that it has the desired
commutator property. We define the scalar Hadamard 2-point
distribution on $\hat D(U)$ by
\ben
\hat
\omega^{\rm s}(x,y) = \sum_{{\bf k} \in
J(S)} \frac{1}{2\nu(S, {\bf k})}
\overline{ u_{\bf k}^{\rm s}(x) } u_{\bf k}^{\rm s}(y) \, .
\een
One can argue as in~\cite{Junker2003} that this is of Hadamard form and
that it satisfies the desired commutator property. The desired
consistency property~\eqref{sv} on the ultrastatic spacetime follows by going through the
definitions. Thus, by the deformation argument, we obtain from this
a pair $(\omega^{\rm v}, \omega^{\rm s})$ on the undeformed
spacetime satisfying also the desired consistency
condition~\eqref{sv}. 

In order to show the general case, we appeal to the deformation argument originally proposed by
Fulling, Narcowich and Wald~\cite{Fulling}, which reduces the statement to the previous case. 
It is easy to see that the desired properties~\eqref{sv} are inherited in the general case
because $d$ and $\delta$ intertwine the action of the wave operators $\delta d$
on 0-forms and $d\delta + \delta d$ on 1-forms, and since
$(\omega^{\rm s}, \omega^{\rm v})$ are bisolutions to the respective
wave equations~\eqref{wveq}. Furthermore, one can show~\cite{Koehler}
using the celebrated ``propagation of
singularities theorem''~\cite{Duistermaat1977} (see Appendix C,E) that the wave front
set condition~\eqref{wfs2} is inherited on $D(U)$, too. We omit the standard details of these arguments. 
\qed

\medskip
\noindent
{\bf Time-ordered products:} The next step is to define within $\W$ the Wick products and time
ordered products satisfying conditions T1--T11. As for the time
ordered products with one factor, we make the same definition as in
the scalar field case, with the only difference that $H$ is replaced by the
matrix valued Hadamard parametrix
\ben
\Big( H_{jk}(x,y) \Big) = (k_{IJ}) \otimes
\left(
\begin{matrix}
H^{\rm v}(x,y) & -i\delta_y H^{\rm v}(x,y) & 0 & 0\\
-i\delta_x H^{\rm v}(x,y) & 0 & 0 & 0\\
0 & 0 & 0 & iH^{\rm s}(x,y)\\
0 & 0 & -iH^{\rm s}(x,y) & 0
\end{matrix}
\right) \, ,
\een
where $j,k=(A^I,B^I,C^I,\bar C^I)$. Using the Hadamard parametrix,
the time ordered products $T_1(\O)$ with one factor $\O \in \P$
are defined by complete analogy with the scalar case, and they satisfy
T1---T11. In particular, it follows from the definition that the Wick
product $T_1(\J_0)$ of the free BRST-current~\eqref{j0def} is conserved,
$dT_1(\J_0) = T_1(d\J_0) = 0$ (modulo $\I$). Hence, we can define a
a conserved BRST-charge (when the Cauchy surfaces are compact, see
above). It also follows directly from the relations in the algebra
$\W$ that $Q_0^2=0$ modulo $\J$. Thus, we can define the algebra of physical
observables, $\hat \F_0$, by the cohomology of $Q_0$
as explained in the previous section. It follows from the Ward
identity (c) below that if $\O \in \P$ is a classically gauge
invariant polynomial expression in $A^I$, i.e., $\O = \prod
\nabla^{s_i} \, dA^{I_i}$ (so that in particular $\hat s_0 \O = 0$),
then the corresponding Wick power $T_1(\O)$ is in the kernel of
$Q_0$ under the graded commutator. Thus, at ghost number 0, the
algebra contains all local covariant quantum Wick powers of
classically gauge invariant observables.

\medskip
\noindent
{\bf Representations with positivity:} 
We must finally show that the algebra $\F_0$ has representations 
which satisfy the conditions of sect. 4.1.2. Thus, we wish to construct a 
Hilbert space representation of the algebra $\F_0=\W/\I$ that gives rise to a corresponding representation of
the algebra of physical observables~\eqref{hatF} on the factor space~\eqref{Hphy0}.

Consider first a domain $U$ of a Cauchy surface as in the previous theorem 4, 
and a metric of the form $\hat g = -dt^2 + h$, where $h$ is a metric on $U$ independent
of $t$.
We let ${\frak h}_b$ be the 1-particle
indefinite inner product space spanned by the orthonormal basis elements
$e_{I,\lambda, {\bf k}}$, with $\lambda = S,L,T$ (notations and setup as in the proof of theorem 4)
and ${\bf k} \in
J(\lambda)$, with indefinite hermitian inner product defined by
$(e_{I,\lambda, {\bf k}}, e_{I',\lambda',{\bf k}'})
= s(\lambda) k_{II'} \delta_{\lambda\lambda'} \delta_{{\bf k}{\bf k}'}$. We
let
\ben
{\frak F}_b = \bigoplus_{n=0}^\infty \bigotimes^n {\frak h}_b
\een
be the corresponding (indefinite metric) standard bosonic Fock space,
with basis vectors
\ben
|I_1 \lambda_1 {\bf k}_1, \dots, I_n \lambda_n {\bf k}_n \rangle
= \frac{1}{n!} \sum_{\pi \in S_n}
e_{I_{\pi 1} \lambda_{\pi 1} {\bf k}_{\pi 1}} \otimes \cdots \otimes
e_{I_{\pi n} \lambda_{\pi n} {\bf k}_{\pi n}}
\een
and we let $a_{I,\lambda, {\bf k}}^+$ be the standard creation
operators associated with the basis vectors, i.e.,
\ben
a_{J,\nu, {\bf p}}^+ |I_1 \lambda_1 {\bf k}_1, \dots, I_n \lambda_n {\bf k}_n
\rangle
= |J \nu {\bf p}, I_1 \lambda_1 {\bf k}_1, \dots, I_n \lambda_n {\bf
  k}_n
\rangle\, .
\een
We let ${\frak h}_f$ be the 1-particle
indefinite inner product space spanned by the orthonormal basis elements
$f_{I,\pm,{\bf k}}$ and ${\bf k} \in
J(S)$, with indefinite hermitian inner product defined by
$(f_{I,s, {\bf k}}, f_{I',s',{\bf k}'})
= i\epsilon_{ss'}k_{II'}\delta_{{\bf k}{\bf k}'}$, where $\epsilon_{ss'}$ is
the anti-symmetric tensor in 2 dimensions. We
let
\ben
{\frak F}_f = \bigoplus_{n=0}^\infty \bigwedge^n {\frak h}_f
\een
be the corresponding (indefinite metric) standard fermionic Fock space,
with basis vectors
\ben
|I_1 s_1 {\bf k}_1, \dots, I_n s_n {\bf k}_n \rangle
= \frac{1}{n!} \sum_{\pi \in S_n} {\rm sgn}(\pi) \,
f_{I_{\pi 1} s_{\pi 1} {\bf k}_{\pi 1}} \otimes \cdots \otimes
f_{I_{\pi n} s_{\pi n} {\bf k}_{\pi n}}
\een
and we let $c_{I,s, {\bf k}}^+$ be the standard creation
operators associated with the basis vectors, i.e.,
\ben
c_{J,r, {\bf p}}^+ |I_1 s_1 {\bf k}_1, \dots, I_n s_n {\bf k}_n
\rangle
= |J r {\bf p}, I_1 s_1 {\bf k}_1, \dots, I_n s_n {\bf k}_n \rangle \, .
\een
The (indefinite) metric space ${\mathcal H}_0$ is defined as the
tensor product ${\mathcal H}_0 = {\frak F}_b \otimes {\frak F}_f$.
We now define the representatives of the fields $\Phi=(A^I, B^I, C^I, \bar
C^I)$ as the following operator valued distributions on ${\mathcal H}_0$:
\bena
\pi_0(A^I(x)) &=& \sum_{\lambda} \sum_{{\bf k} \in
J(\lambda)} \frac{1}{\sqrt{2\nu(\lambda, {\bf k})}}
u^{\rm v}_{\lambda, {\bf k}}(x)  \, a_{I, \lambda, {\bf k}}^+ \quad
+ \text{h.c.} \\
\pi_0(C^I(x)) &=& \sum_{{\bf k} \in J(S)}
\frac{1}{\sqrt{2\nu(S, {\bf k})}} u_{\bf k}^{\rm s}(x) c_{I,+,{\bf k}}^+ \quad
+ \text{h.c.} \\
\pi_0(\bar C^I(x)) &=& \sum_{{\bf k} \in J(S)}
\frac{1}{\sqrt{2\nu(S, {\bf k})}} u_{\bf k}^{\rm s}(x) c_{I,-,{\bf k}}^+ \quad
+ \text{h.c.,}
\eena
where ``h.c.'' is the adjoint defined by the (indefinite) metric structure on ${\mathcal H}_0$. 
We define the representative $\pi_0(B^I(x))$ to be $-i\pi_0(\delta
A^I(x))$ (motivated by the algebraic equation of motion for the field $B$), 
and we define the representative of any anti-field
$\Phi^\ddagger$ to be zero. Finally, we
define the representative of any element $F(u)$ of the
form~\eqref{Fu} by applying a normal ordering on the representatives
(all creation operators to the left or all annihilation
operators). The two-point functions of the vector- and ghost fields
are then precisely given by $\hat \omega^{\rm v}$, resp. by
$\hat \omega^{\rm s}$ defined above in the proof of theorem 4.

It may next be checked that, for compact $G$
(i.e., positive definite Cartan-Killing form $k_{IJ}$) and in the ghost number
0 sector, the positivity requirement of sec.~4.2 is fulfilled. Thus,
the physical Hilbert space~\eqref{Hphy0} inherits a positive
definite inner product. Furthermore, it follows from the
consistency condition~\eqref{sv} that it contains precisely excitation of the longitudinal modes~\eqref{long}. 

In a general, non-static spacetimes of the form $D(U)$, $U$ a bounded 
subset of a Cauchy surface as in theorem 4, a similar
construction can be carried out with the help of a deformation 
as sketched in the proof of theorem~4.

\subsection{Interacting gauge theory}

In this section, we describe in detail how the general
construction of interacting Yang-Mills theory outlined
in sec.~\ref{genoutline} is performed.
To construct perturbatively
the interacting fields in interacting gauge theory,
we need to construct the time-ordered products in the free theory
considered in the previous subsection. For time ordered products
with 1 factor, this was done there. For time ordered products with $n$
factors, this can be done as described in Sect.~3, and these
time ordered products will satisfy the analog of conditions T1--T11.

However, in gauge theory, the time ordered products must satisfy
further constraints related to gauge invariance. As we have argued in
section~4.2, in the gauge fixed formalism, we need to be able to
define an interacting BRST-charge operator, $Q_{\subsc}$, and we need
that operator to be nilpotent, i.e. $Q_{\subsc}^2=0$. In order
to meaningfully construct $Q_{\subsc}$, we need a conserved interacting
BRST-current $\J_{\subsc}$. If our time ordered products only satisfy
T1--T11 [with the symmetry property T6 replaced by graded symmetry
with respect to the Grassmann parity],
then there is in general no guarantee that the interacting
BRST-current is conserved, $d\J_{\subsc}=0$, nor that $Q_{\subsc}^2=
0$, nor that $[Q_\subsc, \Psi_\subsc] = 0$ for strictly gauge
invariant operators $\Psi$ of ghost number 0.

We will now formulate a set of Ward identities in the free theory that
will guarantee that these conditions are satisfied, and which moreover
will guarantee (formally) that the S-matrix---when it exists---is
BRST-invariant. As argued in the
previous section, with such a definition of time-ordered products, the
conditions of gauge invariance of the perturbative interacting quantum
field theory are then satisfied. The Ward identities that we want to
propose are to be viewed as an additional normalization condition
on the time ordered product, and are as follows. Consider a local
operator $\O \in \P$, given by an expansion of the form
\ben
\O = \O_0 + \lambda \O_1 + \dots \lambda^N \O_N \, .
\een
Let $f$ be a smooth compactly supported test function on $M$, and let
\ben\label{Fdef}
F= \int_M [\O_0 + \lambda f \O_1 + \dots + \lambda^N f^N \O_N] \,.
\een
Then the Ward identity that we
will consider is
%\begin{center}
%\begin{tabular}{|c|}
%\hline
%\begin{equation*}
%\\
%\end{equation*}
%\\
%\\
%\hline
%\end{tabular}
%\end{center}
\begin{equation}\label{masterward}
\boxed{
%\hat s_0
\left[Q_0, T\Big(\e^{iF/\hbar}\Big) \right] =
-\frac{1}{2} T\bigg( (S_0+F,S_0+F) \otimes \e^{iF/\hbar}\bigg) \quad
\text{modulo $\I$}\, .
}
\end{equation}
Here, $Q_0$ is the free BRST-charge operator, $(\, . \, , \,
. \,)$ is the anti-bracket~\eqref{bvh}, and $[ \, , \, ]$ is the
graded commutator in the algebra $\W$. $\I \subset \W$ is the ideal 
generated by the free field equation~\eqref{ideal} for the free action $S_0$.\footnote{
Note that the free action contains terms that are linear in $\Phi$ and 
$\Phi^\ddagger$, so the free field equations will have a source given by 
the anti-fields.
}
As with all generating type formulae in this work, this is to be
understood as a shorthand for the hierarchy of identities that
are obtained when the above expression is expanded as a formal power
series in $\lambda$. We now write out explicitly this hierarchy of
identities. For this, it is convenient to introduce some notation.
We denote by $I=\{k_1,\dots,k_r\}$ subsets of $\underline{n}=\{1,\dots,n\}$, and
we write $r=|I|$ for the number of elements. We set
$X_I = (x_{k_1},\dots,x_{k_r})
$, and we put
\ben
\O_r(X_I)=r! \, \O_r(x_{k_1})\delta(x_{k_1},\dots,x_{k_r}).
\een
With these notations, the Ward-identity~\eqref{masterward} can be expressed as
\begin{multline}\label{W}
\sum_{I_1 \cup \dots \cup I_t=\underline{n}}
\bigg(\frac{i}{\hbar}\bigg)^t
\left[Q_0,
T_t(\O_{|I_1|}(X_{I_1}) \otimes \dots \O_{|I_t|}(X_{I_t})
) \right]
=\\ -\sum_{I_1 \cup \dots \cup I_t=\underline{n}} \bigg(\frac{i}{\hbar}\bigg)^{t-1}
  \sum_{k=1}^t (-1)^{\epsilon_k}
T_t(
\O_{|I_1|}(X_{I_1}) \otimes \dots
\hat s_0 \O_{|I_k|}(X_{I_k})
\otimes \dots
\O_{|I_t|}(X_{I_t})
) \\
-\sum_{I_1 \cup \dots \cup I_t=\underline{n}}
\bigg(\frac{i}{\hbar}\bigg)^{t-2}
\sum_{1\le k<l\le n} (-1)^{\epsilon_{k}\epsilon_l}
T_{t-1}(
\O_{|I_1|}(X_{I_1}) \otimes \dots
(\O_{|I_k|}(X_{I_k}), \O_{|I_l|}(X_{I_l}))
\otimes \dots
\O_{|I_t|}(X_{I_t})
)
\end{multline}
modulo $\I$,
where $\epsilon_k =
\epsilon(\O_1)+\dots+\epsilon(\O_{k-1})$.
We will not prove the above Ward identities for arbitrary operators
$\O$ in this work, but only for certain special cases, which are
relevant for our analysis of gauge invariance.
These cases are
\begin{itemize}
\item[(T12a)] $\O$ is given by the interaction Lagrangian,
$\O=\lambda \L_1 + \lambda^2 \L_2$,
\item[(T12b)] $\O$ is given by a linear combination of the
interaction Lagrangian, and the BRST-current
$\O= \lambda \L_1 + \lambda^2 \L_2 + \gamma \wedge (\J_0 + \lambda \J_1)$
(evaluation of the Ward identity to first order in $\gamma \in \Omega^1_0(M)$).
\item[(T12c)] $\O=\lambda \L_1 + \lambda^2 \L_2
+ \gamma \wedge \sum \lambda^k \Psi_k \in \P^4(M)$ is given by a linear combination of the interaction
Lagrangian and a strictly gauge invariant operator $\Psi = \sum_k \lambda^k
\Psi_k \in \P^p(M)$ of ghost number 0, i.e., of the form given by eq.~\eqref{pqd} (evaluation of the Ward identity
to first order in $\gamma \in \Omega^{4-p}_0(M)$).
\end{itemize}
It is only for those cases that
we will prove the Ward-identities~\eqref{W}, and that proof is
provided in section~4.4. For convenience, we
now give explicitly the form of the Ward-identities
in the cases (a), (b), and (c).

{\bf Case (T12a)} The Ward identities in that case are given
explicitly by
\begin{multline}
\sum_{I_1 \cup \dots \cup I_t=\underline{n}}
\bigg(\frac{i}{\hbar}\bigg)^t
\left[Q_0,
T_t(\L_{|I_1|}(X_{I_1}) \otimes \dots \L_{|I_t|}(X_{I_t})
)\right]
=\\ -\sum_{I_1 \cup \dots \cup I_t=\underline{n}}
\bigg(\frac{i}{\hbar}\bigg)^{t-1}
\sum_{k=1}^t
T_t(
\L_{|I_1|}(X_{I_1}) \otimes \dots
\hat s_0 \L_{|I_k|}(X_{I_k})
\otimes \dots
\L_{|I_t|}(X_{I_t})
) \\
-
\sum_{I_1 \cup \dots \cup I_t=\underline{n}}
\bigg(\frac{i}{\hbar}\bigg)^{t-2}
\sum_{1\le j<k\le t}
T_{t-1}(
\L_{|I_1|}(X_{I_1}) \otimes \dots
(\L_{|I_j|}(X_{I_j}), \L_{|I_k|}(X_{I_k}))
\otimes \dots
\L_{|I_t|}(X_{I_t})
) \, ,
\end{multline}
modulo $\I$.
\medskip

{\bf Case (T12b)} The Ward identities in that case are given
explicitly by
\begin{multline}
\sum_{I_1 \cup \dots \cup I_t=\underline{n}}
\bigg(\frac{i}{\hbar}\bigg)^{t-1}
\left[Q_0,
T_t(
\J_{|I_1|}(y,X_{I_1}) \otimes \L_{|I_2|}(X_{I_2}) \otimes
\dots \otimes \L_{|I_t|}(X_{I_t})
)\right]
= \\
\sum_{I_1 \cup \dots \cup I_t=\underline{n}}
\bigg(\frac{i}{\hbar}\bigg)^{t-2}
\sum_{i=2}^t
T_t(
\J_{|I_1|}(y,X_{I_1}) \otimes \L_{|I_2|}(X_{I_2}) \otimes
\dots \hat s_0 \L_{|I_i|}(X_{I_i}) \otimes \dots \L_{|I_t|}(X_{I_t})
) \\
-\sum_{I_1 \cup \dots \cup I_t=\underline{n}}
\bigg(\frac{i}{\hbar}\bigg)^{t-2}
T_t(
\hat s_0 \J_{|I_1|}(y,X_{I_1}) \otimes \L_{|I_2|}(X_{I_2}) \otimes
\dots \otimes \L_{|I_t|}(X_{I_t})
) \\
+
\sum_{I_1 \cup \dots \cup I_t=\underline{n}}
\bigg(\frac{i}{\hbar}\bigg)^{t-3}
\sum_{2\le i<j\le t}
T_{t-1}(
\J_{|I_1|}(y,X_{I_1}) \otimes \L_{|I_2|}(X_{I_2}) \otimes
\dots
(\L_{|I_i|}(X_{I_i}) ,
\L_{|I_j|}(X_{I_j}))
\otimes \dots \L_{|I_t|}(X_{I_t})
) \\
-
\sum_{I_1 \cup \dots \cup I_t=\underline{n}}
\bigg(\frac{i}{\hbar}\bigg)^{t-2}
\sum_{2\le i \le t}
T_{t-1}(
\L_{|I_2|}(X_{I_2}) \otimes \dots (\J_{|I_1|}(y,X_{I_1}),
\L_{|I_i|}(X_{I_i})) \otimes
\dots \L_{|I_t|}(X_{I_t})) \, ,
\end{multline}
modulo $\I$.
Here $\J_1(y,x)=\J_1(y)\delta(x,y)$.
\medskip

{\bf Case (T12c)} Let $\Psi = \Psi_0 + \lambda \Psi_1 + \dots + \lambda^N \Psi_N$
be a strictly gauge invariant local field polynomial of ghost number
zero. Thus, by formula~\eqref{pqd}, up to local curvature terms which
we may ignore, $\Psi = \prod \Theta_{s_i}(F, {\mathcal D} F, {\mathcal
  D}^2 F, \dots)$, where $\Theta_s$ are invariant polynomials of the
Lie-algebra. The Ward identities in that case are given
explicitly by
\begin{multline}
\sum_{I_1 \cup \dots \cup I_t=\underline{n}} \bigg(\frac{i}{\hbar}\bigg)^{t-1}
\left[Q_0,
T_t(
\Psi_{|I_1|}(y,X_{I_1}) \otimes \L_{|I_2|}(X_{I_2}) \otimes
\dots \otimes \L_{|I_t|}(X_{I_t})
) \right]
= \\
-
\sum_{I_1 \cup \dots \cup I_t=\underline{n}}
\bigg(\frac{i}{\hbar}\bigg)^{t-2} \sum_{i=2}^t
T_t(
\Psi_{|I_1|}(y,X_{I_1}) \otimes \L_{|I_2|}(X_{I_2}) \otimes
\dots \hat s_0 \L_{|I_i|}(X_{I_i}) \otimes \dots \L_{|I_t|}(X_{I_t})
) \\
-\sum_{I_1 \cup \dots \cup I_t=\underline{n}} \bigg(\frac{i}{\hbar}\bigg)^{t-2}
T_t(
\hat s_0 \Psi_{|I_1|}(y,X_{I_1}) \otimes \L_{|I_2|}(X_{I_2}) \otimes
\dots \otimes \L_{|I_t|}(X_{I_t})
) \\
-
\sum_{I_1 \cup \dots \cup I_t=\underline{n}}
\bigg( \frac{i}{\hbar} \bigg)^{t-3}
\sum_{2\le i<j\le t}
T_{t-1}(
\Psi_{|I_1|}(y,X_{I_1}) \otimes \L_{|I_2|}(X_{I_2}) \otimes
\dots
(\L_{|I_i|}(X_{I_i}) ,
\L_{|I_j|}(X_{I_j}))
\otimes \dots \L_{|I_t|}(X_{I_t})
) \\
-\sum_{I_1 \cup \dots \cup I_t=\underline{n}}
\bigg( \frac{i}{\hbar} \bigg)^{t-2}
\sum_{i=2}^t
T_{t-1}(
\L_{|I_2|}(X_{I_2}) \otimes \dots (\Psi_{|I_1|}(y,X_{I_1}),
\L_{|I_i|}(X_{I_i})) \otimes
\dots \L_{|I_t|}(X_{I_t})) \, ,
\end{multline}
modulo $\I$.
\medskip
\noindent

We will give a proof of the Ward-identities T12a--T12c
in subsec.~4.4.
We will then show in subsec.~4.6 that
the Ward identities T12a imply the conservation of
the interacting BRST-current, $d\J_{\subsc}=0$.
We will prove in subsec.~4.7
that the Ward identities T12b furthermore imply that $Q_{\subsc}^2 = 0$
and we will show in subsec.~4.8 that the Ward identities
T12c imply $[Q_\subsc, \Psi_\subsc] = 0$ for strictly gauge
invariant operators $\Psi$ at ghost number 0.
The Ward identity T12a also formally implies the BRST-invariance
of the $S$-matrix (see subsec.~4.5), provided the latter exists
(which is not the case in Minkowski space, and appears even more unlikely
in curved spacetime). We will not analyze this existence question here,
so in this sense the BRST-invariance
of the $S$-matrix is not a rigorous result unlike the other results in
our paper.

As an aside,
we note that, the Ward identities T12a,
T12b, and T12c are {\em incompatible} with the identity
\begin{multline}\label{wrong}
\Big[Q_0, T_n(\O_1(x_1) \otimes \cdots \otimes \O_n(x_n)) \Big] \\
= i\hbar \sum_{i=1}^n (-1)^{\epsilon_i} T_n
(\O_1(x_1) \otimes \cdots \hat s_0 \O_i(x_i) \otimes \cdots \O_n(x_n))
\quad \text{mod $\I$ (WRONG!)}\, ,
\end{multline}
unless none of the fields $\O_i$ contains anti-fields.
The above identity has been considered before in the context
of flat spacetime in~\cite{Duetsch2001},
where it has been termed ``Master BRST-identity.'' It appears that
it is impossible to satisfy this identity (even for $n=1$)
when anti-fields are present. It would also not imply either the
conservation of the interacting BRST current $\J_{\subsc}$ nor the nilpotency of the
interacting BRST charge in a framework with anti-fields. Since the
use of anti-fields also appears to be essential in order to derive
sufficiently strong constraints on potential anomalies to the
BRST-Ward identities, we believe that~eq.~\eqref{wrong} is not a good
starting point for the proof of gauge invariance in perturbative
Yang-Mills theory.

\subsection{Inductive proof of Ward identities T12a, T12b, and T12c}

We now show that the Ward identities can be satisfied together with
T1---T11 by making a suitable redefinition of the time-ordered
products if necessary. The Ward identity~\eqref{W} is
an identity modulo $\I$, that is, it is required to hold only on
shell. For the proof of that identity it is actually useful to
consider a more stringent ``off-shell'' version of the identity.
Even though that off-shell version is more stringent, it will in fact turn out
to be easier to prove, as it gives, at the same time, stronger constraints
of cohomological nature on the the possible anomalies than the corresponding
on-shell version.

To set up the off-shell version of our Ward-identity,
we first recall the definition $\hat s_0 = s_0 + \sigma_0$ of
the free Slavnov-Taylor differential, given above in eq.~\eqref{hs01} and~\eqref{hs02}.
As it stands, the differential $\hat s_0$ was defined as a map $\hat s_0: \P(M) \to \P(M)$, i.e.,
it acts on polynomial expressions in the {\em classical} fields
$\Phi, \Phi^\ddagger$. We will now extend the action of $\hat s_0$ to
the non-commutative algebra $\W$. For this, we recall that
the algebra $\W$ may be viewed as the closure of the CCR-algebra $\Wn$, which in turn
is generated by expressions of the form $F_1 \st \dots \st F_n$, where each
$F_i$ is given by $\int f_i \wedge \O_i$, with $f_i$ smooth and of compact support, and
with $\O_i$ given by one of the "basic fields" $\Phi, \Phi^\ddagger$. To define
the action of $\hat s_0$ on such elements of $\Wn$, we set
\ben\label{s0act}
\hat s_0 \bigg(
\O_{1}(x_1) \st \dots \st \O_s(x_n)
\bigg) =
\sum_{i=1}^n (-1)^{\sum_{l<i} \epsilon_l}
\O_1(x_1) \st \dots \hat s_0 \O_i(x_i) \st
\dots  \O_n(x_n) \, ,
\een
where $\O_i$ is either a basic field $\Phi$, or an anti-field
$\Phi^\ddagger$. This defines Slavnov-Taylor differential
$\hat s_0$ as a graded derivation (denoted by the same symbol)
of the algebra $\Wn$. As we have remarked, the subalgebra
$\Wn \subset \W$ is dense (in the H\" ormander topology).
Thus, we can uniquely extend $\hat s_0$ to a graded derivation on $\W$ by
continuity with respect to this topology. We will again denote this graded
derivation $\hat s_0: \W \to \W$ by the same symbol. Actually, we must still
check that the definition~\eqref{s0act} is consistent, i.e., compatible with the
algebra relations in $\Wn$. We formulate this result as a lemma:

\begin{lemma}
The formula~\eqref{s0act} defines a
graded derivation on $\W$.
\end{lemma}

\noindent
{\em Proof:}
The basic algebraic relations in $\Wn$ are the
graded commutation relations
\ben
[\Phi^i(x), \Phi^j(y)] = i\hbar \Delta^{ij}(x, y) \, \myid \, , \quad
[\Phi^i(x), \Phi_j^\ddagger(y)] = 0 = [\Phi_i^\ddagger(x), \Phi_j^\ddagger(y)] \, ,
\een
where $\Delta^{ij}$ is the matrix of commutator functions given by
\ben
\Big( \Delta^{jk}(x,y) \Big) = (k_{IJ}) \otimes
\left(
\begin{matrix}
\Delta^{\rm v}(x,y) & -i\delta_y \Delta^{\rm v}(x,y) & 0 & 0\\
-i\delta_x \Delta^{\rm v}(x,y) & 0 & 0 & 0\\
0 & 0 & 0 & i\Delta^{\rm s}(x,y)\\
0 & 0 & -i\Delta^{\rm s}(x,y) & 0
\end{matrix}
\right) \, ,
\een
where $\Phi^i=(A^I,B^I,C^I,\bar C^I)$, and where $\Delta^{\rm v}, \Delta^{\rm s}$ are the
advanced minus retarded propagators for vectors and scalars, see Appendix~E. To show that
the definition of $\hat s_0$ on $\Wn$ is consistent, we next apply the definition~\eqref{s0act}
to the above graded commutators and check that we get identities. This follows from the relations
\ben\label{svdelta}
d_x \Delta^{\rm s}(x,y) = -\delta_y \Delta^{\rm v}(x,y), \quad
d_y \Delta^{\rm s}(x,y) = -\delta_x \Delta^{\rm v}(x,y),
\een
which in turn a direct consequence of the field equations satisfied by the advanced and
retarded propagators for scalars and vectors. \qed

\medskip

We are now in a position to formulate the desired off-shell version of our (anomalous)
Ward identity that will eventually enable us to prove T12a, T12b, and T12c. We formulate
our result in a proposition:

\medskip
\noindent
\paragraph{\bf Proposition 3:} (Anomalous Ward Identity) For a general
prescription for time-ordered products satisfying T1---T11, the identity
\ben\label{WA}
\boxed{
\\
\hat s_0 T\Big(\e^{iF/\hbar}\Big) = \frac{i}{2\hbar}
T\bigg( (S_0+F,S_0+F) \otimes \e^{iF/\hbar} \bigg) +
\frac{i}{\hbar}
T \bigg( A(\e^{F}) \otimes \e^{iF/\hbar} \bigg)\,
\\
}
\een
holds. Here $F = \int f \wedge \O$ is
any smeared local field with $\O \in \P^p(M), f \in \Omega^{4-p}_0(M)$ and
$A(\e^F)$ is the anomaly, given by
\ben
A(\e^F) = \sum_{n \ge 0} \frac{1}{n!} A_n(F^{\otimes n}) \, ,
\een
and where $A_n: \P^{k_1}(M) \otimes \dots \otimes \P^{k_n}(M) \to
\P^{k_1/\dots/k_n}(M^n)$ are local functionals
supported on the total diagonal. The anomaly satisfies the following
further properties:
\begin{enumerate}
\item[(i)] $A(\e^F)=O(\hbar)$.
\item[(ii)] Each $A_n$ is locally and covariantly constructed out of the metric.
\item[(iii)] Each $A_n$ has ghost number one, in the sense that $\N_{g}^{} \circ A_n - A_n \circ \Gamma_n \N_{g}
= A_n$, where $\N_g$ is the number counter for the ghost fields, see eq.~\eqref{Ngdef} (with
additional terms for the anti-fields), and
\ben
\Gamma_n \N_g = \sum_{i=1}^n id \otimes \cdots \N_g \otimes \cdots id: \P^{\otimes n} \to \P^{\otimes n} \, .
\een
\item[(iv)]  Each $A_n$ has dimension 0, in the sense that $\N_{d} \circ A_n - A_n \circ \Gamma_n \N_{d}
= 0$, where $\N_d := \N_f + \N_r$ is the dimension counter, which is the
sum of the dimensions $\N_f$ of the individual fields and anti-fields
(see the tables above), and the dimensions $\N_r$ of the curvature terms.
\item[(v)] The maps $A_n$ are real in the sense that $A(\e^F)^* = A(\e^{F^*})$.
\end{enumerate}
\medskip
\noindent

Before we come to the proof of this key proposition, we note that, in the absence
of anomalies $A(\e^F) = 0$, the off-shell version of our Ward-identity becomes
\begin{equation}\label{WW}
\boxed{
\hat s_0 T\Big(\e^{iF/\hbar}\Big) =
\frac{i}{2\hbar} T\bigg( (S_0+F,S_0+F) \otimes \e^{iF/\hbar}\bigg) \, .
}
\end{equation}
The difference to~\eqref{masterward} is that on the left side, we
do not have the graded commutator with $Q_0$, but instead we act
with the Slavnov-Taylor map $\hat s_0$, which is the sum
of the standard free BRST-differential $s_0$ generated by $Q_0$,
and the Koszul-Tate differential. The addition of the Koszul-Tate
differential is crucial to obtain an identity that holds {\em off shell},
and not just modulo the free field equations as eq.~\eqref{masterward}.
As already indicated, despite being more
stringent, the sharpened off-shell Ward identity~\eqref{WW} is in fact simpler to prove
than the corresponding on-shell identity~\eqref{masterward}, as
it also allows one to derive more stringent consistency conditions on the possible
anomalies. These consistency conditions rely in an essential way upon
the use of the anti-fields, and this is the principal reason why we
have introduced such fields in our construction.

\medskip
\noindent
{\em Proof of Proposition~3}: The proof of the anomalous
Ward-identity~\eqref{WA} proceeds by induction in
the order $n$ in perturbation
theory, noting that the anomalous Ward-identity holds at order $n$ if it
holds up to order $n-1$, modulo a contribution supported on the total
diagonal. That contribution is defined to be $A_n$.  In more detail,
consider $n$ local functionals $F_1, \dots, F_n$ with $F_i = \int f_i
\wedge \O_i$, with $f_i$ a form of compact support and form degree
complementary to that of $\O_i \in \P(M)$. For definiteness and simplicity,
we assume that all $F_i$ have Grassmann parity 0; in the general case one
proceeds similarly. The anomalous Ward-identity~\eqref{WA}
at order $n$ is then the statement that
\bena\label{componentform}
&& \hat s_0 T_n(F_1 \otimes \dots \otimes F_n) =\non\\
&& \sum_{k=0}^n T_n(F_1 \otimes \dots \hat s_0 F_k \otimes \dots F_n) +
\frac{\hbar}{i}
\sum_{k<j} T_{n-1}(F_1 \otimes \dots (F_j,F_k) \otimes \dots F_n) +\non\\
&&\sum_{t=1}^{n} \sum_{k_1 < \dots < k_t} \sum_{l_1 < \dots < l_{n-t}}
\left( \frac{\hbar}{i} \right)^{t-1} T_{n-t+1}
(A_t(F_{k_1} \otimes \dots F_{k_t}) \otimes F_{l_1} \otimes \dots
F_{l_{n-t}}) \, .
\eena
We now look at the individual terms in this expression. We decompose
$\hat s_0 = s_0 + \sigma_0$ into its pure BRST-part $s_0$ and the Koszul-Tate
differential $\sigma_0$. Letting $\epsilon_i$ be the Grassmann parity of $f_i$
(equal to that of $\O_i$, since $F_i$ is assumed to be bosonic), we have
\bena\label{prep1}
&&\sigma_0 T_n(F_1 \otimes \dots \otimes F_n) \non\\
&=& \sigma_0
\bigg( (-1)^{\sum_{i<j} \epsilon_i \epsilon_j}
\int f_1(x_1) \dots f_n(x_n) \, T_n(\O_1(x_1) \otimes
\dots \otimes \O_n(x_n)) \, dx_1 \dots dx_n \bigg) \non\\
&=& (-1)^{\sum_{i<j} \epsilon_i \epsilon_j}
\int \sum_{k=1}^n (-1)^{\sum_{l<k} \epsilon_l} [f_1(x_1) \dots
\sigma_0 f_k(x_k) \dots f_n(x_n)] \st T_n \Big(\otimes_i \O_i(x_i) \Big) \, dx_1 \dots dx_n \non\\
&=& (-1)^{\sum_{i<j} \epsilon_i \epsilon_j}
\int \sum_{k=1}^n (-1)^{\sum_{l<k} \epsilon_l} [f_1(x_1) \dots
\frac{\delta_R S_0}{\delta \Phi(y)} \frac{\delta_L f_k(x_k)}{\delta \Phi^\ddagger(y)}
\dots f_n(x_n)] \st T_n\Big( \otimes_i \O_i(x_i) \Big) \, dy dx_1 \dots dx_n \non\\
&=& \sum_{k=1}^n \int \frac{\delta_R S_0}{\delta \Phi(y)} \st
T_n
\left(
F_1 \otimes \dots \frac{\delta_L F_k}{\delta \Phi^\ddagger(y)} \otimes
\dots F_n
\right) \, dy \, ,
\eena
and we have
\ben\label{prep2}
\sum_{k=1}^n T_n (F_1 \otimes \dots \sigma_0 F_k \otimes \dots F_n)
=
\sum_{k=1}^n \int T_n \left(F_1 \otimes \dots \frac{\delta_R S_0}{\delta
  \Phi(x)} \wedge
\frac{\delta_L F_k}{\delta
  \Phi^\ddagger(x)} \otimes \dots F_n \right) \, dx \, ,
\een
using the definition of $\sigma_0$, see
eq.~\eqref{hs02} and the following table. We may combine these two identities
into the following identity for the corresponding generating functionals:
\bena\label{koszulmwi}
&&\sigma_0 T(\e^{iF/\hbar}) - \frac{i}{\hbar} T(\sigma_0 F \otimes \e^{iF/\hbar})\\
&=& \frac{i}{\hbar} \int \frac{\delta_R S_0}{\delta \Phi(x)} \st T \bigg(
\frac{\delta_L F}{\delta \Phi^\ddagger(x)} \otimes \e^{iF/\hbar}
\bigg) dx
-\frac{i}{\hbar} \int T \bigg( \Big( \frac{\delta_R S_0}{\delta \Phi(x)} \wedge
\frac{\delta_L F}{\delta \Phi^\ddagger(x)} \Big) \otimes \e^{iF/\hbar}
\bigg) dx \, . \non
\eena
To manipulate this expression,
we now use a proposition formulated and proven first in~\cite{MWD}
[see eq.~(5.48) in lemma~11 of this reference].

\medskip
\noindent
\paragraph{\bf Proposition 4:} ("Master Ward Identity") Let $\psi \in C^\infty_0(M) \cdot \P(M)$ be
arbitrary, i.e., $\psi$ is a local functional of the fields, times a compactly supported cutoff function.
Set
\ben
B = \int_M \frac{\delta S_0}{\delta \Phi(x)} \wedge \psi(x) \, , \quad
\delta_B F = \int_M \frac{\delta F}{\delta \Phi(x)} \wedge \psi(x) \, .
\een
Then we have
\ben\label{dutschid}
T \Big( \Big[ B + \delta_B F + \Delta_B(\e^F) \Big] \otimes \e^{iF/\hbar} \Big)
= \int \frac{\delta S_0}{\delta \Phi(x)} \st T \Big( \psi(x) \otimes \e^{iF/\hbar} \Big)  \, dx \, .
\een
Here $\Delta_B(\e^F) = \sum_n \frac{1}{n!} \Delta_n(F^{\otimes n})$ and each
$\Delta_n: \P^{k_1}(M) \otimes \dots \otimes \P^{k_n}(M) \to
\P^{k_1/\dots/k_n}(M^{n})$ is a linear map that is supported on the total diagonal.
If the $F_i$ do not depend on $\hbar$, then the quantity $\Delta_n(F_1 \otimes \dots \otimes F_n)$ is
of order $O(\hbar)$.

\medskip
\noindent
We will outline the proof of this proposition at the end of the present proof. We now apply
the Master Ward identity to the case when $\psi(x) = \delta_L F/\delta \Phi^\ddagger(x)$. Then we
obtain, for the last term in eq.~\eqref{koszulmwi} the expression
\bena
&&-\frac{i}{\hbar} \int T \bigg( \Big( \frac{\delta_R S_0}{\delta \Phi(x)} \wedge
\frac{\delta_L F}{\delta \Phi^\ddagger(x)} \Big) \otimes \e^{iF/\hbar}
\bigg) = \non\\
&&\frac{i}{\hbar} T \Big( \delta_B F \otimes \e^{iF/\hbar} \Big)
- \frac{i}{\hbar} \int \frac{\delta_R S_0}{\delta \Phi(x)} \st T \bigg(
\frac{\delta_L F}{\delta \Phi^\ddagger(x)} \otimes \e^{iF/\hbar}
\bigg) + \non\\
&&\frac{i}{\hbar} T\Big( \Delta_B(\e^F) \otimes \e^{iF/\hbar} \Big)
\, .
\eena
Now, we have, with our choice $\psi(x) = \delta_L F/\delta \Phi^\ddagger(x)$,
\ben
\delta_B F = \int_M \frac{\delta_R F}{\delta \Phi(x)} \wedge \frac{\delta_L F}{\delta \Phi^\ddagger(x)} = \frac{1}{2} (F, F) \, .
\een
Thus, we altogether obtain the identity
\ben\label{koszulWA}
\boxed{
\\
\sigma_0 T\bigg(\e^{iF/\hbar}\bigg) - \frac{i}{\hbar}
T\bigg( \sigma_0 F \otimes \e^{iF/\hbar} \bigg)
= \frac{i}{2 \hbar} T \bigg( (F, F) \otimes \e^{iF/\hbar} \bigg) +
\frac{i}{\hbar}
T \bigg( \Delta_B(\e^{F}) \otimes \e^{iF/\hbar} \bigg)\,
\\
}
\een
which is in fact just another equivalent way of expressing the Master Ward Identity.
This identity in effect will take care of all terms in eq.~\eqref{componentform} involving the
Koszul-Tate differential.
We now look at the terms involving the pure BRST-differential $s_0$. To deal with these
terms, we now use the following identity:

\begin{lemma}\label{ls0part}
\bena\label{s0part}
&& s_0 T_n(F_1 \otimes \dots \otimes F_n) = \sum_{k=0}^n T_n(F_1 \otimes \dots s_0 F_k \otimes \dots F_n) +
\non\\
&&\sum_{t=1}^{n} \sum_{k_1 < \dots < k_t} \sum_{l_1 < \dots < l_{n-t}}
\left( \frac{\hbar}{i} \right)^{t-1} T_{n-t+1}
(\delta_t(F_{k_1} \otimes \dots F_{k_t}) \otimes F_{l_1} \otimes \dots
F_{l_{n-t}}) \, .
\eena
Here, $\delta_n$ is a map of the same nature as $A_n$, i.e., it is supported on the total diagonal,
and it is of order $O(\hbar)$. A formula generating these identities is
\ben\label{brstWA}
s_0 T\Big(\e^{iF/\hbar}\Big) - \frac{i}{\hbar}
T\bigg( s_0 F \otimes \e^{iF/\hbar} \bigg)
=
\frac{i}{\hbar}
T \bigg( \delta(\e^{F}) \otimes \e^{iF/\hbar} \bigg)\, .
\een
\end{lemma}

\noindent
{\em Proof of Lemma~\ref{ls0part}:}
For $n=1$ the identity says that $s_0 T_1(F) = T_1 (s_0 F) +
T_1(\delta_1(F))$, and we simply define $\delta_1(F)$ in this way. Since there
is no anomaly in the classical limit, it follows that $\delta_1(F)$ is of
order $\hbar$. We now proceed inductively to prove the equation
for all $n$. Assume that it has been shown for any number of
factors up to $n-1$, and the $\delta_1, \dots, \delta_{n-1}$ have consequently
been defined. Take $n$ functionals $F_1,\dots,F_n$ with the property that
the support of the first $l$ functionals is not in the future
of the support of the last $n-l$ functionals, where $l$ is not equal
to $0$ or $n$. Define $M_n$ to be the difference between the left and right
terms in the above equation, with the $n$-th term in the sum (the one containing $\delta_n$)
omitted.
Then, using the causal factorization property of the
time ordered products and the assumed support properties of the $F_i$, it follows
that
\bena
&& M_n(F_1\otimes \dots \otimes F_n) =
-s_0 \Big( T_l(F_1 \otimes \dots \otimes F_l) \st T_{n-l}(F_{l+1} \otimes \dots \otimes F_n) \Big) +\non\\
&& \sum_{k=0}^l T_l(F_1 \otimes \dots s_0 F_k \otimes \dots F_l) \st
T_{n-l}(F_{l+1} \otimes \dots \otimes F_n)
+ \\
&& \sum_{k=l+1}^n T_n(F_1 \otimes \dots \otimes \dots F_l) \st
T_{n-l}(F_{l+1} \otimes \dots s_0 F_k \otimes \dots F_n)
+ \non\\
&&\sum_{t=1}^{l} \sum_{k_1 < \dots < k_t \le l} \sum_{l_1 < \dots < l_{n-t} \le l}
\left( \frac{\hbar}{i} \right)^{t-1} T_{l-t+1}
(\delta_t(F_{k_1} \otimes \dots F_{k_t}) \otimes F_{l_1} \otimes \dots
F_{l_{l-t}}) \st T_{n-l}(F_{l+1} \otimes \dots \otimes F_n) + \non\\
&&\sum_{t=1}^{n-1} \sum_{l<k_1 < \dots < k_t} \sum_{l<l_1 < \dots < l_{n-t}}
\left( \frac{\hbar}{i} \right)^{t-1} T_l(F_1 \otimes \dots \otimes F_l) \st T_{n-l-t+1}
(\delta_t(F_{k_1} \otimes \dots F_{k_t}) \otimes F_{l_1} \otimes \dots
F_{l_{n-l-t}}) \non\, .
\eena
We now apply the inductive hypothesis that eq.~\eqref{s0part} holds at
order $n-1$, together with the fact that $s_0$ is a graded derivation
of $\W$ (we proved this above for $\hat s_0$, the proof for $s_0$ is completely
analogous). If this is done, then it follows that $M_n(F_1 \otimes \dots \otimes F_n)=0$
under the assumed support properties for the $F_i$.
Consequently, $M_n$ must be a functional
valued in $\W$ that is supported on the total diagonal. That
functional must hence be of the form $(\hbar/i)^{n-1}
T_1(\delta_n(F_1 \otimes \dots \otimes F_n))$ for some $\delta_n$,
which we hence take as the definition
of $\delta_n$.

We must next show that $\delta_n(F^{\otimes n})$ is of order $\hbar$.
For this, we pick a quasifree state $\omega$ of $\W$, and we
define, as described in Appendix~B, the "connected time ordered products"
$T^c_\omega$ by the formula
\ben
T_{n,\omega}^c(F_1 \otimes \dots \otimes F_n) := T_n(F_1 \otimes \dots \otimes F_n) -
\sum_P :\prod_{J \in P} T_{|J|} \Big( \otimes_{j \in J} F_j \Big) :_\omega
\een
where $P$ runs over all partitions of $\{1, \dots, n\}$, and where $J$ runs through
the disjoint sets in the given partition. A generating type functional formula
can be obtained using the linked cluster theorem, and is given by eq.~\eqref{conngen}.
The key fact about the connected products is that the $n$-th product
is of order $O(\hbar^{n-1})$ if the $F_i$ themselves are of order $O(1)$.
This will now be used by formulating eq.~\eqref{s0part} in terms of connected
products. Using generating functional expression for the connected time ordered products, and
using the fact that $s_0$ is a derivation with respect to the Wick product (which follows
from eq.~\eqref{sv}),
one can easily see that
\bena\label{s0partc}
&& \left( \frac{i}{\hbar} \right)^{n-1} s_0 T_{n,\omega}^c(F_1 \otimes \dots \otimes F_n) - \sum_{k=0}^n \left( \frac{i}{\hbar} \right)^{n-1} T_{n,\omega}^c(F_1 \otimes \dots s_0 F_k \otimes \dots F_n) -
\non\\
&&\sum_{t=1}^{n-1} \sum_{k_1 < \dots < k_t} \sum_{l_1 < \dots < l_{n-t}}
\left( \frac{i}{\hbar} \right)^{n-t} T_{n-t+1, \omega}^c
(\delta_t(F_{k_1} \otimes \dots F_{k_t}) \otimes F_{l_1} \otimes \dots
F_{l_{n-t}}) \non\\
&& = T_1\Big( \delta_n (F_1 \otimes \dots \otimes F_n) \Big) \, .
\eena
Now, if we inductively assume that $\delta_t$ is of order $O(\hbar)$ for
orders $t<n$, then it follows that the order of the second sum in the above expression
is $O(\hbar)$. Furthermore, the first two terms on the left side in
the above equation precisely cancel up to a term of order $O(\hbar)$.
This follows from the fact that the limit $\lim_{\hbar} T^c_{n,\omega}/\hbar^{n-1}$
correspond to the "tree diagrams", and there are no anomalies at tree level~\cite{NewDuetsch}.
Thus, $\delta_n = O(\hbar)$,
as we desired to show. \qed

We are now in a position to complete the proof. From eqs.~\eqref{brstWA} and~\eqref{koszulWA}
we get the desired Ward identity~\eqref{WA} with
\ben
A(\e^F) := \Delta_B(\e^F) + \delta(\e^F) \,, \quad B =
\int_M \frac{\delta_R S_0}{\delta \Phi(x)} \wedge \frac{\delta_L F}{\delta \Phi^\ddagger(x)}
\, .
\een
We must finally show
that the maps $A_n$ have properties analogous to those of the maps
$D_n$ in sec.~3.6, i.e., properties (i)---(v).
The proof is similar as the proof for the $D_n$ outlined there.
It is again inductive in nature and is based on the expression
\bena\label{componentform1}
&& T_1(A_n(F_1 \otimes \dots \otimes F_n)) = \hat s_0 T_n(F_1 \otimes \dots \otimes F_n) - \non\\
&& \sum_{k=0}^n T_n(F_1 \otimes \dots \hat s_0 F_k \otimes \dots F_n) -
\frac{\hbar}{i}
\sum_{k<j} T_{n-1}(F_1 \otimes \dots (F_j,F_k) \otimes \dots F_n) -\non\\
&&\sum_{t=1}^{n-1} \sum_{k_1 < \dots < k_t} \sum_{l_1 < \dots < l_{n-t}}
\left( \frac{\hbar}{i} \right)^{t-1} T_{n-t+1}
(A_t(F_{k_1} \otimes \dots F_{k_t}) \otimes F_{l_1} \otimes \dots
F_{l_{n-t}}) \,
\eena
for the $n$-th order anomaly. We have already shown that $A_n = O(\hbar)$, because
this is true for $\Delta_n, \delta_n$ to all orders.
The statement (ii) follows because all quantities on the right side of this equation
are locally and covariantly constructed out of the metric. (iii) follows from the fact that
$\hat s_0$ increases the ghost number by 1 unit, and because the anti-bracket increases
the ghost number by 1 unit. (iv) follows because
$\hat s_0$ and the anti-bracket preserve the dimension, and from the known scaling
behavior of the time-ordered products, T2. (v) follows because $\hat s_0$ is compatible with the
*-operation and because the time ordered products are unitary, see T7. For more details on such kinds of
arguments, see again~\cite{Hollands2000}.

To complete the proof, we must still show that Proposition~4 is indeed true.
These arguments are given in detail in thm.~7 and lemma~11
of~\cite{MWD}. For completeness, we here outline a slightly modified
version of these arguments, but we
refer the reader to this work for full details\footnote{The arguments in~\cite{MWD} are
given only for the case of flat spacetime, but the key steps easily generalize to
curved manifolds straightforwardly.}.

\medskip
\noindent
{\em Proof of Proposition~4:} We begin by writing down the $n$-th order part of
eq.~\eqref{dutschid}, given by
\bena\label{t1delta}
&&
-T_1 \Big( \Delta_n(F_1 \otimes \dots \otimes F_n) \Big) =
\bigg( \frac{i}{\hbar} \bigg)^n \int T_{n+1} \bigg(F_1 \otimes \dots \otimes F_n \otimes
\psi(x) \wedge \frac{\delta S_0}{\delta \Phi(x)} \bigg) + \non\\
&&
\bigg( \frac{i}{\hbar} \bigg)^{n-1} \sum_{i=1}^n \int T_{n} \bigg(F_1 \otimes \dots
\psi(x) \wedge \frac{\delta F_i}{\delta \Phi(x)} \otimes \dots F_n \bigg) - \non\\
&&
\bigg( \frac{i}{\hbar} \bigg)^n \int T_{n+1} \Big( F_1 \otimes \dots \otimes F_n \otimes \psi(x) \Big)
\st \frac{\delta S_0}{\delta \Phi(x)}  + \non\\
&&
\sum_{t=1}^{n-1} \bigg( \frac{i}{\hbar} \bigg)^{n-t} \sum_{k_1< ...<k_t} \sum_{l_1<...<l_{n-t}}
T_{n-t+1} \Big(\Delta_t(F_{k_1} \otimes \dots \otimes F_{k_l}) \otimes
F_{l_1} \dots \otimes F_{l_{n-t}} \Big) \, .
\eena
For $n=0$, the identity becomes
\ben
-T_1(\Delta_0) = \int T_1 \Big( \psi(x) \wedge \frac{\delta S_0}{\delta \Phi(x)} \Big) -
\int T_1 \Big( \psi(x) \Big) \st
\frac{\delta S_0}{\delta \Phi(x)} \, .
\een
The function $\Delta_0$ is trivially local in this case. Because the first time ordered product $T_1$
as well as the $\st$-product reduce to the ordinary product in the space of classical local functionals
of the fields when $\hbar \to 0$, it follows that $\Delta_0 = O(\hbar)$, as claimed.

We now proceed iteratively in $n$. We assume that the assertion about $\Delta_n$ in the proposition
has already been proved for $\Delta_k$ up to $k=n-1$. In fact, let us assume for simplicity even that
$\Delta_k = 0$ up to $k=n-1$. We define $M_n(F_1 \otimes \dots \otimes F_n)$ to be the right side of
eq.~\eqref{t1delta}. The aim is to prove that this is a local functional valued in $\W$. To demonstrate this,
consider functionals $F_i$ with the property that
\ben
\left( \bigcup_{i=1}^l {\rm supp} \bigg( \frac{\delta F_i}{\delta \Phi} \bigg) \cup {\rm supp} \psi \right)
\cap J^+ \left( \bigcup_{i=l+1}^n {\rm supp} \bigg( \frac{\delta F_i}{\delta \Phi} \bigg) \right) = \emptyset
\een
for some $l$ not equal to $n$. Then $M_n$ can be written as follows using the
causal factorization properties of the time ordered products:
\bena\label{t1delta1}
&&
M_n \Big(F_1 \otimes \dots \otimes F_n \Big) =
\bigg( \frac{i}{\hbar} \bigg)^n \int T_{l+1} \bigg(F_1 \otimes \dots \otimes F_l \otimes
\psi(x) \wedge \frac{\delta S_0}{\delta \Phi(x)} \bigg) \st T_{n-l} \bigg( F_{l+1} \otimes \dots \otimes F_n \bigg)  + \non\\
&&
\bigg( \frac{i}{\hbar} \bigg)^{n-1} \sum_{i=1}^l \int T_{l} \bigg(F_1 \otimes \dots
\psi(x) \wedge \frac{\delta F_i}{\delta \Phi(x)} \otimes \dots F_l \bigg)
\st T_{n-l} \bigg( F_{l+1} \otimes \dots \otimes F_n \bigg)
- \non\\
&&
\bigg( \frac{i}{\hbar} \bigg)^n \int T_{l+1} \Big( F_1 \otimes \dots \otimes F_l \otimes \psi(x) \Big)
\st \frac{\delta S_0}{\delta \Phi(x)}
\st T_{n-l} \bigg( F_{l+1} \otimes \dots \otimes F_n \bigg) \, ,
\eena
where we have used that, for any $G \in \W$ (of even Grassmann parity), we have the identity
\ben
G \st \frac{\delta S_0}{\delta \Phi(x)} = \frac{\delta S_0}{\delta \Phi(x)} \st G \, ,
\een
which in turn follows from the definition of the star-product given above in sec.~4.2, eq.~\eqref{freegst}, together
with the fact that $\delta S_0/\delta \Phi^i(x) = D_{ij} \Phi^j(x)$, $D_{ij} \omega^{jk}(x,y) = 0$, with the $D_{ij}$
the matrix of linear partial differential operators in the field equation for the free underlying (gauge fixed) theory
with action $S_0$. Using now the inductive assumption in eq.~\eqref{t1delta1}, we conclude that $M_n$ for
the $F_i$ with the assumed support properties. It follows from this that $M_n$ can only be supported on the diagonal,
which is the desired locality property of $\Delta_n(F_1 \otimes \dots \otimes F_n)$.

It remains to be seen that $\Delta_n = O(\hbar)$. For this, we take eq.~\eqref{dutschid} and multiply
from the left with the anti-time ordered products [see eq.~\eqref{atoprod}], to obtain
\bena\label{dutschid1}
&&\bar T\Big( \e^{iF/\hbar} \Big) \st \int T \Big( \e^{iF/\hbar} \otimes \Big[ \psi(x) \wedge \frac{\delta F}{\delta \Phi(x)} +
\psi(x) \wedge \frac{\delta S_0}{\delta \Phi(x)} \Big] \Big) \, dx \\
&=&
\bar T\Big( \e^{iF/\hbar} \Big) \st \int
T \Big( \e^{iF/\hbar} \otimes \psi(x) \Big) \st \frac{\delta S_0}{\delta \Phi(x)} \, dx
+ \bar T\Big( \e^{iF/\hbar} \Big) \st
T \Big( \e^{iF/\hbar} \otimes \Delta_B(\e^F) \Big)
\, .\non
\eena
Using next the definition of the retarded products [see eq.~\eqref{retardeddef}], this may be rewritten in the form
\bena\label{dutschid2}
&&\int R \Big( \psi(x) \wedge \frac{\delta F}{\delta \Phi(x)} +
\psi(x) \wedge \frac{\delta S_0}{\delta \Phi(x)}; \e^{iF/\hbar} \Big) \, dx \non\\
&=&
\int R \Big( \psi(x); \e^{iF/\hbar} \Big) \st \frac{\delta S_0}{\delta \Phi(x)} \, dx +
R \Big( \Delta_B(\e^F) ; \e^{iF/\hbar} \Big)
\, .
\eena
The key point is now the that the retarded products in this equation have a meaningful limit as
$\hbar \to 0$, as proven in~\cite{Duetsch2000}, i.e., the above expressions contain no inverse powers of $\hbar$,
despite the inverse powers of $\hbar$ in the exponentials. This limit is just the classical
limit for the interacting fields as defined by the Bogoliubov formula eq.~\eqref{bogoliubov}.
Furthermore, the classical limit of $\st$ is the usual classical product of classical fields.
Thus, the eq.~\eqref{dutschid2} has a classical limit, the "classical Master Ward Identity" of~\cite{MWD}.
It is shown in this reference that this identity in classical field theory is indeed true with $\Delta = 0$.
Consequently, $\Delta$ itself must be of order $O(\hbar)$, as we desired to show. This concludes our outline
of the proof of Proposition~4. \qed.

Since we have proved Proposition~4, we have proved Proposition~3. \qed

\medskip
\noindent
We next derive a ``consistency
condition'' on the anomaly.

\medskip
\noindent
\paragraph{\bf Proposition 5} ("Consistency condition") The anomaly
satisfies the equation
\begin{equation}\label{anomaly}
\boxed{
\\
\bigg( S_0+F, A(\e^F) \bigg) -\frac{1}{2}
A\bigg( (S_0+F,S_0+F) \otimes \e^F \bigg)
=  A \bigg( A(\e^F)
\otimes \e^F \bigg) \, .
\\
}
\end{equation}

\medskip
\noindent
{\em Proof of Proposition 5}:
We first act with $\hat s_0$ on the anomalous Ward identity
eq.~\eqref{WA} and use that
$\hat s_0^2=0$. We obtain the equation
\ben
0= \hat s_0 T\bigg( A(\e^F) \otimes \e^{iF/\hbar} \bigg) + \frac{1}{2}
\hat s_0 T \bigg(
(S_0 + F, S_0 + F) \otimes \e^{iF/\hbar}
\bigg) = {\rm (I)} + {\rm (II)}
\een
The trick is now to apply the anomalous Ward identity one more
time to each of the terms on the right side. For simplicity,
we assume that $F$ has Grassmann parity 0. We can then write the
first term as
\bena
&&
{\rm (I)} = \frac{\hbar}{i}
\frac{d}{d\tau} \hat s_0 T \bigg( \e^{i(F + \tau A(\e^F))/\hbar}\bigg) \Bigg|_{\tau =
  0} =\non\\
&&
\frac{d}{d\tau} \Bigg[ \frac{1}{2}
T \bigg( \Big(S_0 + F + \tau A(\e^F), S_0 + F + \tau A(\e^F)\Big) \otimes
\e^{i(F + \tau A(\e^F))/\hbar} \bigg) +\non\\
&&
T \bigg( A(\e^{\tau A(\e^F)}) \otimes e^{i(F + \tau A(\e^F))/\hbar}\bigg)
\Bigg]
\Bigg|_{\tau = 0}
= \non\\
&&
T\bigg( (S_0 + F, A(\e^F)) \otimes \e^{iF/\hbar} \bigg)
+ \frac{i}{2\hbar} T\bigg( A(\e^F) \otimes (S_0+F, S_0+F) \otimes
\e^{iF/\hbar} \bigg) -\non\\
&&
T \bigg( A(A(\e^F) \otimes \e^F) \otimes \e^{iF/\hbar} \bigg)
+ T \bigg( A(\e^F) \otimes A(\e^F) \otimes \e^{iF/\hbar} \bigg) \, .
\eena
Since $F$ has Grassmann parity 0, $A(\e^F)$ has Grassmann parity 1,
so by the anti-symmetry of the time-ordered products for such
elements, see~\eqref{Antisym}, the last term vanishes. Next, we apply
the anomalous Ward identity to term (II). We now obtain
\bena
&&
{\rm (II)} = \frac{\hbar}{2i}
\frac{d}{d\tau} \hat s_0 T \bigg( \e^{i(F + \tau (S_0+F,S_0+F))/\hbar}\bigg) \Bigg|_{\tau =
  0} =\non\\
&&
\frac{1}{2}
\frac{d}{d\tau} \Bigg[ \frac{1}{2}
T \bigg( (S_0 + F + \tau (S_0+F,S_0+F), S_0 + F + \tau (S_0+F,S_0+F)) \otimes
\e^{i(F + \tau (S_0+F,S_0+F))/\hbar} \bigg) +\non\\
&&
T \bigg( A(\e^{\tau (S_0+F,S_0+F)}) \otimes \e^{i(F + \tau
  (S_0+F,S_0+F))/\hbar}
\bigg)
\Bigg] \Bigg|_{\tau = 0}
= \non\\
&&
\frac{1}{2} \Bigg[
T\bigg( (S_0 + F, (S_0+F, S_0+F)) \otimes \e^{iF/\hbar} \bigg) + \non\\
&&
\frac{i}{2\hbar} T\bigg( (S_0+F,S_0+F) \otimes (S_0+F, S_0+F) \otimes
\e^{iF/\hbar} \bigg) -\non\\
&&
T \bigg( A((S_0+F,S_0+F) \otimes \e^F) \otimes \e^{iF/\hbar} \bigg)
- \frac{i}{\hbar}
T \bigg( A(\e^F) \otimes (S_0+F,S_0+F) \otimes \e^{iF/\hbar} \bigg) \Bigg]
\, .
\eena
Now, the first term on the right side vanishes due to the graded
Jacobi identity~\eqref{Jacobiid} for the anti-bracket. The second term vanishes due
to the anti-symmetry property of the time ordered products~\eqref{Antisym},
since $(S_0+F,S_0+F)$ has Grassmann parity 1. If we now add up
terms (I) and (II), we end up with the following identity:
\bena
&&T \bigg( \bigg[
(S_0+F,A(\e^F))
-\frac{1}{2} A((S_0+F,S_0+F) \otimes \e^{F})
\bigg] \otimes \e^{iF/\hbar} \bigg) = \non\\
&&T\bigg( A(A(\e^{F}) \otimes \e^F) \otimes \e^{iF/\hbar} \bigg) \, .
\eena
Since a time ordered
product $T(G \otimes \e^{iF/\hbar})$ vanishes if and only if
$G=0$, the desired consistency condition~\eqref{anomaly} follows
\qed

\medskip
\noindent
Let us summarize what we have shown so far:
We first demonstrated that the Ward identity~\eqref{WW}
always holds with an anomaly term of order $\hbar$,
i.e., that eq.~\eqref{WA} holds.
We then showed that the anomaly is not arbitrary, but must
obey the consistency condition~\eqref{anomaly}. This condition
imposes a strong restriction on the possible anomalies, and we will
show in the following subsections using this condition that, when $F$ is as
in the cases T12a, T12b, and T12c, then the anomaly
$A(\e^F)$ can in fact
be removed by a redefinition of the time-ordered products
consistent with T1---T11. Thus, in these cases, we may
achieve that the Ward identity~\eqref{WW}
holds exactly, without anomaly.

To prepare the proof of this statement, we first note that,
since the anomaly itself is of order $\hbar$, the lowest
order in $\hbar$ contribution to the ``anomaly of the
anomaly term'' on the right side of eq.~\eqref{anomaly}
is necessarily of a higher order
in $\hbar$ than the lowest order contribution left side.
An even more stringent consistency condition can therefore be obtained
for the lowest order (in $\hbar$) contribution to the anomaly.
For this, we expand $A(\e^{F})$ in powers of the
coupling, $\lambda$, and $\hbar$,
\ben\label{Anomalyexp}
A(\e^{F})=\sum_{n,m>0} \hbar^m \frac{\lambda^n}{n!} \int \A_n^m(x_1,\dots,x_n)
f(x_1) \dots f(x_n) \, dx_1\dots dx_n \, ,
\een
where $\A_n^m$ is a local, covariant functional of $(\Phi,
\Phi^\ddagger)$, and the metric that is supported on the total diagonal.
Both sums start with positive powers, because the anomaly vanishes
in the classical theory (i.e., $\hbar=0$), and also in the free
quantum theory (i.e., $\lambda=0$). An explicit definition of
$\A_n^m$ is given by
\ben\label{amndefinition}
\A_n^m(x_1, \dots, x_n) =
\frac{1}{m!} \frac{\partial^m}{\partial \hbar^m} \frac{\delta^n}{\delta f(x_1) \cdots \delta f(x_n)}
\, A(\e^F) \Bigg|_{f=0=\hbar} \, .
\een
Let $A^m(\e^{F})$ now be the lowest
order contribution to $A(\e^{F})$ {\em in the $\hbar$-expansion}, that is, $m$ is
the smallest integer for which
\ben\label{Am}
A^m(\e^{F}) := \frac{1}{m!} \frac{\partial^m}{\partial \hbar^m} A(\e^F) \Bigg|_{\hbar = 0}  \,
\een
is not zero.
(Note that the quantity $A^m$ is different from the quantity $A_n$ above!)
Then, from our consistency condition given in Proposition~5, we get the following
version of the consistency condition:

\medskip
\noindent
\paragraph{\bf Proposition 6:} ("$\hbar$-expanded consistency condition")
Let $A$ be the anomaly of the Ward identity in Proposition~3, and let
$A^m$ be the first non-trivial term in the $\hbar$-expansion of $A$.
Then we have
\ben\label{con}
\boxed{
\\
\bigg( S_0+F, A^m \Big(\e^{F} \Big) \bigg)-\frac{1}{2}
A^m \bigg( \Big(S_0+F,S_0+F\Big) \otimes \e^{F}
\bigg) = 0 \, .
\\
}
\een
Here, $( \, . \, , \, . \,)$ is the anti-bracket [see eq.~\eqref{bvh}].

\medskip
\noindent
This stronger form of the consistency condition
is the key relation that will be used in the proofs of T12a, T12b,
and T12c. In those proofs we will actually encounter several
quantities like $\A_n^m$, so it is convenient to use again the
notation from sec.~3.6. As there, $(k_1,\dots,k_n)$ is a set of natural numbers.
We denote by $\P^{k_1/\dots/k_n}(M^n)$ the space of all
local, covariant functionals of $\Phi,\Phi^\ddagger$, and the metric
which are supported on the total diagonal, and which take values in
the bundle~\eqref{bundle}
of antisymmetric tensors over $M^n$. Thus, if $\B_n \in
\P^{k_1/\dots/k_n}(M^n)$,
then $\B_n$ is a (distributional) polynomial,
local, covariant functional of $\Phi,\Phi^\ddagger$ and the
metric taking values in the $k_1+\dots+k_n$ forms over $M^n$,
which is supported on the total diagonal. It is a $k_1$-form
in the first variable $x_1$, a $k_2$-form in the second variable
$x_2$, etc. Concerning such quantities, we have a simple lemma
that we will use below.
\begin{lemma}\label{lemma9}
Let $\B_n \in \P^{k_1/\dots/k_n}(M^n)$, and let
$f_i, i=1, \dots, n$ be closed forms on $M$
of degree $4-k_i$. Assume that for any such forms, we have
\ben\label{smear}
\int \B_n(x_1, \dots, x_n) \wedge \prod_i f_i(x_i) = 0 \, .
\een
Then it is possible to write
\ben
\B_n[\Phi,\Phi^\ddagger] = \sum_{k=1}^n d_k \B_{n/k}[\Phi,\Phi^\ddagger]
+ \B_{n}[0,0] \, ,
\een
where $d_k = dx_k \wedge (\partial/\partial x_k^\mu)$ is the exterior
differential applied to the $k$-th variable.
\end{lemma}
{\em Proof:}
We first consider the case $n=1$. If $k_1=4$, then the assumptions
imply that $F=\int \B_1(x) f_1(x)=0$ for any closed 0-form $f_1$,
i.e., for any constant such as $f_1(x) = 1$. We therefore have
$\delta F/\delta \psi(x) = 0$, using the abbreviation $\psi = (\Phi, \Phi^\ddagger)$.
Consider the path $\psi_\tau = (\tau\Phi, \tau\Phi^\ddagger)$
in field space. Then
\ben
\frac{d}{d\tau} \B_1[\psi_\tau] = \sum_k (\nabla^k \psi) \frac{\partial \B_1[\psi_\tau]}{\partial(\nabla^k \psi)}
= \psi \frac{\delta F[\psi_\tau]}{\delta \psi} + d \vartheta[\psi_\tau] = d \vartheta[\psi_\tau] \, ,
\een
for some locally constructed $3$-form
$\vartheta$. Thus,
\bena
\B_1[\Phi, \Phi^\ddagger] &=& \B_1[0,0] + \int_0^1 \frac{d}{d \tau}
\B_1[\psi_\tau] \, d \tau =
\B_1[0,0] + d \int_0^1 \vartheta[\psi_\tau] \, d \tau \\
&=& \B_1[0,0] + d\B_{1/1}[\Phi,\Phi^\ddagger] \, ,
\eena
which has the desired form. If $k_1=0$, then $f_1$ is a 4-form,
which is always closed. Thus, the assumptions of the lemma imply that
$\B_1[\Phi, \Phi^\ddagger]=0$, which is again of the desired form.
Finally, if $0<k_1<4$, we may choose $f_1 = dh_1$, implying that
$\int d\B_1(x) \wedge h_1(x)=0$ for all $h_1$, and thus that
$d\B_1 = 0$. The statement now follows from the algebraic Poincare lemma.

The proof of the lemma for $n>1$ can now be generalized from the case $n=1$.
Without loss of generality, we may assume
$\B_n[\Phi=0,\Phi^\ddagger=0]=0$, for otherwise, we may simply subtract
this quantity. To reduce the situation to $n=1$,
consider the form on $M$ of degree $k_1$ that is obtained by
smearing $\B_n$ as in~\eqref{smear}, but the smearing over the first
test-form $f_1$ omitted. If $k_1<4$, then this form is a closed
form that is locally and covariantly constructed from
$f_2, f_2, \dots, f_n$ and $\Phi, \Phi^\ddagger$.
This 4-form then by definition obeys the
assumptions of lemma~\ref{lemma2}, so we may write
\ben\label{smear}
0=\int \B_n(x_1, \dots, x_n) \wedge \prod_{i=2}^n f_i(x_i) - d_1
\int \B_{n/1}(x_1, \dots, x_n) \wedge \prod_{i=2}^n f_i(x_i)
\,
\een
for some $\B_{n/1} \in \P^{k_1-1/k_2/\dots/k_n}$. If $k_1=4$ one may
argue similarly. We now repeat this argument, now omitting the
integration over the second test form $f_2$. We then get
\begin{multline}\label{smear2}
0=\int \B_n(x_1, \dots, x_n) \wedge \prod_{i=3}^n f_i(x_i) \\
- d_1
\int \B_{n/1}(x_1, \dots, x_n) \wedge \prod_{i=3}^n f_i(x_i)
- d_2
\int \B_{n/2}(x_1, \dots, x_n) \wedge \prod_{i=3}^n f_i(x_i)
\,
\end{multline}
for some $\B_{n/2} \in \P^{k_1/k_2-1/\dots/k_n}$.
We may continue this procedure, and thus
inductively proceed to construct the remaining $\B_{n/k}$. \qed

\subsubsection{Proof of T12a}

Up to now, we have shown (Proposition~3) that any
prescription for defining time ordered products satisfying properties
T1-T11 satisfies the Ward identity~\eqref{WA} with anomaly.
We shall now prove that we can change the definition of the
time ordered products in such a way that T1-T11 still hold, and
such that in addition the anomaly vanishes in the
case when $F=\int\{\lambda f \L_1 + \lambda^2 f^2 \L_2\}$, where
$f \in C^\infty_0(M)$. Thus, our new prescription will
satisfy~\eqref{masterward} [and in fact even eq.~\eqref{WW}]
for this $F$. This will then enable us to prove that the
new prescription for defining time ordered products will satisfy
property T12a.

The key tool for proving this statement is the consistency
condition on the $\hbar$-expanded anomaly given in Proposition~6.
To take full advantage of this consistency condition, we would like to
put $f=1$, for we then have $S_0+F=S$, and we can take advantage of
BRST-invariance of the full action $S$, see~\eqref{fullaction}.
We note that we cannot simply set $f=1$
in $T(\e^{iF/\hbar})$, for we might encounter infra-red divergences.
However, since the anomaly terms $\A_{n}^m$ are local, covariant
functionals of $\Phi, \Phi^\ddagger$ that are supported on the
total diagonal (taking values in the
$4n$-forms $\wedge^{4n} T^* M^n$ over $M^n$),
we may without any danger set $f=1$ in
eq.~\eqref{con}. As we have already said, in that case we have $F=\lambda S_1+ \lambda^2 S_2$,
and consequently $S_0+F=S$, where $S$
is the full action~\eqref{fullaction}. So from eq.~\eqref{con}
together with $(S,S)=0$ and $\hat s = (S, \,.\,)$ we find
\ben\label{sAm}
%\Big( S, A^m(\e^{\lambda S_1 + \lambda^2
%  S_2}) \Big) =
  \hat s \, A^m \Big( \e^{\lambda S_1 + \lambda^2 S_2} \Big) = 0 \, .
\een
Now, we have
\ben
A^m \Big( \e^{\lambda S_1 + \lambda^2 S_2} \Big) = \int_M a^m(x) = \sum_{n > 0} \frac{\lambda^n}{n!} \, \int_M a^m_n(x) \, ,
\een
where $a^m \in \P^4(M)$ (and likewise for $a^m_n$). Furthermore, from the
properties of the anomaly derived in the previous subsection,
the dimension of $a^m$ must be 4, and the ghost number must be $+1$. Equation~\eqref{sAm} may now
be viewed as saying that $a^m \in H^1(\hat s|d, \P^4)$. From the Lemmas given in sec.~2.2,
we have a complete classification of all the elements in this ring. In fact,
as shown there in theorem~\ref{thm2}, all non-trivial elements in this ring at ghost number +1 and
dimension 4 must be even under parity,
$\epsilon \to -\epsilon$ when the Lie-group has no abelian factors. On the other hand,
it follows from the properties of the anomaly $A$ that $a^m$ is parity odd, i.e.,
$a^m \to -a^m$ under parity $\epsilon \to -\epsilon$. Therefore, $a^m$ must
represent the zero element in the ring $H^1(\hat s|d, \P^4)$, so there are
$b^m \in \P^4_0(M)$ and $c^m \in \P^3_1(M)$ such that
\ben\label{abc}
a^m(x) = \hat s b^m(x) + dc^m(x) \, .
\een
We expand
\ben
b_m(x) = \sum_{n>0} \frac{\lambda^n}{n!} b^m_n(x) \, .
\een
We would like to use the coefficients $b^m_n(x)$ to redefine the time ordered products
in order to remove the anomaly. For this, it is necessary to understand first how the anomaly 
changes if we pass from one prescription $T$ to another prescription $\hat T$. Let $F$
be an arbitrary local function. Eq.~\eqref{unique1} implies that
\bena
\hat T \Big( \e^{iF/\hbar} \Big) &=& T \Big( \e^{i[F + D(\exp_\otimes F)]/\hbar} \Big)\\
\hat T \Big( (S_0 + F, S_0 + F) \otimes
\e^{iF/\hbar} \Big) &=& T \Big( (S_0 + F, S_0 + F) \otimes \e^{i[F + D(\exp_\otimes F)]/\hbar} \Big) +
\\
&& T \Big( D((S_0 + F, S_0 + F) \otimes \e^F) \otimes \e^{i[F + D(\exp_\otimes F)]/\hbar} \Big),\non\\
\hat T \Big( \hat A(\e^F) \otimes \e^{iF/\hbar} \Big) &=& T \Big( [\hat A(\e^F) + D(\hat A(\e^F) \otimes \e^F)]
\otimes \e^{i[F + D(\exp_\otimes F)]/\hbar} \Big) \, ,
\eena
where $\hat A(\e^F)$ is the anomaly~\eqref{WA} in the Ward identity for the modified time ordered
products $\hat T$.  We would now like to relate
the anomaly $A(\e^F)$ of the "old" time ordered products $T$ to the anomaly $\hat A(\e^F)$ of the "new"
time ordered products $\hat T$. We have
\bena
&& \frac{i}{\hbar} T \Big( [\hat A(\e^F) + D(\hat A(\e^F) \otimes \e^F)]
\otimes \e^{i[F + D(\exp_\otimes F)]/\hbar} \Big) \non\\
&=& \frac{i}{\hbar} \hat T \Big( \hat A(\e^F) \otimes \e^{iF/\hbar} \Big) \non \\
&=& \hat s_0 \hat T \Big( \e^{iF/\hbar} \Big) - \frac{i}{2\hbar} \hat T \Big( (S_0 + F, S_0 + F) \otimes
\e^{iF/\hbar} \Big) \non\\
&=& \hat s_0 T \Big( \e^{i[F + D(\exp_\otimes F)]/\hbar} \Big) -
\frac{i}{2\hbar} T \Big( (S_0 + F, S_0 + F) \otimes \e^{i[F + D(\exp_\otimes F)]/\hbar} \Big) \non\\
&-& \frac{i}{2\hbar}  T \Big( D((S_0 + F, S_0 + F) \otimes \e^F) \otimes \e^{i[F + D(\exp_\otimes F)]/\hbar} \Big) \non\\
&=&
\frac{i}{2\hbar} T \Big( (S_0 + F + D(\e^F), S_0 + F+ D(\e^F)) \otimes \e^{i[F + D(\exp_\otimes F)]/\hbar} \Big) \non\\
&-& \frac{i}{2\hbar}  T \Big( D((S_0 + F, S_0 + F) \otimes \e^F) \otimes \e^{i[F + D(\exp_\otimes F)]/\hbar} \Big) \non\\
&-& \frac{i}{2\hbar} T \Big( (S_0 + F, S_0 + F) \otimes \e^{i[F + D(\exp_\otimes F)]/\hbar} \Big) \non\\
&+& \frac{i}{\hbar} T \Big( A(\e^{F+D(\exp_\otimes F)}) \otimes \e^{i[F + D(\exp_\otimes F)]/\hbar} \Big) \non\\
&=&
\frac{i}{\hbar} T \Big( (S_0 + F, D(\e^F)) \otimes \e^{i[F + D(\exp_\otimes F)]/\hbar} \Big)
+ \frac{i}{\hbar} T \Big( A(\e^{F+D(\exp_\otimes F)}) \otimes \e^{i[F + D(\exp_\otimes F)]/\hbar} \Big)\non\\
&+& \frac{i}{2\hbar} T \Big( (D(\e^F), D(\e^F))\otimes \e^{i[F + D(\exp_\otimes F)]/\hbar} \Big) \non \\
&-& \frac{i}{2\hbar}  T \Big( D((S_0 + F, S_0 + F) \otimes \e^F) \otimes \e^{i[F + D(\exp_\otimes F)]/\hbar} \Big) \ 
\, .
\eena
We conclude from the last equation that the following lemma is true:

\paragraph{\bf Lemma 10:}
Let $\hat T$ be a new prescription for time-ordered products related to $T$ by $D$, and $\hat A$ the corresponding 
anomaly. Then there holds:
\begin{eqnarray}
%\boxed{
%\\
\hat A\Big( \e^F \Big) + D\Big(\hat A(\e^F) \otimes \e^F \Big) 
&=& \Big( S_0 + F, D(\e^F) \Big)
+ A\Big( \e^{F+D(\exp_\otimes F)} \Big) \non\\
&+& \frac{1}{2} \Big( D(\e^F), D(\e^F)\Big)
+ \frac{1}{2} D((S_0 + F, S_0 + F) \otimes \e^F)
\, .
%\\
%}
\end{eqnarray}
for any local functional $F$.

\medskip

Similar to the consistency relation for the anomaly, the most useful consequence 
of lemma~10 is obtained by considering the leading order in $\hbar$ (in our case $\hbar^m$). 
We use that, in the case considered, $D(\e^F) = O(\hbar^m)$, and by assumption
$A(\e^F) = O(\hbar^m) = \hat A(\e^F)$. Then it follows
that:

\paragraph{\bf Lemma 11:} If $D$ and $A$ both start at order $\hbar^m, m>0$, we have
\begin{equation}
\boxed{
\\
\hat A^m \Big( \e^F \Big) = A^m \Big(\e^F \Big) + \Big( S_0 + F, D^m(\e^F) \Big) + \frac{1}{2} D^m((S_0 + F, S_0 + F) \otimes \e^F) \, 
\\
}
\end{equation}
for any local functional $F$.

\medskip

We will use lemmas~10 and~11 and similar relations repeatedly in what follows.

Recalling that by thm.~\ref{Tuniqueness}, the changes in the time-ordered products are
parametrized by local, covariant maps $D_n: \P^{p_1}(M) \otimes \cdots \otimes \P^{p_n}(M) \to
\P^{p_1/\dots/p_n}(M^n)$, we define
\ben\label{D1}
D_n(\L_1(x_1) \otimes \dots \otimes \L_1(x_n)) := -\hbar^m b_m^n(x_1) \delta(x_1, \dots, x_n) \, .
\een
It can be shown that this is within the allowed renormalization freedom
for the time-ordered products described in sec.~3.6: First, the locality and
covariance of $D_n$ follows from the corresponding property of $b_n^m$. The scaling
property~\eqref{dimension} follows from the fact that $b_n^m$ has dimension 4, together with the
scaling degree property $sd \, \delta = 4(n-1)$ for the delta function of $n$ spacetime
arguments concentrated on the diagonal in $M^n$. The smooth and analytic dependence
of $D_n$ under changes of the spacetime metric again follows from the corresponding
properties of $b_n^m$, while the symmetry is manifest. The unitarity condition~\eqref{unitary}
follows from the fact that $b_n^m$ is real, which in turn follows from the corresponding
property of the anomaly $A$ derived in the previous subsection. To satisfy the field independence
property~\eqref{fieldindep}, it is furthermore
  necessary to also change the time-ordered products of
  sub-monomials of $\L_1$ in order to be consistent with
  T9. This causes no problems.  The identity~\eqref{Dactionward}
  can be satisfied by defining $D_n$ appropriately for entries $\O_i$ that
  are exterior differentials of $\L_1$. This does not lead to any potential
  consistency problems, because $\L_1$ itself is not the exterior differential of
a locally constructed $3$-form. For details of such kinds of arguments see~\cite{Hollands2005},
where a very similar situation was treated.
Thus, the above $D_n$ (together with the corresponding
$D_n$ for sub-Wick monomials of $\L_1$ and their exterior derivatives) gives a permissible
change in the time ordered products, i.e., the changed time-ordered products
$\hat T$ defined according to~\eqref{unique1} with the above $D$
again satisfy T1--T11. We note also that there are no sub-Wick monomials in $\L_1$ that
are also contained in $(S_0 + F, S_0 + F)$ when $F=\int[\lambda f \L_1 + \lambda^2 f^2 \L_2]$, as one may check explicitly.
Therefore $D^m((S_0 + F, S_0 + F) \otimes \e^F)=0$ in our case. 

For the particular definition of $D$ in eq.~\eqref{D1}, we can immediately see using eq.~\eqref{abc} that $\hat s \, D^m(\e^{\lambda S_1 + \lambda^2 S_2})
= -A^m(\e^{\lambda S_1 + \lambda^2 S_2})$. Therefore, if we now
put  $F=\int[\lambda f \L_1 + \lambda^2 f^2 \L_2]$ with $f=1$ in Lemma~11, then we find
\ben\label{prelimrem}
\hat A^m \Big( \e^{\lambda S_1 + \lambda^2 S_2} \Big) = A^m \Big(\e^{\lambda S_1 + \lambda^2 S_2} \Big)
+ \hat s \, D^m\Big( \e^{\lambda S_1 + \lambda^2 S_2} \Big)
=0 \, .
\een
Thus, by our redefinition of the time ordered products, we have already removed the
anomaly for any constant test function $f$. We will now use this fact to completely
remove the anomaly by a further redefinition of the time ordered products.

To simplify the notation, we will now again use the notations $T$ and $A$ for the redefined
time ordered products and new anomaly, instead of $\hat T$ and $\hat A$. The anomaly
may be expanded in powers of $\hbar$ and $\lambda$ as in eq.~\eqref{Anomalyexp}.
From eq.~\eqref{prelimrem} (remembering that $A$ now denotes $\hat A$), we then have
\ben
\int \A^m_n(x_1, \dots, x_n) \, dx_1 \dots dx_n = 0 \, ,
\een
because we can assume at this stage that the anomaly vanishes for constant $f$.
Consequently, by lemma~\ref{lemma9}, this quantity must be
given by an expression of the form
\ben\label{cohtr}
\A_n^m(x_1, \dots, x_n) = \sum_{k=1}^n d_k {\mathcal C}_{n/k}^m(x_1, \dots, x_n) \, ,
\een
for some ${\mathcal C}_{n/k}^m \in \P^{4/...3/.../4}(M^n)$. It follows from the relation 
$0 = (S_0,S_1)$ that there must exist a local $\O_1 \in \P^3_1(M)$  determined by the
equation
\ben\label{s0L1}
\hat s_0 \L_1 = d\O_1 \, , 
\een
because the left side integrates to 0. 
It is given explicitly by
\ben
\O_1 := f_{IJK} C^I A^J \wedge * dA^K + \frac{1}{2} f_{IJK} C^I C^J \, *d \bar C^K \, .
\een
We now define a set of
$D_n$ by the formula
\ben
D_n(\L_1(x_1) \otimes \dots \O_1(x_k) \otimes \dots \L_1(x_n)) := -\hbar^m
{\mathcal C}_{n/k}^m(x_1, \dots, x_n) \, .
\een
We may again argue that this $D_n$ satisfies all the required properties
for an allowed redefinition of the time ordered products, and we
denote the new time ordered products again by $\hat T$, and the new anomaly
again by $\hat A$. If $F=\int[\lambda f \L_1 + \lambda^2 f^2 \L_2]$ with $f=1$, this redefinition 
has $D^m(\e^F) = 0$, so lemma~11 now gives for this redefinition
\ben
\hat A^m \Big( \e^F \Big) = A^m \Big( \e^F \Big) + \frac{1}{2} D^m \Big( (S_0 + F, S_0 + F) \otimes \e^F \Big) \, .
\een
However, our $D_n$ are designed precisely in such a way that $D^m( (F, F) \otimes \e^F ) = 0$ and
that $D^m(\hat s_0 F \otimes \e^F) = -A^m(\e^F)$, so we find that $\hat A^m(\e^F) = 0$.

In summary, our subsequent definitions of the time ordered products remove the anomaly
$A^m(\e^{F})$ at order $\hbar^m$, and to all orders in $\lambda$.
We now repeat the same argument for $A^{m+1}(\e^F)$,
i.e., order $\hbar^{m+1}$, and we can proceed in just the same way for
any order in $\hbar$. This shows that the anomaly can be removed to
arbitrary orders in $\hbar$ and
$\lambda$ by a redefinition of the time ordered products that is
compatible with T1--T11. The absence of an anomaly in eq.~\eqref{WA} for
our choice of $F$ implies that T12a is satisfied, because eq.~\eqref{WA} is
a generating identity of the identities in T12a. \qed

\subsubsection{Proof of T12b}
The proof that the time ordered products can be adjusted, if necessary,
so that T12b is satisfied is very similar in nature as that
given above for T12a. We therefore only focus on the essential differences.

Consider the local elements $G=\int \gamma \wedge (\J_0 + f\lambda
\J_1)$ and $F=\int(f\lambda\L_1+f^2\lambda^2\L_2)$, where $\gamma$
is a smooth 1-form of compact support, and $f$ is a smooth scalar
function of compact support. The satisfaction of T12b means that the anomaly
in
\bena\label{satsifactiont12b}
&&\hat s_0 T \Big( G \otimes \e^{iF/\hbar} \Big) = T \Big( (S_0 + F, G) \otimes \e^{iF/\hbar} \Big)
+ \frac{i}{2\hbar} T \Big( (S_0 + F, S_0 + F) \otimes G \otimes \e^{iF/\hbar} \Big) \non\\
&+& T \Big( A(G \otimes \e^F) \otimes \e^{iF/\hbar} \Big) + \frac{i}{\hbar} T \Big( A(\e^F) \otimes G \otimes \e^{iF/\hbar}
\Big)
\eena
can be removed by a suitable redefinition of the
time-ordered products. As above, we write
\ben\label{agdf}
A(G \otimes \e^{F}) = \sum_{m,n>0} \hbar^m \frac{\lambda^n}{n!} \int_{M^{n}}
\A_{m,n}(x_1,\dots,x_n)
\gamma(x_1)f(x_1) \dots f(x_n) \, dx_1\dots dx_n \, .
\een
Let $A^m(G \otimes \e^F)$ be the lowest order contribution
in $\hbar$ to the anomaly. Because the anomaly is of order at least $\hbar$, we
have $m>0$. We apply the consistency condition~\eqref{con}
to the element $F+\tau G$ instead of $F$ in that formula,
and we differentiate with
respect to $\tau$ and set $\tau=0$. Then we
obtain the consistency condition
\ben\label{wwwen}
\Big( S_0+F,A^m(G \otimes \e^{F}) \Big) + A^m \Big( (S_0+F,G) \otimes \e^F \Big) -
\frac{1}{2} A^m\Big( (S_0+F,S_0+F) \otimes G \otimes \e^F \Big)= 0 \, .
\een
Now, we put $f=1$ and we take $\gamma$ to satisfy $d\gamma=0$.
Then, $F=\lambda S_1 + \lambda^2 S_2$, and $S_0+F=S$, where $S$
is the full action~\eqref{fullaction} satisfying $(S,S)=0$.
Furthermore, by $\hat s \J = d \K$,
\ben
(S_0+F,G)= (S,G) =
\hat s \int_M \gamma \wedge \J = \int_M \gamma \wedge
d{\bf K} = -\int_M d\gamma \wedge {\bf K} = 0 \, .
\een
Thus, condition~\eqref{wwwen} implies the condition
\ben\label{sGAm}
%\Big(S, A^m (G \otimes
%\e^{\lambda S_1 + \lambda^2 S_2}) \Big)=
\hat s \, A^m \Big(G \otimes
\e^{\lambda S_1 + \lambda^2 S_2} \Big) = 0 \,
\een
when $\gamma$ is closed. Now, we have
\ben\label{Amgs}
A^m \Big( G \otimes \e^{\lambda S_1 + \lambda^2 S_2} \Big) = \int_M \gamma \wedge h^m(x) = \sum_{n > 0} \frac{\lambda^n}{n!} \, \int_M \gamma \wedge h^m_n(x) \, ,
\een
where $h^m \in \P^3(M)$ (and likewise for $h^m_n$). Furthermore, from the
properties of the anomaly derived in the previous subsection,
the dimension of $h^m$ must be 3, and the ghost number must be $+2$. Equation~\eqref{sGAm},
which holds for all closed 1-forms $\gamma$ in the definition of $G$, may now
be viewed as saying that $h^m \in H^2(\hat s|d, \P^3)$. From the Lemmas given in sec.~2.2,
we again have a complete classification of all the elements in this ring. In fact,
as shown there in lemma~\ref{thm2}, all non-trivial elements in this ring at ghost number +2 and
dimension 3 must be even under parity,
$\epsilon \to -\epsilon$ when the Lie-group has no abelian factors. On the other hand,
it follows again from the properties of the anomaly $A$ that $h^m$ is parity odd, i.e.,
$h^m \to -h^m$ under parity $\epsilon \to -\epsilon$. Therefore, $h^m$ must
represent the zero element in the ring $H^2(\hat s|d, \P^3)$, so there are
$j^m \in \P^3_1(M)$ and $k^m \in \P^2_2(M)$ such that
\ben\label{hjk}
h^m(x) = \hat s j^m(x) + dk^m(x) \, .
\een
We again expand $j^m$ in powers of $\lambda$
\ben
j^m(x) = \sum_{n>0} \frac{\lambda^n}{n!} j^m_n(x) \, .
\een
Similar to the proof of T12a,
we would like to use the coefficients $j^m_n(x)$ to redefine the time ordered products
$T_{n}(\J_1(x_1) \otimes \L_1(x_2) \otimes \dots \L_1(x_{n}))$ containing $n-1$ factors of the interaction Lagrangian
and one factor of the free BRST-current.
By thm.~\ref{Tuniqueness}, the changes in the time-ordered products are
parametrized by local, covariant maps $D_n: P^{p_1}(M) \otimes \cdots \otimes P^{p_n}(M) \to
\P^{p_1/\dots/p_n}(M^n)$, and we define
\ben
D_{n}(\J_1(x_1) \otimes \L_1(x_2) \otimes \dots \otimes \L_1(x_{n})) := -
\hbar^m j^m_n(x_1) \delta(x_1, \dots, x_{n}) \, .
\een
This gives changed time-ordered products via~\eqref{unique1}, and one may argue as above in
T12a that these
again satisfy T1--T11.

For a general $D$ and general local functionals $F,G$, lemma~11 implies that
\ben
A^m\Big( G \otimes \e^F \Big) + \Big( S_0 + F, D^m(G \otimes \e^F) \Big)
= \hat A^m\Big( G \otimes \e^F \Big) \, .
\een
Now, if $G = \int \gamma \wedge \J$, and if $F = \lambda S_1+ \lambda^2 S_2$, then
it follows from the above equation that
\ben
A^m\Big( G \otimes \e^{\lambda S_1 + \lambda^2 S_2} \Big) +
\hat s \, D^m\Big( G \otimes \e^{\lambda S_1 + \lambda^2 S_2} \Big)
= \hat A^m\Big( G \otimes \e^{\lambda S_1 + \lambda^2 S_2} \Big) \, .
\een
Furthermore, it follows from the definition of $D$ that $D(G \otimes \e^{\lambda S_1 + \lambda^2 S_2})$
is equal to $-\int \gamma \wedge j^m$. If $\gamma$ is a closed 1-form,
we have shown above that $\hat s \int \gamma \wedge j^m = \int \gamma \wedge h^m$. By eq.~\eqref{Amgs} and our definition of $D$,
we therefore have
\ben
\hat s \, D^m\Big( G \otimes \e^{\lambda S_1 + \lambda^2 S_2} \Big)
%(S, D^m\Big( G \otimes \e^{\lambda S_1 + \lambda^2 S_2} \Big) \Big)
=
-A^m \Big( G \otimes \e^{\lambda S_1 + \lambda^2 S_2} \Big) \, .
\een
Consequently, we have shown that
\ben\label{hatam0}
\hat A^m \Big( G \otimes \e^{\lambda S_1 + \lambda^2 S_2} \Big) = 0 \, .
\een
Therefore, our redefinition of the time ordered products has already removed the
anomaly $\hat A(G \otimes \e^F)$ in the case when $\gamma$ is a closed 1-form, and
$f$ is a constant. We now drop the carret from our notation for the newly defined
time-ordered products and the corresponding anomaly. We may then assume that
eq.~\eqref{hatam0} holds for $A^m$. For the quantities defined in eq.~\eqref{agdf}, this means that
\ben\label{agdf}
0 = \int_{M^{n}}
\A_{n}^m(x_1,\dots,x_n)
\gamma(x_1) \, dx_1\dots dx_n \, .
\een
for any closed 1-form $\gamma$, and any $n$. Lemma~\eqref{lemma9} now implies that
we may write $\A^m_n$ as
\ben
\A_n^m(x_1, \dots, x_n) = d_1 \B^m_{n/1}(x_1, \dots, x_n) + \sum_{k=2}^n
d_k \, \B^m_{n/k}(x_1, \dots, x_n)
\een
Here, the $\B_{m,n/k}$
are now a local covariant functional
of $(\Phi,\Phi^\ddagger)$ in the space
$\P^{2/4/\dots 4}(M^{n})$ for $k=1$, and in the space
$\P^{3/4/\dots 3/ \dots 4}(M^{n})$ for $k \ge 2$. 

Next, we define
for products with $n$ arguments containing 1 factor of ${\bf K}_1 \in
\P^2_2$ [see eq.~\eqref{jladder}] and $n-1$ factors of $\L_1 \in \P^4_0$ by
\ben\label{firstred}
D_{n}({\bf K}_1(x_1) \otimes \L_1(x_2) \dots \otimes \L_1(x_{n})) :=
-\hbar^m {\mathcal B}_{n/1}^m (x_1, \dots, x_{n}) \, .
\een
We redefine the time-ordered products with $n+1$ factors,
containing 1 factor of $\J_1 \in \P^3_1$, one factor of $\O_1 \in \P^3_1$
[see eq.~\eqref{s0L1}],
and $n-2$ factors
of $\L_1 \in \P^4_0$ by
\ben\label{secondred}
D_{n}({\bf J}_1(x_1) \otimes \L_1(x_2) \dots \otimes \O_1(x_{k}) \dots \otimes \L_1(x_{n})) :=
i\hbar^{m+1} {\mathcal B}_{n/k}^m (x_1, \dots, x_{n}) \, .
\een
By going through the same steps as above in T12a,
we find that the new anomaly $\hat A(G \otimes \e^F)$ after the above redefinition
effected by these $D$'s is now
\ben\label{bilanz}
\hat A\Big( G \otimes \e^F \Big) = A\Big( G \otimes \e^F\Big) - D \Big( (S_0+F, G) \otimes \e^F \Big) + \frac{i}{2\hbar}
D \Big( (S_0 + F, S_0 + F) \otimes G \otimes \e^F \Big) \, .
\een
Now, it can be seen that, because of the first redefinition~\eqref{firstred},
\ben
D \Big( (S_0+F, G) \otimes \e^F \Big) = \hbar^m \sum_{n \ge 0} \frac{\lambda^n}{n!} \int
d_1 \, {\mathcal B}_{n/1}^m(x_1, \dots, x_n) \, \gamma(x_1) f(x_1) \dots f(x_n) \, dx_1 \dots dx_n \, ,
\een
using $(S_0, \J_1) = d{\bf K}_1 + \dots$.
It follows from the second redefinition~\eqref{secondred} that
\bena
&&\frac{i}{2\hbar} D \Big( G \otimes (S_0+F, S_0+F) \otimes \e^F \Big) \\
&=& -\hbar^m \sum_{n \ge 0} \sum_{k=2}^n \frac{\lambda^n}{n!} \int
d_k \, {\mathcal B}_{n/k}^m(x_1, \dots, x_n) \, \gamma(x_1) f(x_1) \dots f(x_k) \dots f(x_n) \, dx_1 \dots dx_n \, ,
\non
\eena
using $(S_0, \L_1) = d\O_1$.
Thus, taking the $O(\hbar^m)$-part of eq.~\eqref{bilanz}, using eq.~\eqref{amnx}, we find that
the new anomaly $\hat A^m(G \otimes \e^F)=0$. Thus, the anomaly for the new time-ordered products vanishes at
order $\hbar^m$ and to all orders in $\lambda$. We continue this process by
redefining the time ordered products to the next order in $\hbar$, and
remove the anomaly $A^{m+1}(G \otimes \e^F)$. Since we can do this for all
$m$, we see that we can satisfy T12b above by a suitable redefinition of the time-ordered
products. \qed

\subsubsection{Proof of T12c}
Let $\Psi = \prod \Theta_{s_i}(F, {\mathcal D} F, {\mathcal
  D}^2 F, \dots)$ be the gauge-invariant expression of form-degree
$p$ under
consideration, where $\Theta_s$ are invariant polynomials of the
Lie-algebra, so that in particular $\hat s \Psi=0$. Let $\alpha$
be a $(4-p)$-form, and let $G = \int \alpha \wedge \Psi$.
The satisfaction of the Ward identity T12c means that anomaly in
eq.~\eqref{satsifactiont12b} can be removed, where $G$ in that equation is now
$\int \alpha \wedge \Psi$. As in the proofs of T12a, T12b, one first proves the
consistency condition
\ben
\hat s \, A^m \Big( G \otimes \e^{\lambda S_1 + \lambda^2 S_2} \Big) = 0\, ,
\een
where $m$ is the first order in $\hbar$ where the anomaly occurs, and
where $\alpha$ is now arbitrary.
This
condition is again of cohomological nature. As in T12b, it may be used
to show that the anomaly can be removed, at $n$-th order in $\lambda$, by a redefinition of the time
ordered products with 1 factor of $\Psi_0$ and $n$ factors of $\L_1$,
and by the time ordered products with 1 factor of $\Psi_0$, 1 factor
of $\O_1$ [see eq.~\eqref{s0L1}] and $n-1$ factors of $\L_1$. The details of
these arguments are completely analogous to those given above in the proofs
of T12a and T12b, so we omit them here. \qed

\subsection{Formal BRST-invariance of the $S$-matrix}
We consider the adiabatically switched $S$-matrix ${\mathcal
  S}(F) = T(\e^{iF/\hbar})$ associated with the cut-off interaction
$F = \int_M \{ \lambda f\L_1 + \lambda^2 f^2
\L_2\}$, where $f$ is a smooth switching function of compact support.
Let $Q_0$ be the free BRST-charge operator.
It follows from the definition
${\mathcal S}(F) = T(\e^{iF/\hbar})$ and the Ward-identities T12a
[see eq.~\eqref{masterward}]
that
\ben
[Q_0, {\mathcal S}(F)] =
-\frac{1}{2} T\bigg((S_0 + F, S_0 + F) \otimes \e^{iF/\hbar}\bigg)
\quad \text{mod $\I$} \, .
\een
Now consider a sequence of cutoff functions such that
$f \to 1$ sufficiently rapidly, i.e., the ``adiabatic limit''.
Then it follows
that $S_0+F \to S$, and consequently that
$(S_0+F,S_0+F) \to (S,S)=0$. Thus, formally,
$T((S_0+F,S_0+F) \otimes \e^{iF/\hbar}) \to 0$. Furthermore,
formally, ${\mathcal S}(F)$ converges to the true
$S$-matrix $\mathcal S$. Consequently, assuming that all
these limits exist, we would have
\ben
[Q_0, {\mathcal S}] = 0 \quad \text{mod $\I$ (FORMALLY)} \, .
\een
As we have already said, the adiabatic limit does not appear to exist
for pure Yang-Mills theory in Minkowski spacetime, and there is even less reason to believe that
it ought to exist in generic curved spacetimes. Therefore, the above
statement concerning the BRST-invariance of the $S$-matrix
is most likely only a formal statement, unlike the
other results in this paper. We have nevertheless mentioned it,
because such a condition is often taken to be as the definition of
gauge-invariance at the perturbative level in less rigorous treatments
of quantum gauge field theories in flat spacetime.

\subsection{Proof that $d\J_{\subsc} = 0$}

As above, consider the cutoff interaction
$F = \int_M \{ \lambda f\L_1 + \lambda^2f^2
\L_2\}$, where $f$ is a smooth switching function of compact support,
which is equal to one on some time-slice $M_T=(-T,T) \times \Sigma$. The
desired identity $d\J(x)_{\subsc}$ will follow if we can show that, in the
sense of formal power series,
\ben
0=d\J(x)_F = \sum_n \frac{i^n}{\hbar^n n!} R_n(d\J(x); F^{\otimes n}), \quad x \in M_T
\een
modulo $\I$ for any such cutoff function $f$. Expanding the retarded products
in terms of time ordered products gives the equivalent relation
\ben\label{Jcon1}
T\bigg( d\J(x) \otimes \e^{iF/\hbar} \bigg) = 0
\quad
\text{mod $\I$ forall $x \in M_T$,}
\een
which is again to be understood in the sense of formal power series.
At the level of classical fields, we have
\ben
d\J(x) = (S_0+F, \Phi(x)) \cdot (\Phi^{\ddagger}(x), S_0+F)
\quad
\text{forall $x \in M_T$.}
\een
Hence, \eqref{Jcon1} is equivalent to the equation
\bena\label{Jcon2}
T\bigg( d\J_0(x) \otimes \e^{iF/\hbar} \bigg) &=&
-T\bigg( \left\{
       \hat s_0 \Phi(x) \cdot (\Phi^{\ddagger}(x), F)+
       \hat s_0 \Phi^{\ddagger}(x) \cdot (\Phi(x), F)
\right\} \otimes \e^{iF/\hbar} \bigg) \non\\
&&
-T\bigg( \left\{
(F, \Phi(x)) \cdot (\Phi^{\ddagger}(x), F)
\right\} \otimes \e^{iF/\hbar} \bigg) \quad
\text{mod $\I$.}
\eena
We claim that this equation can be satisfied as a consequence of
our Ward identity T12a by a redefinition of the time-ordered products.
In fact, we shall now show that our Ward identity T12a can even be
used to prove the following stronger identity:
\begin{multline}\label{Jcon3}
\sum_{I_1 \cup \dots \cup I_t=\underline{n}}
\bigg( \frac{i}{\hbar} \bigg)^{t}
T_{t+1}(d\J_0(y) \otimes
\L_{|I_1|}(X_{I_1}) \otimes \dots \L_{|I_t|}(X_{I_t})
)
=\\
-
\sum_{I_1 \cup \dots \cup I_t=\underline{n}}
\bigg( \frac{i}{\hbar} \bigg)^{t-1}
\sum_{i=1}^t
T_t\Bigg(\L_{|I_1|}(X_{I_1}) \otimes \dots \\
\otimes
\left\{
\hat s_0 \Phi(y) \cdot (\Phi^{\ddagger}(y),\L_{|I_i|}(X_{I_i})) +
\hat s_0 \Phi^{\ddagger}(y) \cdot (\Phi(y),\L_{|I_i|}(X_{I_i}))
\right\}
\otimes
\dots \L_{|I_t|}(X_{I_t}) \Bigg) \\
-\sum_{I_1 \cup \dots \cup I_t=\underline{n}}
\bigg( \frac{i}{\hbar} \bigg)^{t-2}
\sum_{1 \le i < j \le t}
T_{t-1}\bigg(\L_{|I_1|}(X_{I_1}) \otimes \dots \\
(\L_{|I_i|}(X_{I_i}), \Phi^{\ddagger}(y)) \cdot (\Phi(y),\L_{|I_j|}(X_{I_j})) \otimes
\dots \L_{|I_t|}(X_{I_t}) \bigg)
\end{multline}
modulo $\I$. This identity
implies~\eqref{Jcon1} as may be seen by multiplying each term
by $\lambda^n/n!$, integrating against $f(x_1), \dots, f(x_n)$,
and summing over $n$. Thus, it remains to be seen that~\eqref{Jcon3}
follows from the Ward identity T12a. For $n=0$, we get the condition
$T_1(d\J_0(y))=0$, which is just the condition of current conservation
in the free theory and hence is satisfied. For $n>0$, we proceed
inductively. This shows that, at the order considered, the failure
of~\eqref{Jcon3} to be satisfied is of the form
$T_1(\alpha_n(y,x_1,\dots,x_n))$, where $\alpha_n(y,x_1,\dots,x_n)$ is a local
covariant functional that is supported on the total diagonal.
We now show that we can set this quantity to 0. To do this, we pick
a testfunction $h \in C^\infty(M)$ with the following properties:
$h(y)=1$ in an open neighborhood of $\{x_1, \dots, x_n\}$,
$h(y)=0$ towards the future of $\Sigma_+$, and towards the past
of $\Sigma_-$, where $\Sigma_\pm$ are Cauchy surfaces in the
future/past of  $\{x_1, \dots, x_n\}$. We may thus write
$dh=\gamma_+-\gamma_-$, where $\gamma_\pm$ are 1-forms that are
supported in the future/past of $\{x_1, \dots, x_n\}$.
Now, from $Q_0=\int_M T_1(\J_0) \wedge \gamma_\pm$, and from the causal
factorization of the time-ordered products, we have
\begin{eqnarray}\label{103}
&&\int_M h(y) T_{t+1}(d\J_0(y) \otimes
\L_{|I_1|}(X_{I_1}) \otimes \dots \L_{|I_t|}(X_{I_t})
) \, dy \non\\
&=&[Q_0, T_{t}(\L_{|I_1|}(X_{I_1}) \otimes \dots \L_{|I_t|}(X_{I_t}))] =i\hbar \hat s_0 T_{t}
(\L_{|I_1|}(X_{I_1}) \otimes \dots \L_{|I_t|}(X_{I_t})) \, ,
\end{eqnarray}
where the last equation is modulo $\I$.
We also have
\ben
\int_M h(y)
(\O(x_i),\Phi^{\ddagger}(y)) \cdot (\Phi(y),\O(x_j)) \, dy = (\O(x_i),\O(x_j))
\een
for any $\O$. It follows from these equations that if we integrate
\eqref{Jcon3} against $h(y)$, then we get an identity follows from the known Ward identity T12a. Stated differently,
because $h(y)=1$ in a neighborhood of $\{x_1, \dots, x_n\}$, and
because the failure $\alpha_n$ of~\eqref{Jcon3} to hold
is supported on the total diagonal, it must satisfy
\ben
\int_M \alpha_n(y, x_1, \dots, x_n) \, dy = 0 \quad
\text{mod $\I$.}
\een
By lemma~\ref{lemma9}, it hence follows that there exists
a local covariant $\beta_n$ supported on the total diagonal
such that $d_y \beta_n(y, x_1, \dots, x_n) =
\alpha_n(y, x_1, \dots, x_n)$, where $\beta_n$ is a 3-form in
the $y$-entry, and a 4-form in each $x_i$-entry, and where
$d_y$ is the exterior differential acting
on the $y$-variable. We may now
redefine time ordered products with one factor of $\J_0(y)$
and $n$ factors of $\L_{1}(x_i), i=1,\dots,n$ by taking
$D_{n+1}(\J_0(y) \otimes \L_1(x_1) \otimes \dots \otimes \L_1(x_n)) := \beta_n(y, x_1,
\dots, x_n)$. Then
the redefined time-ordered products satisfy~\eqref{Jcon3}.

\subsection{Proof that $Q_{\subsc}^2=0$}

We know from the previous subsection that the interacting BRST-current is conserved,
$d\J(x)_{\subsc}=0$ for any $x$, or equivalently,
$d\J(x)_F = 0$ for any $x$ in a domain $M_T = (-T,T) \times \Sigma$
where the function $f$ in $F=\int \{\lambda f \L_1 + \lambda^2 f^2 \L_2\}$
is equal to 1. Thus, the definition of the interacting BRST-charge~
$Q_{\subsc} = \int \gamma \wedge \J_{\subsc}$
is independent of the choice of the compactly supported closed 1-form $\gamma$ dual to
the Cauchy surface $\Sigma$. Using the Bogoliubov formula for the
interacting field operators, the desired equality
$Q_{\subsc}^2=0$ is equivalent to the equation
\bena
&&0=Q_F^2=\bigg( \int \gamma(x) \wedge \J(x)_F \bigg)^2 =\non\\
&&\frac{1}{2}\sum_{n,m} \frac{i^{n+m}}{\hbar^{n+m}n!m!}\int \bigg[
R_n(\J(x); F^{\otimes n}), R_m(\J(y); F^{\otimes m})
\bigg] \, \gamma(x) \gamma(y) \, dx dy
\eena
modulo $\I$, where $\gamma$ is now chosen to be supported in $M_T$. Note that, as usual,
we mean the graded commutator, which is actually the anti-commutator in
the above expression. Now,
because the interacting BRST-charge $Q_F$ as defined using the cutoff
interaction $F$ is independent upon the choice of the compactly supported
closed 1-form in $\gamma$ dual to $\Sigma$, we may
write the interacting BRST-charge either as
$Q_F = \int \gamma^{(1)} \wedge \J_F$, or as
$Q_F = \int \gamma^{(2)} \wedge \J_F$. We may therefore alternatively write
\ben
Q_F^2 = \frac{1}{2} \sum_n \frac{i^n}{\hbar^n n!} \int R_{n+1} \bigg(\J(x); \J(y)
\otimes F^{\otimes n}
\bigg) \gamma^{(1)}(x) \gamma^{(2)}(y) \, dxdy + (1 \leftrightarrow 2) \, ,
\een
where we have also used the GLZ-formula~\eqref{GLZ}. We now make a
particular choice for $\gamma^{(1)}$ and $\gamma^{(2)}$ that will
facilitate the evaluation of this expression. We choose
$\gamma^{(1)} = dh^{(1)} + dh^{(2)}$, where $h^{(1)}$ and $h^{(2)}$
are smooth scalar functions with the following properties:
(a) the support of $h^{(1)}$ is compact, (b) $h^{(1)} = 1$ on the
support of $\gamma^{(2)}$, (c) the support of $h^{(2)}$ is contained
in the causal past of the support of $\gamma^{(2)}$. Due to these
support properties and the causal support properties of the retarded
products, the above expression can then be written as
\ben
Q_F^2 = -\frac{1}{2}
\sum_n \frac{i^n}{\hbar^n n!} \int R_{n+1} \bigg(
d\J(x);
\J(y)
\otimes F^{\otimes n}
\bigg) \, h^{(1)}(x)
\gamma^{(2)}(y)
\, dxdy
\een
Below, we will show that, for any $x,y \in M_T$, the following
identity is a consequence of the Ward-identity T12b:
\begin{multline}
\label{chr}
R \bigg(
d\J(x); \J(y)
\otimes \e^{iF/\hbar}
\bigg) = \\
i\hbar R \bigg(
\left\{
(S_0+F, \Phi(x)) \cdot (\Phi^\ddagger(x),\J(y)) +
(S_0+F, \Phi^\ddagger(x)) \cdot (\Phi(x),\J(y))
\right\}; \e^{iF/\hbar}
\bigg) \quad {\rm mod} \, \I \, .
\end{multline}
We now apply this identity and use that
$h^{(1)}=1$ on the support of $\gamma^{(2)}$. Then we obtain
\ben
Q_F^2 =
\frac{i\hbar}{2} \int R \bigg( (S, \J(x));  \e^{iF/\hbar}
\bigg) \, \gamma^{(2)}(x) \, dx \, ,
\een
again, modulo $\I$.
However, $\hat s \J = d {\bf K}$, so using T11, the right side vanishes
by $d\gamma^{(2)} = 0$. Thus, we have proved
$Q^2_F=0$ modulo $\I$, and it remains to prove eq.~\eqref{chr}.
That equation can be written equivalently in terms of time
ordered products
\bena
\label{chr1}
&&T \bigg(
d\J(x) \otimes \J(y)
\otimes \e^{iF/\hbar}
\bigg)  \\
&& =i\hbar T \bigg(
\left\{
(S_0+F, \Phi(x)) \cdot (\Phi^\ddagger(x),\J(y)) +
(S_0+F, \Phi^\ddagger(x)) \cdot (\Phi(x),\J(y))
\right\}
\otimes \e^{iF/\hbar}
\bigg) \quad {\rm mod} \, \I \, , \non
\eena
using the formulae relating time-ordered and retarded products
given above. We will prove it in this form. Using eq.~\eqref{sjofx},
the eq.~\eqref{chr1} may
be written alternatively as
\bena\label{Jcon4}
&&T\bigg( d\J_0(x) \otimes \J(y) \otimes \e^{iF/\hbar} \bigg) =\\
&&
-T\bigg( \left\{
       \hat s_0 \Phi(x) \cdot (\Phi^{\ddagger}(x), F)+
       (\Phi \leftrightarrow \Phi^\ddagger)
\right\} \otimes \J(y) \otimes \e^{iF/\hbar} \bigg) \non\\
&&
-T\bigg( \left\{
(F, \Phi(x)) \cdot (\Phi^{\ddagger}(x), F)
+(\Phi \leftrightarrow \Phi^\ddagger)
\right\} \otimes \J(y) \otimes \e^{iF/\hbar} \bigg) \non\\
&& +i\hbar T\bigg( \left\{
\hat s_0 \Phi(x) \cdot (\Phi^{\ddagger}(x), \J(y)) +
(F, \Phi(x)) \cdot (\Phi^{\ddagger}(x), \J(y))
+ (\Phi \leftrightarrow \Phi^\ddagger)
\right\} \otimes \e^{iF/\hbar} \bigg) \non
\quad
\text{mod $\I$.}
\eena
We will now show that this equation can be satisfied as a consequence
of our Ward-identity T12b. To prove this identity, we employ the same
technique as in the previous subsection. We first formulate a set of
stronger identities that will imply~\label{Jcon4}. This set of conditions
is completely analogous to eqs.~\eqref{Jcon3}, with the difference
that in eq.~\eqref{Jcon3}, we replace $\L_i(X)$ everywhere by
$\L_i(X)+\tau \J_i(y,X)$, and expand the resulting set of
equations to first order in $\tau$. As in the proof of
eqs.~\eqref{Jcon3}, the resulting equations are established
inductively in $n$. For $n=0$ the identity can be verified directly
using the definitions made in free gauge theory. Inductively, the
resulting equations will then be violated at order $n$ by a potential
``anomaly'' term of the form $T_1(\alpha_n(x,y,x_1, \dots, x_n))$, where
$\alpha_n$ is now an element of $\P^{4/3/4/\dots/4}(M^{n+2})$. As in
the treatment of eq.~\eqref{Jcon3}, the Ward identity T12b then implies
that
\ben\label{a1}
\int_M \alpha_n(x,y,x_1,\dots,x_n) \, dx = 0
\een
while the GLZ-identity, together with the fact that $d\J_{\subsc}=0$ can
be seen to imply the relation
\ben\label{a2}
\int_M d_y \alpha_n(x,y,x_1,\dots,x_n) \, dx_1 \dots dx_n = 0 \, .
\een
Eqs.\eqref{a1} and~\eqref{a2} can now be used to show that the
time-ordered products can be redefined, if necessary, to remove
the anomaly $\alpha_n$. By the same argument as in the previous subsection,
the first identity~\eqref{a1} implies that
\ben\label{adA}
\alpha_n(x,y,x_1,\dots,x_n) =
d_x \delta_n(x,y,x_1,\dots,x_n)
\een
for some $\delta_n \in \P^{3/3/4/\dots/4}(M^{n+2})$. We would like to
redefine the time-ordered products using the quantity $D_n$ (see sec.~3.6)
\ben\label{redef}
D_{n+2}(\J_0(x) \otimes \J_0(y) \otimes \L_1(x_1) \dots
\otimes \L_1(x_n) := \delta_n(x,y,x_1,\dots, x_n) \, .
\een
In view of eq.~\eqref{adA}, this would remove the anomaly. However,
it is not clear that we can make this redefinition, because
the time-ordered products with two free BRST-currents at $x$ and $y$
must be anti-symmetric in $x$ and $y$, and this need not be the case
for $\delta_n$ in \eqref{adA}. We will circumvent this problem by using a
modified $\hat \delta_n$
in eq.~\eqref{redef} to redefine the time-ordered products with 2
currents. To construct the modified $\hat \delta_n$, we consider the
quantity
\ben
\beta(\gamma^{(1)}, \gamma^{(2)}) =
\int \delta_n(x,y,z_1,\dots,z_n) \gamma^{(1)}(x) \gamma^{(2)}(y) \, dxdydz_1
\dots dz_n + (1 \leftrightarrow 2) \, ,
\een
where $\gamma^{(1)},\gamma^{(2)}$ are now arbitrary 1-forms of compact
support. $\beta$ is evidently closely related to the symmetric part of
$\delta_n$, which we would like to be zero. From eq.~\eqref{a2}, we have
$\beta(dh^{(1)}, dh^{(2)})=0$ for any pair of compactly supported
scalar functions $h^{(1)}, h^{(2)}$. As we shall show presently,
this implies that we can write
\ben\label{beta}
\beta(\gamma^{(1)}, \gamma^{(2)}) = C(d\gamma^{(1)}, \gamma^{(2)})
+ (1 \leftrightarrow 2)
\een
where $C$ has a distributional kernel
$C \in \P^{2/3}(M^2)$. We now define
\ben
\hat \delta_n(x,y,z_1,\dots,z_n)=
\delta_n(x,y,z_1,\dots,z_n) - d_x C(x,y) \delta(y,z_1,\dots,z_n) - (x
\leftrightarrow y) \, ,
\een
which is manifestly anti-symmetric in $x,y$.
We use this new $\hat D_n$ in order to redefine the time-ordered products
with 2 currents as in eq.~\eqref{redef} instead of the old $D_n$.
Evidently, the new time ordered product is now
anti-symmetric in $x,y$. Furthermore, as a
consequence of eq.~\eqref{beta}, the new anomaly for the redefined
time-ordered products $\hat \alpha_n$ satisfies
\ben
\int \hat \alpha_n(x,y,z_1,\dots,z_n) \, dz_1 \dots dz_n = 0 \, .
\een
It follows from this equation that
\ben
\hat \alpha_n(x,y,z_1,\dots,z_n)
= \sum_{l=1}^n d_l \delta_{n/l}(x,y,z_1,\dots,z_n) \quad
d_l = dz_l \wedge \frac{\partial}{\partial z_l}
\een
for some $\delta_{n/l} \in \P^{4/3/4/ \dots/3 \dots /4}(M^{n+2})$.
We use these quantities to make a final redefinition of the
time-ordered products. We have
\ben
\hat s_0 \Phi(x_1) \cdot (\Phi^\ddagger(x_1), \L_1(x_2))+
\hat s_0 \Phi^\ddagger(x_1) \cdot (\Phi(x_1), \L_1(x_2))
= d_1\J_1(x_1) \delta(x_1, x_2) + d_2\Sigma_1(x_1,x_2)
\een
for some $\Sigma_1 \in \P^{3/3}(M^2)$. We redefine the
time-ordered products involving these quantities using the quantities
(see sec.~3.6)
\ben\label{redef1}
D_{n+1}(\J_0(x) \otimes \L_1(z_1) \dots
\otimes \Sigma_1(y,z_l) \otimes \dots \L_1(z_n)) :=
\delta_{n/l}(x,y,z_1,\dots, z_n)) \, .
\een
This final redefinition then removes the anomaly $\hat \alpha_n$. \qed

It remains to prove eq.~\eqref{beta}. We formulate this result as
a lemma:

\begin{lemma}
Let $\beta \in \P^{3/3}(M^2)$ such that
$\beta(dh^{(1)}, dh^{(2)})=0$ for any pair of compactly supported
scalar functions $h^{(1)}, h^{(2)}$. Then $\beta$ can
be written in the form~\eqref{beta} for some
$C \in \P^{2/3}(M^2)$.
\end{lemma}

\noindent
{\em Proof:} $\beta$ is of the form
\ben\label{bform}
\beta(\gamma^{(1)}, \gamma^{(2)}) =
\int_M dx \sum_{m=0}^p \beta^{\mu \nu_1 \dots \nu_m \sigma}
\gamma^{(1)}_\mu \nabla_{\nu_1} \cdots \nabla_{\nu_m}
\gamma^{(2)}_\sigma \, ,
\een
where $\beta$ are tensor fields that are locally constructed
out of $g,\nabla$, and $\Phi,\Phi^\ddagger$.
We claim that the condition
$\beta(dh^{(1)}, dh^{(2)})=0$
and the symmetry of $\beta$
implies that $\beta$ can be put into the
form~\eqref{beta}. Since the commutator
of two derivatives gives a Riemann tensor,
we may assume that each tensor $\beta$ in the sum in~\eqref{bform}
is symmetric under the exchange of the indices $\nu_1, \dots, \nu_m$,
\ben\label{sym1}
\beta^{\mu \nu_1 \dots \nu_m \sigma} = \beta^{\mu (\nu_1 \dots \nu_m)
  \sigma} \, .
\een
Now consider the contribution to~\eqref{bform} with the highest number
of derivatives, $m=p$. By varying $\beta(dh^{(1)}, dh^{(2)})=0$
with respect to
$h^{(1)}, h^{(2)}$ there follows the additional symmetry
\ben\label{sym2}
\beta^{(\mu \nu_1 \dots \nu_p \sigma)} = 0 \, .
\een
Consider now the vector field defined by
\ben
B^\mu = \beta^{\mu \nu_1 \dots \nu_p \sigma} \nabla_{\nu_1}
\cdots \nabla_{\nu_p} \gamma_\sigma\, .
\een
Using the symmetry property~\eqref{sym1}, this may be rewritten as
\bena
B^\mu &=& \beta^{\mu \nu_1 \dots \nu_p \sigma} \nabla_{\nu_1}
\cdots \nabla_{[\nu_p} \gamma_{\sigma]}\non\\
&& +\beta^{\mu (\nu_1 \dots \nu_p \sigma)} \nabla_{\nu_1}
\cdots \nabla_{\nu_p} \gamma_{\sigma} \, .
\eena
Then, using the symmetry~\eqref{sym2}, this may further be written as
\bena
B^\mu &=& \beta^{\mu \nu_1 \dots \nu_p \sigma} \nabla_{\nu_1}
\cdots \nabla_{[\nu_p} \gamma_{\sigma]}\non\\
&& -\frac{2}{p+2} \beta^{\sigma (\mu \nu_1 \dots \nu_p)} \nabla_{\nu_1}
\cdots \nabla_{[\nu_p} \gamma_{\sigma]} \non\\
&& -\frac{2(p+1)}{p+2} \nabla_\nu
\left\{\beta^{\mu (\nu \alpha_1 \dots \alpha_{p-1} \sigma)} \nabla_{\alpha_1}
\cdots \nabla_{\alpha_{p-1}} \gamma_{\sigma} - (\mu \leftrightarrow
\nu) \right\} \\
&& +\text{terms with $(p-1)$ derivatives on $\gamma_\sigma$} \, .
\eena
Now put $\gamma = \gamma^{(2)}$ in this equation, contract both sides
with $\gamma^{(1)}$, and integrate, to obtain an expression for the
highest derivative term in $\beta$. Using this expression, we find that
$\beta(\gamma^{(1)}, \gamma^{(2)})$ is given by a sum of terms
each of which contains either $\nabla_{[\mu}\gamma^{(1)}_{\nu]}$
or $\nabla_{[\mu}\gamma^{(2)}_{\nu]}$, or which contains at most
derivative terms of order $p-1$. Consequently, using the
symmetry of $\beta$, we can write
\ben\label{above1}
\beta(\gamma^{(1)}, \gamma^{(2)}) = C(\d \gamma^{(1)}, \gamma^{(2)})
+ C(\d \gamma^{(2)}, \gamma^{(1)}) + R_{p-1}(\gamma^{(1)}, \gamma^{(2)}) ,
\een
where $R_{p-1}$ stands for a remainder term of the form~\eqref{bform}
containing at most $p-1$ derivatives, and where $C$ is also of the
form~\eqref{bform}. If we now take $\gamma^{(1)} = \d h^{(1)}$, and
$\gamma^{(2)} = \d h^{(2)}$ in eq.~\eqref{above1}, and
use $\beta(dh^{(1)}, dh^{(2)})=0$, then we see that $R_{p-1}$
again satisfies $R_{p-1}(dh^{(1)}, dh^{(2)})=0$. Thus, we may repeat the
arguments just given for $R_{p-1}$ and conclude that $\beta$ can be
written as in eq.~\eqref{above1} with a new $C$, and a remainder
$R_{p-2}$ containing at most $p-2$ derivatives. Thus, further repeating this
procedure, we find that \eqref{above1} must hold for some $C$ and a
remainder of the form $R_0(\gamma^{(1)}, \gamma^{(2)}) =
\int \epsilon \gamma^{(1)}_\mu r^{\mu\nu} \gamma^{(2)}_\nu$.

Now, $R_0$ is symmetric, so $r^{[\mu\nu]} = 0$. Furthermore,
we have $R_0(\d h^{(1)}, \d h^{(2)}) = 0$ for all compactly supported
$h^{(1)}, h^{(2)}$. Varying this equation with respect to $h^{(2)}$,
we get $0 = \nabla^\mu(r_{\mu\nu} \nabla^{\nu} h^{(1)})$. Now, pick
a point $x \in M$, and choose $h^{(1)}$ so that $h^{(1)}(x)=0$. Then
it follows that $r_{\mu\nu} \nabla^{\mu} \nabla^{\nu} h^{(1)}=0$ at
$x$. Because $\nabla^\mu \nabla^\nu h^{(1)}$ is an arbitrary symmetric
tensor at $x$, it follows that $r^{(\mu\nu)} = 0$, and therefore
that $r^{\mu\nu} = 0$, thus proving the
desired decomposition~\eqref{beta}. This completes the
proof.
\qed

\subsection{Proof that $[Q_{I}, \Psi_I]=0$ when $\Psi$ is
  gauge invariant}

Here we show that the Ward identity T12c implies
$[Q_I, \Psi_I(x)]=0$ modulo $\I$,
whenever $\Psi \in \P(M)$
is a strictly gauge invariant operator of ghost number 0, i.e.,
$\Psi = \prod
\Theta_{s_i}(F,{\mathcal D} F, \dots, {\mathcal D}^{k_i} F)$. As in
the proof given in the previous subsection, this property will follow
from the identity
\bena\label{Jcon5}
&&T\bigg( d\J_0(x) \otimes \Psi(y) \otimes \e^{iF/\hbar} \bigg) =\\
&&
-T\bigg( \left\{
       \hat s_0 \Phi(x) \cdot (\Phi^{\ddagger}(x), F)+
       (\Phi \leftrightarrow \Phi^\ddagger)
\right\} \otimes \Psi(y) \otimes \e^{iF/\hbar} \bigg) \non\\
&&
-T\bigg( \left\{
(F, \Phi(x)) \cdot (\Phi^{\ddagger}(x), F)
+(\Phi \leftrightarrow \Phi^\ddagger)
\right\} \otimes \Psi(y) \otimes \e^{iF/\hbar} \bigg) \non\\
&& +i\hbar T\bigg( \left\{
\hat s_0 \Phi(x) \cdot (\Phi^{\ddagger}(x), \Psi(y)) +
(F, \Phi(x)) \cdot (\Phi^{\ddagger}(x), \Psi(y))
+ (\Phi \leftrightarrow \Phi^\ddagger)
\right\} \otimes \e^{iF/\hbar} \bigg) \non
\quad
\text{mod $\I$,}
\eena
where again $F=\int (\lambda f \L_1 + \lambda^2 f^2 \L_2)$.
One can now formulate a stronger set of local identities analogous to
eq.~\eqref{Jcon3}, and one can prove these identities using T12c along
the same lines as in the previous subsection, with $\J(y)$ there
replaced everywhere by $\Psi(y)$. The potential anomaly of the
stronger identities (and therefore the possible violation of
eq.~\eqref{Jcon5}) can now be removed by a suitable redefinition of the
time ordered products $T_{n+2}(\J_0(x) \otimes \Psi_0(y) \otimes
\L_1(x_1) \otimes \L_1(x_n))$ at $n$-th order in perturbation theory,
where $\Psi = \Psi_0 + \lambda \Psi_1 +\lambda^2 \Psi_2 +
\dots$. However, contrary to the case in the previous subsection, we
now do not have to worry about potential symmetry issues, that
had to be dealt with there, because $\Psi_0$ is always distinct from
$\J_0$, the latter having ghost number 1.

\subsection{Relation to other perturbative formulations of gauge invariance}

In our approach to interacting quantum gauge theories, the gauge invariance
of the theory was incorporated in the conditions that there
exists a conserved interacting BRST-current operator, and that the
corresponding charge operator be nilpotent. As we demonstrated, this follows
from our Ward identity~\eqref{masterward}, the generating identity
for T12a, T12b, and T12c.
In the literature on perturbative quantum field theory in
flat spacetime, other notions of gauge invariance of the quantum
field theory have been suggested, and other
conditions have been proposed to ensure those. We now briefly discuss
some of these, and explain why these formulations are not suitable in
curved spacetime.

\medskip

{\bf Diagrammatic approaches (dimensional regularization):}
Historically, the first proofs of
gauge invariance of the renormalized perturbation series in
gauge theories on flat $\mr^4$ were performed on the level of Feynman
diagrams. The gauge-invariance of the classical Lagrangian implies
certain formal identities between the diagrams at the unrenormalized
level. At the renormalized level, these identities in turn would
formally\footnote{We say ``formally,'' because amplitudes can have
additional infra-red divergences, which are very hard to treat in a gauge-invariant manner.}
imply the gauge-invariance of amplitudes.
One must thus prove that these identities remain valid at
the renormalized level. For this, it is important to have a
regularization/renormalization scheme that preserves these
identities. Such a scheme was found by 't Hooft and Veltmann~\cite{tHooft1,tHooft2,tHooft3}, namely
dimensional regularization. Because that scheme is also very handy for
calculations (except for certain calculations involving
Dirac-matrices), it has remained the most popular approach among
practitioners. Modern presentations of this approach based on the Hopf-algebra
structure behind renormalization in the BPHZ-approach~\cite{Kreimer1,Kreimer2,Kreimer3}
are~\cite{Suijlekom1,Suijlekom2}.

In curved space, scattering amplitudes are not well-defined, because
there is no sharp notion of particle in general. At a more formal level,
diagrammatic expansions in general are problematic because there does not exist
a unique Feynman propagator, so a given Feynman diagram can mean very different
mathematical expressions depending on one's choice of Feynman propagator.
One may of course expand the theory using any Feynman
propagator. However, then the problem arises that the Feynman propagator is not a
local covariant functional of the metric, but also depends upon boundary/initial conditions,
which are intrinsically non-local. This would interfere with ones ability to
reduce the ambiguity to local curvature terms.
One might be tempted to take the local Feynman parametrix
$H_F$, which is local and covariant. But this has the undesirable property that it is not
a solution of the field equation, but only a Green's function modulo a smooth remainder,
see Appendix~D. This severely complicates the treatment of quantities
that vanish due to field equations, and of the Ward identities.
Finally, in curved space, the Feynman propagator is only well defined
as a distribution in position space,
while techniques such as dimensional regularization seem to work best
in momentum spacetime. Thus,
a diagrammatic proof of quantum gauge invariance of Yang-Mills theory
in curved spacetime seems to be difficult and somewhat unnatural.

\medskip

{\bf Zinn-Justin equation:} In many formal approaches to
perturbative gauge theory in flat
spacetime $\mr^4$, gauge invariance  of the theory is expressed in terms
of an integrated condition involving the so-called ``effective
action'', $\Gamma_{\rm
  eff}(S)$ of the theory associated with the classical action
$S=S_0+\lambda S_1+\lambda^2 S_2$. The effective action is a
generating functional for the 1-particle irreducible Feynman diagrams
of the theory. The condition for perturbative gauge invariance is
simply and elegantly encoded in the relation~\cite{Zinn-Justin1975}
\ben\label{zj}
(\Gamma_\eff(S), \Gamma_\eff(S)) = 0 \, .
\een
Condition~\eqref{zj} is referred to as the ``Slavnov Taylor identity''
in ``Zinn-Justin form''. It is closely related to the ``master
equation'' that arises in the Batalin-Vilkovisky
formalism~\cite{Batalin1981} (see also \cite{Henneaux1995}), and it
reduces to the classical condition $(S,S)=0$ for BRST-invariance
when one puts $\hbar=0$. At the
formal level, the Slavnov-Taylor identity is most straightforwardly
derived from the path integral. It is also in this setting that one can
understand relatively easily that it formally implies the absence of (infinite)
counterterms to the classical action violating gauge invariance.
However, by itself, it does not imply the gauge invariance of physical
quantities such as scattering amplitudes, or identities like $Q_{\subsc}^2 = 0$.

The effective action $\Gamma_\eff(S)$ is
only a formal quantity, since it involves integrations over all of
spacetime. These integrations typically lead to infra-red divergences,
as is in particular the case also in pure Yang-Mills
theory. Therefore, also the Slavnov-Taylor equation~\eqref{zj} is only
a formal identity. If the interaction $\lambda S_1+\lambda^2 S_2$
is replaced by a local interaction, $F=\int \{\lambda f\L_1+ \lambda^2
f^2 \L_2 \}$, with $f$ a smooth cutoff function of compact support,
then the infra-red divergences are avoided, and the
effective action $\Gamma_\eff(S_0+F)$ is well defined. The
precise definition of $\Gamma_\eff(S_0+F)$ within our framework is
given in Appendix~B. However, for the cutoff-interaction, the
Slavnov-Taylor identity no longer holds. Nevertheless, it can be shown that
$\Gamma_\eff(S_0+F)$ satisfies an analogous equation,
given by eq.~\eqref{zjappa}.
That equation can be used to formally ``derive'' eq.~\eqref{zj}, if
one could prove that the anomaly in eq.~\eqref{zjappa} vanishes.
Since the anomaly is closely related to the failure of the
interacting BRST-current to be conserved, one might expect to be able
to remove the anomaly by an argument similar to our proof of
T12a, but this has not been worked out even in flat spacetime.

In curved spacetime, we may still define an effective action,
$\Gamma_\eff(S_0+F)$, which now depends upon the arbitrary choice of
a quasifree Hadamard state $\omega$, see Appendix~E. Hence it is
definitely not a quantity that depends locally and covariantly
upon the metric, but also on the non-local choice of $\omega$,
Therefore, even at the formal level, it is not clear that the
Slavnov-Taylor identity can be viewed as a renormalization condition
that is compatible with the locality and covariance of the
time-ordered products. Also, while the
Slavnov-Taylor identity can again be formally derived from our Ward-Identity
T12a, it does not directly imply the
gauge-invariance of physical quantities such as $n$-point functions,
and it also does not prove (even formally) that the OPE closes
among physical operators. For these reasons, we prefer
to work with the Ward-identities T12a, T12b, T12c in this paper, which
are rigorous, and have a local and covariant character. Despite the above
differences, the Zinn-Justin is probably to be regarded as the closest
analogue to our renormalization conditions expressing local gauge invariance.
The similarities can be made more explicit using our generating formula~\eqref{masterward}
[or eq.~\eqref{WW}] for our Ward identities.

\medskip

{\bf Causal approach:}
A condition expressing perturbative gauge invariance in flat spacetime
that is of a more local nature than~\eqref{zj} has been proposed in
a series of papers by D\" utsch et al.~\cite{Duetsch1994dp,Duetsch1994dq,Duetsch1993bc,Duetsch1994ur,Duetsch1993ee,Scharf1},
see also~\cite{Hurth1,Hurth2,Hurth3,Grigore1,Grigore2,Grigore3}.
These works are also related to the ``quantum Noether
condition''~\cite{Hurth3}.
Let $T_n(x_1,\dots,x_n)$
be the time-ordered product of $T_n(\L_1(x_1) \otimes \dots \otimes
\L_1(x_n))$. (in the above papers, the interaction Lagrangian 4-form
is here identified with a scalar by taking the Hodge dual). Let $Q_0$ be the
free BRST-charge. Then it is postulated that there exists a
set of time-ordered products $T_{n/l}(x_1, \dots, x_n)$ with
the insertion\footnote{Thus in particular, $T_{n/l}(x_1, \dots, x_n)$
should be symmetric in all variables except $x_l$, and it is a
3-form in $x_l$.} of
some (unspecified) 3-form-valued field in the $l$-th
entry such that
\ben\label{scharf}
[Q_0, T_n(x_1, \dots, x_n)] = i\hbar \sum_{l=1}^n
d_l T_{n/l}(x_1, \dots, x_n)
\quad \text{modulo $\I$}
\een
for all $n>0$, where $d_l = dx_l^\mu \wedge \partial/\partial x_l^\mu$
is the exterior derivative acting on the $l$-th entry.
The condition is to be viewed as a normalization
on the time ordered products involving $n$ factors of the
interaction $\L_1$. Note that there are no explicit\footnote{
As explained in the above papers, however, implicit normalization conditions
on time ordered products with factors of $\L_2$ arise
from~\eqref{scharf}. Also,~\eqref{scharf} apparently may even be used
to determine the form of $\L_1$, which is simply given in our
approach.} conditions imposed
on time-ordered products involving $\L_2$. Note also that the
condition is imposed only modulo $\I$, that is, on shell. In fact,
the authors of the above papers always work in a representation, where the field equations
automatically hold (see section~3), rather than at the
algebraic level, where the field equations need not be imposed as a relation.
A related difference is that the above authors do not work with
anti-fields, without which it appears to be very cumbersome to
obtain powerful consistency relations for potential anomalies
of~\eqref{scharf}. (Some aspects of this difference are addressed in~\cite{Barnich1999}.)

The key motivation for condition~\eqref{scharf} is that, as our condition T12a),
it formally implies that the $S$-matrix commutes with $Q_0$ in the
``adiabatic limit,'' see above. Indeed, if we formally
integrate~\eqref{scharf} over $(\mr^4)^n$, then the right hand side
formally vanishes, being a total derivative. This shows that
$\mathcal S$ formally commutes with $Q_0$. However, unlike our Ward
identities, we do not believe
that eq.~\eqref{scharf} would imply $Q_\subsc^2=0$ for the
interacting BRST-charge, or $[Q_\subsc, \Psi_\subsc]=0$ for gauge
invariant operators.

The relation~\eqref{scharf} is apparently different from
our corresponding condition T12a (considered in flat spacetime), so we
now briefly outline how they are related. Consider a prescription for
the time-ordered products satisfying our Ward identity T12a, so that,
in particular, eq.~\eqref{scharf} does not hold for that prescription.
However, let us now make the following redefinition of the
time-ordered products containing two factors of $\L_1$, that is,
\ben
T_2(\L_1(x_1) \otimes \L_1(x_2)) \to
T_2(\L_1(x_1) \otimes \L_1(x_2)) + T_1(\L_2(x_1,x_2)) \, ,
\een
where we recall the notation $\L_2(x_1,x_2) =
2\L_2(x_1)\delta(x_1,x_2)$. Let us further note that
\ben
\hat s_0 \L_2(x_1,x_2) + (\L_1(x_1),\L_1(x_2)) =
d_1 \O_{2/1}(x_1, x_2)
+ d_2 \O_{2/2}(x_1, x_2)
\een
for some fields $\O_{2/1} \in \P^{4/3}$ and $\O_{2/2} \in \P^{3/4}$
supported on the diagonal,
and $\hat s_0 \L_1 = d \O_1$.
Using that $[Q_0,T_n] = i\hbar \hat s_0 T_n$ modulo $\I$, and defining
$T_{n/l}$ by
\bena
T_{n/l}(x_1, \dots, x_n) &=&
\sum_{j=1,2} T_{n-1}\Big(\L_1(x_1)\otimes \dots
\O_{2/j}(x_{l+j-1},x_{l+j}) \otimes
\dots \L_1(x_n) \Big) + \text{cycl. perm.} \non\\
&+& T_n\Big(\L_1(x_1) \otimes \dots \O_1(x_l)\otimes \dots
\L_1(x_{n})\Big) \, ,
\eena
one can then check that eq.~\eqref{scharf} holds. Thus,
our Ward identity implies~\eqref{scharf} if a finite renormalization
change is made, and presumably~\eqref{scharf} may also be used to
deduce our Ward identity T12a. Note, however, that our identities T12b
and T12c are conditions that go definitely beyond
the Ward-identities~\eqref{scharf}.

\section{Summary and outlook}

In this paper, we have given, for the first time, a perturbative construction of non-abelian
Yang-Mills theory on arbitrary globally hyperbolic
curved, Lorentzian spacetime manifolds.
Following earlier work on quantum field theory in curved spacetime, our
strategy was to construct the interacting field operators and the algebra that they
generate. This was accomplished starting from a gauge fixed version of the
theory with ghost and anti-fields, and then defining the algebra of observables of
perturbative Yang-Mills theory as the BRST-cohomology of the corresponding algebra
associated with the gauge fixed theory. To implement this strategy it
was necessary to first find a prescription for defining a {\em conserved}
interacting BRST-current, and for which the corresponding conserved charge
is furthermore {\em nilpotent}. We were able to characterize such a
prescription by a novel set of Ward identities for the time-ordered products
in the underlying free theory. We furthermore showed how to find a
renormalization prescription for which the Ward-identities indeed
hold. In addition, we showed that our renormalization prescription
also satisfies other other important properties,
notably the condition of general covariance. Altogether, these
constructions provide a proof that perturbative Yang-Mills theory can
be defined as a consistent, local
covariant quantum field theory (to all orders in
perturbation theory), for any globally hyperbolic spacetime.

\medskip

A key feature of our approach is that it is entirely local in
nature, in the sense that our renormalization conditions only make
reference to local quantities. A local approach is essential
in a generic curved spacetime in order find the correct renormalization
prescription respecting locality and general covariance. But it
is also advantageous in flat spacetime in many respects
compared to other existing approaches in flat spacetime, such as approaches
focused on the scattering matrix, or approaches based on the
path-integral. The key advantages of our approach are the following:

\begin{itemize}
\item
Because our approach is
completely local, we can completely disentangle the
the infra-red divergences and ultra-violet divergences of the theory.
This is mandatory in Yang-Mills theory, where infra-red divergences
pose a major problem, even in flat spacetime.
\item
Because our approach is algebraic in nature, the objects of primary
interest are the interacting field operators, rather than auxiliary quantities such as
effective actions or scattering matrices. This makes it easy for us to
prove the important result that the operator product expansion of Yang-Mills theory closes
among gauge invariant fields, and that the renormalization group flow does not leave the
space of gauge invariant fields. On the other hand, it tends to be
much more complicated to prove such statements in other formalisms
even in flat spacetime.
\item
Because our approach is local and covariant, we can directly
analyze the dependence of our constructions on the metric. For
example, one can directly obtain the following result: If a non-abelian gauge theory
has trivial RG-flow in flat spacetime (such as the ${\rm N}=4$
super Yang-Mills theory), then it also must have trivial RG-flow
in any spacetime in which possible renormalizable curvature couplings
in the Lagrangian (such as a $R \, {\rm Tr} \Phi^2$-type term) happen to
vanish. Thus, the ${\rm N}=4$ super Yang-Mills theory has trivial
RG-flow in any spacetime with vanishing scalar curvature. Note that,
unlike in flat spacetime, this does by no means imply that the
theory is conformally invariant, because a spacetime with $R=0$
will not in general admit any conformal isometries.
\end{itemize}

A weak point of our constructions, as for most other perturbative
constructions in quantum field theory, is that one does not have any control over
the convergence of the perturbation series. This is in particular a
problem for quantum states such as bound states that are not expected
to have a perturbative description. A partial
resolution of this problem is provided by the operator product
expansion (see sec.~4.2), because it allows one to compute $n$-point
correlation functions in terms of OPE-coefficients and 1-point
functions (``form factors''), which one may regard as additional
phenomenological input.
But a full solution would presumably require to go beyond perturbation theory, which seems a
distant goal even in flat spacetime.

Apart from this problem, there remain a couple of technical questions related to the
perturbation expansion, of which we list a few:

\subsection{Matter fields, anomalies}
In this paper, we have considered only pure Yang-Mills theory for simplicity. Clearly, one would
like to add matter fields, such as fermion fields in a representation R of the gauge group $G$.
In that case, the general strategy and methods of our paper can still be applied.
But it is no longer clear that the
Ward-identities formulated in this paper can still be satisfied, as there can now
be non-trivial solutions to the corresponding consistency conditions in the presence of chiral fermions.
If the Ward-identities cannot be satisfied, one speaks of an anomaly. In our case this would imply
that the interacting BRST-current is no longer conserved, and that a conserved BRST-charge
cannot be defined, meaning that the theory is inconsistent at the quantum level. In flat space, this can happen if
the gauge group contains factors of $U(1)$, for certain representations R. By
the general covariance of our construction, the types of anomalies in flat space must
then also be absent in any curved spacetime. However, in curved space,
a new type of anomaly can also arise in the presence of chiral fermions and abelian
factors in the gauge group. For example, even at the level of free Yang-Mills theory,
one can compute that the divergence $d \J_I$
(exterior differential) of the quantum BRST current operator
is not zero as required by consistency,
but it has a contribution to its divergence proportional of the type
given in eq.~\eqref{chsim}, which cannot be eliminated by
finite renormalization. In particular, one finds a contribution
$d\J_I \propto \A_I + \dots$ at 1-loop order, where
\ben
\A =  {\rm const.} \sum_K {\rm Tr}[{\rm R}(T_K)] \, C^K \, {\rm Tr}(R \wedge R)
\een
and where the sum over $K$ is over the abelian generators of the
Lie-algebra only. In the standard model, with
gauge group $G=SU(3) \times SU(2) \times U(1)$,
the representation of the abelian generator $Y$ (charge
assignments of the fermion fields) is precisely so that $\A=0$,
as also observed by~\cite{Geng,Ramond1990}.
However, we do not know whether the theory remains free of this
kind of anomaly to arbitrary orders in the perturbation
series. This would be important to check.

It is also important to investigate whether the renormalization conditions
considered in this paper can be used to show that a divergence-free
interacting stress tensor $T^{\mu\nu}_I$ can be constructed. Here,
one can presumably use the techniques of~\cite{Hollands2005} to show
that there is no anomaly for this conservation equation, but
it would be important to settle the details. A particularly interesting
question in this connection is to see precisely how the expected
trace anomaly for this quantity arises in the present framework.

\subsection{Other gauge fixing conditions}

In this paper, we have worked with a specific gauge fixing condition (the Lorentz gauge). The important feature of this
condition for our purposes was that the field equation for the spin-1 field then becomes $\square A + \dots = 0$, where
the dots represent terms with less derivatives.
This was important because only in that case are we able
to construct a Hadamard parametrix for the vector field, which is a key ingredient in our constructions.
However, one may wish to consider other types of gauge fixing
conditions, both for practical purposes, as well as a matter of
principle. Even if a Hadamard parametrix could still be defined in
such cases, it is not
a priori clear that the theories defined using different gauge fixing
conditions are equivalent. In our approach,
equivalence would mean that the algebras of observables obtained from
different gauge fixing conditions are
canonically isomorphic. We have not investigated the question whether this is indeed the case.

\subsection{Background independence}

In our constructions (as in all other standard approaches to perturbative Yang-Mills theory), we have
split the Yang-Mills connection ${\mathcal D} = \nabla +i\lambda A$ into the standard flat, non-dynamical background
connection $\nabla$, and a dynamical field $A$. At the level of classical Yang-Mills theory
it is evident that it is immaterial how this split is made, i.e., classical Yang-Mills theory is
background independent in this sense. In particular, the standard choice $\nabla = \partial$
in flat spacetime is just one possibility among infinitely many other ones. In the gauge fixed classical theory
with ghosts and anti-fields, different choices of the background connection give rise to different classical
actions. The difference is, however, only by a BRST-exact term. Since the classical theory is defined as the BRST-cohomology,
such a BRST-exact term does not change the brackets between the physical observables, and hence the theory
is background independent also in the gauge-fixed formalism. Unfortunately, we do not know whether the
same statement is still true in the quantum field theory, i.e., we do not know whether the algebras of physical
observables associated with different choices of the background connection are still isomorphic.
The difficulty is that, in quantum field theory, the background connection $\nabla$ is
treated very differently from the dynamical part $A$: The background connection would enter the
definition of the propagators, e.g., of the local Hadamard parametrices, while $A$ is a
quantum field.

The question whether one is allowed to shift parts of $A$ into $\nabla$ and
vice versa is closely related to the question whether the ``principle of perturbative agreement''
formulated in~\cite{Hollands2005} can be satisfied with respect to the gauge connection.
The satisfaction of this principle is equivalent to certain Ward-identities at the level of the time-ordered products,
but we do not know in the present case whether these Ward identities can be satisfied, i.e., whether there are any anomalies.
In~\cite{Hollands2005}, a potential violation of these identities may be identified with
a certain cohomology class. In our case, when the background structure in question is a
gauge connection, the potential violation would be
represented by a certain 2-cocycle on the space of all gauge potentials. An anomaly of this sort could
arise in theories with chiral fermions.
Thus, the question of background independence in quantum Yang-Mills theory remains an open problem, which
has not been solved, to our knowledge, even in flat spacetime.

\vspace{2cm}

\paragraph{\bf Acknowledgments:} I would like to thank F.~Brennecke,
D.~Buchholz, M. D\" utsch, L.~Faddeev, K.~Fredenhagen, D.~Grigore and
R.~M.~Wald for discussions, and I would also like to thank M.~Henneaux for discussions at an early stage of this work.
Some parts of this work were completed during the 2007 program ``Mathematical and Physical Aspects
of Perturbative Approaches to Quantum Field Theory'' at the Erwin-Schroedinger Institute, Vienna, to which
I express my gratitude for its financial support and hospitality. Later versions have benefitted from improvements/corrections 
suggested by M. D\" utsch, M. Fr\"ob, J. Holland, M. Taslimitherani and J. Zahn.

\appendix

\section{U(1)-gauge theory without vector potential}

In the case of a pure $U(1)$-gauge theory, one may
consider a different starting point for defining the theory,
using as the basic input only the field equations for the 2-form
field strength tensor rather than the action for the gauge potential
$A$. This is because
the field equations may then be written without
reference to the gauge potential as equations for the field
strength $F$, viewed now as the dynamical variable.
The equations are of course Maxwell's equations,
in differential forms notation $d F = 0$ and $d* F = 0$.

On a curved manifold $M$ with nontrivial topology, not every closed
form $F$ need to be exact, so it does not
follow from the field equation
$d F = 0$ that $F$ can be written in terms of a vector potential as
$F=d A$. Thus, using only Maxwell's equations as the
input defines a more general theory classically than the action
$\int dA \wedge {}* dA$,
because cohomologically non-trivial solutions $F$ are possible. In this
section, we briefly indicate how one may quantize such a theory.

A globally hyperbolic spacetime always has topology $M = \Sigma \times
\mr$, so closed but non-exact 2-forms $F$ can exist on $M$ if
$\Sigma$ contains any non-contractible 2-cycles, $C$. Let us cover
$M$ by
\ben
M = \bigcup_i M_i
\een
where each $M_i$ a globally hyperbolic,
connected and simply connected spacetime in its own
right, which does not contain any non-contractible
2-cycles. Consequently on each $M_i$, any closed 2-form is exact, and
the classical theory defined by Maxwell's equations $\d F = 0$, $\d*
F$ is completely equivalent to the theory of a vector potential $A$
with action~\eqref{Sdef}. Thus, by the results of the
previous sections, we can
construct a corresponding algebra of observables $\hat \F_0(M_i)$ for each
$i$, containing gauge-invariant observables such as polynomials of the
field strength.

Each $\hat \F_0(M_i)$ is only given to us as an abstract *-algebra, so
we do not a priori know what is the relation between those algebras
for different $i$. However, if $M_i$ is contained in $M_j$,
then by the general covariance property, there is an
embedding of algebras $\alpha_{i,j} \equiv
\alpha_{\psi(i,j)}: \hat \F_0(M_i) \to \hat \F_0(M_j)$, where
$\psi(i,j): M_i \to M_j$ is the embedding. Thus, following ideas of
Fredenhagen, and K\" usk\" u~\cite{Fredenhagen94,Kusku06,Kuskudiplom}, we may define an
algebra $\A_u(M)$ as the universal algebra
\ben
\A_u(M) \equiv {\rm ind-}\lim_{M_i} \hat \F_0(M_i) \, .
\een
The universal algebra is defined as the unique algebra such that
there exist *-homorphisms $\alpha_i: \hat \F_0(M_i) \to \A_u(M)$ with the
property $\alpha_j \circ \alpha_{j,i} = \alpha_i$. It is characterized
by the fact there are no additional relations in $\A_u(M)$ apart from
the ones in the subalgebras. Thus, $\A_u(M)$ is generated by the
symbols $F_i(f)$ where $\supp f \subset M_i$, which we think of as
smeared field strength tensors
\ben
F_i(f) = \int_{M_i} f \wedge F \, .
\een
Their relations are
\ben\label{relations0}
F_i(f) = F_j(f), \quad \text{if
$\supp f \subset M_i \cap M_j$},
\een
and the $F_i(f)$, with $\supp f \subset M_i$ satisfy
all the relations in $\hat \F_0(M_i)$, which are
\ben\label{relations}
[F_i(f), F_i(h)] =
i\Delta(f, h) \, \myid \, , \quad
F_i(d f) = 0 = F_i(*d f) \, ,
\een
for any 1-forms $f,h$
of compact support in $M_i$. Here, $\Delta: \Omega^2_0(M)
\times \Omega^2_0(M) \to \mr$ denotes the
advanced minus retarded fundamental solution for the
hyperbolic operator $\delta d + d \delta$ acting on
2-forms.

For an arbitrary compactly supported 2-form $f$ on $M$,
we may then define the algebra element $F(f) \in \A_u(M)$ as
\ben
F(f) \equiv \sum_i F_i(\psi_i f) \, ,
\een
where $\supp \psi_i \subset M_i$, and $\sum_i \psi_i = 1$ on $\supp
f$. It is not difficult to show using eq.~\eqref{relations0} that this
definition does not depend upon the particular choice of the covering.
From eq.~\eqref{relations}, it then also follows that $F(\d f) = 0 = F(\d^* f)$
holds for arbitrary compactly supported forms $f$
in $M$. One can also easily show that $F(f) \st F(h) -
F(h) \st F(f) = 0$ for any two test-forms having spacelike related
support. Indeed, after splitting $f,h$ using a suitable
a partition of unity, we may assume that the
supports of $f$ and $h$ are contained in sets $M_i$ and $M_j$.
Since $M$ is assumed to be connected, there exists therefore a
globally hyperbolic spacetime $N \subset M_i \cup M_j$
in which every 2-cycle is contractible, and we may assume that $N$
appears in the covering of $M$. We may then view both $F(f)$ and
$F(h)$ as elements in $\hat \F_0(N)$, where they commute. Since $\Delta$
is uniquely determined by its action on test functions supported in
a neighborhood of a Cauchy surface, it then also follows that
$[F(f),F(h)] = i\hbar \Delta(f,h) \, \myid$.

The universal algebra contains certain central elements that carry
information about the topology of
$M$. They arise as follows. Let $C$ be a 2-cycle in $M$, and let
$\{ \psi_i \}$ be a partition of unity subordinate to the covering
$\{ M_i \}$ of $M$. By Poincare duality, we can find a closed 1-form
$h_C$ on $M$ such that
\ben
\int_M h_C \wedge \alpha = \int_C \alpha
\een
for any closed 2-form $\alpha$, and we may arrange $h_C$ to have
support in a neighborhood of $C$.
The 2-form $\psi_i h_C$ has compact support in $M_i$, and we may define
\ben\label{zedef}
Z_e[C] = F(h_C) \equiv \sum_i F_i(\psi_i h_C)\in \A_u(M) \, .
\een
We claim that $Z_e[C]$ is independent of the particular choice
of $h_C$, and of the partition $\{U_i, \psi_i\}$. Independence
of the partition was already shown above for general 2-forms.
To show independence of $h_C$, consider another $h'_C$ with the same
properties, and let $h_C -h'_C = \omega$. Then $\omega$ is closed, of
compact support and, $\int \omega \wedge \alpha = 0$ for any
closed 2-form $\alpha$. By the well-known fact that the pairing
\ben
\int: H^2(M) \otimes H^2_0(M) \to \mr
\een
is non-degenerate, we
therefore must have that $[\omega]=0$ in $H^2_0(M)$, i.e.,
$\omega = \d \beta$ for some 1-form $\beta$ of compact support.
Independence of $Z_e[C]$ on the particular form of $h_C$ then
follows from $F(\d \beta)=0$.

It then also follows that $Z_e[C]$ only depends upon
the homotopy class of $C$, i.e., $Z_e[C]$ may be viewed as a map
\ben
Z_e: H_2(M; {\mathbb Z}) \to \A_u(M), \quad [C] \mapsto Z_e[C] \, .
\een
In particular $Z_e[C]=0$ for any 2-cycle $C$ that can be deformed into
a point. Because $Z_e[C]$ only depends upon the class $[C]$ of
$C$ in $H_2(M)$, it follows that, given any sufficiently small
compact region $K \subset M$, we may deform $C$ so as to be in
the causal complement of $K$, that is
$C \subset J^+(K) \cup J^-(K)$. By choosing $h_C$ to be supported
in a sufficiently small neighborhood of $C$, it then follows that
\ben
[Z_e[C], F(f)] = 0, \quad \forall f \in \Omega^2_0(K) \, ,
\een
But then this also holds for arbitrary $f$ of compact support,
because $f$ may be written as $\sum \psi_i f$, with each $\supp \psi_i$
so small that $C$ and hence $\supp h_C$
can be deformed so as to lie in the causal complement.
Thus, $Z_e[C]$ is in the center ${\mathcal Z}(\A_u(M))$
of $\A_u(M)$. By taking the dual of
$h_C$ in eq.~\eqref{zedef}, we may similarly define
\ben
Z_m[C] = \sum_i F_i(\psi_i * h_C)\in {\mathcal Z}(\A_u(M)) \, ,
\een
and this quantity has similar properties as $Z_e[C]$.

The center-valued quantities $Z_e[C], Z_m[C]$ correspond to the electric and
magnetic fluxes through a 2-cycle $C$. They are analogous
to the classical quantities $\int_C F$ respectively
$\int_C *F$ and satisfy the same additivity relations under
the addition of cycles. Other interesting derived
quantities may also be defined. For example, let $C_1, C_2, \dots$ be a
basis of 2-cycles in $H_2(M; {\mathbb Z})$, and let
\ben
(Q^{-1})_{jk} = I(C_j,C_k)
\een
be the matrix of their intersection numbers. Then we may define
\ben
q_{top} = \sum_{j,k}^{b_2} Q^{jk} Z_e[C_i] Z_e[C_k] \quad
\in  {\mathcal Z}(\A_u(M)) \,
\een
and this is analogous to the classical topological quantity
\ben
q_{class} = \int_M F \wedge F = \sum_{j,k} Q^{jk} \bigg(
\int_{C_j} F \bigg) \bigg( \int_{C_k} F \bigg)
\een
by the so-called ``Riemann identity'' for closed differential forms.

In any factorial representation $\pi: \A_u(M) \to {\rm End}({\mathcal H})$
on a Hilbert space $\mathcal H$, the representers corresponding to $Z_e[C], Z_m[C]$
are by definition represented by multiples of the identity, i.e.,
\ben
\pi(Z_e[C]) = c_e[C] \cdot I, \quad
\pi(Z_m[C]) = c_m[C] \cdot I \, .
\een
where $c_e, c_m$ are valued in the complex numbers. By DeRahm's
theorem, they can be represented by
2-forms $f_e$ and $f_m$, both of which must be
closed. Choosing a basis $\{\omega^i\}$ of $H^2(M)$,
for example dual to a basis of 2-cycles $\{C_i\}$,
we may thus expand
$f_e = \sum_i q_i \omega^i$, and
$f_m = \sum_i g_i \omega^i$ with numerical constants
$q_i, g_i \in \mr$ depending upon the representation. These constants
are then the (canonically normalized) numerical values of the electric and
magnetic flux through the respective cycle in the representation $\pi$.

The above construction of Maxwell theory (without a vector potential)
is somewhat abstract, and we now discuss an equivalent description. As
above, let $\{\omega^i\}$ be a set of closed forms forming a basis of
$H^2(M)$. Any closed form $F$ may thus be written uniquely as
$F = d A + \sum_i q_i \omega^i$. Substitution into the action $S$
gives
\ben
S = \frac{1}{2} \int dA \wedge *dA + j \wedge *A
\een
where $j = \sum q_i \delta \omega^i$ is considered as an
external (conserved) current coupled to $A$. The quantization of this
theory now proceeds along similar lines as for the action $S$ without
the external current. We correspondingly get an algebra of observables
$\A_q(M)$, which now depends upon the choice of $q \equiv \{q_i\}$ and
$\{\omega^i\}$ through the external current. The algebra is
spanned by generators $\int f \wedge \d A$, and
\ben
\widehat F(f) = \int f \wedge d A +
\sum q_i \bigg( \int \omega^i \wedge f \bigg) \myid \, .
\een
They satisfy the same relations as the generators $F(f)$ above in the
algebra $\A_u(M)$. From this it may be seen that the algebra $\A_q(M)$
only depends upon $q_i$ and the equivalence classes $[\omega^i]$. This
algebra also has further relations not present in $\A_u(M)$,
because the elements $\widehat Z_e[C] \in \A_q(M)$ defined in the same
way as the central elements $Z_e[C] \in \A_u(M)$ above,
are now represented by multiples of the identity, namely
\ben
\widehat
Z_e[C] = \sum q_i \bigg(\int_C \omega^i \bigg) \myid \quad \in
\A_q(M) \, ,
\een
while the elements $Z_e[C] \in \A_u(M)$ are only in the center, but
not necessarily proportional to the identity. Thus, $\A_u(M)$ and $\A_q(M)$
are not isomorphic. Instead, we have
\ben
\A_u(M) \cong \int^{\oplus} \prod_{i=1}^{b_2} \d q_i \, \A_{q}(M) \, .
\een
By contrast, the magnetic fluxes $\widehat Z_m[C]$, defined as above,
are not proportional to the identity but only elements in the center
of $\A_q(M)$. This apparent asymmetry between the electric and
magnetic fluxes arises
from the fact that we have chosen to quantize the theory starting from
a potential for $F$, rather $*F$, which would also be possible. Then
the roles of electric and magnetic fluxes would be reversed.

\medskip

A physically relevant example of a spacetime $M$ with a non-trivial
2-cycle is the Kruskal extension of the Schwarzschild spacetime. It has
line element
\ben
\d s^2 = \frac{32M^3 e^{r/2M}}{r} (-\d T^2 + \d X^2) + r^2 (\d \theta^2 +
\sin^2 \theta \, \d \varphi^2) \quad r>0 \, ,
\een
and topology $M=\mr \times \mr \times S^2$, where $r$ is defined through
$T^2-X^2=(1-r/2M) e^{r/2M}$. It is a globally hyperbolic
spacetime with a non-trivial
2-cycle, homotopic to $S^2$. Hence, the universal algebra
possesses non-trivial central elements $Z_e[S^2], Z_m[S^2]$, and this
gives rise to the possibility of having non-trivial electric and
magnetic fluxes in that spacetime,
as also realized by Ashtekar et al.~\cite{Ashtekar80}.

We now sketch an argument that arbitrary values of the electric and
magnetic charges may be realized in representations $\pi$ carrying a
unitary representation of the time-translation symmetry group.
The spacetime is a solution to the vacuum Einstein-equation
$R_{\mu\nu} = 0$, with static timelike Killing field
$K=\partial/\partial t$, with $t=4M \tanh^{-1}(X/T)$. By
the standard identity $\nabla_{[\mu}
(\epsilon_{\nu\sigma]\alpha\beta}\nabla^\alpha K^\beta) = \frac{2}{3}
R_{\alpha\beta} K^\beta e^\alpha{}_{\mu\nu\sigma}$
valid for any Killing field $K$, $\phi_{\mu\nu} = \frac{1}{4
  \pi} \nabla_{[\mu} K_{\nu]}$ is therefore a static
(meaning $\pounds_K \phi=0$) solution to
the classical Maxwell equations. Given $q,g \in \mr$, we define
$\gamma_{p,q}: F(f) \mapsto F(f) + q\int_{S_2} f \wedge \phi \, \myid+
g \int_{S_2} f \wedge * \phi \, \myid$. This is an automorphism
of $\A_u(M)$. Let us assume that there is a factorial vacuum state
$\langle \, . \rangle_0$ on
$\A_u(M)$ invariant under the action of the time-translation
isometries (which can presumably be constructed by the techniques of
Junker et al.~\cite{Junker2003}), and let us assume that
$\langle Z_e[S^2] \rangle_0 = 0 =
\langle Z_m[S^2] \rangle_0$. Then the states $\langle \, . \,
\rangle_{q,g} = \langle \gamma_{q,g}( \, . \,) \rangle_0$ are also
factorial and the corresponding GNS-representation
carry a unitary representation of the time-translation
symmetries, with invariant vacuum vector.
Furthermore, by $\int_{S^2} * \phi = 1$, we have
\ben
\pi_{q,g}(Z_e[S^2]) = q \, I, \quad
\pi_{q,g}(Z_m[S^2]) = g \, I \, .
\een
in the corresponding GNS-representations
$\pi_{q,g}$ of these states.
Thus, the representations $\pi_{q,g}$ carry electric flux $q$ and
magnetic flux $g$. In this sense, the numbers $q,g$ may be viewed as
superselection charges, as also noted by Ashtekar et al.~\cite{Ashtekar80}.

\section{Effective Actions in curved spacetime}

We here give the
definition of the effective action in our framework
following~\cite{Brennecke2005,MWD} and a derivation of a set of consistency
conditions. We also emphasize that the effective action is a state
dependent quantity, and therefore, unlike the $T$-products,
does not have a local, covariant dependence upon the metric.

In the path integral formulation of quantum field theory, the
effective action in a scalar field theory
is formally defined as follows (see
e.g.,~\cite{Weinberg1996}).
Let $j \in
C^\infty_0(M)$ be an external current density, and define, formally,
\ben
\exp(Z^c(j)) = \int [{\mathcal D} \phi] \exp\bigg(iS/\hbar + \int j\phi
\bigg).
\een
Then the effective action $\Gamma_\eff$ is defined, again formally,
as the Legendre transformation of $Z^c(j)$:
Define $\phi$ through $\phi = \delta
Z^c(j)/\delta j$, and $\Gamma_\eff = \int j\phi - Z^c(j)$. The
quantity $\Gamma_\eff$ is a formal power series in $\hbar$ depending
on $\phi$ (and the action $S$), and may thus be viewed as an element
of $\F$. The above construction is formal in several ways:
The quantity $Z^c(j)$ is typically viewed as the generating
functional for the hierarchy of connected time-ordered $n$-point
functions of the quantum field $\phi$. It thus depends upon a choice
of state, and the same is consequently true for the effective action.
This is obscured in the above functional integral formulation. Here,
the choice of state would enter the precise choice of the formal
path-integral measure $[{\mathcal D} \phi]$. Also, because the path-integral
derivation does not specify the precise definition of the
path-integral measure $[{\mathcal D} \phi]$, it necessarily disregards all issues
related to renormalization. We therefore now give a precise
definition of the effective action in curved spacetime.

For this, we define,
following~\cite{Brennecke2005}, the quantities $T^c_\omega: A^{\otimes
  n} \to \W$ ($A$ the space of local actions) implicitly by
\ben\label{conngen}
T(\exp_\otimes(iF/\hbar)) = \sum_{n \ge 0}
\frac{1}{n!} :T^c_\omega(\exp_\otimes iF/\hbar)
\cdots T^c_\omega(\exp_\otimes iF/\hbar) :_\omega \, ,
\een
where the $n$-th term has $n$ factors. Unlike $T$, the quantity $T^c_\omega$ is
not local and covariant, but depends upon the global choice of
$\omega$. It can be shown that
$\tau^c_\omega(F^{\otimes n}) = \lim_{\hbar \to 0}
T^c_\omega(F^{\otimes n})/\hbar^{n-1} \in A$ exist. Next, define a
functional $\Gamma_\omega: A^{\otimes
  n} \to \W$ implicitly by
\ben
\tau^c_\omega \bigg( \e^{i\Gamma_\omega(\exp_\otimes F)/\hbar}\bigg)
= T^c_\omega \bigg( \e^{iF/\hbar }\bigg) \, .
\een
It can be shown that, for $F \in A$
\ben
\Gamma_\omega(\myid) = 0, \quad \Gamma_\omega(F) = F,
\een
as well as
\ben
\Gamma_\omega(\e^F) = F + O(\hbar) \, .
\een
Given an interaction $F \in A$, we define an ``effective
action'' (with respect to the state $\omega$) associated with $S_0+F$ by
\ben
\Gamma_{\rm eff}(S_0+F) = S_0 + \Gamma_\omega(\e^{F}) = S_0+F + O(\hbar)\, ,
\een
Again, the higher order terms
in $\hbar$ depend upon the state $\omega$, and are not local and
covariant. This property makes the effective action in general
unsuitable to solve the renormalization problem in curved spacetime,
since the local and covariance properties of the renormalization
procedure cannot be controlled.

The effective action obeys a useful identity
that can presumably be used to analyze potential anomalies in the Ward
identities (as an alternative to our approach), at least in flat
spacetime. To formulate this identity, consider
any local field polynomial $\O$, and
the modified action $S_0 + F \to S_0 + F +\int_M h \wedge \O$, where
$h \in \Omega_0(M)$ is a compactly supported smooth form.
Then we have the identity~\cite{Brennecke2005}
\bena\label{zjappa}
&&\int_M
\frac{\delta \Gamma_\eff(S_0+F+\langle h, \O\rangle)}{\delta h(x)} \wedge
\frac{\delta \Gamma_\eff(S_0+F+\langle h, \O\rangle)}{\delta \phi(x)}
\Bigg|_{h=0} \non\\
&=& \int_M \frac{\delta}{\delta h(x)} \Gamma_\eff\Bigg(
S_0 + F + \langle h, \O \delta(S_0+F)/\delta \phi \rangle +
  \Delta_\O\rangle
\Bigg) \Bigg|_{h = 0}\, ,
\eena
where $\Delta_\O(x) = \Delta_\O(\e^{F})(x)\in A$ is the
anomaly corresponding to $\O$ in the corresponding anomalous
``Master Ward Identity'' in sec.~4.4., see also~\cite{Brennecke2005,MWD}.
It is viewed here as a 4-form.

\section{Wave front set and scaling degree}

We here recall the basic definition of the wave front set of
a distribution and some of its elementary properties. For details, see
\cite{Hormander}. If $u$ is a compactly supported {\em smooth} function on
$\mr^n$, then by standard theorems of distribution theory, its Fourier
transform, $\hat u(p)=(2\pi)^{-n/2}u(\exp(ip \, . \,))$ is an analytic function
on $\mr^n$ falling off faster than any inverse power of $p$, i.e.,
\ben
|\hat u(tp)| \le c_N(1+|t|)^{-N}, \quad t \in \mr
\een
for some $c_N$ not depending upon $p$,
and any $N$. Conversely, this bound implies that
a compactly supported distribution $u$ is in fact smooth. The idea of the
wave front set is to use the possible failure of this bound to characterize the
non-smoothness of a distribution. For compactly supported {\em distributions} $u$,
one defines the set of singular directions by
\ben
\Sigma(u) = \{ p \in \mr^n \setminus 0 \mid |\hat u(tp)
\ge c_N(1+|t|)^{-N} \quad
\text{for some $N$, all $t>0$} \} \, .
\een
We define the wave front set of any distribution at a point $x \in \mr^n$ by
\ben
\WF_x(u) = \bigcap_{\psi: x \in \supp \, \psi} \Sigma(\psi u) \, .
\een
where the intersection is over all smooth compactly supported cutoff functions $\psi$.
The wave front set is clearly invariant under dilatation, and therefore a
cone, and it only depends on the behavior of $u$ in an arbitrary small neighborhood of
$x$. For distributions $u$ defined on a smooth $n$-dimensional manifold $X$
one defines the wave front set as follows. Let $\kappa,U$ be a coordinate chart
covering $x$. Then, choosing a smooth cutoff function $\Psi$ supported in $U$ that is 1 near $x$, we can define
$\kappa^*(\psi u)$, which is now a distribution that is defined on $\mr^n$.
We define the wave front set to be the set
\ben
\WF_x(u) = (\kappa^{-1})^* \WF_{\kappa(x)}(\kappa^* (\psi u) ) \subset T^*_x X
\een
It can be proved that this definition does not depend upon the arbitrary choice of $\kappa, \psi$,
and one defines $\WF(u)$ to be the union of all $\WF_x(u)$.
One relevant application of the wave front set in perturbative quantum field theory is
the following theorem~\cite{Hormander} about the product of distributions.
\begin{thm}\label{thm8}
Let $u,v$ be distributions on $X$. If
$0 \notin \WF_x(u) + \WF_x(v)$, then the
pointwise product $uv$ is defined in some neighborhood of $x$, and
$\WF_x(uv) \subset \WF_x(u) + \WF_x(v)$.
\end{thm}
Clearly, if the assumption holds for all $x\in X$,
then the point-wise product is globally
defined on $X$. Another useful theorem about wave front sets is the
following~\cite{Hormander}. Let $K \subset \mr^n$ be a convex
open cone, and let $u(x+iy)$ be analytic in $\mr^n + iK$ for $|y| <
\delta$ and some $\delta$, with the
property that $|u(x+iy)| \le C |y|^{-N}$ for some $N$, and all
$y \in K$ with $|y| < \delta$. Then the boundary value $u(x) = {\rm
  B.V.}_{y \to 0} \, u(x+iy)$, with the limit taken for $y \in K$
defines a distribution on $\mr^n$.
\begin{thm}\label{thm6}
The wave front set of $u(x) = {\rm B.V.}_{y \to 0} \, u(x+iy)$ with the limit taken
within the cone $K$, i.e., $y \in K$, is bounded by
\ben
\WF(u) \subset \mr^n \times K^D \, ,
\een
where $K^D = \{k \in \mr^{n*} \mid \,\, k \cdot y < 0 \,\,
\forall y \in K\}$ is the dual cone.
\end{thm}
In applications, one often deals with distributions that are solutions
to a partial differential equation $Au=0$, where $A$ is partial
differential operator on $X$ (or even a pseudo-differential operator), i.e.,
\ben
A = \sum_{n=0}^N
a^{\mu_1 \dots \mu_n}(x) \nabla_{(\mu_1} \dots \nabla_{\mu_n)} \, .
\een
Under this condition, it can be shown that the wave front set of $u$ must be
restricted to the set
\ben
\WF(u) \subset \{(x,k) \mid
a^{\mu_1 \dots \mu_N}(x) k_{\mu_1} \dots k_{\mu_N} =0 \} \, .
\een
In case when $A$ is the wave operator on a Lorentzian manifold, we hence learn
that any distributional solution $u$ of the wave equation
can only have vectors of the form $(x,k)$ in the wave front set
when $k$ is a null-vector.
Another important application of the wave front set for quantum field theory
in curved spacetime is the propagation of singularities theorem. Consider a
distribution $u$ on a spacetime $(M,g)$ that is a solution to the wave
equation $\square u = f$, with $f$ a {\em smooth} source. The wave operator
defines a 1-particle Hamiltonian on ``phase space'' $T^*M$ by $h(x,p) =
g^{\mu\nu}(x) p_\mu p_\nu$, and Hamilton's equations, defined with
respect to the symplectic structure $dx^\mu \wedge dp_\mu$,
\bena
\dot p_\mu &=& -2\Gamma_{\nu\mu\rho}(x) p^\nu p^\rho\\
\dot x^\mu &=& 2g^{\mu\nu}(x) p_\nu
\eena
define a flow in phase space, $t \mapsto \phi_t$,
which is just the geodesic flow.
The propagation of singularities theorem now states in this example
that this flow must leave the
wave front set $\WF(u)$ invariant, in the sense that $\phi_t^* \WF(u) \subset
\WF(u)$. Thus, the propagation of singularities theorem gives information
how singularities propagate along the bicharacteristic flow.
The theorem as just stated is in fact just a special case of the
celebrated Duistermaat-H\" ormander propagation of singularities
theorem~\cite{Duistermaat1977},
which holds for much more general operators $A$ of real
principal type (including e.g. the massive wave equation). The Hamiltonian
is then given simply by $h(x,k) =
a^{\mu_1 \dots \mu_N}(x) k_{\mu_1} \dots k_{\mu_N}$ in the general case, where $N$ is
the degree of the operator.

Another useful concept in perturbative quantum field theory is that of the scaling degree of a distribution.
Let $u$ be a distribution on $\mr^n$. The scaling degree, $sd_0(u)$ at the origin of $\mr^n$ is defined as
\ben
sd_0(u) = \inf \{\delta \in \mr \mid \lim_{t \to 0+} t^\delta u(tx) = 0\}
\een
where the limit is understood in the sense of distributions, i.e., after smearing with a test function.
One similarly defines the scaling degree $sd_x(u)$ at an arbitrary point $x$ by first translating
$u$ by $x$. On a manifold $X$, the scaling degree is defined by first localizing $u$ with a
cutoff function and then pulling it back with a coordinate chart, $\kappa^*(\psi u)$, as in the
definition of the wave-front set. One again verifies that the definition does not depend upon the
choice of coordinates.

\section{Hadamard parametrices}

In this appendix, we review the definition of the scalar Hadamard parametrix
$H^{\rm s}$, and the vector Hadamard parametrix, $H^{\rm v}$, as well
as the local expressions for the advanced and retarded propagators in
curved spacetime.

\subsection{Scalar Hadamard parametrix}

In a general curved spacetime, it is not possible to find a closed
form expression for $\Delta_{A,R}$, but it is still possible to
present a local expression $H_{A,R}$ involving certain recursively defined
coefficients, which locally coincides with $\Delta_{A,R}$ modulo
$C^\infty$. The distributions $H_{A,R}$ are called ``Hadamard parametrices''
for $\Delta_{A,R}$. To construct them,
let $x,y \in M$, and consider the length functional
\ben
s(x,y) = \int_a^b \Bigg| g_{\mu\nu}(\gamma(t))
\dot \gamma^\mu(t) \dot \gamma^\nu(t) \Bigg|^{1/2} \, dt
\een
for $C^1$-curves $\gamma: [a,b] \to M$ with the property that
$\gamma(a) = x$ and $\gamma(b) = y$, which are either
spacelike, timelike, or null (but do not switch from one
to the other). The functional $s(p,q)$ is invariant under
reparametrizations of the curve, so we may choose a parametrization
so that $g_{\mu\nu} \dot \gamma^\mu \dot \gamma^\nu = 1$ along the
curve when $\gamma$ is either spacelike or timelike (such a parameter
is called an ``affine parameter''). The Euler-Lagrange equations
for the functional are then given by
\ben
\dot \gamma^\mu \nabla_\mu \dot \gamma^\nu = 0,
\een
and curves satisfying this equation are ``geodesics''. If
$\gamma^\mu$ are the components of $\gamma$ in a local chart, then the
geodesic equation reads
\ben
\ddot \gamma^\mu + \Gamma^\mu{}_{\sigma\nu} \dot \gamma^\sigma \dot \gamma^\nu = 0 \, .
\een
Two given points $x,y$ may in general be joined by several geodesics, but
one can show~\cite{Hawking1970} that every point in $M$ has a neighborhood $U$ such
that any pair of points $(x,y) \in U \times U$ may be joined by a
unique geodesic lying entirely within $U$. For $(x,y) \in U \times U$,
we define $\sigma(x,y)$ to be the value of the function $\pm s(x,y)^2$
evaluated on the unique geodesic joining $x$ and $y$, where
$+$ is chosen for a spacelike, and $-$ is chosen for a timelike geodesic.
In Minkowski spacetime, the function $\sigma$ is equal to the invariant
distance between the points $x,y$.
In any spacetime, the function $\sigma$ has the important property that
\ben\label{jacobi}
g^{\mu\nu} \nabla_\mu \sigma \nabla_\nu \sigma = 4\sigma,
\een
where the derivative can act on either the first or second argument.
Now let $T: M \to \mr$ be a time function.
By analogy with flat spacetime, we seek Hadamard parametrices for the advanced
and retarded propagators by the following ansatz:
\ben
H_{A,R}(x,y) = \frac{1}{2\pi} \Theta(\mp t(x,y))\Bigg[
u(x,y) \delta(\sigma(x,y)) - v(x,y) \theta(-\sigma(x,y))
\Bigg],
\een
Here, $u, v$ are as yet unknown smooth, symmetric functions on $U \times U$, $\Theta$ is the step function
supported on the positive axis, 
and $t(x,y) = T(x)-T(y)$. This ansatz is consistent with the support
properties of the advanced and retarded propagators,
and it does not depend on the particular
choice of time function. The unknown functions $u,v$
are to be determined imposing in addition the Klein-Gordon equation,
\bena
(\square -m^2)_x H_{A,R}(x,y) &=& \delta(x,y)
\quad \text{modulo $C^\infty$} \, , \\
(\square -m^2)_y H_{A,R}(x,y) &=& \delta(x,y)
\quad \text{modulo $C^\infty$} \, .
\eena
Using the identity~\eqref{jacobi} one finds that $H_A,H_R$ solve
these equations in $U \times U$ modulo
$C^\infty$ if the following identities hold for $u,v$:
\ben\label{1}
2 \nabla^\mu \sigma \nabla_\mu u = (8-\square \sigma)u
\, .
\een
as well as
\ben\label{2}
(\square - m^2) v= 0,
\een
modulo $C^\infty$, and
\ben\label{3}
2 \nabla^\mu \sigma \nabla_\mu v + (\square \sigma -4)v
= -(\square- m^2)u, \quad \text{on $\partial J^\pm(y)$}
\een
where the derivative operators act on the point $x$.
One can show that the unique smooth solution to the equation for $u$ is
given by $u=D^{1/2}$, where $D(x,y)$ is the
so-called ``VanVleck determinant'', which is defined as follows.
Let $x,y \in U$, and let $A_{\mu\nu} = (\nabla_\mu \otimes
\nabla_\nu) \sigma$, so that $A_{\mu\nu} dx^\mu \otimes dy^\nu$
is a tensor in $T^*_x M \otimes T^*_y M$. We can consider the 4-th
antisymmetric tensor power of this tensor, which may be viewed as a map
\ben
\wedge^4 A : \wedge^4 T_x M \to  \wedge^4 T_y^* M,
\een
where $\wedge^r T_p M$ denotes the space of totally antisymmetric
tensors of type $(r,0)$. Clearly, for $r=4$ this space is
1-dimensional (in 4 dimensions), so if we pick a basis element
at points $x,y$, we can identify $\wedge^4 A$ with a scalar. A
choice of the basis element depending only upon the metric (up to a
sign) is the Levi-Civita tensor $\epsilon$.
With this choice, $D$ is defined as the scalar obtained from $\wedge^4
A$. In local coordinates,
\ben
D = 2^{-4}\frac{1}{4!}
A^{\nu_1}{}_{\mu_1}A^{\nu_2}{}_{\mu_2}A^{\nu_3}{}_{\mu_3}A^{\nu_4}{}_{\mu_4}
\epsilon^{\mu_1\mu_2\mu_3\mu_4}\epsilon_{\nu_1\nu_2\nu_3\nu_4} \, .
\een
where the $\epsilon$ tensors are evaluated at $x$ and
$y$, respectively, and where the factor $2^{-4}$ is inserted to make the 
subsequent formulas simpler. While it is not possible to give a
similarly explicit solution to the equation for $v$, it is possible to
obtain a solution $v$ in the form of a convergent power series
\ben\label{v}
v = \sum_{n=0}^\infty v_n \chi(\sigma/\alpha_n) \sigma^n,
\een
Here, $\chi$ is an arbitrary function of compact support that is
equal to 1 in a neighborhood of $0$, and $\{\alpha_n\}$ is a sequence
growing sufficiently rapidly so as to enforce the convergence of the
series. The coefficients are determined recursively as the solutions
of the ``transport equations''
\ben
2\nabla_\mu \sigma \nabla^\mu v_0 - (\nabla_\mu \sigma
\nabla^\mu \log D -4)v_0
= -(\square - m^2) D^{1/2},
\een
from eq.~\eqref{2} and, for $n>0$
\ben
2\nabla_\mu \sigma
\nabla^\mu v_n - (\nabla_\mu \sigma
\nabla^\mu \log D -4n-4) v_n
= -\frac{1}{n}(\square - m^2)v_{n-1}
\een
from eq.~\eqref{3}.
The solutions to these differential equations
are unique if one assumes, as we have done that $v_n$
are smooth (i.e., in particular regular at $x=y$).
These solutions can be given in integral form as
\ben
v_0 = -\frac{1}{2} D^{1/2} \int_{0}^1
\frac{(\square -
  m^2)D^{1/2}}{D^{1/2}} \lambda^2
\, d\lambda
\een
and, for $n>0$
\ben\label{vn}
v_n = -\frac{1}{2n} D^{1/2} \int_{0}^1
 \frac{(\square -
  m^2)v_{n-1}}{D^{1/2}}  \lambda^{2n+2}
\, d\lambda
\een
where the integrand is evaluated at the point
$(x(\lambda),y)$, where $x(\lambda) = {\rm Exp}_y (\lambda \xi)$,
and where $\xi \in T_y M$ is chosen so that $x(1) = x$. Thus, in terms of the
Riemannian normal coordinates of
$x$ relative to $y$, then the integrand is thought of as
evaluated at the rescaled normal coordinates.
Despite the apparent
asymmetry in the construction of $u,v$, it can be shown that these
functions are symmetric in $x,y$~\cite{Friedlaender,Moretti1999},
and one shows that, indeed,
\ben
H_{A,R}(x,y) = \Delta_{A,R}(x,y) \quad
\text{modulo $C^\infty$}
\een
in $U \times U$. (It can be proved that
exact Greens functions $\Delta_{A,R}$ exist globally, for which the
power series expressions therefore define local asymptotic
expansions.)

From the advanced and retarded parametrices one can define 2 other
parametrices $H_{F,D}$ (for ``Feynman'' and ``Dyson''), given by
\ben\label{Hfdef}
H_{F,D}(x,y) =
\frac{1}{2\pi^2} \bigg(
\frac{u(x,y)}{\sigma \pm i0} + v(x,y) \, \log
(\sigma \pm i0) \bigg) \,
\een
These parametrices are symmetric in $x,y$.
Using the transport equations for $u,v$, one shows that these, too,
are local Green's functions (with $\delta$-function source) modulo
$C^\infty$. The wave-front sets of $H_{A,R,F,D}$ are described by the
following theorem:
\begin{thm}
The wave front set of the 4 Hadamard parametrices are given
by
\bena
\WF(H_{A,R}) &=& \{(x_1, k_1; x_2, k_2) \mid \,\, k_1 \sim -k_2, \, x_1
\in J^\pm(x_2)\}\non\\
&\cup& \{(x, k; x, -k)\}\\
\WF(H_{F,D}) &=& \{(x_1, k_1; x_2, k_2) \mid \,\, k_1 \sim -k_2, \,
k_{1} \in V^*_\pm \, {\rm iff} \,
x_{1} \in J^{\pm}(x_{2})\}\non\\
&\cup& \{(x, k; x, -k)\}
\eena
\end{thm}
The proof of this theorem is similar to that of the next lemma.
It can also be proved that the four parametrices $H_{A,R,F,D}$ are
uniquely characterized by their wave front properties. In fact,
there is a similar classification of parametrices for any operator
or real principal type, as shown by a profound theorem by
Duistermaat and H\" ormander~\cite{Duistermaat1977}.

In the body of the paper, we use a combination, $H$, of the above
Hadamard parametrices, which is called simply
the ``local (scalar) Hadamard parametrix''
for the operator $\square-m^2$. It is the
distribution on $U \times U$ defined by eq.~\eqref{Hdef} in terms
of the same coefficients $u,v$ that appear above in the local
expressions for the advanced and retarded propagators. From
identities like
\ben
\frac{1}{i\pi}
\Im \bigg( \frac{1}{\sigma+i0t} \bigg) = \epsilon(t)\delta(\sigma),
\quad
\frac{1}{i\pi}
\Im \bigg( \log (\sigma+i0t) \bigg) = \epsilon(t)\theta(-\sigma),
\een
we get the relations
\ben
H_F - H_R = -iH = H_A - H_D\, .
\een
In view of the symmetry of $H_{F,D}$, there follows the commutator
property~\eqref{hcom}. Furthermore since $H_{A,R,F,D}$ are local Green's
functions modulo $C^\infty$ with a $\delta$-function source, there
follow the equations of motion
\ben
(\square -m^2)_x H(x,y) = 0 \quad \text{modulo $C^\infty$} \, , \quad
(\square -m^2)_y H(x,y) = 0 \quad \text{modulo $C^\infty$} \, ,
\een
The local Hadamard parametrix $H$ is important because
it characterizes the short distance
behavior of any Hadamard state, see Appendix E.

\subsection{Vector Hadamard parametrix}

The vector Hadamard parametrix $H^{\rm v}(x,y) = H_{\mu\nu}^{\rm v}(x,y)dx^\mu
\wedge dy^\nu$ is constructed by analogy to the scalar case. It now
satisfies the equations
\ben
(d\delta + \delta d)_x H^{\rm v}(x,y) = 0
\quad \text{modulo $C^\infty$} \, , \quad
(d\delta + \delta d)_y H^{\rm v}(x,y) = 0
\quad \text{modulo $C^\infty$} \, ,
\een
where $\delta = *\! d*\!$. In component form, the
equations of motion are given by the operator~\eqref{vec}. The
local vector Hadamard parametrix has an expansion similar to
that of the scalar Hadamard parametrix:
\ben
H_{\mu\nu}^{\rm v}(x,y) = \frac{1}{2\pi^2} \bigg(
\frac{u_{\mu\nu}(x,y)}{\sigma + i0t} + v_{\mu\nu}(x,y)
\log (\sigma + i0t)
\bigg) \, .
\een
The coefficients $u_{\mu\nu}, v_{\mu\nu}$ have expansions that are
analogous to the scalar case. The quantity $u_{\mu\nu}$
is given explicitly by
\ben
u_{\mu\nu} = D^{1/2} I_{\mu\nu}
\een
where $I: T_x M \to T^*_y M$ is the holonomy of the Levi-civita
connection along the unique geodesic connecting $x,y$ (``bitensor of
parallel transport''). The expansion coefficients of $v_{\mu\nu}$
as in eq.~\eqref{v} are again determined by transport equations.
The solutions to these equations take exactly the same form as in the
scalar case, eq.~\eqref{vn}, with the only difference that the
scalar Klein-Gordon operator $\square - m^2$ in those expressions
is replaced by the vector wave-operator
$g_{\mu\nu}\square + R_{\mu\nu}$.

\section{Hadamard states}

In the body of the paper, Hadamard 2-point functions play a key role.
They were introduced in Sec.~3.1 as bidistributions that are
solutions to the wave equation in both entries, that satisfy the
commutator property, and that have a certain wave front set. Here we
show that these conditions allow one to identify the short distance
behavior of any Hadamard 2-point function with that of the local
parametrix $H$ introduced in the previous subsection.

\begin{lemma} Let $\omega(x,y)$ be a 2-point function of Hadamard form,
  i.e., the wave front set $\WF(\omega)$ is given by~\eqref{wfs2}.
Then locally (i.e., where $H$ is defined), $\omega-H$ is smooth, i.e.,
\ben
\omega(x,y) =
\frac{1}{2\pi^2} \bigg(
\frac{u(x,y)}{\sigma + it 0} + v(x,y) \, \log
(\sigma + it0) \bigg)
+ \quad (\text{smooth function in $x,y$}).
\een
Furthermore, any two Hadamard states can at most differ by a globally
smooth function in $x,y$.
\end{lemma}

\noindent
{\em Proof:} We first show that, where it is defined,
$H$ has a wave front set $\WF(H)$ of Hadamard form, i.e.,
is given by eq.~\eqref{wfs2}. Since $v_i$ are smooth functions
on a convex normal neighborhood, it suffices to prove that
$\WF([\sigma + i0t]^{-1})$ and $\WF(\log[\sigma + i0t])$ have the
desired form. To determine the wave front set of such distributions,
we use the above thm.~\ref{thm6}. We apply this theorem to the
distributions in question as follows. First, we pick a local
coordinate system $(\psi, U)$ in a convex normal neighborhood $U$.
Within $U$, we pick a tetrad $e_0, \dots, e_3$ which we use to
identify each $T_x M$ with $\mr^4$ via the map sending $\xi =
(\xi^0, \dots, \xi^3)$ in $\mr^4$
to the point $e_x(\xi) = \xi^0 e_0 |_x + \dots + \xi^3 e_3 |_x$
in $T_x M$. For each given $x \in U$, we can then
write a point $y \in U$ uniquely as $y = \exp_x e_x(\xi)$ for some $\xi \in
\mr^4$. The mapping $(x,y) \in U \times U \mapsto (\psi(x), \xi)$
thus defines a local coordinate chart in $M \times M$,
which we call again $\psi$. Evidently, it then follows that
the pull-back of $(\sigma + i0t)^{-1}$ under $\psi$ is given by the
distribution
\ben
\frac{1}{(y+i0e)^2}=
\bv_{\eta \in V^+, \eta \to 0} \frac{1}{(\xi + i\eta)^2} \, ,
\een
where $e=(1,0,0,0)$,
which is of the form to which we can apply our lemma. Using that
the dual cone of the open future lightcone $V^+$ in Minkowski
spacetime is the closure of the past lightcone $\bar V^-$, it follows
\ben
\WF([\sigma + i0t]^{-1}) \subset \psi^*
[(\mr^4 \times 0) \times (\mr^4 \times \bar V^-)] \, .
\een
 From this, the desired wave front set follows. The logarithmic
term is treated in exactly the same fashion. Consider now the
distribution $d = \omega - H$. The anti-symmetric part of
$\omega$ is given by $i\Delta$, and the anti-symmetric part of
$H$ is given by
\ben\label{hcom}
H(x,y) - H(y,x)= i \epsilon(t) \left\{
u(x,y) \delta(\sigma) + v(x,y)
\theta(\sigma)
\right\}
\, ,
\een
where $\epsilon(t)=1$ for $t>0$, and $\epsilon(t)=-1$ for $t \le 0$.
It can be shown that the right side of the equation is equal to
$i \Delta$ modulo a smooth function. Thus, $d(x,y)$ is symmetric
in $x,y$ modulo a smooth remainder. On the other hand, since we
know that $H$ has the same wave front set as $\omega$, we know that
\bena\label{wfs22}
\WF(d) &\subset& \{(x_1, k_1, x_2, k_2) \in T^* M \times T^*M ; \non \\
                   &&\text{$x_1$ and $x_2$ can be joined by
                     null-geodesic $\gamma$} \\
                   &&\text{$k_1 = \dot \gamma(0)$ and $k_2 = -\dot
                     \gamma(1)$, and $k_1 \in \bar V^+$} \} \, .
\eena
which is evidently not a symmetric set. Thus, the only possibility is
that, in fact, $\WF(d) = \emptyset$, meaning that $d \in C^\infty$, or
equivalently, that $\omega = H$ modulo smooth. This proves the lemma.
\qed

Another proposition about Hadamard 2-point function underlying the
``deformation argument construction'' of Hadamard states given in
subsection~4.2 is the following:

\begin{thm}
Let $\omega$ be a positive definite distributional bi-solution such that $\WF(\omega)$
has the Hadamard wave front property in an open neighborhood of $\Sigma \times \Sigma$,
where $\Sigma$ is a Cauchy surface. Then $\WF(\omega)$ has the Hadamard
form globally on $M \times M$.
\end{thm}

The proof of the theorem is a simple application of the propagation of
singularities theorem for solutions of the Klein-Gordon equation
described in the previous subsection.

A (quasifree) Hadamard state is a 2-point function that is in addition
positive definite, $\omega(\bar f, f) \ge 0$ for any testfunction. The positivity
implies an even stronger ``local-to-global theorem''
than the one given above~\cite{Radzikowski1996b}:

\begin{thm}
Let $\omega$ be a bi-solution to the Klein-Gordon equation in both entries,
with anti-symmetric part $i\Delta$, and with the property that any point
$x \in M$ has a globally hyperbolic neighborhood $N$ such that $\WF(\omega)$ is
of Hadamard form in $N \times N$. Then $\WF(\omega)$ has the Hadamard form
globally in $M \times M$.
\end{thm}


\begin{thebibliography}{99}
\bibitem{Choquet}
  A.~Abrahams, A.~Anderson, Y.~Choquet-Bruhat and J.~W.~.~York,
  ``Einstein And Yang-Mills Theories In Hyperbolic Form Without Gauge Fixing,''
  Phys.\ Rev.\ Lett.\  {\bf 75} (1995) 3377
  [arXiv:gr-qc/9506072].

\bibitem{Witten}
  L.~Alvarez-Gaume and E.~Witten,
  ``Gravitational Anomalies,''
  Nucl.\ Phys.\  B {\bf 234}, 269 (1984).

\bibitem{Ashtekar80}
  A.~Ashtekar and A.~Sen,
  ``On The Role Of Space-Time Topology In Quantum Phenomena: Superselection Of
  Charge And Emergence Of Nontrivial Vacua,''
  J.\ Math.\ Phys.\  {\bf 21}, 526 (1980).

\bibitem{Barnich1999}
G. Barnich, M. Henneaux, T. Hurth, and K. Skenderis:
``Cohomological analysis of gauge-fixed theories,''
hep-th/9910201

\bibitem{Barnich2000}
  G.~Barnich, F.~Brandt and M.~Henneaux,
  ``Local BRST cohomology in gauge theories,''
  Phys.\ Rept.\  {\bf 338}, 439 (2000)

\bibitem{Batalin1981}
I. A. Batalin, G. A. Vilkovisky, Phys. Lett. {\bf B102}, 27 (1981),
Nucl. Phys. {\bf B234}, 106 (1984), J. Math. Phys. {\bf 26}, 172 (1985)

\bibitem{Bayen1977a}
  F.~Bayen, M.~Flato, C.~Fronsdal, A.~Lichnerowicz and D.~Sternheimer,
  ``Deformation Theory And Quantization. 2. Physical Applications,''
  Annals Phys.\  {\bf 111}, 111 (1978).
  %%CITATION = APNYA,111,111;%%

%\cite{Bayen:1977ha}
\bibitem{Bayen1977b}
  F.~Bayen, M.~Flato, C.~Fronsdal, A.~Lichnerowicz and D.~Sternheimer,
  ``Deformation Theory And Quantization. 1. Deformations Of Symplectic
  Structures,''
  Annals Phys.\  {\bf 111}, 61 (1978).

\bibitem{Becci1975}
 C.~Becchi, A.~Rouet and R.~Stora,
  ``Renormalization Of The Abelian Higgs-Kibble Model,''
  Commun.\ Math.\ Phys.\  {\bf 42}, 127 (1975).

\bibitem{Becci1976}
 C.~Becchi, A.~Rouet and R.~Stora,
  ``Renormalization Of Gauge Theories,''
  Annals Phys.\  {\bf 98}, 287 (1976).

\bibitem{Birrell}
  N.~D.~Birrell and P.~C.~W.~Davies,
  ``Quantum Fields In Curved Space,''
%\href{http://www.slac.stanford.edu/spires/find/hep/www?irn=998621}{SPIRES entry}
{\it  Cambridge, Uk: Univ. Pr. ( 1982) 340p}

\bibitem{Bogoliubov1952}
N.N. Bogoliubov and D.V. Shirkov:
``Introduction to the Theory of Quantized Fields.''
John Willey \& Sons, Inc., New York, third edition, 1980.

\bibitem{Hodge}
R. Bott and L. Tu, ``Differential Forms in Algebraic Topology.'' Springer-Verlag, 1981.

\bibitem{Brennecke2005}
F. Brennecke: ``Zum Anomalie-Problem der Master-Ward-Identit\" at,''
Diploma Thesis (in German), Hamburg 2005, available at
{\tt http://www.desy.de/uni-th/lqp/psfiles/dipl-brennecke.ps.gz}

\bibitem{MWD}
F. Brennecke and M. D\" utsch,
``Removal of violations of the Master Ward Identity in perturbative
QFT,'' [hep-th/0705.3160]

\bibitem{Bros}
[4] Bros, J., Iagolnitzer, D., "Causality and local analyticity:a mathematical study", Ann. Inst H. Poincar� {\bf A 18}, 174 (1973), see also
D. Iagolnitzer, Lett. Math. Phys. {\bf 21} 323 (1991), and
references therein.

\bibitem{Brunetti1996}
  R.~Brunetti, K.~Fredenhagen and M.~Kohler,
  ``The microlocal spectrum condition and Wick polynomials of free fields on
  curved spacetimes,''
  Commun.\ Math.\ Phys.\  {\bf 180}, 633 (1996)

\bibitem{Brunetti2000}
  R.~Brunetti and K.~Fredenhagen,
  ``Microlocal analysis and interacting quantum field theories:
  Renormalization on physical backgrounds,''
  Commun.\ Math.\ Phys.\  {\bf 208}, 623 (2000)

\bibitem{Brunetti2003}
 R.~Brunetti, K.~Fredenhagen and R.~Verch,
  ``The generally covariant locality principle: A new paradigm for local
  quantum physics,''
  Commun.\ Math.\ Phys.\  {\bf 237}, 31 (2003)
  [arXiv:math-ph/0112041].


\bibitem{Bunch1981}
  T.~S.~Bunch,
  ``BPHZ Renormalization Of Lambda Phi**4 Field Theory In Curved Space-Time,''
  Annals Phys.\  {\bf 131}, 118 (1981).

\bibitem{Kreimer1}
A. Connes and D. Kreimer: ``Renormalization in quantum field theory and
Riemann-Hilbert problem I: The Hopf algebra structure of graphs and the
main theorem,'' Comm. Math. Phys. {\bf 210}, 249 (2000)

\bibitem{Kreimer2}
A. Connes and D. Kreimer: ``Renormalization in quantum field theory and
Riemann-Hilbert problem II: The beta function, diffeomorphisms, and
the renormalization group,'' Comm. Math. Phys. {\bf 216}, 215 (2001)

\bibitem{Davidychev}
E.E. Boos and A. I. Davydychev:
``A method for calculating massive Feynman diagrams,''
  Theor.\ Math.\ Phys.\  {\bf 89}, 1052 (1991)
  [Teor.\ Mat.\ Fiz.\  {\bf 89}, 56 (1991)].

\bibitem{DeWitt1962}
  B.~S.~DeWitt and R.~W.~Brehme,
  ``Radiation damping in a gravitational field,''
  Annals Phys.\  {\bf 9}, 220 (1960).

\bibitem{deWitt1}
B.~S.~DeWitt: ``Dynamical Theory of Groups and Fields'' (Les Houches
Lectures 1963) (New York: Gordon and Breach)

\bibitem{deWitt2}
B.~DeWitt: ``The Global Approach to Quantum Field Theory,''
Oxford University Press, Oxford 2003

\bibitem{DeWitt2004}
  B.~DeWitt and C.~DeWitt-Morette,
  ``From the Peierls bracket to the Feynman functional integral,''
  Annals Phys.\  {\bf 314}, 448 (2004).

\bibitem{Duistermaat1977}
J.J. Duistermaat and L. Hormander,
``Fourier integral operators II,'' Acta Math., {\bf 128} (1972), 183-269.

\bibitem{Duetsch1994dp}
  M.~D\" utsch, T.~Hurth, F.~Krahe and G.~Scharf,
  ``Causal Construction Of Yang-Mills Theories. 1,''
  Nuovo Cim.\  A {\bf 106} (1993) 1029.

\bibitem{Duetsch1993bc}
  M.~D\" utsch, T.~Hurth, F.~Krahe and G.~Scharf,
  ``Causal construction of Yang-Mills theories. 2,''
  Nuovo Cim.\  A {\bf 107} (1994) 375.

\bibitem{Duetsch1994dq}
  M.~D\" utsch, T.~Hurth and G.~Scharf,
  ``Causal Construction Of Yang-Mills Theories. 3,''
  Nuovo Cim.\  A {\bf 108}, 679 (1995).

\bibitem{Duetsch1994ur}
  M.~D\" utsch, T.~Hurth and G.~Scharf,
  ``Causal construction of Yang-Mills theories. 4. Unitarity,''
  Nuovo Cim.\  A {\bf 108}, 737 (1995).

\bibitem{Duetsch1993ee}
  M.~D\" utsch,
  ``On gauge invariance of Yang-Mills theories with matter fields,''
  Nuovo Cim.\  A {\bf 109}, 1145 (1996).

%  M.~D\" utsch, F.~Krahe and G.~Scharf,
%  ``The Infrared Problem And Adiabatic Switching,''
%  J.\ Phys.\ G {\bf 19} (1993) 503.

%\cite{Duetsch:2004dd}
\bibitem{Duetsch2004}
  M.~D\" utsch and K.~Fredenhagen,
   ``Causal perturbation theory in terms of retarded products, and a proof  of
  the action Ward identity,''
  arXiv:hep-th/0403213, to appear in Rev. Math. Phys.
  %%CITATION = HEP-TH 0403213;%%

%\cite{Duetsch:2002yp}
\bibitem{Duetsch2002}
  M.~D\" utsch and K.~Fredenhagen,
  ``The master Ward identity and generalized Schwinger-Dyson equation in
  classical field theory,''
  Commun.\ Math.\ Phys.\  {\bf 243}, 275 (2003)
  [arXiv:hep-th/0211242].
  %%CITATION = HEP-TH 0211242;%%

%\cite{Duetsch:2001sw}
\bibitem{Duetsch2001}
  M.~D\" utsch and F.~M.~Boas,
  ``The master Ward identity,''
  Rev.\ Math.\ Phys.\  {\bf 14}, 977 (2002)
  [arXiv:hep-th/0111101].
  %%CITATION = HEP-TH 0111101;%%

%\cite{Duetsch:2000de}
\bibitem{Duetsch2000}
  M.~D\" utsch and K.~Fredenhagen,
  ``Perturbative algebraic field theory, and deformation quantization,''
  Published in *Siena 2000, Mathematical physics in mathematics and physics* 151-160, and
  Fields Inst. Commun. {\bf 30}, 151 (2001)
  arXiv:hep-th/0101079.
  %%CITATION = HEP-TH 0101079;%%

%\cite{Duetsch:2000nh}
\bibitem{Duetsch2000b}
  M.~D\" utsch and K.~Fredenhagen,
  ``Algebraic quantum field theory, perturbation theory, and the loop
  expansion,''
  Commun.\ Math.\ Phys.\  {\bf 219}, 5 (2001)
  [arXiv:hep-th/0001129].
  %%CITATION = HEP-TH 0001129;%%

\bibitem{Duetsch1999}
  M.~D\" utsch and K.~Fredenhagen,
   ``A local (perturbative) construction of observables in gauge theories:  The
  example of QED,''
  Commun.\ Math.\ Phys.\  {\bf 203}, 71 (1999)

\bibitem{NewDuetsch}
  M.~D\" utsch,
  ``Proof of perturbative gauge invariance for tree diagrams to all orders,''
  Annalen Phys.\  {\bf 14}, 438 (2005)
  [arXiv:hep-th/0502071].

\bibitem{Dubois-Violette1985}
  M.~Dubois-Violette, M.~Talon and C.~M.~Viallet,
  ``Brs Algebras: Analysis Of The Consistency Equations In Gauge Theory,''
  Commun.\ Math.\ Phys.\  {\bf 102}, 105 (1985).

\bibitem{Dubois-Violette1990}
  M.~Dubois-Violette, M.~Henneaux, M.~Talon and C.~M.~Viallet,
  ``Some results on local cohomologies in field theory,''
  Phys.\ Lett.\ B {\bf 267}, 81 (1991).

\bibitem{Dubois-Violette1992}
M. Dubois-Violette, M. Henneaux, M. Talon, C.M. Viallet,
Phys. Lett. {\bf B289}, 361 (1992)

\bibitem{Epstein1973}
H. Epstein and V. Glaser: ``The r\^ole of locality in
perturbation theory,'' Ann. Inst. H. Poincar\'e Sec. {\bf A XIX},
211--295 (1973)

\bibitem{Fadeev}
  L.~D.~Faddeev and V.~N.~Popov,
  ``Feynman diagrams for the Yang-Mills field,''
  Phys.\ Lett.\  B {\bf 25} (1967) 29.

\bibitem{Fewster2003}
 C.~J.~Fewster and M.~J.~Pfenning,
  ``A quantum weak energy inequality for spin-one fields in curved
  spacetime,''
  J.\ Math.\ Phys.\  {\bf 44}, 4480 (2003)

\bibitem{Fredenhagen94}
K. Fredenhagen: ``The algebraic theory of superselection sectors,''
University of Hamburg lecture notes, available at
{\tt www.desy.de/uni-th/lqp/psfiles/superselect.ps.gz}

\bibitem{Kusku06}
K. Fredenhagen and M. K\" usk\" u: Unpublished notes.

\bibitem{Fulling}
  S.~A.~Fulling,
  ``Aspects of quantum field theory in curved spacetime,''
  London Math.\ Soc.\ Student Texts {\bf 17}, 1 (1989).

\bibitem{FNW}
  S.~A.~Fulling, F.~J.~Narcowich and R.~M.~Wald,
  ``Singularity Structure Of The Two Point Function In Quantum Field Theory In
  Curved Space-Time. Ii,''
  Annals Phys.\  {\bf 136}, 243 (1981), see also
  S.~A.~Fulling, M.~Sweeny and R.~M.~Wald,
  ``Singularity Structure Of The Two Point Function In Quantum Field Theory In
  Curved Space-Time,''
  Commun.\ Math.\ Phys.\  {\bf 63}, 257 (1978).

\bibitem{Friedlaender}
F.G. Friedlaender: ``The Wave Equation on Curved Space-Time,'' Cambridge
University Press, Cambridge 1975.

\bibitem{Woolgar}
  G.~J.~Galloway, K.~Schleich, D.~M.~Witt and E.~Woolgar,
  ``Topological censorship and higher genus black holes,''
  Phys.\ Rev.\  D {\bf 60}, 104039 (1999)
  [arXiv:gr-qc/9902061].

\bibitem{garcia_bondia1}
J. M. Garcia-Bondia: ``Improved Epstein-Glaser renormalization in 
coordinate space I, Euclidean framework'', Math. Phys., Anal. Geom. {\bf 6}, 59 (2003), 

\bibitem{garcia_bondia2}
J. M. Garcia-Bondia and S. Lazzarini: ``Improved Epstein-Glaser renormalization in 
coordinate space II, Lorentz invariant framework'', J. Math. Phys. {\bf 44}, 3863 (2003)

\bibitem{C}
I. M. Gelfand and G. E. Shilov, {\em Les distributions I},
Dunod, Paris 1972

\bibitem{Geng}
  C.~Q.~Geng and R.~E.~Marshak,
  ``Uniqueness of quark and lepton representations
in the standard model from the anomalies viewpoint,''
  Phys.\ Rev.\  D {\bf 39}, 693 (1989).

\bibitem{Geroch2000}
 R.~Geroch,
  ``Partial Differential Equations of Physics,''
  arXiv:gr-qc/9602055.

\bibitem{Grigore1}
D. R. Grigore: ``Ward identities and renormalization of
general gauge theories,'' J. Phys. A: Math. Gen. {\bf 37}, 2803
(2004)

\bibitem{Grigore2}
D. R. Grigore: ``The structure of the anomalies of gauge theories in the
causal approach,'' J. Phys. A: Math. Gen. {\bf 35}, 1665
(2002)

\bibitem{Grigore3}
D. R. Grigore: ``On the uniqueness of the non-abelian
gauge-theories in Epstein-Glaser approach to renormalization,''
Rom. J. Phys.  {\bf 44},
(1999)


\bibitem{Varadarajan}
A. Guichardet: ``Cohomologie des groupes topologiques et des algebres de Lie,''
Textes Mathematiques 2, Cedic/Fernard Nathan, Paris (1980)

\bibitem{Henneaux1995}
 M.~Henneaux and C.~Teitelboim,
  ``Quantization of gauge systems,''
Princeton University Press (1992)

\bibitem{Hawking1970}
 S.~W.~Hawking and G.~F.~R.~Ellis,
  ``The Large scale structure of space-time,''
%\href{http://www.slac.stanford.edu/spires/find/hep/www?irn=6991262}{SPIRES entry}
{\it  Cambridge University Press, Cambridge, 1973}


\bibitem{Hollands2000}
  S.~Hollands and R.~M.~Wald,
   ``Local Wick polynomials and time ordered products of quantum fields in
  curved spacetime,''
  Commun.\ Math.\ Phys.\  {\bf 223}, 289 (2001)

\bibitem{Hollands2001}
  S.~Hollands and R.~M.~Wald,
   ``Existence of local covariant time ordered products of quantum fields in
  curved spacetime,''
  Commun.\ Math.\ Phys.\  {\bf 231}, 309 (2002)

\bibitem{Hollands2003}
  S.~Hollands and R.~M.~Wald,
  ``On the renormalization group in curved spacetime,''
  Commun.\ Math.\ Phys.\  {\bf 237}, 123 (2003)
  %[arXiv:gr-qc/0209029].

\bibitem{Ruan}
%\bibitem{Hollands:2001qe}
  S.~Hollands and W.~Ruan,
  ``The state space of perturbative quantum field theory in curved
  space-times,''
  Annales Henri Poincare {\bf 3}, 635 (2002)
  [arXiv:gr-qc/0108032].

\bibitem{Hollands2005}
  S.~Hollands and R.~M.~Wald,
  ``Conservation of the stress tensor in interacting quantum field theory  in
  curved spacetimes,''
  Rev.\ Math.\ Phys.\  {\bf 17}, 227 (2005)
  [arXiv:gr-qc/0404074].

\bibitem{HollandsPCT}
 S.~Hollands,
  ``A general PCT theorem for the operator product expansion in curved
  spacetime,''
  Commun.\ Math.\ Phys.\  {\bf 244}, 209 (2004)
  [arXiv:gr-qc/0212028].

\bibitem{Hollands2006}
  S.~Hollands,
  ``The operator product expansion for perturbative quantum field theory in
  curved spacetime,'' Commun. Math. Phys., in print,
  arXiv:gr-qc/0605072.

\bibitem{tHooft1}
  G.~'t Hooft and M.~Veltman,
  ``Diagrammar,'' CERN-73-09;

\bibitem{tHooft2}
  G.~'t Hooft and M.~J.~G.~Veltman,
  ``Regularization And Renormalization Of Gauge Fields,''
  Nucl.\ Phys.\  B {\bf 44}, 189 (1972).

\bibitem{tHooft3}
  G.~'t Hooft and M.~J.~G.~Veltman,
  ``Combinatorics of gauge fields,''
  Nucl.\ Phys.\  B {\bf 50}, 318 (1972).

\bibitem{Hormander}
L. Hormander, ``The analysis of linear partial differential operators,
I,'' Springer Verlag (1983)

\bibitem{Hurth1}
T. Hurth: ``Non-abelian gauge symmetry in the causal Epstein-Glaser approach,''
Int. J. Mod. Phys. {\bf A12} 4461 (1995)

\bibitem{Hurth2}
T. Hurth: ``Non-Abelian gauge theories, the causal approach,''
Ann. Phys. {\bf 244} 340 (1995)

\bibitem{Hurth3}
T. Hurth and K. Skenderis: ``Quantum Noether method,''
Nucl. Phys. {\bf B} 566 (1999)

\bibitem{Iyer1994}
 V.~Iyer and R.~M.~Wald,
   ``Some properties of Noether charge and a proposal for dynamical black hole
  entropy,''
  Phys.\ Rev.\ D {\bf 50}, 846 (1994)

\bibitem{Junker1}
W. Junker: ``Hadamard states, adiabatic vacua, and the construction
of physical states for scalar quantum fields in curved space-time,''
Rev. Math. Phys. {\bf 8} (1996) 1091 [Erratum-ibid. {\bf 14} (2002) 511]

\bibitem{Junker2003}
%\cite{Junker:2001gx}
%\bibitem{Junker:2001gx}
  W.~Junker and E.~Schrohe:
   ``Adiabatic vacuum states on general spacetime manifolds: Definition,
  construction, and physical properties,''
  Annales Poincare Phys.\ Theor.\  {\bf 3}, 1113 (2002)

\bibitem{Kallen1950}
 G.~K\" allen: ``Formal integration of the equations of quantum theory
 in the Heisenberg representation,'' Ark. Fysik {\bf 2} 371 (1950)

\bibitem{Kay}
  B.~S.~Kay and R.~M.~Wald,
  ``Theorems on the Uniqueness and Thermal Properties of Stationary,
  Nonsingular, Quasifree States on Space-Times with a Bifurcate Killing
  Horizon,''
  Phys.\ Rept.\  {\bf 207}, 49 (1991).

\bibitem{Koehler}
 M.~K\" ohler:
  ``The Stress energy tensor of a locally supersymmetric quantum field on a
  curved space-time,'' (PhD thesis, Hamburg 1995)
  arXiv:gr-qc/9505014.

\bibitem{Kopper}
  C.~Kopper and V.~F.~M\" uller:
  ``Renormalization proof for massive $\phi_4^4$ theory on Riemannian
  manifolds,''
  arXiv:math-ph/0609089.

\bibitem{Kreimer3}
  D.~Kreimer:
  ``Anatomy of a gauge theory,''
  Annals Phys.\  {\bf 321}, 2757 (2006)
  [arXiv:hep-th/0509135].

\bibitem{Kugo1980}
  T.~Kugo and I.~Ojima:
   ``Local Covariant Operator Formalism Of Nonabelian Gauge Theories And Quark
  Confinement Problem,''
  Prog.\ Theor.\ Phys.\ Suppl.\  {\bf 66}, 1 (1979).
%\bibitem{Kugo:1977zq}
  T.~Kugo and I.~Ojima,
   ``Manifestly Covariant Canonical Formulation Of Yang-Mills Field Theories:
  Physical State Subsidiary Conditions And Physical S Matrix Unitarity,''
  Phys.\ Lett.\ B {\bf 73}, 459 (1978).

\bibitem{Kuskudiplom}
  M.~Kusku,
  ``The free Maxwell field in curved spacetime,'' Diploma thesis (2001), DESY-THESIS-2001-040

\bibitem{GB}
S. Lazzarini and J. M. Garcia-Bondia:
``Improved Epstein-Glaser renormalization. II. Lorentz invariant framework,''
J. Math. Phys. {\bf 44} (2003) 3863-3875

\bibitem{Marolf1994}
 D.~M.~Marolf:
  ``The Generalized Peierls bracket,''
  Annals Phys.\  {\bf 236}, 392 (1994)
  [arXiv:hep-th/9308150].

\bibitem{Ramond1990}
  J.~A.~Minahan, P.~Ramond and R.~C.~Warner:
  ``A comment on the anomaly cancellation in the Standard Model,''
  Phys.\ Rev.\  D {\bf 41}, 715 (1990).

\bibitem{Moretti1999}
 V.~Moretti,
  ``Proof of the symmetry of the off-diagonal Hadamard/Seeley-deWitt's
  coefficients in C(infinity) Lorentzian manifolds by a 'local Wick
  rotation',''
  Commun.\ Math.\ Phys.\  {\bf 212}, 165 (2000)
  [arXiv:gr-qc/9908068].


\bibitem{Piguet1995}
O. Piguet and S. Sorella, ``Algebraic Renormalization,'' Springer
Lecture Notes in Physics, (1995)

\bibitem{Peierls1952}
  R.~E.~Peierls,
  ``The Commutation laws of relativistic field theory,''
  Proc.\ Roy.\ Soc.\ Lond.\ A {\bf 214}, 143 (1952).

\bibitem{pr}
D. Prange: ``Lorentz covariance in Epstein-Glaser renormalization,''
[arXiv:hep-th/9904136]

\bibitem{Radzikowski1996a}
  M.~J.~Radzikowski,
   ``Micro-Local Approach To The Hadamard Condition In Quantum Field Theory On
  Curved Space-Time,''
  Commun.\ Math.\ Phys.\  {\bf 179}, 529 (1996).
  %%CITATION = CMPHA,179,529;%%

\bibitem{Radzikowski1996b}
  M.~J.~Radzikowski,
  ``A Local to global singularity theorem for quantum field theory on curved
  space-time,''
  Commun.\ Math.\ Phys.\  {\bf 180}, 1 (1996).

\bibitem{Sato}
M. Sato, T. Kawai, and M. Kashiwara, "Hyperfunctions and pseudo-differential
equations," Lecture Notes in Mathematics 287, 265-529, Springer Verlag (1973);
M. Sato: "Theory of hyperfunctions I, II," J. Fac. Sci. Univ. Tokyo t 8, 139-193,
387-437 (1959-1960) MR 22:4951, MR24A:2237

\bibitem{Scharf1}
G. Scharf: ``Quantum gauge theories: A true ghost story,''
    Wiley, New York (2001)

\bibitem{Scharf2}
G. Scharf: ``Finite Quantum Electrodynamics,'' Springer, Berlin
Heidelberg New York, 1989 and 1995

\bibitem{Schechter}
M. Schechter, ``General boundary value problems for elliptic partial differential
equations'', Commun. Pure and App. Math., Vol XII, 457-486 (1959)

\bibitem{Speer0}
E.~R. Speer, ``Analytic Renormalization,'' J. Math. Phys.
{\bf 9}, 1404 (1968)

\bibitem{Speer1}
E.~R. Speer, ``On the structure of analytic renormalization,''
Commun. Math. Phys. {\bf 31}, 23 (1971)

\bibitem{Speer2}
E.~R. Speer, ``Analytic renormalization using many
space-time dimensions,'' Commun. Math. Phys. {\bf 37},
83 (1974)

\bibitem{Steinmann1990}
   O. Steinmann: ``Perturbative QED and Axiomatic Field Theory,''
Springer Verlag (2000)

\bibitem{Stora1990}
  R. Stora: ``Pedagogical experiments in renormalized perturbation
  theory,'' in conference {\rm Theory of Renormalization and
    Regularization}, Hesselberg, Germany (2002), at
  {\tt http://wwwthep.physik.uni-mainz.de/~scheck/Hessbg02.html}

\bibitem{Stora1}
R. Stora: ``Continuum gauge theories,'' 1976 Cargese Lectures
published in {\em New developements in quantum field theory
and statistical physics,} eds. M. Levy and P. Mitter NATO ASI
Series B26 (Plenum, 1977)

\bibitem{Stora2}
R. Stora: ``Algebraic structure and Topological Origin of Anomalies,''
Seminar at Cargese Summer Inst. Sept. 1-15 1983, published
in {\em Progress in gauge field theory}, eds. 't~Hooft et al.
(Plenum, 1984)

\bibitem{Suijlekom1}
W. D. van Suijlekom: ``Renormalization of gauge fields: A Hopf algebra approach,''
[arXiv:hep-th/0610137]

\bibitem{Suijlekom2}
W. D. van Suijlekom: ``The Hopf algebra of Feynman graphs in Quantum Electrodynamics,''
Lett. Math. Phys. {\bf 77}, 265 (2006)

\bibitem{zahn}
M. Taslimitherani and J. Zahn: work in progress.

\bibitem{Verch2000}
  R.~Verch,
  ``A spin-statistics theorem for quantum fields on curved spacetime  manifolds
  in a generally covariant framework,''
  Commun.\ Math.\ Phys.\  {\bf 223}, 261 (2001)
  [arXiv:math-ph/0102035].

\bibitem{Wald}
  R.~M.~Wald,
  ``Quantum field theory in curved space-time and black hole thermodynamics,''
%\href{http://www.slac.stanford.edu/spires/find/hep/www?irn=3231020}{SPIRES entry}
{\it  Chicago, USA: Univ. Pr. (1994) 205 p}

\bibitem{Wald1990}
  R.~M.~Wald, ``On identically closed forms locally constructed from a field''
  J. Math. Phys. {\bf 31} (1990) 2378-2384


\bibitem{Weinberg1996}
S. Weinberg: ``The Quantum Theory of Fields, Volume II,''
Cambridge University Press (1996)

\bibitem{Wigner}
E. Wigner, Ann. Math. {\bf 40} 149 (1939)

\bibitem{Zinn-Justin1975}
J. Zinn-Justin, in ``Trends in Elementary Particle Theory ---
International Summer Institute in Theoretical Physics in Bonn 1974''
(Springer-Verlag, Berlin, 1975)
\end{thebibliography}
\end{document}